\pdfoutput=1
\documentclass[pdflatex,sn-mathphys-num]{sn-jnl}

\usepackage{graphicx}%
\usepackage{multirow}%
\usepackage{amsmath,amssymb,amsfonts}%
\usepackage{amsthm}%
\usepackage{mathrsfs}%
\usepackage[title]{appendix}%
\usepackage{xcolor}%
\usepackage{textcomp}%
\usepackage{manyfoot}%
\usepackage{booktabs}%
\usepackage{algorithm}%
\usepackage{algorithmicx}%
\usepackage{algpseudocode}%
\usepackage{listings}%
\usepackage{soul}%
\usepackage{caption}
\usepackage{comment}%
\usepackage{svg}%
\usepackage{marginnote}%
\usepackage{tcolorbox}%
\usepackage{rotating}
\usepackage{hyperref}
\usepackage{xurl}
\AtBeginDocument{\let\burl\url}

\usepackage{geometry}
\usepackage{mathpazo}          
\selectfont   

\definecolor{inkdeep}{RGB}{18,52,86}      
\definecolor{inkteal}{RGB}{13,102,110}    
\definecolor{inkbrick}{RGB}{150,54,42}    
\definecolor{inkmute}{RGB}{110,118,128}   

\makeatletter
\def\sectionfont{\reset@font\fontfamily{\rmdefault}\fontsize{14bp}{16bp}%
  \bfseries\selectfont\raggedright\boldmath\color{inkdeep}}
\def\subsectionfont{\reset@font\fontfamily{\rmdefault}\fontsize{12bp}{14bp}%
  \bfseries\selectfont\raggedright\boldmath\color{inkteal}}
\def\subsubsectionfont{\reset@font\fontsize{11bp}{13bp}%
  \bfseries\selectfont\raggedright\boldmath\color{inkdeep}}
\makeatother

\newcommand{\rhtext}{Meteorology-driven causal nowcasting of fugitive landfill emissions}
\makeatletter
\def\ps@headings{%
  \let\@oddfoot\@empty  
  \let\@evenfoot\@empty
  \def\@oddhead{%
    \parbox[b]{\textwidth}{%
      \footnotesize\color{inkmute}\rhtext\hfill
      \raisebox{0.1ex}{\color{inkteal}\thepage}\\[2.5pt]
      {\color{inkteal}\rule{\textwidth}{0.4pt}}}}%
  \let\@evenhead\@oddhead
  \let\@mkboth\markboth}
\makeatother

\usepackage{microtype}
\hypersetup{
  colorlinks = true,
  linkcolor  = inkdeep,
  citecolor  = inkteal,
  urlcolor   = inkbrick,
  breaklinks = true
}

\DeclareRobustCommand{\chadd}[1]{#1}
\DeclareRobustCommand{\chdelete}[1]{}
\DeclareRobustCommand{\chreplace}[2]{#1}
\DeclareRobustCommand{\chrev}[1]{}

\DeclareRobustCommand{\chstack}[2]{#1}
\providecommand{\linenumbers}{}

\theoremstyle{thmstyleone}%
\theoremstyle{thmstyletwo}%
\theoremstyle{thmstylethree}%
\newcommand{\degree}{\ensuremath{^\circ}}

\usepackage{tikz}
\usetikzlibrary{arrows.meta,positioning,decorations.pathmorphing}
\newif\ifgraphicalabstract
\graphicalabstracttrue
\ifgraphicalabstract
\makeatletter
\let\GA@printabstract\printabstract
\def\printabstract{%
  \GA@printabstract
  \par\vspace{10pt}%
  \centerline{\resizebox{\linewidth}{!}{
\definecolor{gaSky}{RGB}{226,236,242}
\definecolor{gaGround}{RGB}{198,190,172}
\definecolor{gaPlume}{RGB}{150,160,150}
\definecolor{t0}{RGB}{ 60,140,105}
\definecolor{t1}{RGB}{224,178, 60}
\definecolor{t2}{RGB}{219,124, 48}
\definecolor{t3}{RGB}{190, 60, 48}

\begin{tikzpicture}[x=1mm,y=1mm,font=\sffamily\scriptsize,
  stage/.style={rounded corners=1.2pt,draw=inkteal!55,fill=white,line width=0.5pt},
  hdr/.style={inkdeep,font=\sffamily\bfseries\scriptsize},
  ganote/.style={inkmute,font=\sffamily\fontsize{6}{7}\selectfont,align=center},
  flow/.style={-{Stealth[length=2mm,width=1.6mm]},inkteal,line width=0.9pt}]

\fill[gaSky!55,rounded corners=2pt] (-2,-39) rectangle (162,29);

\begin{scope}[xshift=5.5mm]

\node[stage,minimum width=33mm,minimum height=27mm] at (16,11) {};
\node[hdr] at (16,21.3) {Landfill source};
\begin{scope}
  \clip (2,0.5) rectangle (30,18);
  \fill[gaGround] plot[smooth,domain=2:30] (\x, {3.4+2.6*exp(-((\x-16)/7)^2)}) -- (30,0.5) -- (2,0.5) -- cycle;
  \foreach \o/\op in {0/0.55, 1.9/0.42, 3.8/0.30, 5.7/0.20}{
    \draw[gaPlume,opacity=\op,line width=1.6pt]
      plot[smooth,domain=14:31] (\x, {6.0+\o+0.085*(\x-14)^1.4});}
  \foreach \hx in {23.5,26.5}{
    \fill[inkdeep!55] (\hx-1.1,0.5) rectangle (\hx+1.1,2.5);
    \fill[inkdeep!75] (\hx-1.4,2.5) -- (\hx,3.9) -- (\hx+1.4,2.5) -- cycle;}
\end{scope}
\draw[inkmute!60,line width=0.4pt] (2,0.5) -- (30,0.5);
\node[ganote] at (16,-5.8) {H$_2$S and CH$_4$ at monitored\\ receptors, set by weather};

\node[stage,minimum width=33mm,minimum height=27mm] at (55,11) {};
\node[hdr] at (55,21.3) {Meteorological drivers};
\foreach \i/\lab/\core in {0/{wind direction}/1, 1/{wind speed}/1, 2/{pressure}/1, 3/{temperature}/0}{
  \ifnum\core=1
    \node[rounded corners=1pt,fill=inkteal!22,draw=inkteal!60,line width=0.4pt,
          minimum width=25mm,minimum height=3.4mm,inner sep=0pt,
          font=\sffamily\fontsize{6}{7}\selectfont,text=inkdeep]
          at (55,16.6-\i*4.3) {\lab};
  \else
    \node[rounded corners=1pt,fill=inkteal!6,draw=inkteal!25,line width=0.3pt,
          minimum width=25mm,minimum height=3.4mm,inner sep=0pt,
          font=\sffamily\fontsize{6}{7}\selectfont,text=inkmute]
          at (55,16.6-\i*4.3) {\lab};
  \fi}
\node[ganote] at (55,-5.8) {causal core and timescales\\ from multiscale transfer entropy};

\node[stage,minimum width=33mm,minimum height=27mm] at (94,11) {};
\node[hdr] at (94,21.3) {CAIRN nowcaster};
\draw[inkmute!45,line width=0.3pt] (80.5,5.5) -- (107.5,5.5);
\draw[t2,line width=0.9pt] plot[smooth,domain=0:27] ({80.5+\x}, {5.5+9*exp(-\x/3.1)});
\draw[inkteal,line width=0.9pt] plot[smooth,domain=0:27] ({80.5+\x}, {5.5+9*exp(-\x/13)});
\node[font=\sffamily\fontsize{6}{7}\selectfont,text=t2]      at (87.5,16.4) {fast · 1\,h};
\node[font=\sffamily\fontsize{6}{7}\selectfont,text=inkteal] at (102,11.6) {slow · 6\,h};
\node[ganote] at (94,-5.8) {memory anchored on the measured\\ timescales; weather and calendar only};

\node[stage,minimum width=33mm,minimum height=27mm] at (133,11) {};
\node[hdr] at (133,21.3) {Fused exposure tier};
\foreach \i/\c/\lab in {0/t3/{Tier 3}, 1/t2/{Tier 2}, 2/t1/{Tier 1}, 3/t0/{Tier 0}}{
  \fill[\c,rounded corners=0.8pt] (121.5,15.2-\i*4.2) rectangle (135,17.9-\i*4.2);
  \node[anchor=west,font=\sffamily\fontsize{6}{7}\selectfont,text=inkdeep]
        at (136,16.55-\i*4.2) {\lab};}
\node[ganote] at (133,-5.8) {four nowcasters, WHO-referenced;\\ agrees with sensor tiers, $\kappa_w = 0.67$};

\foreach \a/\b in {33/38, 72/77, 111/116} \draw[flow] (\a,10) -- (\b,10);
\end{scope}

\draw[inkteal!35,line width=0.4pt] (2,-11.5) -- (158,-11.5);
\node[hdr,anchor=west] at (2,-15.3) {External check: community odour reports \textmd{\textcolor{inkmute}{(schematic)}}};

\draw[-{Stealth[length=1.8mm,width=1.4mm]},inkmute,line width=0.6pt] (34,-31) -- (126,-31);
\node[ganote,anchor=east,text=inkmute] at (33,-31) {days};

\foreach \x/\h/\c in {38/1/t0, 44/1/t0, 50/3/t1, 56/6/t2, 62/8/t3, 68/7/t3, 74/4/t2, 80/2/t1, 86/1/t0, 92/1/t0}{
  \fill[\c,rounded corners=0.4pt] (\x-2.2,-30.5) rectangle (\x+0.2,-30.5+\h);}
\foreach \x/\h in {38/0.4, 44/0.6, 50/2.2, 56/4.8, 62/7.4, 68/6.2, 74/3.0, 80/1.2, 86/0.5, 92/0.3}{
  \fill[inkbrick!70] (\x+0.5,-30.5) rectangle (\x+2.3,-30.5+\h);}
\node[anchor=west,font=\sffamily\fontsize{6}{7}\selectfont,text=inkdeep] at (98,-22.5) {\textcolor{t3}{\rule{2mm}{1.6mm}} fused tier (daily mean)};
\node[anchor=west,font=\sffamily\fontsize{6}{7}\selectfont,text=inkdeep] at (98,-26.5) {\textcolor{inkbrick!70}{\rule{2mm}{1.6mm}} odour reports per day};

\node[anchor=west,ganote,align=left,text=inkdeep] at (128,-17.5)
      {not used in training;\\ same-day peak (lag 0),\\ $r = 0.71$};
\end{tikzpicture}
}}%
  \par\vspace{2pt}}
\makeatother
\fi

\usepackage{docmute}

\begin{document}

\documentclass[pdflatex,sn-mathphys-num,referee]{sn-jnl}

\usepackage{graphicx}%
\usepackage{multirow}%
\usepackage{amsmath,amssymb,amsfonts}%
\usepackage{amsthm}%
\usepackage{mathrsfs}%
\usepackage[title]{appendix}%
\usepackage{xcolor}%
\usepackage{textcomp}%
\usepackage{manyfoot}%
\usepackage{booktabs}%
\usepackage{algorithm}%
\usepackage{algorithmicx}%
\usepackage{algpseudocode}%
\usepackage{listings}%
\usepackage{hyperref}
\usepackage{xurl}
\AtBeginDocument{\let\burl\url}%

\usepackage{xr}
\makeatletter
\let\ORIG@externaldocument\externaldocument
\renewcommand*{\externaldocument}[1]{%
  \begingroup\let\bibcite\@gobbletwo\ORIG@externaldocument{#1}\endgroup}
\makeatother
\externaldocument{cee_supp}

\theoremstyle{thmstyleone}

\theoremstyle{thmstyleone}%
\newtheorem{theorem}{Theorem}%
\newtheorem{proposition}[theorem]{Proposition}%

\theoremstyle{thmstyletwo}%
\newtheorem{example}{Example}%
\newtheorem{remark}{Remark}%

\theoremstyle{thmstylethree}%
\newtheorem{definition}{Definition}%

\raggedbottom
\usepackage[mathlines]{lineno}

\usepackage{changes}
\usepackage{marginnote}
\makeatletter\@mparswitchfalse\makeatother
\setlength{\marginparsep}{8pt}\setlength{\marginparwidth}{1.6cm}
\newif\ifshowchanges
\showchangestrue
\ifshowchanges
  \DeclareRobustCommand{\chadd}[1]{\added{#1}}
  \DeclareRobustCommand{\chdelete}[1]{\deleted{#1}}
  \DeclareRobustCommand{\chreplace}[2]{\replaced{#1}{#2}}
  \DeclareRobustCommand{\chrev}[1]{\textsuperscript{\textcolor{red}{\bfseries\sffamily #1}}}
  \newenvironment{chaddblock}{\par\begingroup\color{blue}}{\par\endgroup}
  \DeclareRobustCommand{\chstack}[2]{\shortstack{\added{#1}\\\deleted{#2}}}
\else
  \DeclareRobustCommand{\chadd}[1]{#1}
  \DeclareRobustCommand{\chdelete}[1]{}
  \DeclareRobustCommand{\chreplace}[2]{#1}
  \DeclareRobustCommand{\chrev}[1]{}
  \newenvironment{chaddblock}{}{}
  \DeclareRobustCommand{\chstack}[2]{#1}
\fi

\newcommand{\degree}{\ensuremath{^\circ}}

\begin{document}
\linenumbers

\title{\centering\fontsize{15}{17.7}\selectfont Meteorology-driven Causal Nowcasting of Fugitive Landfill\\ Emissions \texorpdfstring{from Measured Coupling Timescales}{from Measured Coupling Timescales}}

\author*[1]{\fnm{Timothy C.} \sur{Pearce}}\email{t.c.pearce@le.ac.uk}
\author[2]{\fnm{David J. T.} \sur{Smith}}\email{toby.smith@ukhsa.gov.uk}
\author[2]{\fnm{Alec} \sur{Dobney}}\email{alec.dobney@ukhsa.gov.uk}
\author[2]{\fnm{Alessia} \sur{Freddo}}\email{alessia.freddo@ukhsa.gov.uk}

\affil*[1]{\orgdiv{Biomedical Engineering Research Group, School of Engineering}, \orgname{University of Leicester}, \orgaddress{\city{Leicester}, \country{United Kingdom}}}
\affil[2]{\orgdiv{Environmental Hazards and Emergency Department}, \orgname{UKHSA}, \orgaddress{\city{Nottingham}, \country{United Kingdom}}}

\abstract{Which meteorological processes control exposure to fugitive gases downwind of a source, and on what timescales, have largely been inferred from dispersion theory and partial field evidence. Here we show that the meteorological drivers of elevated hydrogen sulphide (H$_2$S) exposure at a long-monitored European landfill, and the timescales over which each acts, can be identified directly from monitoring data. Wind direction, wind speed and atmospheric pressure form the causal core, with the share of directed information carried by pressure increasing with aggregation scale. The recovered timescales are consistent with those expected from the underlying atmospheric processes. We use these driver timescales to initialise CAIRN (Causal-Anchored Inference for Receptor Nowcasting), a machine-learning nowcaster with fast and slow memory components. Trained on past exceedances of WHO guideline levels, CAIRN nowcasts them from surface weather measurements and the calendar alone, without hand-engineered features.  Combining four such nowcasters produces a site-level, tiered alert that agrees substantially with that generated by a direct sensor network and tracks an independent record of community odour reports. Meteorological variables can therefore serve as an inference-time proxy for exposure relative to WHO guideline levels, and they link atmospheric dynamics to community impact as an episode unfolds.}

\maketitle

\section*{Introduction}

Communities living adjacent to regulated industrial sources of fugitive
odorous and toxic gases (for example, landfills, sewage works, composting
facilities, intensive livestock operations and biogas plants) bear a
disproportionate burden of respiratory, neurological and mental-health
morbidity that current response frameworks manage primarily through
retrospective intervention, once impacts in surrounding populations are
already documented~\citep{carson2024malodors, ukhsa2025walleys}. The
scale of the underlying challenge is growing: municipal solid waste
alone is projected to nearly double globally by
2050~\citep{kaza2018what}, and landfill methane emissions are
substantially under-reported worldwide~\citep{wang2024methane}. Moreover, evidence indicates that odour-emitting facilities are more likely to be located in less privileged communities characterised by lower income, lower levels of education attainment, and higher proportions of minority populations~\citep{desouza2026evaluating, Martuzzi2010InequalitiesWaste}. This unequal distribution of exposure contributes to environmental health inequalities, as the odour-related impacts on health and wellbeing are compounded by a reduced capacity to influence decision-making. Odour from such sites is often treated as a nuisance issue; its unequal distribution makes it also a matter of environmental justice, with persistent health burdens that fall unequally.

Among the gases these odorous industrial sites release, hydrogen sulphide (H$_2$S) is a
sentinel: generated when sulphate-reducing bacteria metabolise sulphate
in decomposing waste under anaerobic conditions, it carries a
characteristic ``rotten-egg'' odour detectable at concentrations far
below those of toxicological concern~\citep{atsdr2001landfillgas}. Its
health burden is correspondingly graded. Acute, high-level exposure is
overtly toxic, but the dominant impact on communities near regulated
sites can be chronic  -- persistent malodour that degrades
wellbeing through sleep disturbance, headache, irritation and
psychological stress~\citep{shusterman1992health, hirasawa2019subjective}.

Odour is a substantial environmental nuisance with measurable population-level impacts. Across Europe, environmental odours are recognised as a major source of public concern and are among the leading causes of environmental complaints to regulatory authorities, second only to noise~\citep{RevisitingOdourPollution2021}. The prevalence of odour annoyance varies substantially across contexts, with relatively low levels reported in general population surveys and markedly higher levels (often exceeding 40--50\%) in populations residing near agricultural, industrial, or waste-related emission sources~\citep{BlanesVidal2012OdorAnnoyance, Wroniszewska2020OdorAnnoyance}. The regulatory threshold for intervention in the UK is defined under Part III of the Environmental Protection Act 1990, where odour may constitute a statutory nuisance if it is “prejudicial to health or a nuisance.” Local authorities have a duty to investigate complaints and, where satisfied that a statutory nuisance exists, must serve an abatement notice requiring the responsible party to stop or mitigate emissions~\citep{epa1990sec79}. However, determining nuisance relies on professional judgement and witnessed evidence, meaning intermittent odours, one-off events, or highly sensitive receptors may fall into “grey areas” where impacts are real but do not meet the statutory threshold. In these situations, the Environmental Permitting Regulations (EPR) provide an important complementary control. Environmental permits often include conditions to manage emissions such as odour, and breaches can be enforced by regulators independently of the statutory nuisance regime. For permitted facilities, even impacts below the statutory nuisance threshold may still breach permit conditions and require action, so EPR operates alongside statutory nuisance controls~\citep{DEFRA2017}.

Assessment of odour impacts in the UK is structured around the FIDOL framework, comprising Frequency, Intensity, Duration, Offensiveness and Location, which provides a practical basis for evaluating how odours affect receptors. FIDOL is embedded in the Institute of Air Quality Management (IAQM) guidance on odour assessment, widely applied in planning and regulatory decision-making. It recognises that odour impact is not solely concentration-based but depends on the interaction of these perceptual and contextual factors. This approach helps standardise subjective assessments, but it can still introduce variability and uncertainty, particularly where monitoring data are limited and reliance is placed on human sensory judgement.

Operationally, statutory nuisance cases are managed in the UK through a multi‑stakeholder system to support joint risk assessment, communication, and response. This system spans environmental regulation, public health science, and emergency planning. Within this framework, air quality guideline levels for H$_2$S, such as WHO 24-hour recommendations for health protection ($150\,\mu\mathrm{g\,m^{-3}}$) and 30 minute odour annoyance guideline ($7\,\mu\mathrm{g\,m^{-3}}$), are used to assess exposures to the gas and whether odour episodes are likely to affect health, wellbeing or quality of life. Assessing exposure against these levels depends on receptor concentrations, obtained either from monitors or from dispersion models that rely on steady-state, well-mixed approximations and typically require site-specific calibration.

Three evidential gaps stand between this monitoring infrastructure and a community-protective intervention. \emph{First}, the precise links are missing: we
lack an empirical, data-derived account of how routine meteorological
variables map onto fugitive emission and exposure, as opposed to a
mechanistic account assumed in advance. \emph{Second}, the drivers are
multiscale: exposure is governed by processes acting from minutes to
days that interact non-linearly, and the canonical timescales of those
processes have not been recovered from observation and used to structure
a predictive model. \emph{Third}, and most consequential for public
health, the validation is missing: to our knowledge, no learnt classifier of fugitive-emission exposure has been tested against an independent record of
community wellbeing -- the very outcome that fugitive-emission
regulation exists to protect~\citep{eykelbosh2021elucidating}.

The first two gaps share a common root in the physics of the boundary
layer. At a monitoring receptor, H$_2$S concentration is set by a
hierarchy of meteorological controls operating across distinct
timescales: advection sets which receptor lies downwind; mechanical
dilution at sub-hourly scales sets the instantaneous concentration
through an inverse dependence on wind
speed~\citep{pasquill1983atmospheric}; barometric pumping potentially modulates the
source flux as multi-hour soil--atmosphere pressure gradients drive gas
out of the waste mass~\citep{xu2014impact, young2003relating,
forde2019barometric}; and the diel growth and collapse of the mixing
layer governs the depth of the volume into which emissions are
diluted~\citep{parolari2021multiscale}. Each process leaves a signature
on a different temporal horizon, so the meteorology--exposure
relationship is inherently multiscale and compounding. Engineered-feature machine-learning pipelines, which now dominate air-quality classification~\citep{wang2019prediction, machinelearning2025realtime}, represent this memory through derivative windows, lag depths and stagnation intervals whose horizons are fixed in advance by the analyst; they capture the physics the analyst anticipates, and so their performance depends on those choices. Structured state-space
models~\citep{gu2022parameterization} can in principle learn long-memory
representations directly from raw meteorology, but their timescale priors
are initialised data-agnostically; resolving this would let a model
autonomously capture the compounding physics of weather over time
without constant human recalibration, yet whether the canonical
boundary-layer horizons can be recovered from data and used as
architectural priors has not, to our knowledge, been demonstrated for fugitive emissions.

The third gap is of a different kind. Uncovering the mathematical links between weather and emission
is only half of an implementation problem: a model validated solely
against the sensor signal it was trained on remains a technical exercise,
blind to whether its outputs track the real-world outcome (i.e. human
exposure and the distress it causes). Demonstrating real-world effectiveness, as distinct
from in-sample efficacy, requires confronting model predictions with an
external, independently generated record of community impact. Such
records exist, in the form of daily odour reports lodged by affected
residents, but we are not aware of their use to externally validate a learnt
exposure model operating in an uncontrolled, real-world setting.

Here we address all three gaps using post-calibration-adjustment
monitoring~\citep{ea2024dataadjustment} from a long-observed European
landfill site associated with persistent odour-related impacts and community concern. For the first two, information-theoretic causal
inference~\citep{runge2019inferring} recovers the missing links
empirically, identifying wind direction, wind speed and atmospheric
pressure as the causal core of receptor exposure and ranking their
influence across ten temporal scales spanning fifteen
minutes to ninety-six hours -- the multiscale structure made explicit, from observation
rather than assumption. We then introduce CAIRN (Causal-Anchored
Inference for Receptor Nowcasting), a dual-pathway state-space
classifier whose memory kernels are anchored on these data-derived
scales. Supplied only with raw meteorology, CAIRN matches the
high-exposure detection that engineered pipelines reach only through
multi-stage hand-coded derivative, stagnation and recirculation
features, and it transfers without modification across two
receptor geometries and to co-emitted methane.

For the third gap, we couple the model's output directly to lived
community experience. Fusing four per-channel classifiers under a
Bayesian log-odds accumulator yields a four-tier alert system, aligned
with WHO odour-annoyance guidance, that agrees with a deterministic
raw-sensor ground truth and tracks an independent
record of daily community odour reports at lag zero~\citep{aldegunde2024pollutionalerts}. This external validation tests whether the exposure signal tracks community outcomes in an uncontrolled setting.
Trained on a single year of routinely collected meteorology at one regulated UK landfill, CAIRN yields an inference-time proxy for exposure relative to WHO guideline levels that tracks community impact as an episode unfolds.

\section*{Results}

\subsection*{Causal hierarchy of meteorological drivers}

\begin{figure}[htp!]
    \centering
    \makebox[\textwidth][c]{\includegraphics[width=1.02\textwidth]{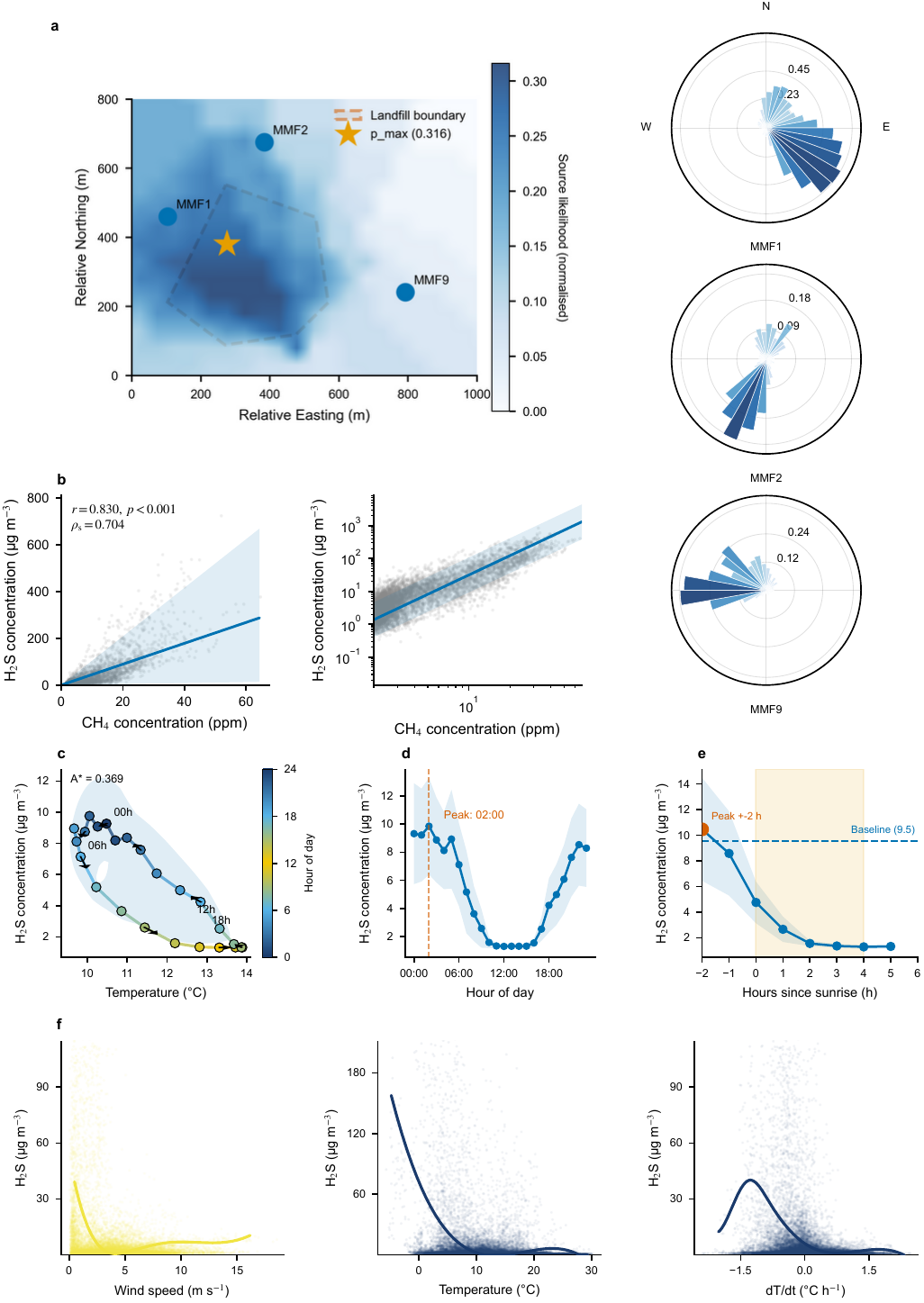}}
  \caption{Caption continued on next page.}
  \label{fig:env_characterisation}
  \end{figure}

We analysed a year of post-calibration-adjustment 2024 monitoring data
from the principal receptor site (MMF9, $n = 34{,}015$ valid 15-minute
records of co-located H$_2$S, CH$_4$ and four meteorological channels)
together with concurrent records from two further downwind sites
(MMF1, MMF2; data and processing in Methods, ``Data and study sites'').
H$_2$S monitoring at this site is exceptional rather than routine: it was instituted in response to a community-impact emergency at the site, not as a standard regulatory requirement. We use the resulting record to derive a model
whose inference-time inputs depend only on the four routinely
collected meteorological channels. Two complementary analyses
establish the meteorological control on receptor-level exposure: a
multi-method environmental characterisation of the
source--pathway--receptor system
(Fig.~\ref{fig:env_characterisation}), and an information-theoretic
ranking of the candidate causal drivers across ten temporal scales
(Fig.~\ref{fig:causal_links}; Methods, ``Causal analysis'').

Conditional probability function (CPF) rose plots at the three
co-located stations (Fig.~\ref{fig:env_characterisation}a) converge on
a common source area: at the principal receptor the exceedance
probability peaks in the westerly sector
(peak bearing $265^{\circ}$, CPF $\approx 0.37$, peak/opposing-sector
ratio $4.09\times$; circular--linear correlation $r = 0.149$,
$p < 10^{-3}$; spike-conditional Fisher's odds ratio $1.43$,
$p = 2.9 \times 10^{-4}$). The modest circular--linear $r$ reflects sectoral rather than
sinusoidal dependence on bearing; the peak/opposing-sector and
spike-conditional odds ratios are the stronger directional evidence. This anisotropy is consistent with a
continuous ground-level fugitive source upwind of the receptor and is
the basis on which a receptor-modelling triangulation
\citep{ashbaugh1985principal}, whose intersection falls inside
the permitted landfill boundary, narrows the emitting area to within a
single upwind sector. Bivariate analysis of simultaneously measured
H$_2$S and CH$_4$ (Fig.~\ref{fig:env_characterisation}b) yields a
strong positive correlation ($r = 0.830$, $p < 10^{-15}$; Spearman
$\rho = 0.683$; power-law exponent $b = 1.95$) and a Jaccard spike
co-occurrence index $J = 0.654$ ($\chi^2 = 21{,}298$, $p < 10^{-15}$).
PM$_{10}$ negative controls give $r = -0.028$, ruling out mechanical
dust resuspension as a shared pathway, while a moderate NO$_x$
correlation ($r = 0.377$) is accounted for by joint nocturnal trapping
(a 6-hour diel phase offset between the two species; Supplementary
Note 2). Together these results identify the source as anaerobically
generated landfill gas \citep{atsdr2001landfillgas}.

The remaining four panels resolve the boundary-layer mechanisms through
which meteorology modulates exposure. The temperature--H$_2$S phase loop
(Fig.~\ref{fig:env_characterisation}c) traces a strong counter-clockwise
hysteresis (normalised area $A^{*} = 0.369$; cooling/warming-limb slopes
$\beta_{\mathrm{cool}} = -2.83$, $\beta_{\mathrm{warm}} = -2.45$),
which indicates that the surface-layer mixing state depends on recent thermal history as well as on the instantaneous temperature. The mean diel profile
(Fig.~\ref{fig:env_characterisation}d) peaks at 02:00 LST
($9.69\,\mu\mathrm{g\,m^{-3}}$ mean; $51.6\,\mu\mathrm{g\,m^{-3}}$ at
P95) and reaches an early-afternoon minimum across 12:00--15:00 LST
($1.31\,\mu\mathrm{g\,m^{-3}}$), giving peak/trough ratios of
$7.4\times$ in the mean and $12.3\times$ at the upper tail
(Kruskal--Wallis $H = 338.5$, $p = 7 \times 10^{-58}$). Aligning every
calendar day to astronomical sunrise
(Fig.~\ref{fig:env_characterisation}e) collapses 365 independent
trajectories onto a single reproducible dispersal curve: a pre-sunrise
baseline of $9.49\,\mu\mathrm{g\,m^{-3}}$ falls to
$1.30\,\mu\mathrm{g\,m^{-3}}$ within four hours
($7.3\times$ reduction, one-sided Mann--Whitney $p < 10^{-15}$).
Bivariate regression (Fig.~\ref{fig:env_characterisation}f) confirms
all three theoretically predicted negative associations
(WS $r = -0.122$; TEMP $r = -0.173$; $\mathrm{d}T/\mathrm{d}t$
$r = -0.143$; all $p < 10^{-3}$); concentration-spike events cluster
in the low-wind, low-temperature, negative-tendency regime
corresponding to Pasquill class E--F stable boundary layers
\citep{pasquill1983atmospheric, stull1988introduction}. Pressure is
excluded from this bivariate panel because its causally significant
horizon is multi-hour and not resolvable in instantaneous 15-minute
scatter; the pressure-tendency response is established below by the
multiscale transfer-entropy analysis
(Fig.~\ref{fig:causal_links}a).

Multiscale transfer entropy (MSTE) across seven candidate drivers
and ten temporal scales (15\,min to 96\,h, the nine estimable of which
are shown in Fig.~\ref{fig:causal_links}a; Methods, ``Causal analysis'')
ranks these
drivers quantitatively. Among the four directly measured
meteorological variables, three form a robust core of directed
information flow, significant after Benjamini--Hochberg correction
across a single family of all 70 driver--scale tests: atmospheric
pressure (effective transfer entropy, $\mathrm{ETE} = 0.005$--$0.058$ nats, significant at every
scale from 15\,min to 12\,h, its share of total positive ETE rising
from $11.5\%$ at 15\,min to $31.4\%$ at 12\,h, the range over which barometric responses to passing synoptic systems become resolvable),
wind direction ($0.021$--$0.058$ nats, dominant at every scale up to
6\,h, $38$--$50\%$ of total ETE, and not significant beyond), and wind
speed ($0.011$--$0.029$ nats, significant at $\tau \leq 3$\,h and
again at 12\,h). The scale-by-scale composition of that total is
shown in Fig.~\ref{fig:causal_links}b. The core is robust to the dependence assumption:
under a correction valid under arbitrary dependence between the tests,
wind direction retains all six scales and pressure all seven, and 18 of
the 19 cells belonging to these three drivers survive, the exception
being wind speed at 12\,h (Fig.~\ref{fig:causal_links}a, filled versus
open symbols; Methods). Temperature enters through its rate of change rather than
its level. The raw variable reaches significance at the 15\,min scale
only and under Benjamini--Hochberg alone, and is not interpreted; the
six-hour temperature tendency is significant under both corrections
($\mathrm{ETE} = 0.022$, $q = 0.009$ and $q = 0.042$ respectively) and
is the only cell of the temperature family to survive the
arbitrary-dependence correction. The tendency variables are
deterministic transforms of drivers already in the family and are not
counted as separate discoveries, but the contrast between the pair is
informative: at six hours the rate of cooling carries more directed
information than the temperature itself ($0.022$ against $0.018$
nats). Six hours is the interval over which nocturnal radiative
cooling establishes the surface inversion, and the result corroborates,
by an independent route, the thermal-history dependence that the
phase-loop hysteresis of Fig.~\ref{fig:env_characterisation}c
shows. The analysis does not localise where in the source--pathway--receptor chain a driver acts, so a thermal signature consistent with inversion trapping is equally consistent with temperature modulating the rate at which the waste mass generates gas; the two are not separable here.

Directed information from every driver falls away beyond
12\,h: no cell is significant at 24 or 48\,h, both of which are
estimable, and 96\,h is not estimable ($n = 89$ blocks).
That fall-away is a property of the test rather than an absence
of coupling: the surrogate null widens approximately twentyfold as
coarse-graining reduces the sample from $34{,}015$ blocks at 15\,min to
$180$ at 48\,h, so a given effective transfer entropy is progressively
harder to distinguish from chance at coarser scales. Raw ETE is
consequently not comparable across scales. The coarse scales are reported as underpowered at
this record length rather than as uncoupled.

The picture that emerges is a five-phase exposure lifecycle. H$_2$S
and CH$_4$ are co-emitted from the surface
(Fig.~\ref{fig:env_characterisation}b) under barometric modulation
(Fig.~\ref{fig:causal_links}a, raw pressure, whose share of
directed information rises with aggregation scale and peaks at
12\,h); advection sets which receptor is exposed
(Fig.~\ref{fig:env_characterisation}a; wind direction dominant at
every scale up to 6\,h and not significant beyond in Fig.~\ref{fig:causal_links}a); mechanical dilution at
sub-hourly scales sets the instantaneous concentration through the
inverse-wind-speed dependence of the Gaussian plume
(Fig.~\ref{fig:env_characterisation}f); under stable nocturnal
boundary layers a shallow inversion traps emissions over successive
hours (Fig.~\ref{fig:env_characterisation}c--d), recorded as the
strong counter-clockwise thermal hysteresis and, independently,
as the six-hour temperature tendency that is the one thermal cell to
survive the arbitrary-dependence correction in
Fig.~\ref{fig:causal_links}a; and post-sunrise convective mixing
erodes the inversion and entrains the accumulated reservoir into a
much larger volume (Fig.~\ref{fig:env_characterisation}e). The two
analyses agree: the variables that information-theoretic causality
identifies as significant at a given timescale are the variables that
classical dispersion physics identifies as operative on that
timescale. We note that the pressure signal will enter the
engineered-feature classifier in the next subsection through its
derivative $\mathrm{d}P/\mathrm{d}t$ and the stagnation index rather
than as raw pressure: the engineered features already encode the
synoptic information that pressure carries, a point that the ALE
analysis confirms below.

\subsection*{Engineered-feature benchmark and physical interpretability}

To establish whether the causal hierarchy can be operationalised by a
model that uses it explicitly, and to provide a strong baseline against
which a learnt representation can be tested, we trained four seasonal
XGBoost classifiers on a feature set that encodes the causal hierarchy
explicitly: causal Butterworth derivatives of T, P and WS, each
fitted over a strictly backward window $[t-6\,\mathrm{h},\,t]$ and
verified causal by truncation invariance (values computed on a
truncated series are identical to those computed on the full series at
every retained timestamp). The backward window spans 6\,h with its effective centre at $t-3\,\mathrm{h}$. The remaining features are 2- and
6-hour stagnation indices; a recirculation index; cyclic time and
direction encodings; and four annual Fourier harmonics (Methods,
``Ensemble classifier-tree seasonal nowcaster''). Three ordinal classes follow World
Health Organization guideline thresholds (Low $< 2$, Medium $2$--$7$,
High $\geq 7\,\mu\mathrm{g\,m^{-3}}$). Aggregated across the
calendar year 2024 ($n = 33{,}617$ evaluation samples), the classifier
attains a sample-weighted accuracy of $79.9\%$, with per-season
accuracy of $76.4$--$83.6\%$ and per-season High-class recall of
$0.583$--$0.749$ (Fig.~\ref{fig:classification_performance}a;
Supplementary Tables~\ref{tab:si_xgb_hyperparams}--\ref{tab:si_xgb_hyperopt_vs_eval}). Hyperparameter-search validation and seasonal-evaluation
F$_1$ agree within $|\Delta\mathrm{F}_1| \leq 0.009$ across all four
seasons, as expected if the temporally isolated holdout prevents hyperparameter leakage \citep{cawley2010over}.  A 3-hour majority-vote
ensemble (Fig.~\ref{fig:classification_performance}d) suppresses
classification noise; rolling accuracy is sustained
across each seasonal window (Fig.~\ref{fig:classification_performance}c),
and High-class predictions co-locate in time with daily community
odour reports (Fig.~\ref{fig:classification_performance}b), the first indication that the classifier captures an operationally
meaningful exposure signal.

Accumulated local effects (ALE) computed on held-out validation data
(Fig.~\ref{fig:classification_xai}a; Methods, ``Ensemble classifier-tree
seasonal nowcaster'') identify a physical hierarchy that overlaps with
the one the causal analysis recovered from the data alone.
The leading features are the diel cosine encoding
(class-mean importance $\bar{I} = 0.138$), the
2-hour stagnation index
($\bar{I} = 0.045$) and wind speed ($\bar{I} = 0.029$); the
temperature tendency $\mathrm{d}T/\mathrm{d}t$ ranks twelfth
($\bar{I} = 0.001$).
The dependence curves (Fig.~\ref{fig:classification_xai}b)
are physically interpretable: wind speed enters with the negative slope
of the inverse-wind-speed dilution law, and stagnation contributes
monotonically positively.
Notably, raw atmospheric
pressure has zero ALE importance despite being among the strongest drivers in the MSTE analysis:
the 2-hour stagnation index already encodes the synoptic
information that pressure carries, and the raw value adds no further
signal.
The benchmark thus recovers the same meteorological drivers
that the causal analysis identifies. It does so, however, chiefly
through raw wind channels and a diel encoding rather than through
the hand-coded derivative kernels: of the engineered features, only the
stagnation index carries appreciable attribution. Detailed seasonal interpretation of each
feature is given in Supplementary Note~3.

\subsection*{Physics-anchored learnt nowcasting}

To test whether the same physics is discoverable from raw meteorology
alone, we constructed CAIRN, a dual-pathway diagonal structured state
space (S4D) model whose only structural prior is a channel-wise split
into a fast lane (kernels initialised at a 1-hour timescale, matching
the wind-driven advective band of Fig.~\ref{fig:causal_links}a) and a
slow lane (initialised at a 6-hour timescale, which lies
within the band of scales over which pressure remains causally
significant in
Fig.~\ref{fig:causal_links}a)~\citep{gu2022parameterization}. The
input is restricted to four raw meteorological channels and minimal
calendar encodings (Fig.~\ref{fig:s4_arch}a); no derivative, lag,
rolling statistic, stagnation index or recirculation index is supplied
(Methods, ``CAIRN: structured state-space dual-pathway nowcaster'').
Strict causality is preserved at every layer: the SSM kernel is
one-sided, the convolution is explicitly truncated, and the readout is
the final-timestep slice. We evaluate the model under a weekly
expanding-window walk-forward protocol (13 retrains over the autumn
2024 quarter), the conservative choice because autumn was the worst
season for the engineered benchmark (Macro F$_1 = 0.608$) and the
season in which long-memory drivers carry the largest predictive load.

A six-condition initialisation ablation, holding optimiser, loss,
features and walk-forward protocol fixed (Table~\ref{tab:ed_s4_six_conditions}),
isolates the architectural contribution. Sweeping the slow-lane anchor
across two orders of magnitude
($\tau^{\star}_{\mathrm{slow}} \in \{6, 48, 147\}$\,h -- respectively
a scale lying within the causally significant pressure band
identified by the MSTE analysis, an intermediate sub-synoptic value
lying outside it, and the anchor of the baseline configuration, condition 1a of Table~\ref{tab:ed_s4_six_conditions})
identifies the MSTE-aligned anchor as the only condition
that outperforms all the others. Each condition was run three times from
independent random seeds; Table~\ref{tab:ed_s4_six_conditions} reports mean $\pm$
s.d., and the pooled within-condition standard deviation is 0.013 in
F$_1$-High (12 d.f.). The 6\,h anchor attains $0.57 \pm 0.01$ and
outperforms every alternative initialisation on a paired comparison
across the 13 weekly folds (two-sided Wilcoxon signed-rank,
Benjamini--Hochberg-corrected $q \leq 0.048$ in all five
comparisons).
The optimum lies inside the causally significant pressure
band ($\tau \leq 12$\,h; Fig.~\ref{fig:causal_links}a), whereas both
larger anchors lie outside it.  The sweep
did not test anchors between 6 and 48\,h, so the optimum is identified
within the tested set rather than located within the band, and the band extends to 12\,h, which was not tested. The 12\,h scale is the one at which pressure carries its largest share of directed information (Fig.~\ref{fig:causal_links}a), so the causal analysis itself identifies an untested anchor as its strongest candidate. Because 12\,h lies both inside the band and within the learnt kernels' reachable-memory regime, testing it would also separate the band-membership and kernel-reachability explanations, which the present sweep cannot.
Independently, the learnt slow-lane kernels accumulate 99\% of their
sensitivity mass by 27--45\,h and retain only 0.04--0.23\% of peak
sensitivity at 48\,h, so the two larger anchors initialise memory the
model cannot use. Band membership and kernel reachability therefore both predict the
observed ordering across the tested anchors; a 12\,h anchor, which lies
inside both the band and the learnt kernels' reachable-memory regime,
would separate the two explanations.

Only the unbounded condition, in which the upper $\Delta$ clamp
is lifted and the optimiser is free to migrate the slow lane, falls
clearly below every bounded condition ($0.48 \pm 0.01$; 10 of 13 folds against the
best bounded alternative, Wilcoxon $p = 0.011$). Because that
condition differs from the 147\,h baseline only in the removal of the
clamp, bounding the slow lane is consequential independently of where
within the bound it is anchored.

Within the bound, the conditions resolve into two tiers
rather than a graded ordering. The 147\,h, 48\,h, overlapping-lane (fast and slow anchors at 3 and 6\,h) and random-initialisation
conditions are mutually indistinguishable ($0.50$--$0.52$; one-way
ANOVA over three seeds each, $F = 1.75$, $p = 0.23$), so their
apparent ranking within any single run is not interpretable: the 147\,h condition, for example, ranks second on one seed but fifth on the three-seed mean.
Two further points qualify this result. First, random
initialisation ($0.52 \pm 0.01$) outperforms both the 48\,h and
147\,h anchors, so these data do not establish physics anchoring as a
general mechanism; they support the narrower claim that the
MSTE-derived 6\,h anchor is superior to every alternative tested,
including no anchor at all. Second, the comparison against random
initialisation is the weakest of the five: the 6\,h anchor wins on 9
of 13 folds, which reaches significance on the signed-rank test
($q = 0.048$) but not on a two-sided sign test ($p = 0.267$); when every fold is instead
scored without per-fold checkpoint selection, at the final retained epoch
or at a common fixed epoch, the same
comparison improves to $q \leq 0.013$. The
skill retained under random initialisation indicates that the
channel-wise lane partition itself provides a structural benefit even
without anchored sampling.

A complementary lane-zeroing ablation (Supplementary Methods, ``Walk-forward validation'')
does not resolve the two pathways.
Under the kernel-path mask of Eq.~\eqref{eq:s_lane_mask} the paired
per-fold difference between them is $-0.002$ (fast leads on 8 of 13
folds; sign test $p = 0.58$), the ordering reverses between
aggregation rules, and the interaction term exceeds either marginal
contribution while ranging from $-0.468$ to $+0.474$ across folds.
The lanes therefore contribute jointly rather than separably, and the
marginal values reported in Supplementary Methods, ``Walk-forward validation'' do not attribute
High-class skill to either lane.

The trained model reproduces phenomenology it was not given
directly.
Hour-of-day and 10$^{\circ}$ wind-sector aggregates of the predicted
High-class rate (Fig.~\ref{fig:s4_arch}b) reproduce the empirical
nocturnal accumulation peak at 02:00 LST. The directional
response tracks the empirical one closely (sector-profile correlation
$r = 0.93$) and concentrates predicted High probability in the source
sector, which carries $0.291$ against $0.077$ outside it -- an
enrichment of $3.8\times$. The predicted peak sector is nonetheless
displaced approximately $34^{\circ}$ clockwise of the empirical peak,
so the model recovers the source direction as a broad sector rather
than resolving its precise bearing. Per-feature, per-lane sensitivities
(Fig.~\ref{fig:s4_arch}c) show qualitatively distinct integration
kernels: the fast lane decays smoothly from $k = 0$ over the first
$\sim$5\,h with wind variables retained longest, while the slow lane
retains appreciable sensitivity across the whole displayed
window. That extent reflects the
input-sensitivity construction, not the learnt kernel timescales, which
move only marginally during training (slow-lane mean
$7.55 \rightarrow 7.56$\,h under the MSTE-aligned configuration).
The marginal response
of the predicted High probability to the unprovided pressure tendency
$\mathrm{d}P/\mathrm{d}t$ (Fig.~\ref{fig:s4_arch}d) and temperature
tendency $\mathrm{d}T/\mathrm{d}t$ (Fig.~\ref{fig:s4_arch}e) is
non-monotonic and physically interpretable in both cases, with the
empirical High-class frequency tracking the model curve within
bootstrap confidence:
from raw $P$ and $T$ alone, the model reproduces marginal
responses that engineered pipelines obtain only by hand-coding
explicit derivatives. The corresponding lane-resolved association is
weak ($\rho_s = +0.14$) and is reported as consistency with the
joint synoptic signature rather than as evidence that the signature
is encoded in any particular lane.

\subsection*{Walk-forward performance and community-impact validation}

Spike detection is the operational target of CAIRN: training uses
focal cross-entropy with explicit High-class weighting to drive
rare-event sensitivity (Methods, ``CAIRN: structured state-space
dual-pathway nowcaster''). Across the 13 weekly walk-forward retrains
over Oct--Dec 2024 ($n = 8{,}040$ evaluation timesteps, of which
872 are High-class; Methods,
``Walk-forward validation''),
CAIRN leads the engineered XGBoost benchmark on
High-class detection under a matched protocol in which both models
select their stopping point on the same inner temporal probe:
F$_1$-High $0.533$ against $0.501$, a gap of $+0.032$ ($+6.3\%$
relative). The paired per-fold difference does not reach
significance (two-sided Wilcoxon $p = 0.110$; CAIRN leads in 8 of 13
weeks), although the paired High-class decision over all 8,040
predictions does (McNemar, 344 against 292, $p = 0.043$, uncorrected for
the eight metrics compared). Both figures pool the confusion counts over all 8,040
evaluation timesteps before computing the metric. The alternative
convention, an unweighted mean of the 13 weekly values, gives
$0.528$ against $0.465$ and a larger gap of $+0.063$; the pooled rule
is reported because it matches the operational question the alert
system poses and because it is the more conservative of the two here.
The difference between the rules is itself informative: High-class
support ranges from 8 to 150 samples across the 13 weeks, so an
unweighted mean gives a week with eight positives the same influence
as one with a hundred and fifty.
Recall-High is $0.618$ against $0.575$ ($+0.044$) and
precision-High $0.468$ against $0.445$ ($+0.024$), so the detection
advantage holds on both components. The probabilistic terms do not
follow it. Brier-High marginally favours CAIRN ($0.075$ against
$0.077$), but the ranked probability score and the multiclass log-loss
both favour the benchmark, the latter substantially ($0.139$ against
$0.102$, and $0.900$ against $0.537$). CAIRN's advantage is therefore
in discrimination rather than calibration: it separates High-class
events more sharply while returning a less well-calibrated posterior.
This is the expected consequence of the selection protocol. Selecting each weekly checkpoint on High-class F$_1$ measured on the evaluation block itself would optimise the decision boundary rather than the probability estimate; selecting on the inner probe instead yields a better-ranked and less well-calibrated model (full per-week breakdown in Table~\ref{tab:ed_s4_period_aggregate} and
Supplementary Table~\ref{tab:si_s4_xgb_weekly}).
Under mismatched selection protocols, with CAIRN selecting its checkpoint on the evaluation block and the benchmark using a fixed iteration budget, the gap would be $+0.086$; with both selecting on a common inner probe it is $+0.032$. The benchmark itself is essentially unaffected by inner-probe selection (change in F$_1$-High under inner-probe selection $-0.004$ with a 7-day probe,
$+0.001$ with a 28-day probe), so the difference between the two gaps is attributable to CAIRN's selection protocol rather than to the benchmark.
Discrimination is higher than the benchmark
(AUC $0.898$ vs $0.877$;
Fig.~\ref{fig:S4_performance_H2S}b, left), although the per-fold
paired difference does not reach significance (CAIRN leads in 8 of 13
weeks, sign test $p = 0.59$) and the claim is therefore not made across the
operating range.
It detects 38 more High-class events than the
benchmark while producing 14 fewer false alarms (612 against 626).
Because both models decide by the argmax of the three-class posterior,
neither has a free operating point, so this is not a threshold trade-off;
the higher pooled AUC across $8{,}040$ timesteps is consistent with a better
ordering of events by predicted probability, although the per-fold
difference does not reach significance. Rolling 24-hour metrics
(Fig.~\ref{fig:S4_performance_H2S}a) show the gain is sustained rather
than concentrated in single weeks, and the spike-detection track at the
top of the panel shows that residual false negatives concentrate near
the High-class threshold rather than on the largest plumes.

Comparing CAIRN with the engineered benchmark varies two things at once, the input representation and the model, so a third model isolates them. An identically configured tree ensemble was trained on
CAIRN's own 16 raw inputs, none of which carries a lag, rolling
window or derivative; it is therefore memoryless by construction, and
it reaches F$_1$-High $0.463$. Against that common floor, the
24-feature engineered set, whose Butterworth derivatives and stagnation indices supply memory by hand, recovers $+0.040$ in pooled
F$_1$-High, while the state-space architecture, given only the raw
16, recovers $+0.073$. \emph{Both} increments reach significance
on the paired per-fold test that the endpoint comparison fails
(two-sided Wilcoxon $p = 0.027$ and $p = 0.022$; each leads in 10 of 13
folds), and the learnt route recovers roughly $1.8$ times as much.
Hand-coded and learnt memory therefore both recover skill that is absent from the memoryless floor; the learnt route recovers more of it, and does so without an analyst specifying the physics in advance. The ordering
holds on every High-class metric and on false alarms, which fall
monotonically across the three models (717, 626, 612); it does not extend
to overall accuracy or multiclass log-loss, where CAIRN
remains last because of its less well-calibrated posterior. Both comparisons are sensitive to fold
composition, and unequally so. Two folds have an inner probe containing no High-class sample, and between them they carry 201 of the 872 High-class samples, including the largest single week. Excluding them reduces the endpoint difference from $+0.032$ to $+0.012$, while the architecture
increment falls only from $+0.073$ to $+0.042$ and the engineering
increment from $+0.040$ to $+0.030$; the ordering survives every
jackknife (Supplementary Table~\ref{tab:si_three_arm}). The
decomposition against the memoryless floor is therefore the more stable
of the two comparisons.

A second model with an architecturally identical specification,
retrained without modification on co-emitted CH$_4$ under the same
walk-forward protocol, attains a daily-mean predicted High probability
that correlates with that of the H$_2$S model at
$r = 0.777$
($R^2 = 0.604$, slope $= 1.00$, $n = 88$ days,
$p = 5.5 \times 10^{-19}$; Fig.~\ref{fig:S4_performance_H2S}d, right)
and a Brier-High score ($0.0724$) within
$0.003$ of the H$_2$S model
(Supplementary Table~\ref{tab:si_s4_h2s_ch4_weekly}).
This comparison is cross-protocol: the H$_2$S series is the
matched inner-probe model and the CH$_4$ series selects its checkpoint on the evaluation week (Methods, ``Walk-forward
validation''). The difference between the two series is therefore not attributable to the
models alone. The near-unit slope and
small intercept indicate that the same physics-anchored architecture
transfers to a second co-emitted species despite differing absolute concentration scales. This is the structural pre-condition for a multi-species ensemble
whose redundancy rests on two independent
\emph{measurement} chains, though not on independent labels: the CH$_4$
class boundaries are derived from the H$_2$S boundaries by ordinary
least squares
($\mathrm{H_2S} = 5.55\,\mathrm{CH_4} - 8.29$), so the two targets are
a linear transform of one another and the redundancy is instrumental
rather than definitional for two co-emitted gases sharing a common source
(Supplementary Table~\ref{tab:si_s4_h2s_ch4_weekly}).

The final test couples the model output to an independent record of
community wellbeing. The lagged Pearson cross-correlogram between the
day-mean predicted High probability and the daily community odour
complaint count over the same Oct--Dec 2024 window
(Fig.~\ref{fig:S4_performance_H2S}c) shows a single pronounced peak at
lag 0 for both species: H$_2$S $r = 0.585$
($R^2 = 0.342$, $p = 2.2 \times 10^{-9}$, $n = 88$ days) and CH$_4$
$r = 0.517$ ($R^2 = 0.267$, $p = 2.5 \times 10^{-7}$); the correlation
falls to roughly half its peak within $\pm 1$~day on either side of the
lag-zero maximum (Fig.~\ref{fig:S4_performance_H2S}c).
This single-receptor correlation must be read against what
meteorology alone achieves on the same days. A single predictor, the negated daily mean temperature, reaches $r = 0.563$ when fitted out of
sample under blocked cross-validation, and across four independent
runs of the identical protocol the model attains $0.581 \pm 0.028$.
The margin over that floor, $+0.018$, is within run-to-run variation
(one-sided $t = 1.28$, $p = 0.145$). Correlation with odour reports at a
single receptor is therefore not, on its own, evidence that the model
captures a community-impact signal beyond ambient seasonal
meteorology.
Two features of the result are nonetheless informative. First, the lag structure is
informative even where the magnitude is not: the absence of a dominant
positive lag indicates that the strictly causal architecture is
behaving as a nowcaster of present community exposure, not a
forecaster of future odour reports. Second, the association strengthens
substantially when the monitoring channels are fused across the network
(next section), and the community-impact claim rests on that fused result
rather than on the single-receptor correlation.

  \subsection*{Multi-site network tier alerting}

  CAIRN extends to a multi-receptor setting in which independently
  trained nowcasters, one per species and receptor (hereafter channels),
  are fused into a single tiered indicator at the timescale of community impact.
  This extension tests three properties that go beyond the
  single-receptor analyses above: (i)~whether the CAIRN architecture
  transfers without retuning to a second receptor geometry;
  (ii)~whether classification skill is maintained when each constituent
  classifier is driven by the meteorology of its own receptor rather
  than by the principal MMF9 record; and (iii)~whether the fused
  output maps onto an operationally interpretable tiered-alert
  framework matched to the public-health response framework of Supplementary Note~1. MMF1 is excluded from this analysis: it was
  relocated mid-record, and the record at the post-move receptor (MMF1A) is too short to support an independent walk-forward retrain.

  Four constituent CAIRN nowcasters -- H$_2$S and CH$_4$ at MMF9 and
  MMF2, each retrained on the co-located meteorology of its own
  receptor under the protocol of
  Sec.~\ref{sec:methods_walkforward} -- feed a Bayesian log-odds
  accumulator (Sec.~\ref{sec:methods_multisite},
  Eq.~\eqref{eq:multisite_fusion}). The predicted tier is constructed
  \emph{purely from meteorology}: no raw target-species concentration
  enters the predicted pipeline at inference time. Per-channel debouncing matches the WHO 30-minute averaging
  convention, and the channel-state likelihood ratios are discounted
  below each classifier's uncorrected precision-to-prevalence ratio to absorb the
  inter-channel dependence implicated by the shared advection bearing
  and synoptic barometric modulation of
  Figs.~\ref{fig:env_characterisation}a and~\ref{fig:causal_links}a
  (Supplementary Note~6, ``Bayesian fusion of multi-site,
  multi-species evidence''; Supplementary Table~\ref{tab:si_lr}).
  The continuous posterior is mapped to four sequential ordinal tiers
  at cut-points $(\theta_1, \theta_2, \theta_3) = (0.15, 0.50, 0.92)$
  aligned with the Supplementary Note~1 public-health response: Tier~1
  at the FIDOL nuisance-assessment regime; Tier~2 at multi-channel
  corroboration; Tier~3
  at the near-maximum attainable posterior under coordinated four-channel meteorological evidence (analytical ceiling $P_{\max} \approx 0.95$).

  Across 13 weekly walk-forward folds spanning Jan--Mar 2025
  ($n = 8{,}536$ synchronous 15-min timesteps;
  Fig.~\ref{fig:multisite_tier}a--c), the fused predicted tier agrees
  with the deterministic raw-concentration ground-truth tier at
  quadratic-weighted Cohen's kappa $\kappa_w = 0.669$ (block-bootstrap
  95\% CI $0.565, 0.755$), strict accuracy $75.7\%$ (Supplementary Table~\ref{tab:ed_multisite_tier}).

  Because the tier distribution is dominated by Tier~0 (normal
  background), a majority-class predictor attains strict accuracy $80.0\%$;
  the accuracy figure alone therefore does not demonstrate skill, and
  the agreement claim rests on $\kappa_w$, which is chance-corrected.
  The unweighted $\kappa$ is $0.397$: the quadratic weighting
  materially improves the reported agreement because most errors are
  off-by-one tier.

  The hysteresis counts, likelihood ratios and tier cut-points
  that parameterise the fusion were selected by grid search on nine of the
  13 weeks (Supplementary Note~6). Refitting these parameters
  leave-one-week-out, so that every timestep is scored out of sample with
  respect to them, gives $\kappa_w = 0.661$ against $0.669$ and a complaint correlation of $0.698$ against $0.711$: the fusion-parameter optimism is $0.008$ in agreement and $0.014$ in correlation (Supplementary Table~\ref{tab:ed_multisite_tier}, Supplementary
  Table~\ref{tab:si_tier_decomposition}). The insensitivity is structural: the accumulator
  consumes latched binary channel states after debouncing, so the fused
  output has coarse leverage over the parameters being tuned. The four constituent nowcasters select their checkpoints on an inner temporal probe that precedes each evaluation week (Methods). Selecting the checkpoint on the evaluation week instead, which lets the scored week influence training, gives $\kappa_w = 0.709$ and an odour-report correlation of $0.729$ with the same fusion, an inflation of $0.040$ in agreement and $0.018$ in correlation (Supplementary Note~6, Supplementary Table~\ref{tab:si_tier_decomposition}).
  Performance is concentrated at the extremes: Tier~0 and Tier~3 are
  recovered (F$_1 = 0.89$ and $0.58$, with Tier~3 recall $0.53$, roughly half of
  ground-truth Tier~3 support, from meteorology alone) while the
  intermediate tiers are not (F$_1 \approx 0.30$; Supplementary Table~\ref{tab:ed_multisite_tier}). Tier~2 is the operationally consequential
  intermediate state, and recalibration of the constituent nowcasters
  would be required before this layer of the alert system is
  deployable for statutory-nuisance confirmation
  (Supplementary Note~6, Discussion).
  The following properties of the fusion configuration follow directly
  from its arithmetic and bound how it can be used. Predicted Tier~3
  requires all four channels active, because the largest three-channel
  posterior is $0.903$ against $\theta_3 = 0.92$; a single channel
  missing therefore caps the posterior at that value and makes Tier~3
  unreachable, whereas the ground-truth tier requires only three latched
  channels, so the two scales are not aligned at the top. Both H$_2$S
  receptors active with both CH$_4$ channels quiet gives $P = 0.40$ and
  reaches Tier~1 only. These follow from the likelihood-ratio set and
  prior of Supplementary Table~\ref{tab:si_tier_params} and account for
  the low intermediate-tier F$_1$ reported above. A same-architecture vote-count
  baseline returns $\kappa_w^{\mathrm{vote}} = 0.578$;
  the Bayesian fusion delivers $\Delta\kappa_w = +0.09$ over
  baseline, and under a \emph{paired} block bootstrap, which resamples the same blocks for both estimators so that the shared fold-to-fold variation cancels, the improvement is significant (95\% CI
  $0.048, 0.127$; $p = 0.001$). An unpaired bootstrap, which resamples the two $\kappa$ values
  independently and so ignores that both estimators are evaluated on
  identical timesteps, gives an interval approximately three times wider. Beyond the agreement gain, the fusion produces a continuous
  posterior $P_t$ that the vote-count baseline cannot deliver.
  The predicted tier nonetheless coincides with the
  vote count minus one on $92.5\%$ of timesteps, so the fusion should
  be understood as a calibrated refinement of channel counting rather
  than as a categorically different decision rule.

  External validation against an independent record of daily community
  odour reports (Fig.~\ref{fig:multisite_tier}d), which was not used in training, hysteresis or LR selection, provides an external check on the fused tier. The daily-mean CAIRN tier correlates with daily complaint
  count at Pearson $r = 0.711$ (95\% CI $0.40$--$0.84$ by a
  7-day moving-block bootstrap, which respects the strong day-to-day
  autocorrelation of both series (lag-1 $+0.68$ and $+0.59$);
  Spearman $\rho = 0.656$,
  $n = 89$ days; days on which no complaint record was returned
  are excluded rather than treated as zero counts; Fig.~\ref{fig:multisite_tier}e).
  Under the leave-one-week-out refit the same correlation is
  $0.698$, and it sits $0.095$ below the deterministic ground-truth
  reference ($r = 0.793$); a paired 7-day moving-block bootstrap on the
  difference, in which the same resampled blocks are used for both series
  so that shared day-to-day variation cancels, gives
  $\Delta r = +0.095$ (95\% percentile CI $-0.211$ to $+0.187$;
  $B = 4{,}000$). The paired difference is not statistically
  distinguishable from zero, though the interval is wide and asymmetric
  because a small number of high-count complaint days dominate the
  resampled blocks.
  Measured against the best naive meteorological predictor on
  the same 89 days (daily stagnation fraction, $r = 0.475$), the fused
  tier, scored with the in-sample fusion parameters ($r = 0.711$), improves on ambient meteorology by $+0.236$ (paired 7-day
  moving-block bootstrap, 95\% CI $+0.038$ to $+0.347$, $P(\Delta \leq 0) = 0.007$; the same resampled blocks are
  used for both series so that shared seasonal variation cancels). The
  corresponding margin of the single-receptor H$_2$S nowcaster over its best naive meteorological predictor (the negated daily mean temperature) is $+0.018$
  and does not reach significance. It is this multi-channel result that
  supports the community-impact claim: fusing four independently
  trained channels recovers a signal that no single channel separates
  from seasonal meteorology.
  The lagged cross-correlogram peaks
  at lag~0 (Fig.~\ref{fig:multisite_tier}f, Supplementary
  Table~\ref{tab:si_xcorr}). The lag-0 bar is the unique global
  maximum of a broad structure around lag zero; its location supports
  interpretation as a nowcaster rather than a delayed
  reflector of past events or a forecaster of future ones. The same architecture, retrained without retuning, serves both receptors and both species.
Figure~\ref{fig:complaint_distribution} maps the same complaint record in space: weekly postcode-level counts over January--March 2025 are greatest in the postcodes nearest the landfill and decline with distance. The temporal agreement of Fig.~\ref{fig:multisite_tier}e--f and this spatial gradient are both consistent with plume transport and dilution from the site, although the spatial pattern may also reflect population density and reporting propensity.

\section*{Discussion}
\label{sec:discussion}

Responses to fugitive landfill emissions remain largely retrospective: exposure is typically characterised only after community
impact has occurred, reflecting the source--pathway--receptor framework
that underpins chemical-incident response. This posture understates both
the scale and distribution of impact. Satellite inversions indicate that
landfill methane emissions exceed reported inventories~\citep{cusworth2024quantifying,
nesser2024high, wang2024methane}, associated health and wellbeing
burdens are increasingly recognised~\citep{guadalupe2021industrial,
heaney2011relation, carson2024malodors}, and these burdens fall
disproportionately on less advantaged populations~\citep{desouza2026evaluating,
Martuzzi2010InequalitiesWaste}.

The results presented here demonstrate
that receptor-scale exposure can be inferred directly from
meteorological conditions, represented in a causal nowcaster, and
validated against independent records of community impact.
The controls exerted by meteorology on receptor hydrogen sulphide have largely been inferred from dispersion theory and partial field evidence~\citep{ko2015review, njoku2025landfill}. Here, a year of continuous co-located gas and meteorological monitoring allows them to be identified directly, across timescales from minutes to days that short field campaigns cannot resolve. Multiscale transfer entropy identifies a
consistent hierarchy of drivers, with wind direction, wind speed and
atmospheric pressure forming the core of directed information flow
(Fig.~\ref{fig:causal_links}).
The hierarchy is found to be stable: the ordering of the three drivers is unchanged under six estimator configurations (Supplementary Note~5) and identical in two disjoint half-years, with the sign of the effective transfer entropy agreeing in 19 of 21 driver--scale cells. Pressure carries an increasing share of directed information as the aggregation scale coarsens, up to 12\,h, the coarsest scale at which the record resolves it. Two mechanisms could carry a pressure signal on this multi-hour horizon, pumping of gas from the source and suppression of the mixing depth~\citep{xu2014impact, stull1988introduction}, and directed information alone does not separate them.

The hierarchy is a statement about control rather than about where in the chain that control is exerted. The more likely reading, given the dispersion physics, is that meteorology sets the conditions under which emissions reach a receptor, but a contribution from temperature- or pressure-dependent emission rates at the source is not excluded by these data. Three analyses of the record, system characterisation, transfer entropy and model attribution, point to consistent drivers (Figs.~\ref{fig:env_characterisation}, \ref{fig:causal_links}, \ref{fig:classification_xai}), although they share the same data and the attribution model was given features built around the hierarchy; their agreement makes an artefact of any single method less likely.

These insights extend to model design. Air-quality prediction systems
typically impose temporal structure through engineered lags and
aggregation windows~\citep{mendez2023machine, houdou2024interpretable}.
In contrast, CAIRN links its temporal receptive field to
the coupling timescales the causal analysis
measures. The tested anchor inside the causally significant pressure
band outperformed every tested anchor outside it (Table~\ref{tab:ed_s4_six_conditions}). Bounding the slow pathway
to physically plausible timescales is consequential in its own right:
releasing the bound and allowing the optimiser to migrate the lane
freely lowers F$_1$-High by $0.04$ relative to the best alternative bounded condition, with lower scores on 10 of 13 folds ($p = 0.011$,
Table~\ref{tab:ed_s4_six_conditions}). This alignment allows the model to reconstruct
responses to pressure and temperature tendencies directly from raw
meteorological inputs (Fig.~\ref{fig:s4_arch}), although the marginal
contributions of the two lanes are not separately identifiable
(Supplementary Methods, ``Walk-forward validation'').

Odour reports are typically used for operational support rather than as a health-relevant endpoint~\citep{wang2019prediction, prudenza2023implementation, brancher2014odour}. Here, odour reports withheld from all model development provide a validation endpoint. Their independence is procedural rather than physical: reports and concentrations both follow from the same emission and dispersion events, so the comparison tests whether a meteorology-only signal tracks a human-reported endpoint, not whether unrelated series agree. At a single receptor the model's exposure probability correlates with daily reports at a modest $r = 0.585$ ($n = 88$ days), peaking at zero lag at the daily resolution of the record (Fig.~\ref{fig:S4_performance_H2S}), but no more than ambient meteorology does; the evidence for the community-impact link strengthens when two receptors and two gases are fused.

The zero-lag relationship is central to interpretation. A delayed peak
would indicate reconstruction of exposure after it has been reported,
while a leading relationship would imply the use of future information.
Instead, both the single-receptor signal and the fused tier peak at lag zero (Figs.~\ref{fig:S4_performance_H2S}c and \ref{fig:multisite_tier}f), the signature expected of a nowcast: an estimate of present exposure from concurrent meteorology, produced by an architecture that is strictly causal and receives no future input. At the daily resolution of the odour-report record, this places the meteorology--exposure link on the day of exposure.

The fused meteorological signal agrees substantially with the WHO-referenced exposure tiers derived from the sensors ($\kappa_w = 0.669$; $0.661$ under a leave-one-week-out refit; Fig.~\ref{fig:multisite_tier}), and because those tiers follow the FIDOL/WHO odour-annoyance framework used in UK statutory assessment~\citep{nicell2009assessment}, exposure inferred from meteorology is expressed in the same terms in which it is assessed. These findings bear on public-health practice, regulation and environmental policy because they show that exposure relative to WHO guideline levels can be inferred from meteorology once a model has been trained at a site, so that exposure is understood in terms of the conditions that drive it rather than only as a measured concentration. Whether expressing exposure in this way improves outcomes is not tested here, and evidence that short-term alerts improve outcomes is mixed~\citep{lyons2016airaware, dai2026significant}. Because the predictors are general meteorological and temporal variables rather than site-specific operational inputs, the approach may apply to other fugitive sources, such as wastewater and industrial facilities, although this remains to be validated.

Taken together, these results establish the link between meteorology and exceedance of WHO guideline levels against an exceptional ground truth: continuous receptor H$_2$S measurements, in place at this site because of the community-impact emergency, which define the exposure classes on which the model is trained. The meteorology-only signal learnt against that ground truth also tracks the independent record of odour reports. The fused tier correlates with daily reports at $r = 0.698$ under the leave-one-week-out refit, against $r = 0.793$ for the raw-sensor ground truth scored the same way ($n = 89$ days; Fig.~\ref{fig:multisite_tier}e), a difference not distinguishable from zero (paired $\Delta r = +0.095$, 95\% CI $-0.211$ to $+0.187$). Because the meteorology-only tier tracks odour reports at a correlation not distinguishable from that of the direct measurements, although the interval on the difference is wide, the relationship established here could support prediction of community impact from meteorology directly, combining routinely available meteorological data, including weather-prediction products~\citep{lam2023learning, lang2024aifs, price2024probabilistic}, with odour-report histories and, where available, surface gas measurements.

\section*{Methods}

\subsection*{Data and study sites}
\label{sec:methods_data}

Continuous ambient air-quality monitoring data were collected at three
downwind monitoring stations (denoted MMF9, MMF1, MMF2) and one
supplementary receptor (denoted MMF1A) throughout the calendar year
2024 (1 January--31 December). Records at 15-minute sampling interval
yielded $34{,}609$, $23{,}104$, $28{,}331$ and $11{,}374$
observations with valid H$_2$S respectively, against a complete 2024 grid of
35,136 15-minute intervals. The MMF1 and MMF1A records are short not because
of instrument downtime: MMF1 ceased operation on 31 August 2024 and the
supplementary receptor began on 2 September 2024, a relocation rather
than an outage, and the two are therefore not concurrent. Each station recorded H$_2$S
($\mu\mathrm{g\,m^{-3}}$), CH$_4$ (ppm), wind direction
(WD, deg), wind speed (WS, m\,s$^{-1}$), air temperature
(TEMP, $^\circ$C) and barometric pressure ($P$, hPa); the temperature
tendency $\mathrm{d}T/\mathrm{d}t$ ($^\circ$C\,hr$^{-1}$) was computed
as a first-order finite difference at the 15-minute sampling interval
(TEMP and $T$ are used interchangeably hereafter). The
receptor-modelling analysis in Fig.~\ref{fig:env_characterisation}a
uses all three downwind stations; per-panel chemical and
boundary-layer analyses (Fig.~\ref{fig:env_characterisation}b--f,
Fig.~\ref{fig:causal_links}) and all classifier training and
evaluation use the principal receptor MMF9, which provides the
longest contiguous valid record during the 2024 evaluation window.
The 2024 record uses post-calibration-adjustment data following the
independent reanalysis of the H$_2$S analyser by the regulator
\citep{ea2024dataadjustment}; no pre-2024 H$_2$S
observation enters any evaluation set. Training windows for the
walk-forward protocol begin on 1 September 2023, so pre-adjustment
data contribute to model fitting but never to any reported
metric. Pasquill--Gifford stability
classes were approximated using measured wind speed combined with a
solar-insolation proxy from time-of-day and time-of-year (pyranometer
data unavailable). All analyses were conducted in Python 3.10
(NumPy, SciPy, pandas, scikit-learn, hyperopt, PyTorch and Matplotlib).

\subsection*{Environmental characterisation}
\label{sec:methods_envchar}

The six panels of Fig.~\ref{fig:env_characterisation} use standard
statistical procedures, specified panel by panel in Supplementary
Methods, ``Environmental characterisation'':
conditional probability function (CPF) rose plots over $n_s = 36$
ten-degree wind-direction sectors at an H$_2$S exceedance threshold
$C_{\mathrm{thr}} = 5\,\mu\mathrm{g\,m^{-3}}$
\citep{ashbaugh1985principal} (\textbf{a}); Pearson and Spearman
correlation, a log--log OLS power-law fit and a Jaccard spike
co-occurrence index tested by $\chi^2$ for the H$_2$S--CH$_4$ pair
(\textbf{b}); the closed diel (TEMP, H$_2$S) phase loop, whose
enclosed area is computed by the Shoelace formula and normalised by
the mean ranges, with a hysteresis-asymmetry index from
cooling/warming-limb slopes (\textbf{c}); the mean diel profile
with hourly variation tested by Kruskal--Wallis (\textbf{d}); the
sunrise-aligned composite over all 365 days, pre- versus post-sunrise
concentrations compared by a one-sided Mann--Whitney $U$ test
(\textbf{e}); and bivariate OLS regressions of H$_2$S on wind speed,
temperature and $\mathrm{d}T/\mathrm{d}t$ with two-tailed $t$-tests on
the slopes (\textbf{f}). The quantitative summary across
all panels is in Supplementary Note~2 and Supplementary Table~\ref{tab:si_envchar_summary}.

\subsection*{Causal analysis}
\label{sec:methods_causal}

Multiscale transfer entropy (MSTE) quantifies directed nonlinear
information flow from each candidate driver to H$_2$S across
ten temporal scales $\tau \in \{15\,\mathrm{min},\,
30\,\mathrm{min},\, 1\,\mathrm{h},\, 2\,\mathrm{h},\, 3\,\mathrm{h},\,
6\,\mathrm{h},\, 12\,\mathrm{h},\, 24\,\mathrm{h},\, 48\,\mathrm{h},\,
96\,\mathrm{h}\}$,
following the framework of \citet{runge2019inferring}. For each
driver--scale combination we (i)~coarse-grain by non-overlapping
block averaging at aggregation factor
$m = \tau/(15\,\mathrm{min})$; (ii)~compute conditional transfer entropy
$T_{X \to Y \mid \mathbf{Z}} = I\!\left(Y_{t+1};\,X_t \mid Y_{t}^{(h)}, \mathbf{Z}_t\right)$
\citep{schreiber2000measuring} using the Frenzel--Pompe
$k$-nearest-neighbour estimator \citep{frenzel2007partial} conditioned on a
single greedily selected confounder; (iii)~apply the rank-uniform
marginal transform, tie-breaking jitter, Theiler exclusion window
$W = 4$ \citep{theiler1986spurious} and an adaptive neighbour count
$k$, all specified in Supplementary Methods, ``Multiscale transfer
entropy''; (iv)~assess significance using $B = 2{,}000$ circular-shift
surrogates that preserve the marginal distribution, periodicity and
autocorrelation of the source while breaking its temporal alignment
with the target; (v)~compute effective transfer entropy by
subtracting the surrogate-mean baseline; (vi)~apply Benjamini--Hochberg
false-discovery-rate correction at $\alpha = 0.05$ across
a single family of all 70 driver--scale tests, with
surrogate $p$-values computed as
$p = (1 + \#\{b : \mathrm{TE}_b \geq \mathrm{TE}_{\mathrm{obs}}\})/(B+1)$
so that $p = 0$ is unattainable and the correction is well
posed \citep{benjamini1995controlling}.
To test whether the directed information reflects shared diel
periodicity rather than transport, the analysis was repeated with the
circular-shift offsets restricted to integer multiples of 24\,h. This
null preserves diel co-oscillation between driver and target while
destroying their day-to-day correspondence, and is an exact
permutation test over the 365 distinct day shifts the record admits.
The directed information is essentially unchanged: the median
proportion of effective transfer entropy attributable to shared
diel structure is $0.027$ across the four directly measured
drivers, and no cell exceeds $0.75$. This test is uninformative at
aggregation scales of 24\,h and coarser, where every circular shift is
already a day multiple and the day-shift null coincides with the
ordinary one; those scales are reported as not tested for diel
confounding rather than as unconfounded.
The scale grid, driver set, embedding depths, conditioning
strategy and surrogate scheme were fixed before the reported analysis
was run and are reported in full; the estimator-configuration
comparison of Supplementary Note~5 is exploratory and is reported as a
robustness check rather than as a set of confirmatory tests. Only the
single pre-specified configuration enters the Benjamini--Hochberg
family.
Benjamini--Hochberg controls the false discovery rate under
independence or positive regression dependence. These tests are dependent by construction: the scales are nested coarse-grainings of a single series, and the three tendency variables are deterministic transforms of drivers already in the family. Positive regression dependence is therefore plausible but not demonstrated. The correction was
therefore repeated as a sensitivity analysis under
Benjamini--Yekutieli, which is valid under arbitrary dependence at the
cost of a harmonic-sum penalty. Benjamini--Hochberg remains the reported
procedure; cells surviving both corrections are drawn with filled
symbols in Fig.~\ref{fig:causal_links} and cells significant under
Benjamini--Hochberg alone with open symbols, so the effect of the dependence
assumption can be read directly from the figure.
A scale is reported as estimable only if it retains
$n \geq 100$ coarse-grained blocks and the adaptive neighbour count
satisfies $k \leq n/10$; a scale failing either test is reported as
``not estimable'' rather than ``not significant''. Under this rule the
96\,h scale ($n = 89$) is not estimable, and the coarsest estimable
scale is 48\,h. Blocks were retained only when at least 50\% of their
nominal sub-samples survived.
The history
embedding depth $h$ is set adaptively and capped at $h \leq 2$ to keep
the conditioning dimension within the regime where the KSG estimator
retains adequate sensitivity. Pseudo-code for the full algorithm and the
Frenzel--Pompe KSG-CMI subroutine is given in Supplementary Methods,
``Multiscale transfer entropy''. A six-configuration ablation study
confirms that the reported causal hierarchy is robust to estimator
hyperparameter choice (Supplementary Note~5; Supplementary
Tables~\ref{tab:si_mste_configs}--\ref{tab:si_mste_summary}).
That ablation varies the estimator settings; it does not vary
the sample. Stability with respect to the sample was assessed
separately by split-half resampling on January--June and
July--December 2024 with every estimator parameter held fixed: the
ordering of the three directly measured drivers is identical in both
halves and in the full-year analysis, and the sign of the effective
transfer entropy agrees in 19 of 21 driver--scale cells. Per-half
significance retention, the two cells that disagree, and the
limitations of the test are reported in Supplementary Note~5,
``Split-half stability of the recovered hierarchy''.

\subsection*{Ensemble classifier-tree seasonal nowcaster}
\label{sec:methods_xgb}

Four independent XGBoost classifiers \citep{chen2016xgboost} were
trained, one per calendar quarter of 2024 (Winter Q1, Spring Q2,
Summer Q3, Autumn Q4), each using an expanding-window protocol rooted
at the start of the monitoring record (September 2023) and extending
to the boundary of the target evaluation quarter. Three ordinal
classes use WHO guideline values, namely the odour-detection and odour-annoyance guidance levels for H$_2$S:
Low $< 2$, Medium $2$--$7$, High $\geq 7\,\mu\mathrm{g\,m^{-3}}$. The
feature set encodes the causal hierarchy explicitly: causal
Butterworth derivatives of $T$, $P$ and WS \citep{oppenheim1999discrete};
$2$-h and $6$-h stagnation indices (fraction of 15-minute intervals
with $\mathrm{WS} < 1.5\,\mathrm{m\,s^{-1}}$); a scalar/vector
recirculation index; cyclic time and wind-direction encodings;
and four annual
Fourier seasonality harmonics.
The resolved set contains 24 features and includes no lagged
or rolling-window terms. One of the 24, the station identifier, is
constant in the single-receptor configuration reported here and
carries no information; it is retained only for compatibility with
the multi-site pipeline.
With a strictly causal derivative window the classifier's skill is essentially the same as with a centred window (Macro F$_1$ $0.612$ against $0.608$) while its attribution differs substantially, indicating that the engineered temperature tendency matters for interpretation rather than for predictive performance. Class imbalance ($\sim 9.5{:}1$
Low:High) is corrected by inverse prior-probability weighting via
balanced sample weights computed in scikit-learn \citep{elkan2001foundations}; SMOTE
\citep{chawla2002smote} and scaled-exponent weighting alternatives
were rejected because the former produces physically implausible
feature combinations (interpolation between temporally distant
samples) and the latter degraded Low-class precision below
acceptable levels. Bayesian hyperparameter optimisation uses the
Tree-structured Parzen Estimator \citep{bergstra2011algorithms} over
an eight-dimensional search space (\texttt{max\_depth}, learning
rate, \texttt{n\_estimators}, \texttt{subsample},
\texttt{colsample\_bytree}, $\alpha$, $\lambda$,
\texttt{min\_child\_weight}) with up to 40 trials and patience-10
early stopping. To prevent leakage between hyperparameter selection
and seasonal evaluation, each season uses a temporally isolated
one-week hyperparameter-selection holdout consumed before the season boundary,
separated from the evaluation quarter by a one-week gap
\citep{cawley2010over}. The final model uses the
\texttt{multi:softprob} objective with \texttt{mlogloss} as the
training metric. The benchmark reported in Table~\ref{tab:ed_s4_period_aggregate} was trained for a fixed 200 boosting rounds with early
stopping disabled, so no evaluation-window data influenced its
stopping point. A 3-hour majority-vote ensemble (12 consecutive 15-minute
samples per block) post-processes predictions to align with the
30-minute WHO averaging convention; ties are broken in favour of the
highest-prior class (Low) for conservative false-alarm control. Model
interpretability uses Accumulated Local Effects (ALE) computed on
held-out validation data independently per seasonal classifier with
PyALE at $K = 20$ quantile-based grid points \citep{apley2020visualizing};
the full ALE derivation, the per-class importance metric, and the
multiclass aggregation procedure are given in Supplementary Methods,
``Accumulated Local Effects''. Hyperparameter values, per-season
performance, sample counts and the comparison of hyperparameter-selection and
evaluation performance are in Supplementary Tables~\ref{tab:si_xgb_hyperparams}--\ref{tab:si_xgb_hyperopt_vs_eval}; per-season ALE
interpretation is in Supplementary Note~3.

\subsection*{CAIRN: structured state-space dual-pathway nowcaster}
\label{sec:methods_s4}

The CAIRN architecture removes the engineered-feature inductive bias
and learns meteorological memory directly from a minimal
16-dimensional input: four raw meteorological channels
($\mathrm{WD},\,\mathrm{WS},\,T,\,P$) augmented only by the sin/cos
decomposition of WD (resolving the $0/360^{\circ}$ angular
discontinuity), an hour-of-day sin/cos pair, and four annual Fourier
seasonality harmonics represented as sin/cos pairs (eight
channels).
Raw WD is retained alongside its sin/cos decomposition, so the
$0/360^{\circ}$ discontinuity the decomposition removes is
reintroduced on one of the 16 channels; the decomposed pair carries the
directional information, and the ablation of Table~\ref{tab:ed_s4_six_conditions}
holds this input set fixed across all conditions, so no reported
comparison is affected. No derivative, lag, rolling statistic,
stagnation index or recirculation index is supplied. The model is a
stack of $n_{\mathrm{layers}} = 3$ pre-norm S4D blocks
\citep{gu2022parameterization, gu2021efficiently} of width $H = 128$
and state size $N = 256$, applied to a sequence of length
$L = 384$ steps (96\,h at 15-minute sampling interval). Each S4D
layer convolves its input against a length-$L$, one-sided convolution
kernel
\begin{equation}
\bar{\mathbf{K}}
\;=\;\bigl(\mathbf{C}\bar{\mathbf{B}},\;
            \mathbf{C}\bar{\mathbf{A}}\bar{\mathbf{B}},\; \dots,\;
            \mathbf{C}\bar{\mathbf{A}}^{\,L-1}\bar{\mathbf{B}}\bigr)
\;\in\; \mathbb{R}^{L},
\label{eq:ssm_kernel}
\end{equation}
which is parameterised in S4D form with diagonal $\mathbf{A}$,
complex eigenvalues $\lambda_n = -\exp(\log A^{\mathrm{real}}_n) + \mathrm{i}\,\pi(n-1)$,
and learnable timescale $\Delta$. The full continuous-time SSM,
zero-order-hold discretisation, complex-Vandermonde diagonal-kernel
construction, and S4D block specification (LayerNorm, gated-linear-unit
fusion, FFN, DropPath, residual; \citealp{xiong2020layer,
dauphin2017language, hendrycks2016gaussian, huang2016deep}) are given
in Supplementary Methods, ``S4D state space model''.

The architectural innovation is a channel-wise split into a fast lane
(large decay magnitude $|A^{\mathrm{real}}_{\mathrm{fast}}| = 10.0$,
short-$\Delta$ window) covering sub-hour to few-hour timescales, and
a slow lane (small decay magnitude
$|A^{\mathrm{real}}_{\mathrm{slow}}| = 0.5$, long-$\Delta$ window)
covering multi-hour to multi-day timescales, with 50/50 channel split
($f = 0.5$). Per-channel $\Delta$ values within each lane are sampled
in log-space around a physics-anchored centre that biases coverage
onto the canonical horizons identified by the MSTE analysis
(Fig.~\ref{fig:causal_links}a): a fast anchor
$\tau^{\star}_{\mathrm{fast}} = 1$\,h matching the wind-driven
advective band, and a slow anchor
$\tau^{\star}_{\mathrm{slow}} = 6$\,h -- a scale within the
band over which atmospheric pressure remains causally significant
(15\,min to 12\,h; Fig.~\ref{fig:causal_links}a). Given the
15-minute sampling base $\Delta t_{\mathrm{base}} = 0.25$\,h,
\begin{equation}
\Delta^{\star}_{\mathrm{lane}}
\;=\; \frac{\Delta t_{\mathrm{base}}}
           {|A^{\mathrm{real}}_{\mathrm{lane}}| \cdot \tau^{\star}_{\mathrm{lane}}},
\qquad
\log\Delta^{(h)}_{\mathrm{lane}}
\;\sim\;
\mathcal{N}\!\bigl(\log\Delta^{\star}_{\mathrm{lane}},\,s^2\bigr),
\label{eq:dt_anchor}
\end{equation}
clamped to log-uniform per-lane bounds with log-space spread
$s = 0.5$. Strict causality is preserved at every layer: the SSM
kernel is one-sided, the FFT convolution is explicitly truncated to
length $L$, the readout is the final-timestep slice, and no
bidirectional, attention-based or global-pool operation is used. The
training objective is focal cross-entropy \citep{lin2017focal} with
focusing parameter $\gamma = 2.62$ and per-class weights
$(\alpha_{\mathrm{Low}}, \alpha_{\mathrm{Med}}, \alpha_{\mathrm{High}})
= (1.00, 11.26, 14.07)$ imposed manually rather than derived from
inverse frequencies, complemented by inverse-prior balanced minibatch
sampling. The optimiser is AdamW \citep{loshchilov2019decoupled} with
five differential learning-rate groups, a six-epoch warmup during
which $\log A^{\mathrm{real}}$ and $A^{\mathrm{imag}}$ are frozen,
and a reduce-on-plateau learning-rate scheduler monitoring validation
High-class F$_1$ with patience-2 reduction and floor $10^{-5}$. Full
optimiser configuration is given in Supplementary Methods,
``Optimiser schedule''.

\subsection*{Walk-forward validation and metrics}
\label{sec:methods_walkforward}

CAIRN models are evaluated under a weekly expanding-window walk-forward
protocol covering 1 October--31 December 2024 (13 non-overlapping
weekly blocks; $n = 8{,}040$ evaluation timesteps). For each weekly
block, training data comprise all observations up to the previous
Sunday with the training cutoff retreated by $L - 1 = 383$ steps
($\approx 96$\,h) before the first validation timestamp, reserving a
causal context window for the input convolution that contains no
training labels.
The engineered benchmark applies no such reserve: it is a
pointwise classifier with no context window and therefore no
corresponding leakage mode, so it trains to the week boundary and sees
approximately four additional days per fold. The matched comparison
retains this difference, since removing it would handicap the
benchmark for a risk it does not carry. A fresh standardisation (zero mean, unit variance) is fitted on
training data alone and applied to the validation week and its
context reserve. A fresh model is trained from scratch (no
carry-over of weights, optimiser state or learning-rate schedule)
for up to 60 epochs with patience-7 early stopping on
an inner temporal probe: the seven days immediately
preceding each evaluation week are withheld from training and used
solely to select the epoch at which training stops.
Selecting instead on the evaluation block itself, with early stopping, best-epoch restoration, the learning-rate schedule and checkpoint selection all keyed on the week being scored, lets that week influence training. Comparing the two across all 13 folds, together with a matched control that withholds the same quantity of training data but selects on the evaluation block and so isolates the effect of the reduced training set, shows that evaluation-block selection inflates F$_1$-High
by $0.031$ (inner-probe selection lower in 12 of 13 folds; sign test $p = 0.0017$, Wilcoxon
$p = 0.0085$). This protocol governs the H$_2$S nowcaster
evaluated over Oct--Dec 2024, which is the model reported in
Fig.~\ref{fig:S4_performance_H2S}a,b, Table~\ref{tab:ed_s4_period_aggregate} and
Supplementary Table~\ref{tab:si_s4_xgb_weekly}. The cross-species CH$_4$ model uses checkpoint selection on the evaluation week; its reported values therefore carry selection optimism of the order of the $0.031$ measured above (Supplementary Table~\ref{tab:si_s4_h2s_ch4_weekly}; Fig.~\ref{fig:S4_performance_H2S}c,d). The four nowcasters fused in Fig.~\ref{fig:multisite_tier} select on the inner probe. In a comparison whose protocol and analysis were fixed before the runs, over 13 folds per nowcaster, selecting on the inner probe rather than on the evaluation week lowered nowcaster F$_1$-High by $0.028$--$0.080$, significantly for three of the four nowcasters, and the fused-tier $\kappa_w$ by $0.040$ (Supplementary Note~6). Because the inner probe also removes the most recent training days from each window, each of these nowcasters is then refitted: the epoch count and learning-rate schedule are chosen on the seven-day probe, and the model is then retrained on the full training window for that number of epochs, with the probe run's learning-rate schedule replayed, early stopping disabled and the final-epoch weights retained, so that no stage of training or selection reads the evaluation week. This refitting protocol was chosen from candidate protocols on four Oct--Dec 2024 weeks that precede the evaluation window and was fixed before the single evaluation of the Jan--Mar 2025 window.
The inflation is not a consequence of class sparsity: it is
uncorrelated with the number of High-class samples in the evaluation
week ($r = -0.19$, $p = 0.54$), and arises instead from selecting the
maximum of a volatile epoch-wise metric (mean absolute epoch-to-epoch
change $0.048$) over approximately 18 epochs.
At this site's event rate a seven-day block occasionally
contains no High-class exceedance at all: evaluation weeks span
8--150 High samples and probe blocks 0--180. Any weekly-blocked
protocol monitoring a rare-class metric at this event rate is
therefore operating on a selection signal that is intermittently
near-degenerate, and folds in which the probe contains no positive
class are reported separately. In the matched comparison against the
engineered benchmark, two of the 13 inner probes contain no
High-class sample and four more contain fewer than thirty; in those
folds the stopping point is selected on a metric computed over a block
containing none, or almost none, of the class it measures. Where the probe holds no High-class sample the monitored F$_1$-High evaluates to zero at every epoch rather than to an undefined value, so no epoch registers an improvement, the early-stopping counter runs from the first post-warmup epoch and the weights restored are that epoch's; the learning-rate scheduler receives the same zero and reduces the rate on its plateau schedule, falling back to validation loss only when the probe block is empty, not when the class is merely absent. The alternative, a 28-day probe ending four days before the evaluation
week, removes the degeneracy but doubles the apparent inflation relative
to the adjacent 7-day probe, because a block up to a month old selects a
checkpoint suited to conditions that have since changed; the 7-day probe
is reported because its selection block immediately precedes the period
being predicted. The consequence is
visible in one week of the benchmark comparison, whose inner block
contains no High-class sample and whose F$_1$ is correspondingly
degraded despite the evaluation week itself holding 150 positives; this
reflects the selection protocol and is a further reason for pooling rather than weighting weeks
equally.
Period-wide aggregates in Table~\ref{tab:ed_s4_period_aggregate} are
computed by pooling the confusion counts over all 8,040 evaluation
timesteps and deriving one metric from the pooled counts; the
per-week table (Supplementary Table~\ref{tab:si_s4_xgb_weekly}) reports the unweighted mean of
the 13 per-week values. The two conventions give F$_1$-High of 0.533 and 0.528
respectively and are stated separately wherever both appear.
Rows with any missing feature are dropped listwise; labels are
forward-filled over gaps of at most 30 minutes; channels absent at a
given timestep contribute their prior rather than an imputed
probability.
Under the reported protocol the MSTE-aligned configuration (condition 2a)
reproduces to $\pm 0.006$ F$_1$-High across three independent random
seeds, and the pooled within-condition standard deviation across all
six ablation conditions is $0.013$ (12 d.f.). The metric suite spans hard
classification (per-class and macro F$_1$, and conditional-rolling
F$_1$/precision/recall and false-alarm rate over a 4-h moving window)
and probabilistic calibration (Brier-High,
\citealp{brier1950verification}; binary ranked probability score,
\citealp{epstein1969scoring}; log-loss). Their mathematical
definitions, the per-week pipeline enumeration and the no-leakage
controls (label, feature, model, hyperparameter, retrain and checkpoint selection) are given in Supplementary Methods, ``Walk-forward
validation''. Per-week
breakdown is reported in Supplementary Table~\ref{tab:si_s4_xgb_weekly}; period-wide
aggregates are in Table~\ref{tab:ed_s4_period_aggregate}.

\subsection*{Cross-species CH$_4$ nowcaster and complaint validation}
\label{sec:methods_cross}

A second CAIRN model with architecturally identical specification was
trained on co-emitted CH$_4$ at the principal receptor under the same
weekly walk-forward protocol, with checkpoint selection on the evaluation week rather than on the inner temporal probe, so its reported values are optimistic to the extent
quantified in ``Walk-forward validation''. CH$_4$ class boundaries were derived
from the H$_2$S thresholds via an ordinary least-squares regression
$\mathrm{H_2S} = 5.55\,\mathrm{CH_4} - 8.29$ fitted on the principal
receptor record (giving Low $< 1.854$, Medium $1.854$--$2.756$,
High $\geq 2.756$\,ppm). External validation against community
wellbeing uses an independent record of daily community
odour-complaint counts ($n = 88$ days, Oct--Dec 2024,
excluding days lacking complaint data). Daily-mean predicted
High-class probabilities for both species were computed by averaging
$\hat{p}_{\mathrm{High}}$ within calendar-day windows; lagged
Pearson cross-correlations between $\hat{p}_{\mathrm{High}}$ and the
daily complaint count were evaluated at lags $-7$ to $+7$ days, with
the lag-0 correlation reported as the primary
endpoint. The complaint record is not used during training and is
unavailable to the classifier at inference time. Per-week
performance for both species is in Supplementary Table~\ref{tab:si_s4_h2s_ch4_weekly}; the
underlying co-located 15-minute time series is shown in
Supplementary Figure~1.

  \subsection*{Multi-site Bayesian alert-tier classifier}
  \label{sec:methods_multisite}
  A joint analysis fuses four CAIRN nowcasters -- H$_2$S and CH$_4$ at
  receptors MMF9 and MMF2, each trained independently under
  Sec.~\ref{sec:methods_s4} and the walk-forward protocol of
  Sec.~\ref{sec:methods_walkforward} -- into a continuous network-event
  probability and an ordinal alert tier $\tau_t \in \{0,1,2,3\}$ at the
  native 15-minute sampling interval. The predicted tier is constructed
  \emph{purely from meteorology} at inference time: no raw target-species
  concentration is read by any stage of the predicted pipeline. We refer to each of the four species--receptor nowcasters as a channel $c \in \mathcal{C}$. Each
  channel's per-step High-class probability is thresholded at $p^\star
  = 0.5$ and passed through an asymmetric
  $(N_{\mathrm{on}}, N_{\mathrm{off}}) = (3, 2)$ latch (45-minute onset,
  30-minute clearance -- one full WHO 30-minute averaging interval
  plus a 15-minute safety margin, and one full interval, respectively).
  The four latched states $\ell_{c,t}$ then contribute additive
  log-likelihood-ratio updates to a prior log-odds at base-rate
  $\pi = 0.10$:
  \begin{equation}
  \log O_t \;=\; \log\!\frac{\pi}{1-\pi}
  \;+\; \sum_{c \in \mathcal{C}} \log \mathrm{LR}_{c,t},
  \qquad P_t \;=\; O_t/(1+O_t),
  \label{eq:multisite_fusion}
  \end{equation}
  with channel-state likelihood ratios calibrated below each
  nowcaster's uncorrected precision-to-prevalence ratio to absorb inter-channel
  chemical and meteorological dependence (Supplementary Note~6,
  Supplementary Table~\ref{tab:si_lr}). H$_2$S-quiet states contribute
  mild negative log-LR (silence at a direct receptor is informative);
  CH$_4$-quiet and all missing-data states contribute
  $\log \mathrm{LR} = 0$ (neutral). The continuous posterior is mapped
  to four sequential ordinal tiers by hard thresholding alone -- no
  raw-sensor override on the predicted side -- at cut-points
  $(\theta_1, \theta_2, \theta_3) = (0.15, 0.50, 0.92)$ aligned with
  the FIDOL public-health framework of Supplementary Note~1: Tier~1
  at the FIDOL nuisance-assessment regime, Tier~2 at multi-channel
  corroboration, and Tier~3 at near-maximum attainable posterior (analytical ceiling
  $P_{\max} \approx 0.949$). Robustness of the parameter choice is
  established by a $\pm 25\%$ sensitivity analysis on the channel
  likelihood ratios (maximum $|\Delta \kappa_w| = 0.003$;
  Supplementary Note~6, Supplementary Table~\ref{tab:si_lr_sensitivity}).

  Performance is assessed against a deterministic ground-truth tier
  $\tau^{\mathrm{GT}}_t$ constructed from raw concentrations by a
  channel-vote rule on the same debounced threshold exceedances
  (Supplementary Note~6, Eq.~\eqref{eq:s_gt_tier_mapping}). The two
  pipelines share their per-channel hysteresis but differ in the
  multi-channel fusion stage (Bayesian log-odds versus hard
  channel-vote count) and in the target signal used as the per-channel
  input (model probability versus raw concentration). The primary
  statistic is the quadratic-weighted Cohen's
  kappa~\citep{cohen1968weighted, landis1977measurement} on the
  synchronous Jan--Mar 2025 walk-forward window ($n = 8{,}536$ shared
  15-min timesteps; 13 weekly retrains per constituent nowcaster).
  Confidence intervals are obtained by block bootstrap ($B = 1{,}000$
  resamples, 24-hour block length) to accommodate within-episode
  autocorrelation. A vote-count baseline that replaces the Bayesian
  accumulator with a hard count of latched channels -- equivalent to
  applying the ground-truth rule directly to the predicted activations
  -- isolates the contribution of the soft-evidence fusion stage from
  the per-channel nowcasting stage. External validation against an
  independent record of daily community odour reports is reported as
  a held-out endpoint that was not used in any stage of training,
  parameter selection or hysteresis design. Per-tier metrics with
  bootstrap CIs, the confusion matrix, the per-week kappa breakdown,
  the LR sensitivity analysis, and the complaint cross-correlation are
  in Supplementary Tables~\ref{tab:si_lr_sensitivity}--\ref{tab:si_xcorr}.

\subsection*{Statistical analyses}
\label{sec:methods_stats}

Tests are two-sided at $\alpha = 0.05$ unless stated otherwise; full
specifications are given in the subsections above. Environmental
characterisation (Fig.~\ref{fig:env_characterisation}) uses correlation and
regression $t$-tests, a $\chi^2$ test of independence, Kruskal--Wallis and a
one-sided Mann--Whitney $U$ test. Directed information flow
(Fig.~\ref{fig:causal_links}) is tested against $B = 2{,}000$ circular-shift
surrogates with Benjamini--Hochberg correction across the 70 driver--scale
tests (Benjamini--Yekutieli as a sensitivity analysis), a day-shift
permutation null for diel confounding and split-half resampling for
stability. Paired per-fold comparisons over the 13 walk-forward folds use
two-sided Wilcoxon signed-rank and sign tests, with Benjamini--Hochberg
correction across the five ablation comparisons
(Table~\ref{tab:ed_s4_six_conditions}); the benchmark comparison adds a
McNemar test on the 8,040 paired High-class decisions
(Table~\ref{tab:ed_s4_period_aggregate}). Tier agreement is
quadratic-weighted Cohen's $\kappa$ with block-bootstrap 95\% confidence
intervals ($B = 1{,}000$, 24-h blocks), and the fusion-versus-baseline
difference uses a paired block bootstrap; complaint agreement is the lag-0
Pearson correlation of daily means, with intervals from a paired 7-day
moving-block bootstrap. Sample sizes: $n = 8{,}040$ timesteps and 88
complaint days (Oct--Dec 2024); $n = 8{,}536$ timesteps and 89 complaint days
(Jan--Mar 2025). Model training used three fixed seeds (42, 43, 44); analyses
used Python 3.10 with NumPy, SciPy, pandas, scikit-learn and PyTorch.

\section*{Code availability}

A reference implementation of the CAIRN nowcaster is publicly
available at \url{https://github.com/tcpearce/CAIRN}. It contains the
dual-pathway state-space model definition, the 13 weekly
walk-forward checkpoints for each species, the per-week standardisation
parameters, the feature construction and labelling code, and an
test suite that reproduces the per-timestep predictions
reported here.
The multiscale transfer-entropy estimator, the gradient-boosted
benchmark and the scripts generating Figs.~1--7 are available from the
corresponding author and will be deposited in the same archive on
acceptance.
Fig.~\ref{fig:complaint_distribution} is not reproducible
from the released data, since it rests on the community complaint
record, which is withheld as personal data (Data availability). It is a
descriptive map of the spatial distribution of that record and no
quantitative result in this work is derived from it.
A versioned snapshot with a persistent DOI will be deposited at Zenodo
on acceptance.

\section*{Data availability}

The 15-minute meteorological and gas-concentration record underlying
the nowcasting results is released with the reference implementation
above as a single Parquet file covering 1 September 2023 to 31
December 2024 at the principal receptor, comprising wind direction,
wind speed, air temperature, barometric pressure, H$_2$S and CH$_4$.

Community odour-complaint records are \emph{not} released. They are
personal data collected under a regulatory complaints process, are not
required to reproduce any model result reported here, and are excluded
from the released dataset and its file-level metadata. The complaint
analyses of Figs.~\ref{fig:S4_performance_H2S}c--d,
\ref{fig:multisite_tier}d--f and \ref{fig:complaint_distribution}
therefore cannot be regenerated from the public release; the
derived daily counts used in those panels are available from the
corresponding author on reasonable request, subject to the
information-governance conditions under which they were obtained.

The monitoring site is identified by station code only.

\section*{Author contributions}
\begin{itemize}
    \item Conceptualization: A.F., T.C.P., D.J.T.S., A.D.
    \item Methodology: T.C.P. and A.F.
    \item Investigation: T.C.P. and A.F.
    \item Visualization: T.C.P. and A.F.
    \item Writing -- original draft: T.C.P. and A.F.
    \item Writing -- review \& editing: A.F., T.C.P., D.J.T.S., A.D.
\end{itemize}

\section*{Funding acknowledgement}
This study is part-funded by the National Institute for Health and Care Research (NIHR) Health Protection Research Unit in Chemical Threats and Hazards. T.C.P., D.J.T.S., A.D. and A.F. are supported by grant award NIHR207293. The views expressed are those of the author(s) and not necessarily those of the NIHR or the Department of Health and Social Care.

\section*{Acknowledgements}
The authors gratefully acknowledge Vince Jenner (Environmental Public Health Scientist, Environmental Hazards and Emergency Department, UK Health Security Agency) for valuable insights on the meteorological correlations associated with fugitive emissions. Sincere appreciation is also extended to Aphrodite Niggebrugge (Senior Geospatial Analyst, Geospatial Team, UK Health Security Agency) for developing the high-resolution GIS outputs used to visualise the findings.

\section*{Conflicting interests}
The authors declare no conflicting interests.

\bibliography{sn-bibliography}

\clearpage

  \clearpage
  \clearpage
  \begin{figure}[t]
   \small\textbf{Environmental characterisation of receptor-level
    hydrogen sulphide exposure.}
    \textbf{a}, Conditional probability function (CPF) rose plots at the
    three downwind monitoring stations (peak bearing $265^{\circ}$ at the
    principal receptor; CPF threshold $5\,\mu\mathrm{g\,m^{-3}}$);
    dashed polygon, source area; $n = 34{,}609$ 15-minute observations
    at the principal receptor.
    \textbf{b}, Bivariate H$_2$S--CH$_4$ relationship at the principal
    receptor, linear (left) and log--log (right) scales; Pearson
    $r = 0.830$ ($p < 0.001$); no-intercept OLS linear slope
    $\hat{\beta}_0 = 4.47\,\mu\mathrm{g\,m^{-3}\,ppm^{-1}}$;
    power-law fit $y = 0.356\,x^{1.95}$ (exponent $b = 1.95$).
    \textbf{c}, Hourly-mean diel trajectory in (temperature, H$_2$S)
    phase space; counter-clockwise loop with normalised area $A^{*} = 0.369$.
    \textbf{d}, Mean diel H$_2$S concentration profile (line) with
    95\% confidence interval of the hourly mean (shading; $\pm$1.96\,SEM
    from daily composites); peak 02:00\,LST, flat early-afternoon minimum across 12:00--15:00\,LST.
    \textbf{e}, Sunrise-aligned mean concentration profile; pre-sunrise
    baseline $9.5\,\mu\mathrm{g\,m^{-3}}$ (dashed), fumigation window
    0--4\,h (shading); concentrations are means $\pm$ 95\% CI across
    all available calendar days.
    \textbf{f}, Bivariate scatter of H$_2$S against wind speed (yellow),
    temperature and temperature tendency (navy); curves show smoothed
    90th-percentile spline fitted to 20 equal-width bins in $x$,
    capturing upper-tail dependence on each driver.
    Methods, ``Environmental characterisation''. Full numerical detail in
    Supplementary Note~2 and Supplementary Table~\ref{tab:si_envchar_summary}.
  \end{figure}

  \begin{figure}[htp!]
    \centering
    \includegraphics[width=1\textwidth]{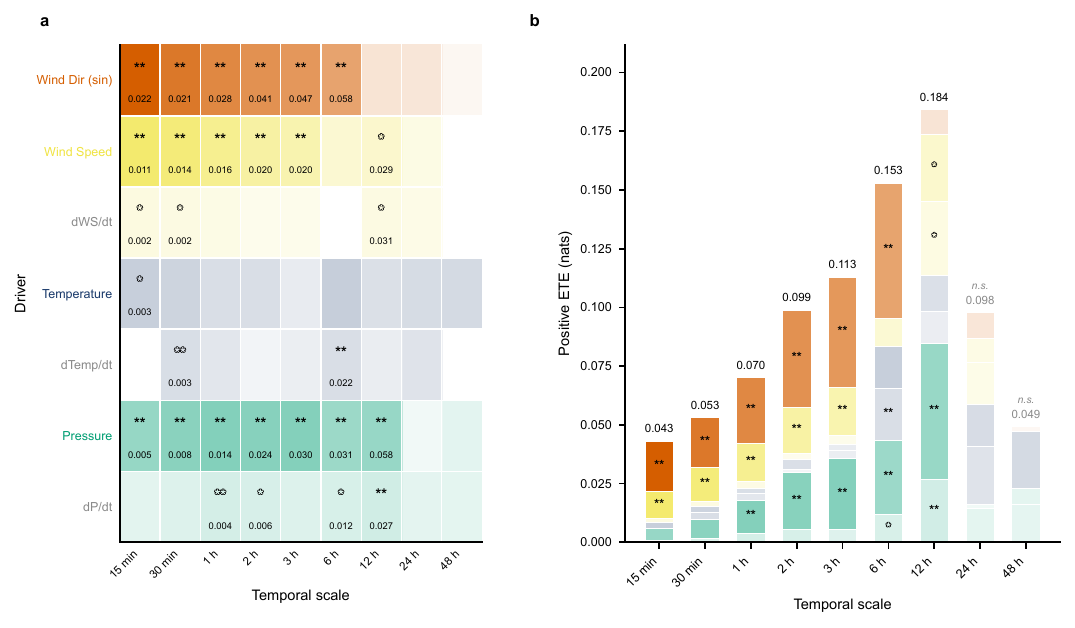}
  \caption{\textbf{Multiscale information flow from meteorology to
  hydrogen sulphide.}
   \small\textbf{a}, Effective transfer entropy (ETE; nats) from each of seven
    candidate meteorological drivers to H$_2$S, evaluated at ten temporal
    scales ($\tau = 15$\,min to $96$\,h), of which the nine estimable scales
    ($\tau = 15$\,min to $48$\,h) are displayed; row colours identify each driver
    (shared with panel~b). Cell values give ETE in nats. Colour
    intensity encodes ETE relative to that cell's own surrogate standard
    deviation, because raw ETE is not comparable across scales (Methods); asterisks denote
    significance after Benjamini--Hochberg false-discovery-rate correction
    across a single family of all 70 tests ($^{**}\,p_{\mathrm{adj}} \leq 0.025$;
    $^{*}\,0.025 < p_{\mathrm{adj}} < 0.05$).
    Filled symbols mark cells that additionally survive
    Benjamini--Yekutieli correction, which is valid under arbitrary
    dependence; open symbols mark cells significant under
    Benjamini--Hochberg alone.  Of the 29 cells reaching significance, 20 are filled
    and 9 open; 18 of the 19 cells belonging to the three
    directly measured core drivers are filled, the exception being wind
    speed at 12\,h. Seven of the nine open cells are meteorological
    tendency variables and the remaining two are wind speed at 12\,h and
    temperature at 15\,min. The
    seven cells of the 96\,h row are retained in the correction family
    although not estimable and not displayed, so the figure shows 63 of the
    70 tested cells while the multiplicity penalty is applied across all
    70.
    The three tendency-variable rows are shown in grey: they are deterministic transforms of drivers already present in the family, retained so that the Benjamini--Hochberg correction is applied across all 70 tests, but not interpreted in the main text (Supplementary Note~5).
    \textbf{b}, Stacked-bar decomposition of total positive ETE by driver
    at each temporal scale. Segment fills and stacking order match panel~a; numerals give the
    cumulative total per scale; asterisks follow the same FDR threshold as
    panel~a and are shown only for segments tall enough to carry a
    marker. Scales at which no cell reaches significance are labelled
    n.s.
    Estimator and confounder selection: Methods, ``Causal analysis'' ($B = 2{,}000$ circular-shift surrogates).
    Sensitivity to estimator hyperparameters in Supplementary Note~5.}\label{fig:causal_links}

  \end{figure}

  \begin{figure}[htp!]
    \centering
    \makebox[\textwidth][c]{\includegraphics[width=1.1\textwidth]{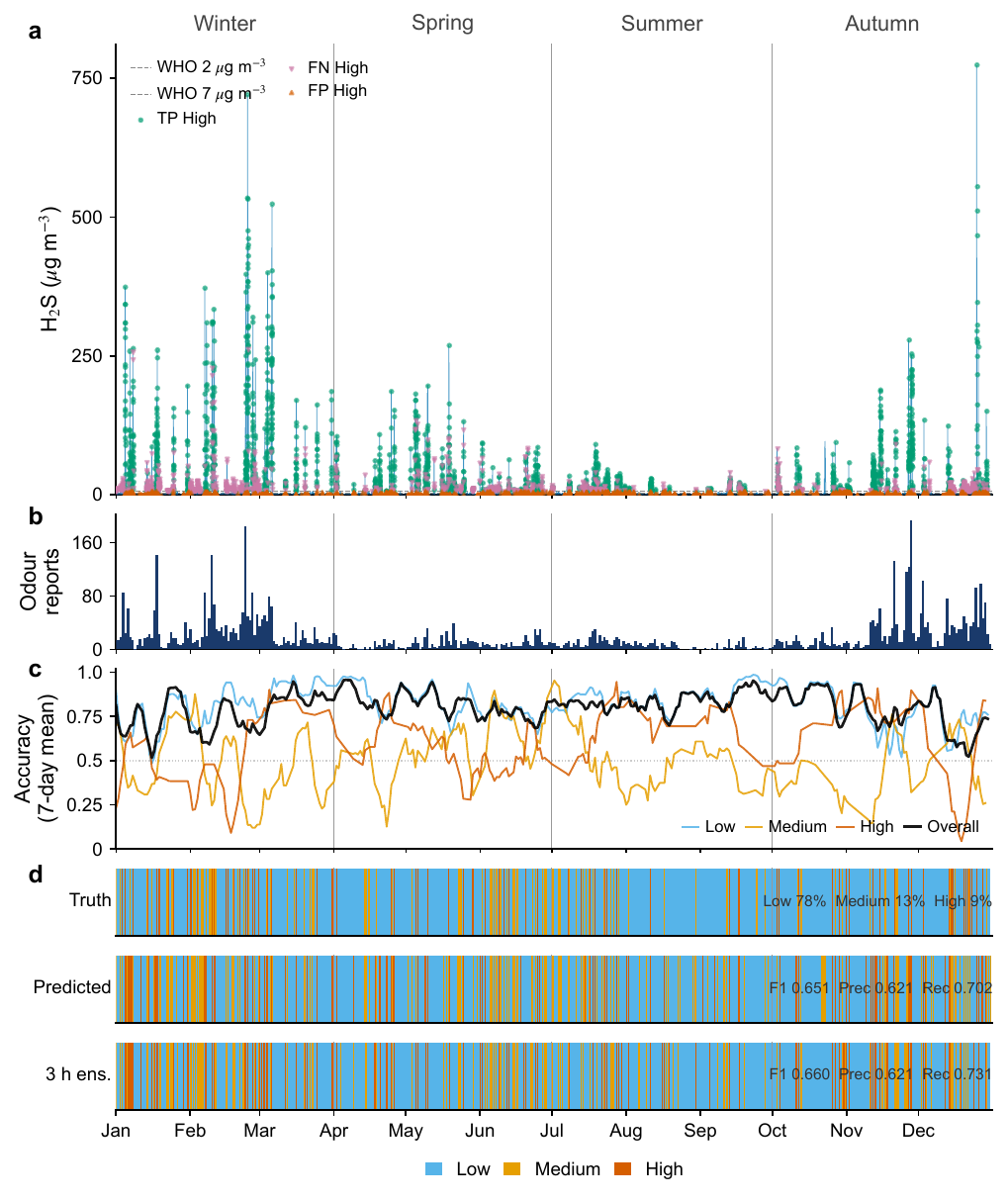}}
    \caption{Caption continued on next page.}
    \label{fig:classification_performance}
  \end{figure}
  \clearpage
  \begin{figure}[t]
  \small \textbf{Three-class XGBoost classifier of receptor exposure.}
  WHO-anchored class boundaries: Low $< 2$, Medium $2$--$7$, High
  $\geq 7\,\mu\mathrm{g\,m^{-3}}$.
  \textbf{a}, Seasonal expanding-window classification of 15-minute
  H$_2$S concentrations across 2024; WHO thresholds (dashed). True
  positives (TP$= 1{,}914$), false positives (FP$= 2{,}395$) and false
  negatives (FN$= 1{,}147$) for High exceedance are overlaid.
  \textbf{b}, Co-located daily community odour-complaint counts,
  overlaid in time with the High-class predictions in panel (a).
  \textbf{c}, 7-day rolling classification accuracy across the full 2024
  validation period.
  \textbf{d}, 3-hour majority-vote ensemble (12 consecutive 15-minute
  predictions per block) aligning the model output with the 30-minute
  WHO averaging convention; per-sample ground truth and class
  probabilities also shown.
  Period-wide macro-averaged performance: 15-min $F_1 = 0.651$
  (precision $= 0.621$, recall $= 0.702$); 3-h ensemble
  $F_1 = 0.660$ (precision $= 0.621$, recall $= 0.731$).
  Methods, ``Ensemble classifier-tree seasonal nowcaster''. Per-season performance,
  tuned hyperparameters, sample counts and a comparison of
  hyperparameter-selection and evaluation performance in Supplementary Tables~\ref{tab:si_xgb_hyperparams}--\ref{tab:si_xgb_hyperopt_vs_eval}.
  \end{figure}

  \begin{figure}[htp!]
    \centering
    \includegraphics[width=1\textwidth]{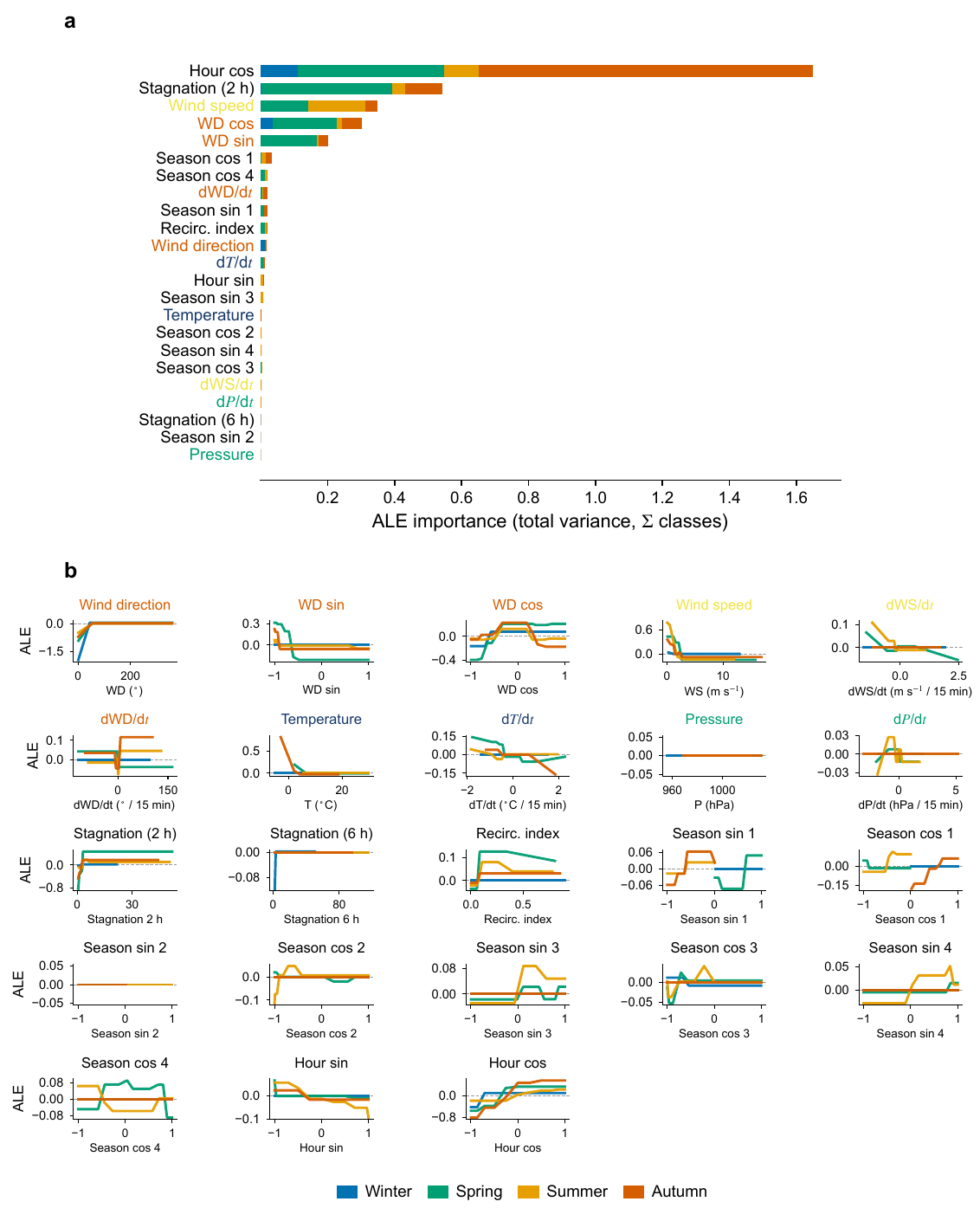}
  \caption{Caption continued on next page.}
  \label{fig:classification_xai}
  \end{figure}
  \clearpage
  \begin{figure}[t]
  \small \textbf{Accumulated local effects (ALE) interpretation of the XGBoost classifier.}
  \textbf{a}, Class-averaged ALE feature importance grouped by season
  (Methods, ``Ensemble classifier-tree seasonal nowcaster''; full
  derivation in Supplementary Methods, ``Accumulated Local Effects'');
  features sorted by descending mean importance across seasons.
  The leading features are the diel cosine encoding,
  the 2-hour stagnation index and wind speed; the wind channels coincide
  with the multiscale transfer entropy core of
  Fig.~\ref{fig:causal_links}a, while the diel term reflects the
  nocturnal accumulation regime resolved in
  Fig.~\ref{fig:env_characterisation}d,e.
  \textbf{b}, Class-averaged ALE dependence curves, one subplot per
  feature (23 features); curves show four seasons overlaid. Positive
  ALE shifts predictions towards Medium/High; negative ALE shifts
  towards Low.
  ALE computed on held-out validation data for each seasonal classifier.
  Per-season interpretation in Supplementary Note~3.
  \end{figure}

  \begin{figure}[htp!]
        \centering
        \includegraphics[width=1\textwidth,keepaspectratio]{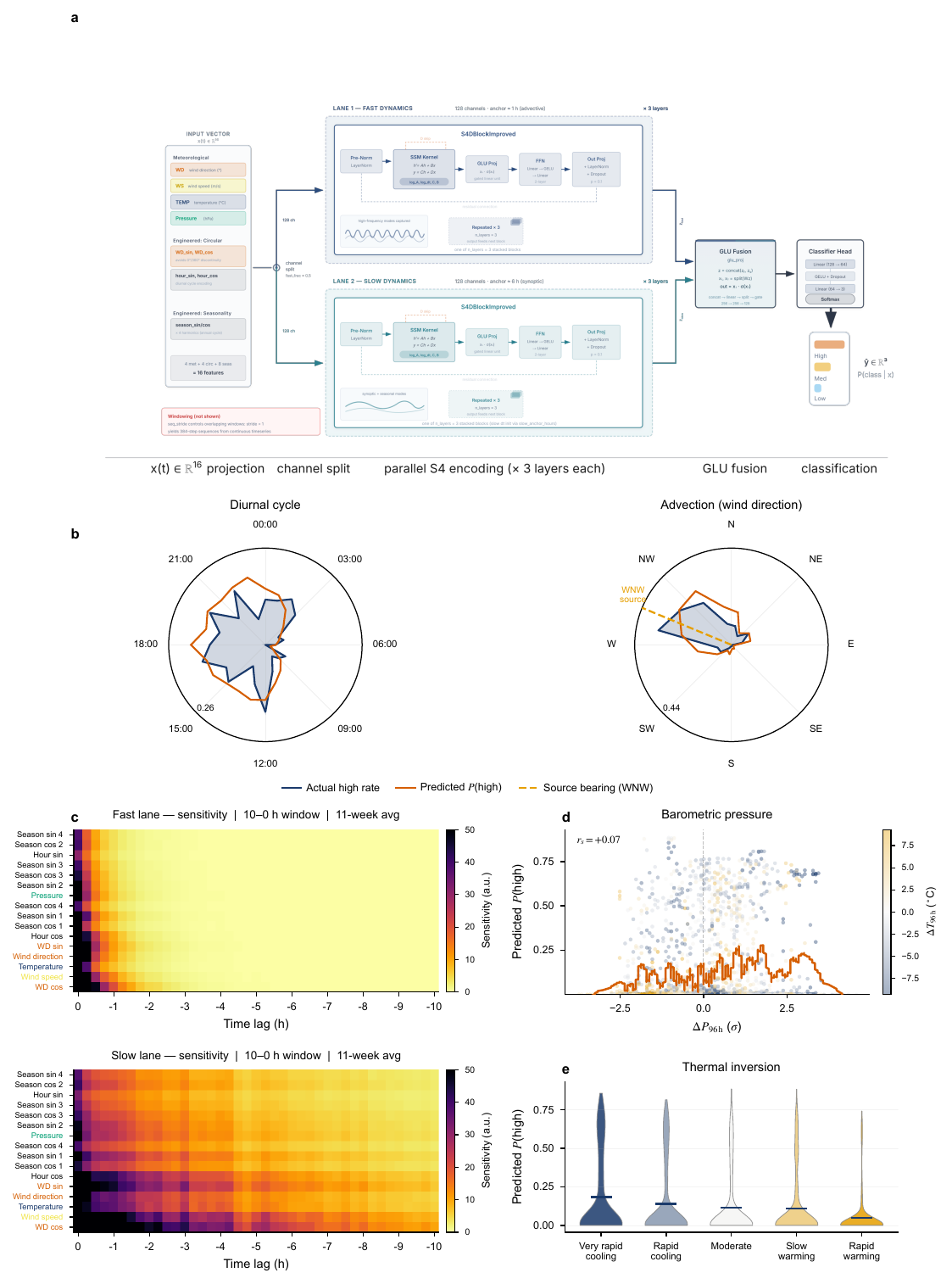}
  \caption{Caption continued on next page.}
  \label{fig:s4_arch}
  \end{figure}
  \clearpage
  \begin{figure}[t]
  \small \textbf{Dual-pathway state space architecture and learnt
  representations.}\textbf{a}, Schematic of the dual-pathway diagonal S4 model (Methods,
  ``CAIRN: structured state-space dual-pathway nowcaster''). The
  16-feature input (four raw meteorological channels and
  circular/Fourier calendar encodings; no derivative, lag, rolling,
  stagnation or recirculation feature) is linearly projected to model
  width $H = 128$ and split channel-wise into equal fast and slow
  lanes, each processed by three pre-norm S4D blocks with a gated
  linear unit, the two lanes then combined by a single GLU fusion.
  Fast lane: 1\,h advective anchor; slow lane: 6\,h synoptic anchor
  (a scale within the causally significant pressure band
  of
  Fig.~\ref{fig:causal_links}a).
  \textbf{b}, Empirical (ground-truth) versus model-predicted High-class
  rates against hour of day (left) and $10^{\circ}$ wind-direction
  sector (right); 13-week walk-forward validation (Oct--Dec 2024).
  \textbf{c}, Column-normalised per-feature, per-lane sensitivity over
  the most recent 10\,h of the 96\,h input window; fast lane (top) and slow lane (bottom),
  averaged across the 11 of 13 weekly checkpoints for which
  sensitivity maps were computed.
  The slow lane's extended response reflects the sensitivity
  construction rather than the learnt kernel timescales, which change
  only marginally during training, and should not be read as memory
  acquired by the model.
  \textbf{d}, Scatter of predicted $\hat{p}_{\mathrm{High}}$ against the
  96-hour normalised barometric-pressure change
  $\Delta P_{96\,\mathrm{h}}\,/\,\sigma_P$, coloured by the concurrent
  96-hour temperature change $\Delta T_{96\,\mathrm{h}}$; red curve,
  Savitzky--Golay smoothed trend; Spearman $r_s$ annotation; both
  quantities are derived from raw $P$, $T$ traces and are not supplied
  as input features.
  \textbf{e}, Violin distributions of $\hat{p}_{\mathrm{High}}$ across
  five equal-frequency quintile bins of the 96-hour temperature change
  (very rapid cooling to rapid warming); horizontal tick, within-quintile
  mean; violin colour reflects empirical High-class rate in that quintile
  (darker $=$ higher rate).
  Six-condition initialisation ablation in Table~\ref{tab:ed_s4_six_conditions};
  lane-ablation in Supplementary Methods, ``Walk-forward validation''.
  \end{figure}

  \begin{figure}[htp!]
  \centering
  \includegraphics[width=1\textwidth,keepaspectratio]{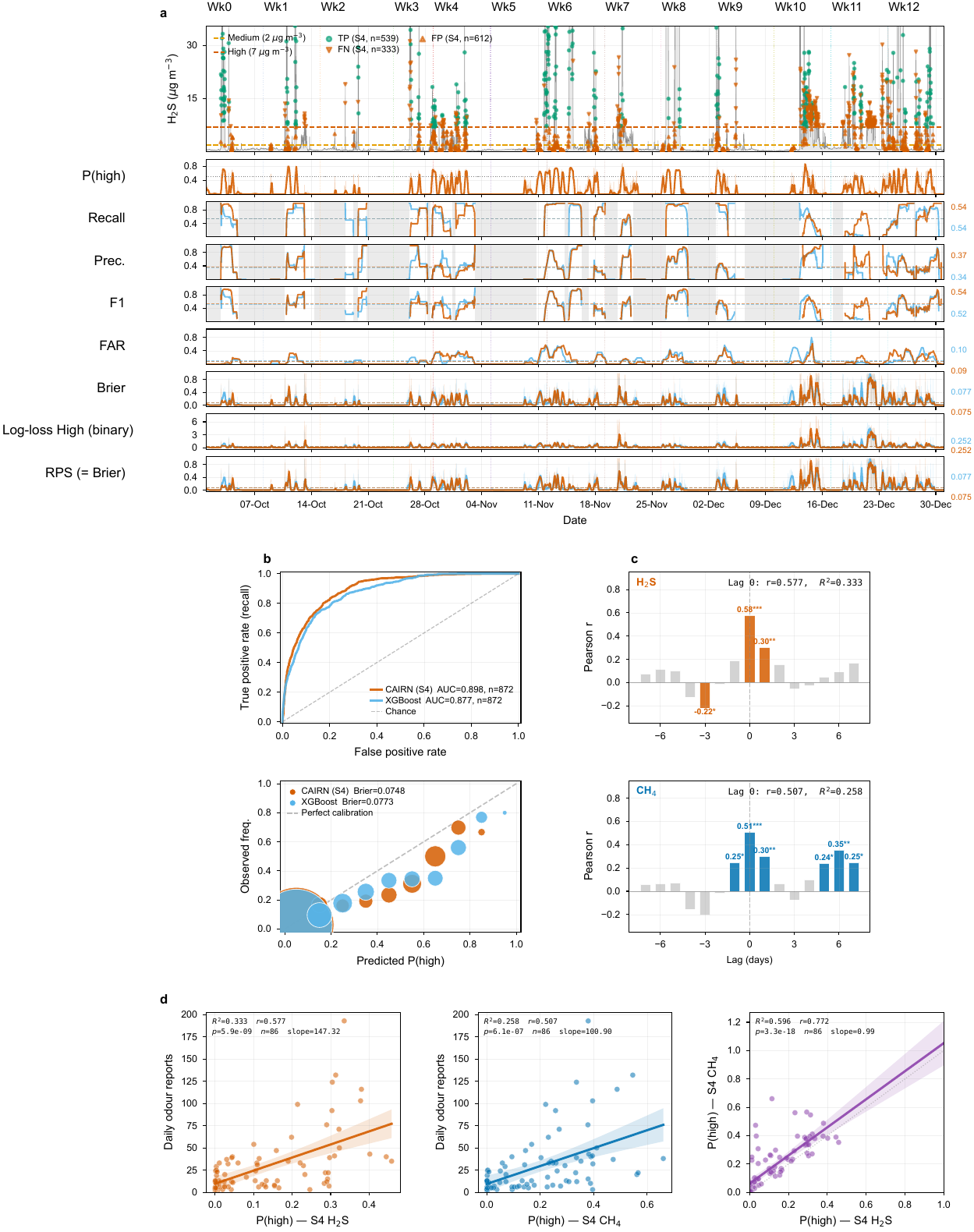}
  \caption{Caption continued on next page.}
  \label{fig:S4_performance_H2S}
  \end{figure}
  \clearpage
  \begin{figure}[t]
  \small \textbf{Walk-forward performance and external
  community-impact validation.} 13 weekly expanding-window retrains over Oct--Dec 2024 (Methods,
  ``Walk-forward validation''); $n = 8{,}040$ evaluation timesteps.
  \textbf{a}, Top: 15-minute H$_2$S concentration with WHO Medium
  ($2\,\mu\mathrm{g\,m^{-3}}$) and High ($7\,\mu\mathrm{g\,m^{-3}}$)
  thresholds; per-sample S4 outcomes (true positives, false positives,
  false negatives) overlaid; weekly retrain boundaries marked. Lower
  rows: predicted High-class probability and rolling 24\,h recall,
  precision, F$_1$, false-alarm rate and 4\,h moving-average Brier-High,
  log-loss-High and binary RPS-High for both models. Horizontal dashed
  lines: period-wide means.
  \textbf{b}, Receiver-operating characteristic (left) and 10-bin
  reliability diagram (right) over the full validation window;
  CAIRN AUC $= 0.898$, XGBoost AUC $= 0.877$. The
  paired per-fold difference is not significant (CAIRN leads in 8 of 13
  weeks, sign test $p = 0.58$; Wilcoxon $p = 0.11$).
  \textbf{c}, Lagged Pearson cross-correlogram between day-mean
  $\hat{p}_{\mathrm{High}}$ and the daily community odour-complaint
  count ($n = 88$ days; lags $-7$ to $+7$ days; negative lag, model
  leads odour reports). Top: H$_2$S model; bottom: CH$_4$ model.
  \textbf{d}, Daily-aggregate scatter relationships ($n = 88$ days):
  odour reports versus H$_2$S model probability (left) and CH$_4$ model
  probability (middle); cross-model agreement
  $\hat{p}_{\mathrm{High}}^{\mathrm{H_2S}}$ versus
  $\hat{p}_{\mathrm{High}}^{\mathrm{CH_4}}$ (right; OLS trend
  extrapolated to $\hat{p} = 1$); ordinary least-squares fits with
  95\% confidence intervals for the mean response.
  Both nowcasters share the dual-pathway architecture of Fig.~\ref{fig:s4_arch}a and
  Methods (``CAIRN: structured state-space dual-pathway nowcaster'').
  Panels (a) and (b), and the H$_2$S series in (c) and (d), use
  the matched inner-probe protocol reported throughout the Results. The
  CH$_4$ series in (c) and (d) selects its checkpoint on the evaluation week, so the
  CH$_4$ correlation and the cross-model agreement panel are descriptive
  rather than like-for-like comparisons. The same asymmetry is noted in
  Supplementary Table~\ref{tab:si_s4_h2s_ch4_weekly}.
  Period-wide aggregates in Table~\ref{tab:ed_s4_period_aggregate}; weekly breakdown in
  Supplementary Tables~\ref{tab:si_s4_xgb_weekly}--\ref{tab:si_s4_h2s_ch4_weekly}.
  \end{figure}

  \begin{figure}[htbp]
    \centering
    \makebox[\textwidth][c]{\includegraphics[width=1.1\textwidth]{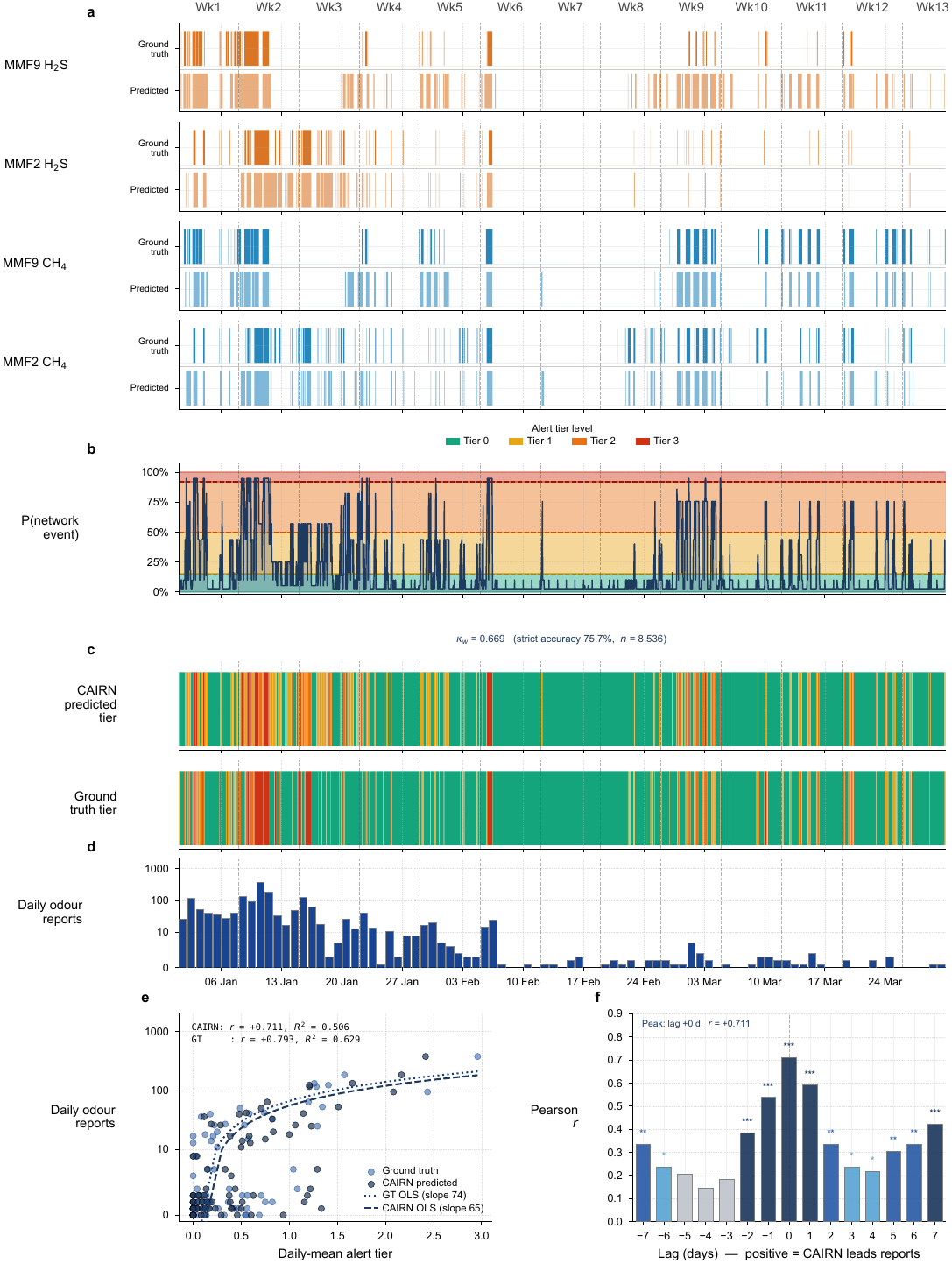}}
    \caption{Caption continued on next page.}
    \label{fig:multisite_tier}
  \end{figure}
  \clearpage
  \begin{figure}[t]
    \small \textbf{Multi-site Bayesian tier classifier agrees with the deterministic
    raw-sensor ground truth and tracks community odour reports.}
    \textbf{a},~Per-channel binary High-class activation for the four
    independently trained CAIRN nowcasters (H$_2$S and CH$_4$ at receptors MMF9
    and MMF2), each weekly retrained under the walk-forward protocol of
    Sec.~\ref{sec:methods_walkforward}, with checkpoints selected on an inner temporal probe (Methods). Within each channel row, the upper
    strip shows the ground-truth raw-concentration exceedance and the lower
    strip shows the model's hysteresis-debounced High-class prediction; vertical
    dashed lines mark the 13 weekly walk-forward folds.
    \textbf{b},~The fused continuous network-event probability $P_t$
    (Eq.~\eqref{eq:multisite_fusion}); coloured horizontal bands shade the four
    ordinal-tier zones and the three dashed lines mark the Tier~1
    ($\theta_1 = 0.15$), Tier~2 ($\theta_2 = 0.50$) and Tier~3
    ($\theta_3 = 0.92$) cut-points. \textbf{c},~CAIRN-predicted (top strip) versus deterministic raw-sensor
    ground-truth (bottom strip) ordinal alert tier
    $\tau_t \in \{0, 1, 2, 3\}$. Quadratic-weighted Cohen's $\kappa_w = 0.669$, strict element-wise accuracy 75.7\% ($n = 8{,}536$ synchronous 15-min timesteps).
    \textbf{d},~Daily community odour-complaint counts (symlog $y$), dark blue bars
    on the shared calendar axis of panels (a)--(c). The complaint record is
    fully external -- never used during training or hyperparameter
    selection.
    \textbf{e},~Daily-mean alert tier versus daily complaint count
    ($n = 89$ days carrying a complaint record, including the 19 on
    which the recorded count was zero). Dark navy circles and dashed
    line: CAIRN predicted tier versus odour reports (Pearson $r = 0.711$, $R^2 = 0.506$). Semi-transparent blue circles and dotted line:
    deterministic raw-sensor ground-truth tier versus odour reports
    ($r = 0.793$, $R^2 = 0.629$). The two ordinary least-squares fits have similar slopes. \textbf{f},~Lagged Pearson cross-correlation of daily-mean predicted tier
    with daily complaint count across a $\pm 7$-day window. Bars: Pearson $r$ at each lag, shaded by significance; stars denote thresholds ($^{***}\,p < 0.001$, $^{**}\,p < 0.01$, $^{*}\,p < 0.05$).
    The lag-0 bar is the unique global maximum at $r = 0.711$, with roll-off at $\pm 1$~day to $r \approx 0.54$--$0.59$.
    Correlations at $-7$, $-6$ and $+2$ to $+7$ days also reach
    significance, so the roll-off is not strictly monotonic and the peak
    is the global maximum of a broad structure, heavier at positive lags, rather
    than an isolated one.
    The lag-0 maximum is consistent with a nowcast of concurrent conditions rather than a lagged reflection of past events or a forecast of future ones.
    \end{figure}

\begin{sidewaysfigure}[htp!]
\centering
\makebox[\textwidth][c]{\includegraphics[width=1.1\textwidth,keepaspectratio]{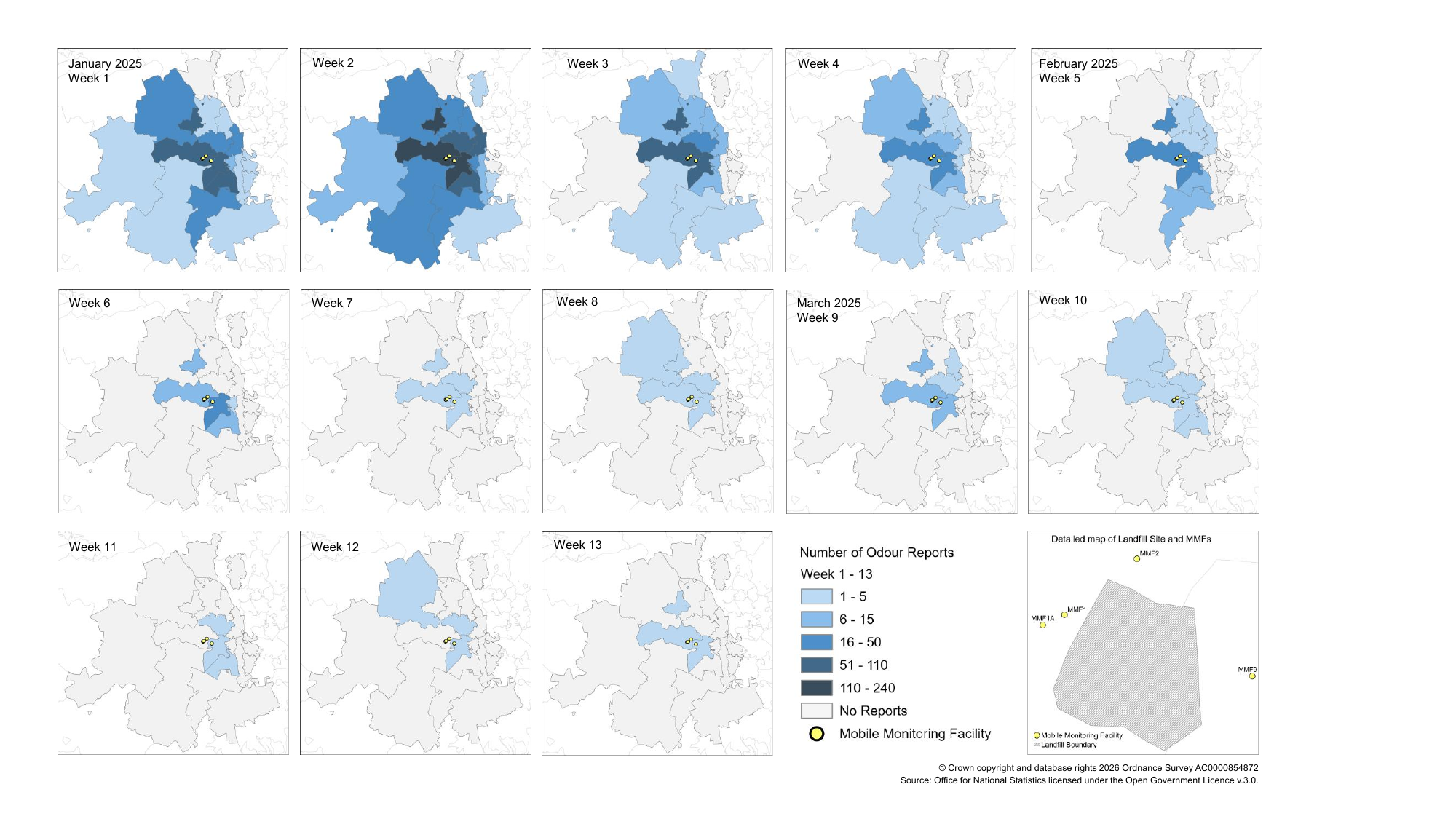}}
 \caption{Caption continued on next page.}
 \label{fig:complaint_distribution}
 \end{sidewaysfigure}
  \clearpage
  \begin{figure}[t]
    \small \textbf{Spatial reconstruction of weekly odour-complaint activity around the landfill site.}
    Each panel shows the postcode-level distribution of daily odour reports aggregated by week between January and March 2025. Colour intensity is proportional to the number of reports received, with darker shading indicating greater complaint activity. A persistent concentration of odour reports is observed in communities closest to the landfill, with reporting intensity progressively declining with distance from the source. The stable spatial gradient is consistent with the exposure patterns inferred from H$_2$S measurements and meteorologically driven dispersion processes.
\end{figure}

\clearpage
\begin{sidewaystable}[htbp]
\centering
\caption{\textbf{Six-condition initialisation ablation isolating the
architectural contribution of physics-anchored timescale priors.} All
conditions share the dual-pathway diagonal S4 architecture, optimiser,
focal-cross-entropy loss, 16-feature input, walk-forward protocol
(13 weekly retrains over Oct--Dec 2024) and hyperparameter set; they
differ only in the initialisation of the SSM kernel parameters
$\{\Delta_{\mathrm{lane}}, A^{\mathrm{real}}_{\mathrm{lane}}\}$ and
the per-lane $\Delta$ bounds. The MSTE-aligned condition (\textbf{2a},
bold) is the configuration used throughout the main text. Period-wide
F$_1$-High is the primary metric for High-class spike
detection. Each condition was run three times from
independent random seeds (42, 43 and 44); values are mean $\pm$ s.d.\ over
those three runs, quoted to two decimal places because the third
decimal lies below the run-to-run noise floor. The pooled
within-condition standard deviation is $0.013$ (12 d.f.). Conditions
1a, 1b, 1c and 2b are mutually indistinguishable (one-way ANOVA
$F = 1.75$, $p = 0.23$); only 2a and 3b separate from that tier.
Paired per-fold comparisons of 2a against each alternative use the
seed-mean value per fold across the three seeds, giving 13 paired
differences. Benjamini--Hochberg-corrected $q$ ranges from $0.003$ to
$0.048$ on two-sided Wilcoxon signed-rank tests; the comparison against
random initialisation (1b) is the weakest at $q = 0.048$, and does not
reach significance on a two-sided sign test (9 of 13 folds,
$p = 0.267$). Under two protocols that remove the checkpoint selection
--- scoring every fold at the final retained epoch (patience expiry) and
at a common fixed epoch, the last two columns --- the same comparison
improves to $q \leq 0.013$ and every paired margin is preserved or
widened. Headline values are reported in Results, ``Physics-anchored
learnt nowcasting''.}
\label{tab:ed_s4_six_conditions}
\footnotesize
\setlength{\tabcolsep}{4pt}
\begin{tabular}{@{}clp{2.6cm}ccp{2.7cm}ccc@{}}
\toprule
ID & Name & Init.\ scheme
   & $\tau^{\star}_{\mathrm{fast}}$
   & $\tau^{\star}_{\mathrm{slow}}$
   & Hypothesis tested
   & F$_1$-High (selected checkpoint)
   & Patience expiry & Common epoch \\
\midrule
1a & Baseline    & physics-anchored, hard-clamped & 1\,h & 147\,h
   & Hard-clamped reference, single $\Delta$ value & $0.50 \pm 0.02$ & 0.41 & 0.41 \\
1b & Random      & log-uniform (no anchor) & --- & ---
   & Anchor irrelevant beyond bounds & $0.52 \pm 0.01$ & 0.43 & 0.43 \\
1c & Overlap     & physics-anchored, wide spread & 3\,h & 6\,h
   & Lane separation matters & $0.52 \pm 0.01$ & 0.44 & 0.45 \\
\textbf{2a} & \textbf{MSTE-aligned} & \textbf{physics-anchored} & \textbf{1\,h} & \textbf{6\,h}
   & \textbf{Slow anchor at MSTE 6\,h scale} & $\mathbf{0.57 \pm 0.01}$ & \textbf{0.50} & \textbf{0.52} \\
2b & Mid-scale   & physics-anchored & 1\,h & 48\,h
   & Intermediate slow anchor & $0.51 \pm 0.02$ & 0.44 & 0.44 \\
3b & Unbounded   & physics-anchored, free upper bound & 1\,h & 147\,h (free)
   & Optimiser-free slow timescale & $0.48 \pm 0.01$ & 0.40 & 0.41 \\
\bottomrule
\end{tabular}
\end{sidewaystable}

\clearpage
\begin{table}[htbp]
\centering
\caption{\textbf{Period-wide walk-forward
performance of CAIRN versus the engineered XGBoost
benchmark under a matched selection protocol.} Pooled across the 13
weekly walk-forward retrains over Oct--Dec 2024 ($n = 8{,}040$
evaluation timesteps) on the case-study site; both models select their
stopping point on the same 7-day inner temporal probe, and the decision
rule is the argmax of the three-class posterior. CAIRN leads on
every High-class detection metric, on discrimination and on false
alarms, contributing 38 additional true positives while producing 14
fewer false positives. The individual margins are modest and none
reaches significance on a paired per-fold test over the 13 folds
(F$_1$-High $p = 0.110$); the paired High-class decision over all 8,040
predictions does (McNemar, 344 against 292, $p = 0.043$). The
probabilistic terms separate from the detection terms: Brier-High
marginally favours CAIRN, while the ranked probability score and
log-loss favour the benchmark, reflecting that inner-probe checkpoint selection yields a better-ranked but less well-calibrated posterior (Results, ``Walk-forward performance''). Whole-distribution
metrics are reported for completeness and are an operational target of
neither model. Weekly per-fold breakdown in Supplementary
Table~\ref{tab:si_s4_xgb_weekly}.}
\label{tab:ed_s4_period_aggregate}
\small
\setlength{\tabcolsep}{6pt}
\begin{tabular}{@{}lccr@{}}
\toprule
Metric & XGBoost & CAIRN & $\Delta$ \\
\midrule
\multicolumn{4}{@{}l}{\textit{High-class detection metrics (argmax of the three-class posterior)}} \\
F$_1$-High            & 0.501 & \textbf{0.533} & $+0.032$ ($+6.3\%$) \\
Recall-High           & 0.575 & \textbf{0.618} & $+0.044$ ($+7.6\%$) \\
Precision-High        & 0.445 & \textbf{0.468} & $+0.024$ ($+5.3\%$) \\
False-alarm rate      & 0.087 & \textbf{0.085} & $-0.002$ \\
\midrule
\multicolumn{4}{@{}l}{\textit{Discrimination}} \\
AUC (High class) & 0.877 & \textbf{0.898} & $+0.021$ \\
\midrule
\multicolumn{4}{@{}l}{\textit{Probabilistic metrics}} \\
Brier score (High class)   & 0.077 & \textbf{0.075} & $-0.003$ \\
Ranked probability score   & \textbf{0.102} & 0.139 & $+0.038$ \\
Log-loss                   & \textbf{0.537} & 0.900 & $+0.362$ \\
\midrule
\multicolumn{4}{@{}l}{\textit{Whole-distribution metrics (operational target of neither model)}} \\
Accuracy      & \textbf{0.787} & 0.694 & $-0.092$ \\
Macro-F$_1$   & \textbf{0.615} & 0.576 & $-0.039$ \\
\midrule
\multicolumn{4}{@{}l}{\textit{Detection tallies}} \\

  & \multicolumn{2}{c}{} & \\
True positives, High class
  & 501 & \textbf{539} & $+38$ \\
False positives, total
  & 626 & \textbf{612} & $-14$ \\
False negatives, High class
  & 371 & \textbf{333} & $-38$ \\
Weekly F$_1$-High wins
  & 5/13 & \textbf{8/13} & --- \\
Total evaluation timesteps
  & \multicolumn{2}{c}{$n = 8{,}040$} & \\
\bottomrule
\end{tabular}
\end{table}

\clearpage

\end{document}

\clearpage

\thispagestyle{empty}
\vspace*{0.24\textheight}
\begin{tcolorbox}[colback=inkdeep!4,colframe=inkteal,boxrule=1pt,arc=3pt,
                  left=16pt,right=16pt,top=16pt,bottom=16pt]
  \centering
  {\color{inkteal}\scshape\large Supplementary Information}\\[0.9em]
  {\color{inkdeep}\Large\bfseries Meteorology-driven Causal Nowcasting\\ of Fugitive Landfill Emissions\\ from Measured Coupling Timescales}\\[1.4em]
  {\color{inkteal}\rule{0.34\textwidth}{0.8pt}}\\[1.2em]
  \begin{minipage}{0.84\textwidth}\centering\small
    Supplementary Notes 1--6, Supplementary Methods, Supplementary Tables
    S1--S19 and Supplementary Figures.\\[0.7em]
    References are listed once, at the end of the main article.
  \end{minipage}
\end{tcolorbox}
\clearpage

\renewcommand{\thefigure}{S\arabic{figure}}
\renewcommand{\thetable}{S\arabic{table}}
\renewcommand{\theequation}{S\arabic{equation}}
\renewcommand{\thesection}{S\arabic{section}}
\setcounter{figure}{0}
\setcounter{table}{0}
\setcounter{equation}{0}
\setcounter{section}{0}

\renewcommand{\bibliography}[1]{}

\makeatletter\let\newgeometry\@gobble\makeatother

\renewcommand{\rhtext}{Supplementary Information\, \textbullet\, Meteorology-driven causal nowcasting}

\documentclass[pdflatex,sn-mathphys-num]{sn-jnl}

\usepackage{graphicx}
\usepackage{multirow}
\usepackage{amsmath,amssymb,amsfonts}
\usepackage{booktabs}
\usepackage{xcolor}
\usepackage{textcomp}
\usepackage{hyperref}
\AtBeginDocument{\let\burl\url}
\usepackage{caption}
\usepackage{algorithm}
\usepackage{algpseudocode}
\let\comment\relax\let\endcomment\relax
\usepackage{xr}
\makeatletter
\let\ORIG@externaldocument\externaldocument
\renewcommand*{\externaldocument}[1]{%
  \begingroup\let\bibcite\@gobbletwo\ORIG@externaldocument{#1}\endgroup}
\makeatother
\externaldocument{cee_article}

\newcommand{\degree}{\ensuremath{^\circ}}

\usepackage{changes}
\usepackage{marginnote}
\makeatletter\@mparswitchfalse\makeatother
\setlength{\marginparsep}{8pt}\setlength{\marginparwidth}{1.6cm}
\newif\ifshowchanges
\showchangestrue
\ifshowchanges
  \DeclareRobustCommand{\chadd}[1]{\added{#1}}
  \DeclareRobustCommand{\chdelete}[1]{\deleted{#1}}
  \DeclareRobustCommand{\chreplace}[2]{\replaced{#1}{#2}}
  \DeclareRobustCommand{\chrev}[1]{\textsuperscript{\textcolor{red}{\bfseries\sffamily #1}}}
  \newenvironment{chaddblock}{\par\begingroup\color{blue}}{\par\endgroup}
  \DeclareRobustCommand{\chstack}[2]{\shortstack{\added{#1}\\\deleted{#2}}}
\else
  \DeclareRobustCommand{\chadd}[1]{#1}
  \DeclareRobustCommand{\chdelete}[1]{}
  \DeclareRobustCommand{\chreplace}[2]{#1}
  \DeclareRobustCommand{\chrev}[1]{}
  \newenvironment{chaddblock}{}{}
  \DeclareRobustCommand{\chstack}[2]{#1}
\fi

\geometry{left=3cm,right=3cm,top=2.5cm,bottom=2.5cm}

\renewcommand{\thefigure}{S\arabic{figure}}
\renewcommand{\thetable}{S\arabic{table}}
\renewcommand{\theequation}{S\arabic{equation}}
\renewcommand{\thesection}{S\arabic{section}}

\setcounter{figure}{0}
\setcounter{table}{0}
\setcounter{equation}{0}
\setcounter{section}{0}

\usepackage{caption}

\raggedbottom

\begin{document}
\newgeometry{left=3cm,right=3cm,top=2.5cm,bottom=2.5cm}
\linespread{1}\selectfont

\begin{center}
    \vspace*{1em}

    {\LARGE \textbf{Meteorology-driven Causal Nowcasting of\\[0.5em]  Fugitive Landfill Emissions from\\[0.8em] Measured Coupling Timescales}}\\[1.5em]

    {\Large \textbf{Supplementary Information}}\\[2em]
\end{center}

\tableofcontents
\vspace{1em}

\noindent This document accompanies the main manuscript. It consists of six
Supplementary Notes, Supplementary Methods in six subsections, the
Supplementary Tables and one Supplementary Figure, which hold the
contextual, per-panel, sensitivity and full-procedural material
referenced from the main text, figure captions and Methods.
Supplementary Note~1 provides the regulatory and public-health
context that frames the case-study site. Supplementary Note~2 gives the
full per-panel quantitative detail underlying Fig.~1 of the main text.
Supplementary Note~3 reports the per-season interpretation of the
Accumulated Local Effects (ALE) hierarchy of the XGBoost classifier.
Supplementary Note~4 provides an extended comparison with prior
literature on landfill hydrogen sulphide, machine-learning nowcasting,
and community-impact validation. Supplementary Note~5 documents the multiscale transfer
entropy (MSTE) estimator ablation that supports the robustness of the
causal hierarchy reported in main Fig.~2. Supplementary Note~6 expands the joint multi-site, multi-species analysis and the ordinal alert tiers summarised in main Fig.~7. \textbf{Supplementary
Methods} holds the full mathematical and procedural detail for the
methods summarised in main Methods, organised in six subsections: ``Environmental characterisation'' (per-panel mathematical detail for Fig.~1); ``Multiscale transfer entropy'' (full procedural specification, including temporal coarse-graining and surrogate testing, the KSG--Theiler estimator, and pseudo-code for Algorithm~1 and the KSG-CMI subroutine); ``Accumulated Local Effects'' (ALE
derivation, multiclass aggregation and importance metric); ``S4D state
space model'' (continuous-time SSM, ZOH discretisation, S4D diagonal
kernel, block specification, end-to-end architecture and causality
guarantees); ``Optimiser schedule'' (joint focal-cross-entropy/MSE
loss, AdamW with five differential learning-rate groups, warmup-freeze
and reduce-on-plateau scheduler); and ``Walk-forward
validation'' (per-week pipeline, leakage controls, lane-ablation
and feature-sensitivity attribution, and metric definitions).
The Supplementary Tables and Supplementary Figure~1 hold the
numerical detail referenced from the main text and from the Notes
and Methods below.

\section{Regulatory and public-health context}
\label{sec:si_note1}

\subsection*{Waste, landfills and fugitive gas emissions}

Global municipal solid waste generation reached an estimated 2.01 billion
tonnes per year, representing approximately 0.74~kg per person per day on
average, and is projected to rise to 3.40 billion tonnes by 2050
\citep{kaza2018what}. Even
under modern engineered designs that incorporate phased cell construction,
liners, leachate extraction and capping, the long-term anaerobic
decomposition of organic waste continues to release methane (typically
45--60\% by volume), carbon dioxide (40--60\%), water vapour, and a complex
trace fraction including volatile organic compounds and reduced sulphur
species, principally hydrogen sulphide (H$_2$S)
\citep{atsdr2001landfillgas, parker2002gaseous}. Landfill gas migration is
nominally controlled through extraction wells supported by monitoring
networks, but cap permeability, extraction rate, and the integrity of
sub-cell containment together determine whether emissions remain on-site
or migrate laterally to nearby receptors \citep{nastev2001gas}. Where these controls are
incomplete, fugitive emissions to air pose persistent environmental and
public-health risks for surrounding communities. Recent regulatory data
indicate that the share of poor-performing regulated industrial sites in
the waste sector is at the highest level on record, with such sites
disproportionately implicated in pollution incidents and fugitive
emissions \citep{ends2024, ea2025chief}.

\subsection*{Hydrogen sulphide: toxicology and odour}

H$_2$S is generated in waste cells by sulphate-reducing bacteria converting
sulphate (SO$_4^{2-}$) under anaerobic conditions
\citep{lee2006reduced, townsend2004heavy, townsend2005cd}; in mixed waste streams
the dominant sulphate source is gypsum (CaSO$_4$) from construction and
demolition debris \citep{jang2001sulfate, asakura2015sulfate}. Its odour threshold is cited as between approximately $0.75\,\mu\mathrm{g\,m^{-3}}$ and $11\,\mu\mathrm{g\,m^{-3}}$ \citep{ATSDR2016HydrogenSulfideToxGuide},
making it one of the most readily identifiable environmental odours, but
prolonged or high-concentration exposure causes olfactory fatigue that
masks levels potentially dangerous to health \citep{govuk_nd_h2s}. Acute inhalation of low
concentrations irritates the eyes and respiratory tract, producing sore
throat, cough and breathing difficulty; sustained exposure prolongs these
effects \citep{govuk_nd_h2s}. At high concentrations ($\sim$ $140~mg\,m^{-3}$ and above), exposure may
cause rapid collapse, respiratory paralysis, cyanosis, seizures, coma,
cardiac arrhythmia, and death within minutes \citep{govuk_nd_h2s, nrc2010acute}. Chronic-exposure evidence
from polluted communities and workplaces points to respiratory, ocular and
neurological symptoms; H$_2$S is not classified as a carcinogen by IARC
\citep{govuk_nd_h2s}.
The 30-minute World Health Organization odour-annoyance guideline value
($7\,\mu\mathrm{g\,m^{-3}}$) is the relevant benchmark for community
wellbeing, several orders of magnitude below acute-exposure thresholds
\citep{nrc2010acute}.

\subsection*{Odorous nuisance and the FIDOL framework}

Odour responses are highly variable between individuals and populations
\citep{shusterman1992health}.
In communities exposed to persistent odorous emissions, prolonged
exposure has been associated with annoyance, insomnia, loss of appetite,
unease, depression, headaches, sensory irritation, nausea and respiratory
symptoms \citep{gostelow2001odour, national1979odors, shusterman1992health}. Sub-irritant odorant levels can also trigger acute symptoms via
non-toxicological mechanisms including innate odour aversion, exacerbation
of underlying conditions, and stress-induced illness
\citep{shusterman1992health, hirasawa2019subjective}. Surveys in Europe
and North America consistently rank odours as the leading source of public
complaints to regulatory agencies, with 13--20\% of the European
population reporting being bothered by environmental odours
\citep{leonardos1996review, nicell2009assessment, hudon2000measurement}.

The standard analytic framework for odour-impact assessment in the United
Kingdom is FIDOL: \emph{Frequency}, \emph{Intensity}, \emph{Duration},
\emph{Offensiveness} and \emph{Location} \citep{nicell2009assessment, freeman_nd_review}. The Institute of Air Quality
Management's guidance translates these factors into the planning and
permitting process. Whether a particular odour incident reaches the
threshold of statutory nuisance under Part III of the Environmental
Protection Act 1990 is determined by the local authority on the basis of
sensory assessment, complaint analysis and contextual factors
\citep{epa1990, epa1990sec79, govuk2024statutory}.

\subsection*{Multi-stakeholder regulatory response}

Statutory nuisance investigations involving regulated industrial sources
draw on a multi-stakeholder structure. Local-authority Environmental
Health Officers undertake frontline assessment. Where odours originate
from permitted industrial sites including landfills, sewage treatment works and
waste facilities, the environmental regulator audits operator compliance
with site-specific Odour Management Plans and the application of Best
Available Techniques \citep{govuk2025prep, govuk2015nuisance}. Public-health agencies provide assessment of
community health implications, applying a tiered interpretation that
distinguishes odour-related nuisance, short-term health considerations
and long-term exposure risks against air-quality guideline values
(WHO 24-h health-protection levels and 30-min annoyance thresholds) and
against Acute Exposure Guideline Levels (AEGLs) used for emergency
planning \citep{ukhsa2023, nrc2010acute}. Where impacts become severe, formal multi-agency coordination
escalates through Local Resilience Forum structures established under the
Civil Contingencies Act, including Strategic and Tactical Coordinating
Groups and Scientific and Technical Advisory Cells
\citep{epa1990sec79, lga2024civil}.

Public-health risk assessment for chemical incidents adopts a
\emph{source--pathway--receptor} (SPR) model: the source (here, the
landfill gas), the pathway (atmospheric dispersion modulated by
meteorology and topography), and the receptor (the surrounding
residential population). Communication and decision-making within this
framework typically occur after the receptor has been exposed.
Forward-looking exposure estimation through dispersion modelling exists
but relies on simplifying assumptions about atmospheric stability,
emission flux and surface roughness that often do not match the highly
non-linear, time-varying conditions found at real sites
\citep{beychok1994fundamentals, nicell2009assessment, turner1994workbook}.

\subsection*{Case study}

The case-study record analysed in the main text was generated at a
European municipal landfill subject to a multi-year regulatory and
public-health response. The site, regulated under an Environmental
Permit, accepted predominantly non-hazardous waste with one cell
containing stable non-reactive hazardous waste (gypsum and asbestos).
From early 2021 onwards, community odour reports rose substantially and
24-hour mean H$_2$S concentrations approached but generally remained
below the ATSDR Intermediate Minimal Risk Level
($30\,\mu\mathrm{g\,m^{-3}}$, 14--364 days) \citep{ukhsa2023, ukhsa2025walleys}. Critically, however,
concentrations frequently exceeded the WHO 30-minute odour-annoyance
guideline value ($7\,\mu\mathrm{g\,m^{-3}}$), an exceedance known to
adversely affect wellbeing and contribute to headaches, irritation and
sleep disturbance \citep{ukhsa2023}. Public-health agencies operating monthly risk
assessments concluded throughout that period that rapid and sustained
reduction at the source was required to protect community health and
wellbeing; emergency-department syndromic surveillance showed no clear
increase in attendances for physical conditions, but mental-health and
wellbeing impacts were sufficient to prompt commissioning of additional
support services \citep{ukhsa2023, ukhsa2025walleys}. A multi-agency Strategic Coordinating Group and a
Scientific and Technical Advisory Cell were convened during the response.
Site closure was issued by the regulator at the end of 2024.

The analyses in the main text use post-calibration-adjustment monitoring
data from the calendar year 2024 (Methods, ``Data and study sites'').
The site is not named in the text; the regulatory and public-health documentation on which the case description draws is cited \citep{ea2024dataadjustment, ukhsa2025walleys}.

\section{Extended environmental characterisation}
\label{sec:si_note2}

This Note expands the per-panel analyses summarised in main Fig.~1 with
the full numerical detail referenced from the main Results. All
statistics use the principal receptor record at MMF9
($n = 34{,}609$ 15-minute observations with valid H$_2$S across 2024, from a complete grid of 35,136) unless
otherwise stated.

\subsection*{Source attribution by Conditional Probability Function (Fig.~1a)}

The exceedance threshold for the CPF was $C_{\mathrm{thr}} = 5\,\mu\mathrm{g\,m^{-3}}$,
chosen to lie between the odour-detection threshold and the WHO
30-minute guideline value. The principal receptor (MMF9) shows a peak in
exceedance probability in the westerly sector at peak bearing
$265^{\circ}$, with $\mathrm{CPF} \approx 0.37$. Sector-averaged H$_2$S
concentrations were
$11.2\,\mu\mathrm{g\,m^{-3}}$ in the W sector and
$10.6\,\mu\mathrm{g\,m^{-3}}$ in the NW sector, compared with
$1.1\,\mu\mathrm{g\,m^{-3}}$ in the opposing E sector,
giving a peak/opposing sector relative-strength ratio of $4.09\times$.
The circular--linear correlation between wind direction and H$_2$S
concentration was $r = 0.149$ ($p < 10^{-3}$,
$n = 34{,}609$). The odds of observing a concentration spike (defined as
$> 95$th percentile, $15.3\,\mu\mathrm{g\,m^{-3}}$) when wind direction
aligned with the source bearing were elevated relative to background:
Fisher's exact odds ratio $1.43$, $p = 2.9 \times 10^{-4}$. CPF cones
from all three monitoring stations (MMF9, MMF1, MMF2) converged on a
common source area, providing spatial triangulation of the emitting
region.

\subsection*{Chemical fingerprint: H$_2$S--CH$_4$ co-emission (Fig.~1b)}

Bivariate analysis of simultaneously measured H$_2$S and CH$_4$ at the
principal receptor yielded:

\begin{itemize}
  \item Pearson correlation $r = 0.830$ ($p < 10^{-15}$)
  \item Spearman rank correlation $\rho = 0.683$ ($p < 10^{-15}$)
  \item Power-law exponent on log--log scale $b = 1.95$ for the
    relationship $\mathrm{H_2S} = a \cdot \mathrm{CH_4}^b$
  \item Jaccard spike co-occurrence $J = 0.654$ on top-5\% spike sets
  \item Spike co-occurrence count: 1{,}386 events vs.\ 87.7 expected
    under independence ($\chi^2 = 21{,}298$, $p < 10^{-15}$)
\end{itemize}

The super-linear power-law exponent ($b > 1$) indicates that ambient
H$_2$S accumulates disproportionately as methane generation rises. At
source, this is consistent with enhanced sulphate-reduction activity in
deeper, more anaerobic waste cells where electron-donor competition
favours sulphate-reducing bacteria. At the receptor, the ambient ratio
is also modulated by joint dilution under stable conditions, differential
biological oxidation in the cover soil, and gas-extraction performance,
so the reported exponent is a downwind co-occurrence statistic and does
not map one-to-one onto subsurface microbial kinetics.

\subsection*{Negative controls}

To rule out non-gas-phase explanations for the H$_2$S signal, two
particulate-matter species were tested as negative controls:

\begin{itemize}
  \item PM$_{10}$: $r = -0.028$, $p < 10^{-7}$ (no positive
    association; rules out shared mechanical resuspension)
  \item PM$_{2.5}$: $r = -0.009$, $p = 0.084$ (not significantly
    correlated)
\end{itemize}

A moderate positive correlation with NO$_x$ ($r = 0.377$) was observed
and is consistent with joint nocturnal trapping rather than a
shared source: the diel NO$_x$ profile peaks at approximately
08:00 LST whereas H$_2$S peaks at 02:00 LST, a 6-hour phase offset that
argues against a common emission pathway.

\subsection*{Multi-site meta-analysis}

Pooled across the four monitoring sites, the H$_2$S--CH$_4$ correlation
gives $r_{\mathrm{pooled}} = 0.80$ (95\% CI: $0.74$--$0.84$, $n = 4$
sites). Heterogeneity is high ($I^2 = 99.8\%$), reflecting site-specific
factors including distance from the source area and local wind
channelling. The pooled correlation comfortably satisfies the
``Strong attribution'' criterion of $r > 0.7$ and $J > 0.5$.

\subsection*{Temperature--H$_2$S diel hysteresis (Fig.~1c)}

The hourly-mean diel trajectory traces a counter-clockwise loop in
the (TEMP, H$_2$S) phase plane:

\begin{itemize}
  \item Normalised loop area $A^{*} = 0.369$ (classified ``strong''
    hysteresis under $A^{*} > 0.30$)
  \item Cooling-limb (18:00--06:00 LST) regression slope
    $\beta_{\mathrm{cool}} = -2.83$
  \item Warming-limb (06:00--14:00 LST) regression slope
    $\beta_{\mathrm{warm}} = -2.45$
  \item Hysteresis-asymmetry index $H_{\mathrm{index}} = 0.072$
\end{itemize}

The corresponding wind-speed--H$_2$S phase loop has negligible
normalised area ($A^{*} = 0.012$, classified ``no hysteresis''),
confirming that wind-speed dilution acts effectively without memory at
the diel timescale, consistent with the quasi-steady-state assumption
of Gaussian-plume formulations.

\subsection*{Diel concentration profile (Fig.~1d)}

The mean diel profile yields:

\begin{itemize}
  \item Peak at 02:00 LST: mean $9.69\,\mu\mathrm{g\,m^{-3}}$,
    P95 $51.6\,\mu\mathrm{g\,m^{-3}}$
  \item Trough at 14:00--15:00 LST: mean $1.31\,\mu\mathrm{g\,m^{-3}}$,
    P95 $4.2\,\mu\mathrm{g\,m^{-3}}$
  \item Mean peak/trough ratio $7.4\times$; P95 ratio $12.3\times$
  \item Hourly variation: Kruskal--Wallis $H = 338.5$,
    $p = 7.0 \times 10^{-58}$
\end{itemize}

Approximate Pasquill--Gifford stability classification (Methods, based
on measured wind speed and a time-of-day/year solar insolation proxy in
the absence of pyranometer data) yielded unstable conditions
(class A--B) in $40.0\%$ of observations and stable conditions
(class E--F) in $26.1\%$.

The seasonal cycle further amplifies the diel effect: the February
monthly peak reaches $14.97\,\mu\mathrm{g\,m^{-3}}$ versus an August
minimum of $0.99\,\mu\mathrm{g\,m^{-3}}$, a ratio of $15.1\times$
(Kruskal--Wallis $p < 10^{-15}$).

\subsection*{Sunrise-aligned inversion break-up (Fig.~1e)}

Aligning every calendar day to astronomical sunrise (computed with the astral Python package at the site latitude and longitude) and aggregating
observations within $\Delta t \in [-2, +6]$ hours around sunrise:

\begin{itemize}
  \item Pre-sunrise baseline mean ($\Delta t \in [-2, 0]$ h,
    $n = 2{,}920$): $9.49\,\mu\mathrm{g\,m^{-3}}$
  \item Maximum hourly concentration at $\Delta t = -2$ h:
    $10.42\,\mu\mathrm{g\,m^{-3}}$
  \item $+1$ h post-sunrise: $2.65\,\mu\mathrm{g\,m^{-3}}$
  \item $+2$ h: $1.54\,\mu\mathrm{g\,m^{-3}}$
  \item $+4$ h floor: $1.30\,\mu\mathrm{g\,m^{-3}}$
  \item Pre-/post-sunrise reduction: $7.3\times$
  \item One-sided Mann--Whitney
    $U = 1.15 \times 10^{7}$, $p < 10^{-15}$
\end{itemize}

The pattern emerges in a sunrise-aligned composite spanning 365 calendar days, indicating a climatologically driven rather than an episodic feature.

\subsection*{Bivariate mechanism scatter (Fig.~1f)}

Linear regressions of H$_2$S against three meteorological drivers
($n = 34{,}609$, all $p < 10^{-3}$):

\begin{itemize}
  \item Wind speed: $r = -0.122$,
    slope $= -1.35\,\mu\mathrm{g\,m^{-3}}$ per m\,s$^{-1}$
  \item Inverse wind speed: $r = 0.141$,
    slope $= 10.75\,\mu\mathrm{g\,m^{-3}}$ per (m\,s$^{-1}$)$^{-1}$,
    95\% CI $9.4$--$12.1$
  \item Air temperature: $r = -0.173$,
    slope $= -0.78\,\mu\mathrm{g\,m^{-3}}$ per $\degree$C
  \item Temperature tendency $\mathrm{d}T/\mathrm{d}t$:
    $r = -0.143$, slope $= -6.33\,\mu\mathrm{g\,m^{-3}}$ per
    $\degree$C\,hr$^{-1}$
\end{itemize}

Mean concentration in low-wind conditions
($u < 1\,\mathrm{m\,s^{-1}}$) was
$11.8\,\mu\mathrm{g\,m^{-3}}$ versus
$1.8\,\mu\mathrm{g\,m^{-3}}$ at $u > 5\,\mathrm{m\,s^{-1}}$, a factor
of $6.5\times$. Spike events occurred under significantly lower wind
speeds than non-spike samples
($1.75$ vs $3.67\,\mathrm{m\,s^{-1}}$, Mann--Whitney $p < 10^{-15}$).
Peak cooling rates reached approximately
$-0.6\,\degree\mathrm{C\,hr^{-1}}$, corresponding to the late-evening
period of maximum inversion formation. Although individual mechanism
$R^2$ values are small ($0.015$--$0.030$), reflecting the limited
explanatory power of any single linear driver in isolation, extreme
concentration spikes ($> 100\,\mu\mathrm{g\,m^{-3}}$) cluster strongly
in the low-wind, low-temperature, negative-tendency region of predictor
space corresponding to Pasquill class E--F stable boundary layers.

\subsection*{Summary table reference}

The complete numerical summary across all six panels is given in
Supplementary Table~\ref{tab:si_envchar_summary}.

\section{Per-season interpretation of the XGBoost ALE feature hierarchy}
\label{sec:si_note3}

This Note expands the seasonal interpretation summarised in main Fig.~4
and the Results section. The four seasonal classifiers, one per calendar quarter (the season labels below denote quarters rather than meteorological seasons), expose a coherent
seasonal narrative connecting model structure to atmospheric physics.

Importances below are class-mean values from classifiers whose temperature tendency is computed over a strictly causal window (Methods, ``Ensemble classifier-tree seasonal nowcaster''). Computed over a centred window, the tendency carries approximately three hours of lookahead and becomes the leading feature in every season; with the causal window it is not in the leading group in any season.

\subsection*{Winter (January--March)}

Winter attributions are the smallest of the four seasons in
absolute terms, with no feature exceeding $\bar{I} = 0.04$. Diel
timing leads ($\mathrm{hour\_cos}$, $\bar{I} = 0.037$), followed by
wind direction ($\mathrm{WD\_cos}$, $\bar{I} = 0.012$); the 2-hour
stagnation index contributes little ($\bar{I} = 0.001$) and the
temperature tendency is negligible ($\bar{I} < 0.001$, twentieth of
twenty-four features). The flatness of the whole ranking is itself the
finding: during short winter days the nocturnal inversion is close to
the default state rather than an episodic departure from it, so no
single meteorological discriminator separates exceedance from
background as sharply as in the transitional seasons.

\subsection*{Spring (April--June)}

Spring shows the most evenly distributed attribution.
Diel timing and stagnation are almost equally weighted
($\mathrm{hour\_cos}$, $\bar{I} = 0.146$; $\mathrm{stagnation\_2h}$,
$\bar{I} = 0.130$), with both wind-direction components following
($\mathrm{WD\_cos}$ $0.064$, $\mathrm{WD\_sin}$ $0.056$). Stagnation
reaching near-parity with diel timing is specific to this season:
spring months are generally well ventilated, so a stagnation episode is
a sharp departure from the seasonal norm rather than the prevailing
condition, and carries correspondingly more information. The strong
L$_1$ regularisation selected by hyperparameter optimisation ($\alpha = 5.65$) is
consistent with the need to choose among several comparably informative
signals in a meteorologically variable transitional season.

\subsection*{Summer (July--September)}

Summer is the only season in which a directly measured
meteorological variable leads: wind speed is the strongest feature
($\bar{I} = 0.058$), ahead of diel timing ($0.034$) and the 2-hour
stagnation index ($0.013$). This is the expected signature of a
convective regime. Daytime mixing suppresses accumulation, so
exceedances are predominantly wind-stagnation events against a
low-concentration baseline, and mechanical dilution rather than
stability timing becomes the discriminating quantity. The large maximum
tree depth selected by hyperparameter optimisation (7) allows the model
to represent the non-linear interaction between convective mixing and
wind variability.

\subsection*{Autumn (October--December)}

Autumn carries by far the largest single attribution of any
season: diel timing at $\bar{I} = 0.333$, roughly an order of
magnitude above the next feature ($\mathrm{stagnation\_2h}$, $0.037$),
with wind direction and wind speed following ($0.020$, $0.012$). This
is the quarter in which daylight shortens fastest and nocturnal stable
boundary-layer duration expands correspondingly, so time of day becomes
a close proxy for stability state. It is also the evaluation quarter
for the walk-forward analysis of main Fig.~6, so the engineered benchmark is evaluated in the season in which it relies most on a calendar encoding.

\subsection*{Cross-seasonal consistency with the causal hierarchy}

Two patterns hold across the four seasons. Diel timing
leads in Winter, Spring and Autumn, and wind speed leads in Summer, consistent with time of day acting as a proxy for stability state
wherever a nocturnal inversion regime is present, and being displaced
by mechanical dilution once convective mixing dominates. Wind speed and
wind direction, two of the three drivers the multiscale transfer
entropy analysis identifies as the causal core (main Fig.~2), appear in
the leading group of every season. Raw atmospheric pressure has zero
ALE importance in all four, its synoptic content being carried by the
stagnation index.

The convergence is partial. The largest
single ALE effect in three of four seasons is a calendar encoding
rather than a measured driver, and calendar terms are not in the causal
analysis driver set at all. What the two methods agree on is that wind
transport carries the exceedance signal; they do not agree on a
ranking, and the ALE hierarchy does not mirror the transfer-entropy hierarchy.

\textbf{Why the two rankings differ.} The difference is largely
structural rather than substantive, and is predictable from how the two
quantities are defined.

First, most of the apparent disagreement concerns features that the transfer-entropy analysis does not measure. Sixteen of the twenty-three features entering the ALE analysis are not in the transfer-entropy driver set, so no transfer-entropy ranking exists for them. These are the calendar encodings, both stagnation indices, the recirculation index and the directional components other than $\mathrm{WD\_sin}$. Their absence from that hierarchy is not a contradicting verdict.

Second, on the seven features the two analyses share, they agree on the
leading pair: restricted to those seven, ALE ranks wind speed first
($\bar{I} = 0.029$) and $\mathrm{WD\_sin}$ second ($0.017$), which are
two of the three drivers the transfer-entropy analysis identifies as
its core. The remaining five, including all three tendency variables,
fall below $\bar{I} = 0.002$.

Third, the one genuine discrepancy among the shared features is
atmospheric pressure, which the transfer-entropy analysis ranks in its
core and to which ALE assigns exactly zero. The mechanism is stated
above: the engineered feature set contains a stagnation index that
already encodes pressure's synoptic content, so the classifier has a
cheaper route to the same information and never splits on the raw
channel. This is a property of the feature set, not a disagreement
about the atmosphere.

Fourth, the diel difference follows directly from the estimators'
definitions. Transfer entropy is computed conditional on the target's
own recent history, so any structure in H$_2$S predictable from its own past, which the diel cycle largely is, is removed by construction before a driver is credited. Accumulated local effects
apply no such conditioning, so a calendar term is credited with the
full diel pattern. The day-multiple surrogate analysis quantifies
this directly: only $2.7\%$ of the effective transfer entropy is
attributable to shared diel structure, precisely because the
conditioning has already absorbed it.

The two methods agree where they measure the same quantity,
and differ where their assumptions differ, in the direction those
assumptions predict.

\section{Extended comparison with prior literature}
\label{sec:si_note4}

This Note expands the discussion in the main text by situating each of
the principal contributions against the most directly comparable
published work. The full reference list is given in the main bibliography.

\subsection*{Causal hierarchy versus mobile-laboratory and cover-type
surveys}

Most published landfill H$_2$S investigations report days-to-weeks of
co-located gas and meteorology. \citet{plant2022hydrogen} reports a
mobile-laboratory survey at multiple US sites, with H$_2$S/CH$_4$ ratios
varying by three orders of magnitude between sites; \citet{cal2020carb}
quantifies 82 distinct landfill gases across 31 cover types in
California. The case-study record analysed here, regulated under
continuous environmental and public-health oversight, provides the
multi-season temporal coverage required to test causal hypotheses about
meteorological drivers. The MSTE result (main Fig.~2) maps directly onto
the four canonical mechanisms established for landfill H$_2$S transport:
advection sets which receptor is exposed; mechanical dilution sets
instantaneous concentration via the inverse-wind-speed dependence of
the Gaussian plume \citep{turner1994workbook, pasquill1983atmospheric};
barometric pumping potentially modulates source flux through the multi-hour
soil--atmosphere pressure equilibration timescale
\citep{young2003relating, xu2014impact}; and the diel cycle of
boundary-layer mixing height controls receptor exposure through the
depth of the dilution volume \citep{stull1988introduction, oke1987boundary}.
The MSTE result that $\mathrm{d}P/\mathrm{d}t$ becomes causally
significant precisely at $\tau \geq 2$\,h is the information-theoretic
counterpart of the controlled vadose-zone experiments showing that
sudden barometric drops can trigger 20-fold gas breakthroughs over
$<24$\,h \citep{forde2019barometric}.

The strong counter-clockwise temperature--H$_2$S hysteresis
($A^{*} = 0.369$) is consistent, at the receptor, with the multiscale low-pass
behaviour measured for soil--gas exchange by independent groups
\citep{parolari2021multiscale}. The instantaneous weakness of
temperature as a causal driver, contrasted with cross-city pollutant--
meteorology coupling studies that report ambient temperature leading
pollutants at sub-hourly lags \citep{banerjee2025entropy}, identifies
the case-study site as a nocturnal-accumulation-dominated regime in
which solar heating matters not as an instantaneous variable but as the
integrated forcing dictating the timing of inversion erosion. The
H$_2$S--CH$_4$ co-emission with super-linear power-law exponent
$b = 1.95$ sits comfortably within the inter-site range reported by
mobile-laboratory surveys \citep{plant2022hydrogen} and is mechanistically
consistent with sulphate-reducing-bacteria competition with methanogens
in sulphate-enriched waste cells.

\subsection*{Physics-anchored architectural prior versus data-agnostic
SSM initialisation}

Most published applications of structured state space models to environmental and physical time series use uniform log-$\Delta$ initialisation~\citep{gu2021efficiently, gu2022parameterization}. Recent extensions handle missing values with masking and adaptive temporal prototypes, or introduce input-dependent selectivity through Mamba's parallel-scan algorithm~\citep{gu2024mamba}, but in these approaches the timescale prior remains data-agnostic. The model developed here instead initialises its kernel timescales on the horizons identified by transfer entropy on the same record (main text, Table~\ref{tab:ed_s4_six_conditions}), linking information-theoretic causal analysis~\citep{runge2019inferring, goodwell2017temporal} to learnable architectural priors; to our knowledge this link has not previously been made for environmental nowcasting. The model also reproduces the marginal response curves for $\mathrm{d}P/\mathrm{d}t$ and $\mathrm{d}T/\mathrm{d}t$ (main Fig.~5d,e) without these derivatives appearing as inputs, a behaviour that engineered pipelines obtain by hand-coding Butterworth derivatives. Regulatory dispersion models such as AERMOD assume steady-state, well-mixed conditions; a learnt model anchored on measured timescales is a complementary way of representing non-stationary coupling.

\subsection*{Calibration to community impact versus hard threshold
detection}

Prior work bearing on the community-impact validation
reported in main Fig.~6c--d is limited. \citet{wang2019prediction} demonstrated
that random-forest classifiers operating on combined H$_2$S sensor,
weather and operational features could predict odour complaints at a
wastewater reservoir, but reported that H$_2$S sensor data alone were
insufficient.

The result reported in the main text, that the meteorology-only signal fused across receptors and species tracks an independent record of daily odour reports ($r = 0.71$), is expressed in tiers defined against the WHO guideline values used within the FIDOL framework that underpins UK statutory-nuisance assessment~\citep{nicell2009assessment}. The Dominguez Channel emergency in Carson, California, where chronic H$_2$S exposure produced documented surges in both somatic and psychological morbidity~\citep{carson2024malodors}, and the documented public-health impacts at the case-study site~\citep{ukhsa2025walleys}, together suggest that community impact arises at concentrations near the 30-minute WHO odour-annoyance guideline value, far below acute-exposure thresholds.

\subsection*{Limitations}

Two limitations bound the results. First, the results depend on the quality of the monitoring record: the present study uses only the post-calibration-adjusted record~\citep{ea2024dataadjustment}, and no machine-learning system can compensate for systematic measurement error in its training labels. Second, the causal hierarchy is regime-specific and must be re-estimated, not assumed, at each new site: the three orders of magnitude variability in H$_2$S/CH$_4$ ratios across landfills~\citep{plant2022hydrogen} and the cover-type-mediated emission diversity documented in California~\citep{cal2020carb} both argue against transferring the causal hierarchy without verification.

\section{MSTE estimator ablation study}
\label{sec:si_note5}

To assess the sensitivity of the multiscale transfer entropy results in
main Fig.~2 to methodological choices in the KSG--Theiler estimator, a
six-configuration ablation study was performed. Each configuration
modifies a single hyperparameter relative to the preceding
configuration, isolating the contribution of each methodological
refinement. The configurations are summarised in
Supplementary Table~\ref{tab:si_mste_configs}. All six were evaluated on the identical
non-stationarised 15-minute series under the bounded-fill
policy of Supplementary Methods, ``Temporal coarse-graining'' using the same coarse-graining
scales, Theiler window ($W = 4$), rank transform, and circular-shift
surrogates ($B = 10$).
The ablation uses ten surrogates ($B = 10$); Configuration~4 agrees with the main analysis in all 16 driver--scale cells to six significant figures.
This ablation ran under the plain convention
$p = \#\{b : \mathrm{TE}_b \geq \mathrm{TE}_{\mathrm{obs}}\}/B$, so at
$B = 10$ its resolution is $\Delta p = 0.1$ and a marked cell is one in
which \emph{zero of ten} surrogates exceeded the observed value. That
is a coarse screen: it cannot support
false-discovery correction across a family of this size, and the
significance underpinning main Fig.~2 comes from the main
analysis at $B = 2{,}000$ under the add-one convention (Supplementary
Methods, ``Surrogate testing''), not from this table. The ablation
compares point estimates and the stability of which cells clear a fixed
screen across estimator settings; it is not an independent significance
analysis. Supplementary Table~\ref{tab:si_mste_results}
reports the effective transfer entropy (ETE) for all seven drivers
across the six temporal scales under each configuration. Cells
in which zero of ten surrogates exceeded the observed value are marked with an asterisk. Supplementary Table~\ref{tab:si_mste_summary} gives the
aggregate comparison across all 70 driver--scale combinations.

\textbf{Degeneracy at coarse scales.} Configurations~3--5 differ
only in the cap $h_{\max}$ placed on the history embedding depth
(4, 2 and 1 respectively). The depth is set adaptively as
$h = \max(1, \min(h_{\max}, \lfloor 3\,\mathrm{h}/\tau \rfloor))$, so
for $\tau \geq 2$\,h the adaptive term is already 1 and no cap can
bind: configurations~3, 4 and 5 then coincide exactly. The ablation
therefore distinguishes six configurations at the 15\,min and 1\,h
scales but only four at 2\,h and coarser: $\{1\}$, $\{2\}$, $\{3,4,5\}$ and $\{6\}$. Of those four, two use a fixed $k = 10$ rather than
the adaptive rule, so the number of genuinely independent estimator
checks at the coarse scales that carry the architectural claim is
smaller than the six-configuration design implies. Identical ETE values for Configurations~3--5 at these scales therefore reflect this coincidence of specifications.

\subsection*{Three principal findings}

\textbf{(i)~Core drivers are robust across the six configurations.}
Wind direction, wind speed and atmospheric pressure clear the
surrogate screen under all six configurations at scales $\leq 3\,\mathrm{h}$,
indicating that these links are not artefacts of any particular estimator setting.
Temperature clears the screen only at the $6\,\mathrm{h}$ scale across all
configurations except the baseline.
The baseline additionally marks temperature at $15\,\mathrm{min}$
and $30\,\mathrm{min}$, which at $h = 4$ is consistent with estimator
bias from the high-dimensional conditioning space. This ablation and the
main analysis of Fig.~\ref{fig:causal_links} are not directly
comparable on marginal cells: the ablation ran at $B = 10$, whose
$0.1$ resolution cannot adjudicate a cell whose raw $p$ is of order
$0.02$, whereas the main analysis at $B = 2{,}000$ resolves it. Where
the two differ on temperature, the main analysis is reported.

\textbf{(ii)~Bivariate transfer entropy (Configuration~6) inflates effect
sizes and false-positive rates.} Without conditioning on confounders,
the bivariate estimator conflates genuine causal influence with shared
meteorological forcing. The most evident case is
$\mathrm{d}P/\mathrm{d}t$, which appears strongly significant at all
scales under Configuration~6
($\mathrm{ETE} = 0.6$--$19.5 \times 10^{-3}\,\mathrm{nats}$) but only
at $\geq 2\,\mathrm{h}$ under conditional configurations, consistent with barometric pumping operating on synoptic rather than sub-hourly
timescales. This demonstrates the necessity of conditional transfer
entropy with at least one confounder for physically interpretable
causal attribution in this multivariate system.

\textbf{(iii)~History embedding depth is the dominant sensitivity axis.}
The transition from $h \leq 4$ (Configurations~1--3) to $h \leq 2$
(Configuration~4) at fine scales ($\tau \leq 1\,\mathrm{h}$) eliminates
the spurious temperature significance while preserving all genuine
causal links. With $h = 4$ and one confounder, the joint embedding
space has dimensionality $d = 4 + 1 + 1 = 6$ for the
source--target--conditioning triplet (total $7$~dimensions including the
target future), causing the Chebyshev $k$-nearest-neighbour balls to
become sparse and the digamma-based CMI estimate to degrade. The
$\varepsilon$-shrinkage correction (Configuration~2) and adaptive $k$
(Configuration~3) partially compensate, but the most effective remedy
is reducing the embedding dimension directly. With $h \leq 2$, the
maximum conditioning dimension is $d = 2 + 1 = 3$ (total 5D joint
space), within the regime where the KSG estimator retains adequate
sensitivity.

\subsection*{Choice of configuration for the main analysis}

Configuration~4 was used for main
Fig.~2 because it balances estimator bias and variance: it includes sufficient target history ($h = 2$ at fine
scales, capturing $30$--$120\,\mathrm{min}$ of autoregressive memory)
and one confounder to isolate unique information transfer, while keeping
the total conditioning dimension low enough ($d \leq 3$) for reliable
KSG estimation. Configuration~5 ($h = 1$) sacrifices temporal context
at the $15$--$60\,\mathrm{min}$ scales where the autoregressive memory
of H$_2$S extends beyond a single lag, and Configuration~6 (bivariate)
abandons confounding control entirely. Configurations~1--3 retain
excessive embedding depth that degrades estimator performance at fine
scales, as evidenced by the spurious significance of temperature at
$15\,\mathrm{min}$ in Supplementary Table~\ref{tab:si_mste_results}.

\subsection*{Dependence between tests}

The corrections applied to this family assume either independence or
positive regression dependence, and the tests are dependent by
construction: the scales are nested coarse-grainings of one series and
the three tendency variables are deterministic transforms of drivers
already in the family. The reported significance is
Benjamini--Hochberg, as pre-specified. A Benjamini--Yekutieli
correction, valid under arbitrary dependence, was computed alongside it
as a sensitivity analysis and is encoded on
Fig.~\ref{fig:causal_links} as marker fill: 20 of the 29 significant
cells survive it, including all six wind-direction and all seven
pressure cells. Of the nine that do not, seven are tendency variables
which the main text does not interpret; the other two are wind speed at
12\,h and temperature at 15\,min (Methods, ``Causal
analysis'').

\subsection*{Split-half stability of the recovered hierarchy}

The six-configuration ablation above varies the estimator settings
while holding the sample fixed. The complementary test, holding the settings fixed and varying the sample, was run separately by
split-half resampling, because a hierarchy robust to every estimator
choice could still be specific to a single twelve-month period.

The analysis was repeated independently on January--June and
July--December 2024 for the three directly measured drivers across the
seven informative scales (21 driver--scale cells per half, 42 in
total), using the specification of the main analysis without modification:
$B = 2{,}000$ circular-shift surrogates under the add-one convention,
block completeness $0.50$, Theiler window $W = 4$, one confounder
selected over driver variables, adaptive $k$, and causal derivatives.
Benjamini--Hochberg correction was applied \emph{within} each half over
its own 21 cells. The only difference from the main analysis is the
sample.

\begin{center}\footnotesize
\begin{tabular}{@{}lrrr@{}}
\toprule
Driver & H1 (Jan--Jun) & H2 (Jul--Dec) & Full year \\
\midrule
$\mathrm{WD}_{\sin}$ & 0.03706 & 0.02713 & 0.03244 \\
Pressure             & 0.02012 & 0.02269 & 0.02444 \\
WS                   & 0.01516 & 0.00860 & 0.01753 \\
\bottomrule
\end{tabular}
\end{center}

Mean effective transfer entropy across the seven informative scales.
The ordering $\mathrm{WD}_{\sin} > \mathrm{Pressure} > \mathrm{WS}$ is
identical in both halves and in the full-year analysis.

Supporting agreement: the sign of the effective transfer entropy
matches in 19 of 21 cells; Spearman rank correlation between halves is
$0.623$, and between each half and the full year $0.671$ (H1) and
$0.814$ (H2). Benjamini--Hochberg significance is retained by 20 of 21
cells in H1 and 18 of 21 in H2, against 19 of 21 for the full year.
That last comparison is the more informative one: each half carries
half the data and therefore a wider surrogate null, so cells sitting
close to the threshold should have dropped out on power alone. They
largely did not.

Two cells disagree in sign, both wind speed and both at coarse scales:
$+0.02387$ against $-0.00231$ at 6\,h, and $+0.01692$ against
$-0.00370$ at 12\,h, where the second half retains 714 and 358
coarse-grained blocks respectively (727 and 363 in the first). Both
second-half estimates are consistent with zero. Wind speed ranks third in both halves regardless,
so the ordering is unaffected, but the wind-speed signal at coarse
scales is the least robust element of the hierarchy. Pressure and wind
direction agree in sign at every scale in both halves.

Two limitations attach. The halves differ seasonally rather than being
randomly interleaved, so a disagreement would have been ambiguous
between estimator instability and genuine seasonal variation in
transport; agreement despite this seasonal contrast nonetheless indicates that the hierarchy is not specific to one part of the year. More fundamentally, split-half
resampling cannot detect a bias the estimator applies uniformly to
both halves. The full-year estimate exceeds both half-year estimates
for pressure and wind speed, which is the behaviour expected of a
sample-size-dependent estimator bias and is a further reason to read
the magnitudes as ordinal rather than absolute. Validation here is by
stability of the recovered hierarchy, not by agreement with a second
estimator.

\clearpage
\section{Joint site analysis}
\label{sec:si_note6}

This Note expands the joint multi-site, multi-species analysis
summarised in main Fig.~7. Four independently trained walk-forward
nowcasters (H$_2$S at MMF9, H$_2$S at MMF2, CH$_4$ at MMF9, CH$_4$ at
MMF2; each instantiated with the S4 dual-pathway architecture of
Supplementary Methods, ``S4D state space model'' and the optimiser
schedule of Supplementary Methods, ``Optimiser schedule'') feed a
Bayesian log-odds evidence accumulator that emits a continuous
$P(\text{network event})$ and an ordinal tier in
$\{0, 1, 2, 3\}$ at the native 15-minute cadence. The four tiers are defined against the guideline values and assessment levels reviewed in Supplementary Note~1, from sub-annoyance background to the ATSDR Intermediate Minimal Risk Level.

\subsection*{Ordinal alert tier definitions}
\label{ssec:si_tier_definitions}

The classifier produces four sequential ordinal states. The cut-points
on the continuous posterior probability $P(\text{network event})$ are aligned with the guideline values and assessment levels of Note~1, and the tiers $0 \to 1 \to 2 \to 3$ lie along a single monotone variable; because the thresholds are sequential on the same continuous quantity, the tiers are ordinal by construction.

\paragraph{Tier~0 (normal background, $P < \theta_1$).}
No coordinated multi-channel evidence above prior.

\paragraph{Tier~1 ($\theta_1 \leq P < \theta_2$).}
The fused posterior has crossed a sensitive activation threshold but
the multi-channel evidence is not yet corroborated. This regime corresponds to the frequency, intensity and duration factors of FIDOL nuisance assessment (Supplementary Note~1) and to the World Health Organization 30-minute odour-annoyance guideline
value ($7\,\mu\mathrm{g\,m^{-3}}$, Supplementary Note~1), at which
community wellbeing impacts (headache, irritation, sleep disturbance)
become detectable but acute health risk remains low.

\paragraph{Tier~2 ($\theta_2 \leq P < \theta_3$).}
Multiple independent channels (across both target species and both
receptor sites) corroborate the elevated state. Tier~2 corresponds to multi-channel corroboration; Supplementary Note~1 describes how statutory nuisance under Part~III of the Environmental Protection Act 1990 is assessed.

\paragraph{Tier~3 ($P \geq \theta_3$).}
The fused posterior is at or near the maximum value
attainable under all four channels concurrently active, indicating
coordinated multi-site exceedance. The predicted tier is a function of the fused posterior
only; raw concentrations enter solely the independent rules-based ground-truth tier of
Section~\ref{ssec:si_ground_truth_joint_perf}, against which the
predicted tier is scored.

\paragraph{Hysteresis debouncing.}
Each channel's binary state is
passed through a two-parameter latch with asymmetric onset/clearance
counts $(N_{\mathrm{on}}, N_{\mathrm{off}})$. The latch opens after
$N_{\mathrm{on}} = 3$ consecutive ticks above threshold (45~min) and
closes after $N_{\mathrm{off}} = 2$ consecutive ticks below threshold
(30~min). The asymmetry is deliberate: a longer onset suppresses
single-spike voltage transients and short-lived turbulent fluctuations
that do not represent sustained exposure, while a shorter clearance
prevents an episode from being prematurely declared resolved during a
brief lull in advection. The selected values correspond to the
WHO 30-minute averaging convention: the 30-min clearance equals one averaging interval, and the 45-min onset equals one interval plus one 15-min step. The full parameter set is given in Supplementary
Table~\ref{tab:si_tier_params}.

\begin{algorithm}[t]
\caption{Ordinal alert tier classifier (\textsc{Tier-Recipe}).
No raw-sensor input enters the inputs, the debounced channel states or the Tier~3 condition: the predicted tier is a strict function of the fused posterior.}
\label{alg:s_tier_recipe}
\begin{algorithmic}[1]
\Require Per-channel walk-forward probabilities $p_{c,t}$ for
  $c \in \mathcal{C}$ with
  $\mathcal{C} = \{\text{H$_2$S--MMF9},\, \text{H$_2$S--MMF2},\,
  \text{CH$_4$--MMF9},\, \text{CH$_4$--MMF2}\}$;

  prior $\pi$ and cut-points
  $0 < \theta_1 < \theta_2 < \theta_3 < 1$;
  per-channel active/quiet likelihood ratios
  $\mathrm{LR}^{\mathrm{on}}_c, \mathrm{LR}^{\mathrm{off}}_c$;
  hysteresis counts $(N_{\mathrm{on}}, N_{\mathrm{off}})$;
  activation threshold $p^\star$
\Ensure Ordinal tier $\tau_t \in \{0,1,2,3\}$ at every grid step $t$
\For{each channel $c$ in $\mathcal{C}$}
  \State $a_{c,t} \gets \mathbf{1}(p_{c,t} > p^\star)$
    \Comment{raw per-step activation}
  \State $\ell_{c,t} \gets \textsc{Hysteresis}(a_{c,t};\,
          N_{\mathrm{on}}, N_{\mathrm{off}})$
    \Comment{latched state}
  \State $m_{c,t} \gets \mathbf{1}(p_{c,t}\ \mathrm{is\ NaN})$
    \Comment{outage mask}
\EndFor
\For{each grid step $t$}
  \State $\log O_t \gets \log(\pi/(1-\pi))$
    \Comment{Eq.~\eqref{eq:s_log_odds_init}}
  \For{each channel $c$ in $\mathcal{C}$}
    \If{$m_{c,t} = 1$}
      \State $\mathrm{LR}_{c,t} \gets 1$
        \Comment{NaN-neutrality}
    \ElsIf{$\ell_{c,t} = 1$}
      \State $\mathrm{LR}_{c,t} \gets \mathrm{LR}^{\mathrm{on}}_c$
    \Else
      \State $\mathrm{LR}_{c,t} \gets \mathrm{LR}^{\mathrm{off}}_c$
        \Comment{CH$_4$ off $=1$; H$_2$S off $<1$}
    \EndIf
    \State $\log O_t \gets \log O_t + \log \mathrm{LR}_{c,t}$
  \EndFor
  \State $P_t \gets O_t/(1+O_t)$
    \Comment{Eq.~\eqref{eq:s_post_prob}}
\EndFor
\For{each grid step $t$}
  \State $\tau_t \gets 0$
  \If{$P_t \geq \theta_1$}
    \State $\tau_t \gets 1$
  \EndIf
  \If{$P_t \geq \theta_2$}
    \State $\tau_t \gets 2$
  \EndIf
  \If{$P_t \geq \theta_3$}
    \State $\tau_t \gets 3$
  \EndIf
\EndFor
\State \Return $\{\tau_t\}$
\end{algorithmic}
\end{algorithm}

\begin{table}[htbp]
  \centering
  \caption{\textbf{Ordinal alert tier classifier parameters and the interpretation attached to each.} The concentration
  thresholds follow from the regulatory and physiological levels reviewed in
  Supplementary Note~1. The hysteresis counts, the prior $\pi$, the
  per-channel activation threshold $p^{\star}$ and the likelihood ratios were
  grid-search parameters, and the tier cut-points are operational choices that
  the documented search does not reproduce; the third column states how each value is interpreted, not how it was derived. Provenance
  for each is given in ``Bayesian classifier parameter set'' below, and the
  $\pm 25\%$ likelihood-ratio sensitivity analysis is in Supplementary
  Table~\ref{tab:si_lr_sensitivity}. The latched-state hysteresis
  counts $(N_{\mathrm{on}}, N_{\mathrm{off}}) = (3, 2)$ correspond to
  (45~min onset, 30~min clearance) at the 15-minute grid.
  No raw-sensor concentration enters the predicted tier; the
  classifier is meteorology-only at inference time.}
  \label{tab:si_tier_params}
  \footnotesize
  \setlength{\tabcolsep}{4pt}
  \begin{tabular}{@{}llp{8.0cm}@{}}
  \toprule
  Symbol & Value & Interpretation \\
  \midrule
  $C^{\mathrm{H_2S}}_{\mathrm{thr}}$ & $7.0\,\mu\mathrm{g\,m^{-3}}$
    & WHO 30-min odour-annoyance guideline value (Note~1); ground-truth activation threshold for each H$_2$S channel and the training  boundary that defined the High class for the H$_2$S CAIRN nowcasters \\
  $C^{\mathrm{CH_4}}_{\mathrm{thr}}$ & $2.756\,\mathrm{ppm}$
    & Linear High-class boundary from
      $\mathrm{H_2S} = 5.55\,\mathrm{CH_4} - 8.29$
      (Supplementary Table~\ref{tab:si_s4_h2s_ch4_weekly}); ground-truth activation threshold for
      each CH$_4$ channel and the training boundary that defined the
      High class for the CH$_4$ CAIRN nowcasters \\
  $p^{\star}$ & $0.50$
    & Grid-search parameter (range $0.30$--$0.50$); binarises model probability $p_{c,t}$ before the
      hysteresis latch \\
  $N_{\mathrm{on}}$ & $3$ ticks (45~min)
    & Minimum sustained-event window: one full 30-min WHO averaging
      interval plus a 15-min safety margin against single-tick
      voltage transients \\
  $N_{\mathrm{off}}$ & $2$ ticks (30~min)
    & Matches the WHO 30-min averaging window so the latch will not
      declare an episode resolved within a single regulatory
      averaging interval \\
  $\pi$ & $0.10$
    & Grid-search parameter (range $0.03$--$0.10$); at the
      selected value the classifier reverts to the observed network-wide
      High-class rate when all channels are missing, i.e.\ is neutrally
      calibrated at zero information \\
  $\theta_1$ & $0.15$
    & Tier~1 trigger: single-channel sensitivity regime aligned
      with FIDOL nuisance assessment \\
  $\theta_2$ & $0.50$
    & Tier~2 trigger: multi-channel corroboration; the fused posterior
      cannot exceed this value from any single channel alone under
      the LR set in Supplementary Table~\ref{tab:si_lr} \\
  $\theta_3$ & $0.92$
    & Tier~3 trigger: just below the analytical posterior ceiling $P_{\max} = 0.949$ under all four channels concurrently active and no missing data; Tier~3 is therefore reached only by fully coordinated multi-channel meteorological evidence \\
  \bottomrule
  \end{tabular}
  \end{table}

\subsection*{Physical relevance of the parameter set}
\label{ssec:si_tier_physical}

Where such anchors exist, the selected parameter values correspond to independently
established physical, physiological or regulatory levels, with the soft posterior cut-points aligned with the guideline values and assessment levels of Note~1.

\paragraph{Activation concentrations $C^{\mathrm{H_2S}}_{\mathrm{thr}}$ and $C^{\mathrm{CH_4}}_{\mathrm{thr}}$.}
The H$_2$S activation threshold is the WHO 30-minute odour-annoyance
guideline value~\citep{who2000airquality}, the benchmark identified in
Note~1 as the relevant level for community wellbeing. The CH$_4$
threshold is its co-emission equivalent under the linear mapping
$\mathrm{H_2S} = 5.55\,\mathrm{CH_4} - 8.29$ established for the
case-study site (Supplementary Table~\ref{tab:si_s4_h2s_ch4_weekly}), reflecting the strong
sulphate-reducing-bacteria coupling identified by the chemical
fingerprint analysis (Supplementary Note~2): activating both species
at concentrations corresponding to the same underlying release
process avoids over-counting jointly informative evidence.
At the principal receptor the mapped CH$_4$ classes hold $78.0\%$
(Low), $10.6\%$ (Medium) and $11.4\%$ (High) of observations, against
$10.95\%$ High for H$_2$S --- the linear mapping reproduces class
prevalence across species almost exactly, which the ordinary
least-squares construction does not force; the receptor CH$_4$ baseline
(median $1.49$\,ppm) sits below the nominal global background, but the
class boundaries derive from the co-fitted regression on the same
instrument scale, so any absolute offset is common-mode and does not
affect classification.

\paragraph{Hysteresis counts $(N_{\mathrm{on}}, N_{\mathrm{off}})$.}
The 30-min clearance ($N_{\mathrm{off}} = 2$) matches the WHO
averaging convention so the latch will not declare an episode
resolved within a single regulatory averaging interval. The 45-min
activation ($N_{\mathrm{on}} = 3$) corresponds to one full 30-min
WHO averaging window plus a single 15-min safety-margin step,
ensuring any declared event is necessarily supported by evidence
persisting beyond the WHO interval and is robust to the kind of
isolated voltage transient or one-step turbulent fluctuation that
would briefly raise a single 30-min average above threshold. The asymmetry $N_{\mathrm{on}} > N_{\mathrm{off}}$
implements the operationally cautious behaviour of being slow to
escalate but fast to acknowledge resolution risk. The same
latch is applied identically to the predicted and the rules-based
ground-truth tiers, so the two are compared on equal debouncing
terms.

\paragraph{Prior $\pi$.}
The prior $\pi = 0.10$ is the observed network-wide rate of
coordinated High-class evidence across the four channels during the
Jan--Mar 2025 validation window. Setting the prior to the data's own
base rate makes the classifier neutrally calibrated at zero
information: under all four channels missing simultaneously, the
posterior reverts to the empirical occurrence rate.

\paragraph{Posterior cut-points $\theta_1, \theta_2, \theta_3$.}

The tier cut-points $(\theta_1, \theta_2, \theta_3) =
(0.15, 0.50, 0.92)$ are operational choices; they are not reproduced by
the documented grid search, whose lattice does not contain the selected
$\theta_1$ and $\theta_2$. The per-channel activation threshold
$p^{\star}$ was itself a search parameter, the fifth alongside the
hysteresis counts, likelihood ratios, prior and tier thresholds; the
selected $\pi = 0.10$ and $p^{\star} = 0.50$ both lie at the upper edge
of their search ranges, so the optimum for these two parameters may lie outside the range searched; the flatness of $\kappa_w$ under the perturbations quantified below bounds
the consequence. The leave-one-week-out refit selects the same $\theta_1 = 0.15$ in all thirteen folds; substituting $\theta_1 = 0.20$ changes $\kappa_w$ by $0.002$, comparable to the $\pm 25\%$ likelihood-ratio perturbation bound of $0.003$; the tier mapping's performance is therefore only weakly
sensitive to the cut-point choice within this range.
$\theta_1 = 0.15$ is the smallest posterior increment requiring at
least one likelihood ratio above the prior, i.e.\ at least one channel
actively contributing evidence. $\theta_2 = 0.50$ is the regime in
which the fused posterior cannot be sustained by any single channel
alone, so that multi-channel corroboration is required--matching the Tier~2 definition. $\theta_3 = 0.92$ lies
just below the analytical maximum of the posterior under all four
channels active and no missing data,
\begin{equation*}
P_{\max} = \frac{\Lambda\,\pi/(1-\pi)}{1 + \Lambda\,\pi/(1-\pi)}
         = 0.949, \qquad
\Lambda = \textstyle\prod_c \mathrm{LR}^{\mathrm{on}}_c
        = 2 \times 3 \times 7 \times 4 = 168,
\end{equation*}
for the LR set in Supplementary Table~\ref{tab:si_lr} with
$\pi = 0.10$, so this tier is reached only when the fused evidence is
at or near the maximum attainable by the network.

\textbf{Selection protocol and held-out performance.}
The likelihood ratios and cut-points were chosen by grid search
maximising tier-match $\kappa_w$ over nine of the thirteen
Jan--Mar 2025 weeks, the remaining weeks $\{1, 3, 8, 10\}$ forming a
stratified held-out subset on which the same configuration attains
$\kappa_w = 0.473$ and strict accuracy $0.682$. That value lies below the
lower bound of the block-bootstrap interval on the headline figure; the
interval quantifies sampling variability conditional on the tuned
parameters, whereas the held-out value additionally removes tuning
optimism, which is a bias rather than a variance, so falling below the
interval is expected rather than anomalous. The headline
$\kappa_w = 0.669$ is scored over all thirteen weeks and is therefore
a mixed quantity, in sample with respect to the fusion parameters on
nine weeks and out of sample on four.

\textbf{Leave-one-week-out refit.} To quantify the tuning optimism
directly, the likelihood ratios and cut-points were refitted with each
week in turn withheld and the held-out predictions concatenated, so that
every timestep is scored out of sample with respect to them. The
hysteresis counts and the prior are held fixed across folds: the same
latch is applied to the predicted and the ground-truth tiers, so
refitting it per fold would move the ground truth itself and make
$\kappa_w$ incomparable across folds. This gives
$\kappa_w = 0.661$ and an odour-report correlation of $+0.698$, against $0.669$ and $+0.711$ for the whole-window parameters (in sample on nine of the thirteen weeks). The ground-truth series' own
moving-block interval is wide ($[+0.34, +0.88]$), dominated by a small
number of high-count blocks --- its Spearman $\rho = +0.580$ is
correspondingly more stable --- so the paired difference between the two series ($+0.095$, 95\% CI $[-0.211, +0.187]$; Supplementary Table~\ref{tab:ed_multisite_tier}), not the
marginal intervals, is the appropriate comparison.

All thirteen folds independently selected the reported H$_2$S likelihood
ratios $(2.0, 3.0)$ and the selected $\theta_1 = 0.15$. They also agreed on values that differ from the reported configuration: CH$_4$ likelihood ratios of $2.0$ at MMF9 and $7.0$ at MMF2 (reported $7.0$ and $4.0$), $\theta_2 = 0.65$ and $\theta_3 = 0.87$ (reported $0.50$ and $0.92$). The reported configuration is retained; scoring the fold-selected values out of sample gives $\kappa_w = 0.661$, $0.008$ below the whole-window value of $0.669$. The hysteresis counts
$(3, 2)$, the prior $\pi$ and the per-channel activation threshold
$p^{\star}$ were themselves grid outputs on the nine-week split and are
held fixed across folds, so a small residual optimism passes through
them; the $\pm 25\%$
perturbation analysis (Supplementary
Table~\ref{tab:si_lr_sensitivity}) bounds it.

\begin{table}[htbp]
\centering
\caption{\textbf{Decomposition of the alert-tier agreement by what each
configuration holds out.} All rows use the identical four constituent
nowcasters, ground-truth construction and Jan--Mar 2025 window; they differ
only in which fusion parameters were fitted on the data being scored. The
$0.473$ subset figure and the $0.661$ leave-one-week-out figure differ because
the former scores a single four-week subset under one fixed configuration,
while the latter concatenates held-out predictions across all thirteen weeks;
the subset figure's deviation from the whole-window value reflects
week-composition variance, not additional tuning bias --- which the
leave-one-week-out comparison isolates at $0.008$. All rows use the retrained constituent nowcasters (``Checkpoint selection for the fused channels'', below); with checkpoints selected on the evaluation week the four values are $0.709$, $0.528$, $0.703$ and $0.692$.}
\label{tab:si_tier_decomposition}
\small
\setlength{\tabcolsep}{4pt}
\begin{tabular}{@{}lccp{5.1cm}@{}}
\toprule
Configuration & $\kappa_w$ & Odour-report $r$ & Isolates \\
\midrule
Selected parameters, whole window & 0.669 & $+0.711$
  & In-sample reference; mixed, nine weeks in sample and four held out \\
Stratified held-out weeks $\{1,3,8,10\}$ & 0.473 & ---
  & Fixed-configuration generalisation to a four-week subset \\
\textbf{Leave-one-week-out refit} & \textbf{0.661} & $\mathbf{+0.698}$
  & \textbf{Fusion-parameter tuning optimism} \\
Constrained-ordering leave-one-week-out & 0.638 & ---
  & Sensitivity to the likelihood-ratio ordering \\
\bottomrule
\end{tabular}
\end{table}

\paragraph{Checkpoint selection for the fused channels.}
Selecting checkpoints on the evaluation week, with early stopping,
best-epoch restoration and the learning-rate schedule keyed on the
week being scored, lets that week influence training. The four constituent nowcasters were therefore compared under two checkpoint-selection protocols, each fixed before the runs and applied over the same 13 weekly folds for every channel: selection on the inner probe, the seven days immediately preceding each evaluation week, withheld from training; and a control that withholds the same seven training days but still selects on the evaluation week. Relative to the control, selecting on the inner
probe lowered mean weekly F$_1$-High by $0.028$ for H$_2$S at MMF9 (lower
in all 8 folds that differed; sign test $p = 0.008$), $0.080$ for H$_2$S
at MMF2 ($p = 0.006$), $0.032$ for CH$_4$ at MMF9 ($p = 0.34$) and $0.066$
for CH$_4$ at MMF2 ($p < 0.001$). With the fusion parameters held fixed,
the fused-tier $\kappa_w$ was $0.709$ with evaluation-week selection, $0.698$
with the control channels and $0.659$ with the inner-probe channels, so
$0.040$ is attributable to selection and $0.011$ to the shorter training
window; the odour-report correlation was $0.729$, $0.709$ and $0.690$.
Because the inner probe removes the most recent training days, the
channels reported in the main text are retrained after selection: the epoch count
and learning-rate schedule are chosen on the inner probe, and the model
is then retrained on the full training window for that number of epochs
with the same learning-rate schedule, early stopping disabled and the final-epoch weights retained; no stage of training or selection reads the evaluation week. This protocol was selected from
candidate protocols on four Oct--Dec 2024 weeks that precede the
evaluation window, as the candidate with the largest mean gain in weekly F$_1$-High over inner-probe selection among those whose retraining was computationally feasible, and was fixed before its single evaluation on the Jan--Mar 2025 window. With
these retrained channels, $\kappa_w = 0.669$ and the odour-report correlation is
$0.711$ ($0.661$ and $0.698$ under the leave-one-week-out refit).

\paragraph{Likelihood ratios.}
Each channel-state likelihood ratio is a Bayes factor expressed in the standard log-odds-update form~\citep{kass1995bayesfactors},
\begin{equation}
\mathrm{LR}^{\mathrm{on}}_c
= \frac{\Pr(\ell_{c,t}=1 \mid \text{network event})}
       {\Pr(\ell_{c,t}=1 \mid \neg\text{network event})},
\label{eq:s_lr_definition}
\end{equation}
i.e.\ the relative likelihood of a latched ``active'' state under the
event versus non-event hypotheses.
The likelihood-ratio values are grid outputs; the dependence-discounting rationale given below is a design intention, not the selection mechanism. A monotonicity-constrained
refit imposing $\mathrm{LR}(\mathrm{H_2S}) \geq \mathrm{LR}(\mathrm{CH_4})$
inside the same leave-one-week-out loop changes $\kappa_w$ by $0.023$ on the concatenated held-out series, and the paired per-fold comparison favours the unconstrained ordering (9 of 13 folds; $p = 0.03$, Wilcoxon signed-rank over thirteen folds). The CH$_4$-heavy ordering therefore performs modestly better; this is an empirical property of the fitted ratios rather than a physical ordering of the species, and no result in this work rests on the ordering.
The bare precision-prevalence
ratio of each constituent nowcaster lies in the range
$10$--$19$ for the four channels; the integer values in
Supplementary Table~\ref{tab:si_lr} are smaller than that ratio, which is consistent with three sources of
inter-channel dependence that violate the conditional-independence
assumption underlying a naive product of channel-specific Bayes
factors:
\begin{itemize}
\item the two H$_2$S nowcasters share a sulphate-reducing chemical
  source term and a partially overlapping advection field
  (westerly principal bearing, Supplementary Note~2);
\item the two CH$_4$ nowcasters share the same primary chemistry
  (sulphate-reducer/methanogen co-emission, Supplementary Note~2);
\item the H$_2$S and CH$_4$ channels at the same receptor share local
  meteorology, so a high-pressure inversion event registers
  simultaneously on both species at that receptor.
\end{itemize}
The values in Supplementary Table~\ref{tab:si_lr} therefore function
as calibrated multipliers chosen to keep the multi-channel posterior
in a well-conditioned range under positive inter-channel correlation,
rather than as a direct application of the precision-prevalence ratio.
A copula or empirical-Bayes treatment of the joint channel-state
distribution would refine the calibration but is left to future work;
the present integer values are deliberately conservative.

H$_2$S quiet states ($\mathrm{LR}^{\mathrm{off}} < 1$) provide mild
\emph{negative} evidence because silence at the most exposed receptor
(MMF9) is itself informative: a long-running
plume that is not detected at MMF9 is unlikely to constitute a
network event. CH$_4$ quiet states carry $\mathrm{LR}^{\mathrm{off}} = 1$
(no negative evidence) because CH$_4$ exhibits a higher baseline rate
of background activation than H$_2$S; allowing CH$_4$ silence to
suppress the posterior risks falsely clearing an H$_2$S-led event.
Missing data (\textsc{NaN}) is treated as $\mathrm{LR} = 1$ at every
channel in keeping with the principle that an absent sensor provides
no evidence in either direction.

\begin{table}[htbp]
  \centering
  \caption{\textbf{Channel-state likelihood ratios for the Bayesian
  fusion stage.} Each ratio is the multiplicative update applied to
  the prior odds when the corresponding channel is in the indicated
  state (Eq.~\eqref{eq:s_channel_update}). Active-state values are
  calibrated multipliers, substantially smaller than the bare
  precision-prevalence ratio of each constituent nowcaster
  (column 2; the bare ratio is in the range $10$--$19$ across the four
  channels at $\pi = 0.10$) to account for inter-channel dependence
  (shared chemistry between H$_2$S receptors, between CH$_4$ receptors,
  and between species at the same receptor). H$_2$S-quiet states carry
  moderate negative evidence because the H$_2$S channels are the most
  direct receptor measurements of the public-health-relevant species;
  CH$_4$-quiet states carry $\mathrm{LR} = 1$ because the higher
  CH$_4$ background would make suppressive quiet evidence
  operationally unsafe. Missing data is always neutral. Robustness of
  these values is established by the $\pm 25\%$ sensitivity analysis
  of Supplementary Table~\ref{tab:si_lr_sensitivity}. Native precision is that of the constituent nowcasters on which the likelihood ratios were selected; the same likelihood ratios are used for the retrained nowcasters reported in the main text.}
  \label{tab:si_lr}
  \footnotesize
  \setlength{\tabcolsep}{6pt}
  \begin{tabular}{@{}lcccc@{}}
  \toprule
  Channel & Native precision &
    $\mathrm{LR}^{\mathrm{on}}_c$ &
    $\mathrm{LR}^{\mathrm{off}}_c$ &
    $\mathrm{LR}^{\mathrm{NaN}}_c$ \\
  \midrule
  H$_2$S @ MMF9 (broad-coverage)        & 0.526 & 2.00 & 0.50 & 1.00 \\
  H$_2$S @ MMF2 (tight-coverage confirmer)     & 0.636 & 3.00 & 0.50 & 1.00 \\
  CH$_4$ @ MMF9 (chemical corroboration)       & 0.632 & 7.00 & 1.00 & 1.00 \\
  CH$_4$ @ MMF2 (chemical corroboration)       & 0.673 & 4.00 & 1.00 & 1.00 \\
  \bottomrule
  \end{tabular}
  \end{table}

  \subsection*{Bayesian fusion of multi-site, multi-species evidence}
  \label{ssec:si_bayesian_fusion}
  The fusion stage maps the four meteorology-trained walk-forward
  nowcaster outputs to a continuous network-event probability and an
  ordinal alert tier on the synchronous 15-minute grid. The classifier
  is purely meteorology-driven at inference time: the predicted tier
  $\tau_t$ never reads any raw target-species concentration; only the
  ground-truth label $\tau^{\mathrm{GT}}_t$ uses raw concentrations.

  \paragraph{Per-channel state.}
  For channel $c \in \mathcal{C}$ at timestep $t$, the latched binary
  state $\ell_{c,t}$ and the missing-data mask $m_{c,t}$ are computed
  from the model probability $p_{c,t}$ by
  \begin{equation}
  \ell_{c,t} = \textsc{Hysteresis}\bigl(
  \mathbf{1}(p_{c,t} > p^\star);\, N_{\mathrm{on}}, N_{\mathrm{off}}\bigr),
  \qquad
  m_{c,t} = \mathbf{1}(p_{c,t}\;\text{is missing}).
  \label{eq:s_state_hysteresis}
  \end{equation}
  with $p^\star = 0.5$, $N_{\mathrm{on}} = 3$ (45~min onset) and
  $N_{\mathrm{off}} = 2$ (30~min clearance, matched to the WHO 30-min
  averaging convention).

  \paragraph{Log-odds initialisation.}
  The prior log-odds is anchored at the empirical network-event base
  rate:
  \begin{equation}
  \log O_0 = \log\!\frac{\pi}{1-\pi},
  \qquad \pi = 0.10.
  \label{eq:s_log_odds_init}
  \end{equation}
  \paragraph{Channel-state log-odds update.}
  Each channel contributes additively in log-odds:
  \begin{equation}
  \log \mathrm{LR}_{c,t} =
  \begin{cases}
  0, & m_{c,t} = 1 \quad (\text{sensor missing}) \\[1pt]
  \log \mathrm{LR}^{\mathrm{on}}_c, & m_{c,t} = 0 \land \ell_{c,t} = 1 \\[1pt]
  \log \mathrm{LR}^{\mathrm{off}}_c, & m_{c,t} = 0 \land \ell_{c,t} = 0.
  \end{cases}
  \label{eq:s_channel_update}
  \end{equation}

  CH$_4$-quiet contributions are exactly neutral
  ($\mathrm{LR}^{\mathrm{off}}_{\text{CH}_4} = 1$); only H$_2$S-quiet
  states deliver $\log\mathrm{LR} < 0$ (mild suppression).

  \paragraph{Posterior probability.}
  The fused log-odds and posterior are
  \begin{align}
  \log O_t &= \log O_0 + \sum_{c \in \mathcal{C}} \log \mathrm{LR}_{c,t},
  \label{eq:s_log_odds_sum} \\
  P_t &= \sigma(\log O_t) = O_t/(1 + O_t),
  \qquad \sigma(x) = (1 + e^{-x})^{-1}.
  \label{eq:s_post_prob}
  \end{align}

  \paragraph{Tier mapping (pure meteorology).}
  The continuous posterior is mapped to the ordinal tier by sequential thresholding only, with no raw-sensor override on the predicted side:
  \begin{equation}
  \tau_t = \sum_{k=1}^{3} \mathbf{1}\bigl(P_t \geq \theta_k\bigr).
  \label{eq:s_tier_mapping}
  \end{equation}
  with $(\theta_1, \theta_2, \theta_3) = (0.15, 0.50, 0.92)$. Because
  $\theta_1 < \theta_2 < \theta_3$ and the indicator sum is monotone in
  $P_t$, the trajectory $0 \to 1 \to 2 \to 3$ cannot skip a tier on
  the soft path. The Tier~3 cut-point sits just below the analytical
  posterior ceiling under all four channels concurrently active
  ($P_{\max} \approx 0.949$ for the LR set in
  Supplementary Table~\ref{tab:si_lr}), so Tier~3 is reached only by
  fully coordinated multi-channel meteorological evidence.

\paragraph{Bayesian classifier parameter set} The hysteresis counts, prior,
likelihood ratios and tier cut-points in Supplementary
Table~\ref{tab:si_tier_params} were selected to maximise tier-match
$\kappa_w$ against the deterministic ground-truth tier of
Eq.\eqref{eq:s_gt_tier_mapping} on nine of the thirteen Jan--Mar 2025
validation weeks (``Selection protocol and held-out performance'' above), so the
window-wide figure is a mixed quantity. A $\pm 25\%$
perturbation of each $\mathrm{LR}^{\mathrm{on}}_c$ in turn (eight perturbations, all other
parameters held fixed) changes the window-wide $\kappa_w$ by at most 0.003 (mean $|\Delta\kappa_w| = 0.001$ across the eight perturbations, Supplementary
Table~\ref{tab:si_lr_sensitivity}). The framework's agreement with the ground
truth is therefore not contingent on the precise LR values within the range tested; it reflects the structural choice to fuse four
meteorology-trained channels under a log-odds accumulator with H$_2$S-quiet
suppression and CH$_4$-quiet neutrality.

\begin{table}[htbp]
\centering
\caption{\textbf{$\pm 25\%$ perturbation
sensitivity of the window-wide tier-match $\kappa_w$ to each active-state
likelihood ratio.} Each row scales one $\mathrm{LR}^{\mathrm{on}}_c$ by the
indicated factor with all other parameters held fixed (other LRs, hysteresis,
prior, tier cut-points). $\Delta\kappa_w$ is the deviation from the baseline
$\kappa_w = 0.669$. The maximum absolute deviation is 0.003 (one row), indicating
that the headline performance is robust to the precise LR choice.}
\label{tab:si_lr_sensitivity}
\footnotesize
\begin{tabular}{@{}lcrr@{}}
\toprule
Perturbed parameter & Factor & $\kappa_w$ & $\Delta\kappa_w$ \\
\midrule
H$_2$S MMF9 $\mathrm{LR}^{\mathrm{on}}$ & $\times 0.75$ & 0.669 & \phantom{$+$}0.000 \\
H$_2$S MMF9 $\mathrm{LR}^{\mathrm{on}}$ & $\times 1.25$ & 0.669 & \phantom{$+$}0.000 \\
H$_2$S MMF2 $\mathrm{LR}^{\mathrm{on}}$ & $\times 0.75$ & 0.669 & \phantom{$+$}0.000 \\
H$_2$S MMF2 $\mathrm{LR}^{\mathrm{on}}$ & $\times 1.25$ & 0.666 & $-0.003$ \\
CH$_4$ MMF9 $\mathrm{LR}^{\mathrm{on}}$ & $\times 0.75$ & 0.668 & \phantom{$+$}0.000 \\
CH$_4$ MMF9 $\mathrm{LR}^{\mathrm{on}}$ & $\times 1.25$ & 0.669 & \phantom{$+$}0.000 \\
CH$_4$ MMF2 $\mathrm{LR}^{\mathrm{on}}$ & $\times 0.75$ & 0.667 & $-0.002$ \\
CH$_4$ MMF2 $\mathrm{LR}^{\mathrm{on}}$ & $\times 1.25$ & 0.669 & \phantom{$+$}0.000 \\
\bottomrule
\end{tabular}
\end{table}

\paragraph{Inter-channel dependence.} The conditional-independence assumption
underlying Eq.\eqref{eq:s_log_odds_sum} is violated in practice: the two H$_2$S
channels share a chemical source and a partially overlapping advection field,
the two CH$_4$ channels share the same primary chemistry, and the H$_2$S/CH$_4$
channels at a single receptor share local meteorology. A formally correct fusion
under dependent channel states would require a copula model of the joint
$\{\ell_{c,t}\}_{c \in \mathcal{C}}$ distribution or an empirical-Bayes treatment
from joint state frequencies. We do not adopt either here; the LR values in
Supplementary Table~\ref{tab:si_lr} absorb the positive inter-channel correlation
by remaining substantially below the bare precision-prevalence ratio of each
constituent classifier. The sensitivity analysis above shows that $\kappa_w$ changes by at most $0.003$ under a $\pm 25\%$ perturbation of each likelihood ratio; a dependence-aware fusion is left to future work.

\subsection*{Ground-truth tier definition and headline performance}
\label{ssec:si_ground_truth_joint_perf}

\paragraph{Ground-truth tier construction.} The ground-truth tier
$\tau^{\mathrm{GT}}_t$ is constructed from the raw 15-min sensor record by a
deterministic channel-vote rule on debounced threshold exceedances. For each
channel $c$ define the raw indicator and its latched form:
\begin{equation}
b^{\mathrm{GT}}_{c,t} = \mathbf{1}(r_{c,t} \geq C^{\mathrm{thr}}_c), \qquad
\ell^{\mathrm{GT}}_{c,t} = \textsc{Hysteresis}(b^{\mathrm{GT}}_{c,t};\,
N_{\mathrm{on}}, N_{\mathrm{off}}), \label{eq:s_gt_indicator}
\end{equation}
where $C^{\mathrm{H_2S}}_{\mathrm{thr}} = 7\,\mu\mathrm{g\,m^{-3}}$ and
$C^{\mathrm{CH_4}}_{\mathrm{thr}} = 2.756\,\mathrm{ppm}$, the \emph{same} class boundaries used to define each CAIRN nowcaster's training target. Missing data
is treated conservatively ($b^{\mathrm{GT}}_{c,t} = 0$) so that an absent sensor
cannot raise the ground-truth tier. The number of latched channels is
$n^{\mathrm{GT}}_t = \sum_{c \in \mathcal{C}} \ell^{\mathrm{GT}}_{c,t}$ and the
ground-truth tier is
\begin{equation}
\tau^{\mathrm{GT}}_t = \sum_{k=1}^{3}
\mathbf{1}\bigl(n^{\mathrm{GT}}_t \geq k\bigr), \label{eq:s_gt_tier_mapping}
\end{equation}
i.e.\ Tier~1 when at least one channel is latched, Tier~2 when at
least two, Tier~3 when at least three. This channel-vote rule encodes the tier definitions for a multi-receptor network: a single
isolated exceedance warrants Tier~1, multi-channel corroboration is
required for Tier~2, and convergent evidence across both species or
both receptors for Tier~3. The same hysteresis
$(N_{\mathrm{on}}, N_{\mathrm{off}})$ is applied to both sides, so $\tau_t$ and
$\tau^{\mathrm{GT}}_t$ live on the same ordinal scale. The ground-truth label is
deterministic and reproducible given the threshold and hysteresis declaration
above, but not independent of those choices: a different threshold scheme would yield a different labelling.

\paragraph{Headline tier-match performance.} On the synchronous Jan--Mar 2025
validation window ($n = 8{,}536$ shared-grid timesteps), the meteorology-only
predicted tier agrees with the deterministic raw-sensor ground truth at
quadratic-weighted Cohen's kappa
\begin{equation*}
\boxed{\kappa_w = \text{0.669} \quad (\text{95\% CI: 0.565, 0.755}), \quad \text{strict accuracy} = \text{75.7\%}}
\end{equation*}
where the 95\% confidence interval is obtained by a block
bootstrap with $B = 1{,}000$ resamples and a 24-hour block length (96 time steps at
15-minute resolution) to accommodate within-episode autocorrelation. The
matched-block bootstrap has a $\kappa_w$ standard error of 0.050. The headline
$\kappa_w$ value sits in the ``substantial agreement'' band of the Landis--Koch
interpretive scale~\citep{landis1977measurement}, although the lower bound of its interval falls in the ``moderate'' band. Per-tier metrics with
bootstrap 95\% CIs are in Supplementary Table~\ref{tab:si_tier_per_tier}; the
full confusion matrix is in Supplementary Table~\ref{tab:si_tier_confusion}.

\paragraph{Comparison to a vote-count baseline.} A natural baseline applies the
same channel-vote rule of Eq.~\eqref{eq:s_gt_tier_mapping} directly to the four
predicted latched activations $\{\ell_{c,t}\}$, replacing the Bayesian log-odds accumulator with a hard count of latched channels and holding all other
pipeline stages fixed. On the same window this baseline returns
$\kappa_w^{\mathrm{vote}} = 0.578$ (95\% CI: 0.480, 0.667). The Bayesian fusion improves on it by $\Delta\kappa_w = +0.090$; under a paired block
bootstrap, in which the same resampled blocks are used for both estimators so
that shared block-to-block variation cancels, the improvement is significant
(95\% CI $0.048$, $0.127$; one-sided $p = 0.001$). An unpaired bootstrap, which
resamples the two $\kappa$ values independently, approximately triples the
interval width and is not the appropriate test for two estimators evaluated on
identical timesteps. The Bayesian fusion also produces a continuous posterior $P_t$, exposes channel-specific evidence weights, and accommodates missing data and CH$_4$-quiet neutrality natively. The vote-count baseline delivers none of these.

\begin{table}[htbp]
\centering
\caption{\textbf{Confusion matrix of predicted
versus ground-truth ordinal alert tiers across the Jan--Mar 2025 validation
window} ($n = 8{,}536$ synchronous 15-min timesteps). Rows are the ground-truth
tier $\tau^{\mathrm{GT}}_t$; columns are the meteorology-only predicted tier
$\tau_t$. Diagonal cells are typeset in bold. Parenthesised values are row
percentages.}
\label{tab:si_tier_confusion}
\footnotesize
\setlength{\tabcolsep}{6pt}
\begin{tabular}{@{}lcccc|c@{}}
\toprule
& \multicolumn{4}{c|}{Predicted $\tau_t$} & \\
Ground truth $\tau^{\mathrm{GT}}_t$ & Tier~0 & Tier~1 & Tier~2 & Tier~3 & Support \\
\midrule
Tier~0 & \textbf{5{,}762 (84\%)} & 791 (12\%) & 230 (3\%) & 43 (1\%) & 6{,}826 \\
Tier~1 & 261 (37\%) & \textbf{274 (38\%)} & 168 (24\%) & 10 (1\%) & 713 \\
Tier~2 & 113 (19\%) & 197 (33\%) & \textbf{213 (36\%)} & 70 (12\%) & 593 \\
Tier~3 & 17 (4\%) & 39 (10\%) & 134 (33\%) & \textbf{214 (53\%)} & 404 \\
\bottomrule
\end{tabular}
\end{table}

\begin{table}[htbp]
\centering
\caption{\textbf{Per-tier and macro
classification metrics for the meteorology-only multi-channel Bayesian tier
classifier.} Block-bootstrap 95\% CIs ($B = 1{,}000$, block = 24~h) in brackets.
Window-wide $\kappa_w = 0.669$ (CI 0.565, 0.755); strict element-wise accuracy 0.757.}
\label{tab:si_tier_per_tier}
\footnotesize
\setlength{\tabcolsep}{4pt}
\begin{tabular}{@{}lccccc@{}}
\toprule
Tier & Precision & Recall & F$_1$ & Support & Population share \\
\midrule
Tier~0 & 0.936 [0.916, 0.958] & 0.844 [0.793, 0.888] & 0.888 [0.855, 0.916] & 6{,}826 & 79.9\% \\
Tier~1 & 0.211 [0.136, 0.288] & 0.384 [0.297, 0.459] & 0.272 [0.191, 0.346] & 713 & 8.4\% \\
Tier~2 & 0.286 [0.189, 0.387] & 0.359 [0.229, 0.511] & 0.318 [0.219, 0.411] & 593 & 6.9\% \\
Tier~3 & 0.635 [0.363, 0.801] & 0.530 [0.275, 0.751] & 0.578 [0.320, 0.722] & 404 & 4.7\% \\
\midrule
Macro avg & 0.517 & 0.529 & 0.514 & 8{,}536 & --- \\
Accuracy & \multicolumn{2}{c}{0.757} & --- & 8{,}536 & --- \\
$\kappa_w$ & \multicolumn{2}{c}{0.669 [0.565, 0.755]} & --- & 8{,}536 & --- \\
\bottomrule
\end{tabular}
\end{table}

\paragraph{Effective sample size at the extremes.} The bootstrap CI on Tier~3
F$_1$ is wide ([0.32, 0.72]) because Tier~3 episodes have an effective sample size
much smaller than the nominal count of 404 timesteps. Episodes are temporally
autocorrelated on 1--3 hour scales (4--12 timesteps), giving an effective
independent count of roughly 30--100 episodes. Per-tier point estimates at the
extremes should be read against this CI.

\paragraph{Per-week tier-match performance.} Supplementary
Table~\ref{tab:si_tier_weekly_qwk} reports the within-week $\kappa_w$ breakdown
with single-week bootstrap CIs. Single-week $\kappa_w$ values are unstable when
the support of any tier falls below ${\sim}10$: the statistic's variance
grows by an order of magnitude under these conditions, and weekly values in
those regimes should be read as descriptive rather than as calibrated indicators
of tier-match quality. Week~7 contains no ground-truth deviation from Tier~0 (all
672 timesteps GT~$= 0$) and yields a degenerate $\kappa_w$ that is undefined because the
ground-truth labels do not vary; the strict accuracy of 0.981 in that week is informative as a specificity check on the classifier. The headline
$\kappa_w = 0.669$ is the pooled estimate over the entire $n = 8{,}536$ synchronous
timeseries, not a sample-size-weighted average of the per-week values.

\begin{table}[htbp]
\centering
\caption{\textbf{Per-week tier-match performance
with bootstrap 95\% CIs.} Each row reports the within-week quadratic-weighted
Cohen's kappa, strict element-wise accuracy, and the per-tier support counts in
the ground-truth labels. Bold marks weekly blocks with point estimate
$\kappa_w \geq 0.70$. Bootstrap CIs ($B = 1{,}000$, within-week resampling) capture
sampling variation within the week and do not account for cross-week
autocorrelation. Week~7 has degenerate ($\kappa_w$ undefined) tier variation.}
\label{tab:si_tier_weekly_qwk}
\footnotesize
\setlength{\tabcolsep}{5pt}
\begin{tabular}{@{}ccccccccc@{}}
\toprule
& & & & \multicolumn{4}{c}{Ground-truth tier support} & \\
Wk & $\kappa_w$ & 95\% CI & Acc.\ & Tier~0 & Tier~1 & Tier~2 & Tier~3 & $n_{\mathrm{wk}}$ \\
\midrule
1 & 0.461 & [0.376, 0.530] & 0.509 & 423 & 82 & 132 & 27 & 664 \\
2 & 0.613 & [0.565, 0.662] & 0.423 & 314 & 19 & 120 & 219 & 672 \\
3 & 0.370 & [0.310, 0.432] & 0.448 & 538 & 33 & 26 & 75 & 672 \\
4 & 0.353 & [0.255, 0.444] & 0.771 & 609 & 21 & 42 & 0 & 672 \\
5 & 0.359 & [0.260, 0.441] & 0.720 & 604 & 50 & 18 & 0 & 672 \\
6 & \textbf{0.958} & [0.927, 0.981] & 0.972 & 604 & 7 & 11 & 50 & 672 \\
7 & --- & --- & 0.981 & 672 & 0 & 0 & 0 & 672 \\
8 & 0.231 & [0.115, 0.349] & 0.893 & 608 & 64 & 0 & 0 & 672 \\
9 & \textbf{0.709} & [0.664, 0.753] & 0.676 & 418 & 109 & 128 & 17 & 672 \\
10 & 0.441 & [0.333, 0.546] & 0.878 & 587 & 49 & 24 & 12 & 672 \\
11 & \textbf{0.772} & [0.714, 0.822] & 0.881 & 545 & 91 & 36 & 0 & 672 \\
12 & \textbf{0.730} & [0.674, 0.781] & 0.836 & 495 & 127 & 46 & 4 & 672 \\
13 & 0.644 & [0.529, 0.733] & 0.892 & 409 & 61 & 10 & 0 & 480 \\
\midrule
Pooled & 0.669 & [0.565, 0.755] & 0.757 & 6{,}826 & 713 & 593 & 404 & 8{,}536 \\
\bottomrule
\end{tabular}
\end{table}

\subsection*{External validation against community odour reports}
\label{ssec:si_complaint_validation}

The Jan--Mar 2025 validation window includes a co-registered daily record of
community odour-complaint counts at the principal receptor (n = 89 days; mean 58
odour reports/day, max 1{,}137 on 2025-01-13). \emph{This record was not used in
training, in the hysteresis or LR selection, or in any tier-mapping decision.}
It provides an external validation endpoint that is independent of model fitting, though not of the emission events that also generate the sensor labels (main Discussion).

\paragraph{Daily-mean alert tier versus daily odour reports.} Aggregating
the 15-min predicted tier $\tau_t$ to daily means and correlating with daily
complaint count gives Pearson $r = 0.711$, $R^2 = 0.51$, Spearman $\rho = 0.656$, $p = 5.6 \times 10^{-15}$; the same aggregation on the deterministic ground-truth tier $\tau^{\mathrm{GT}}_t$ gives $r = 0.793$, $R^2 = 0.63$ (intervals in Supplementary Table~\ref{tab:ed_multisite_tier}). The CAIRN classifier therefore captures approximately 80\% of the ground-truth's explained variance in odour reports
($0.51 / 0.63 = 0.80$), within 0.082 of the correlation achieved by the raw-sensor ground truth.

\paragraph{Lagged cross-correlation.} The lagged Pearson cross-correlation
between daily-mean predicted tier and daily complaint count has its unique
global maximum at lag~0 ($r = 0.711$), with roll-off to $r = +0.539$ at lag~$-1$ day and $r = +0.592$ at lag~$+1$ day, and $r \leq 0.42$ at
$|{\rm lag}| \geq 2$ days (Supplementary Table~\ref{tab:si_xcorr}). At the daily resolution of the report record, the dominant lag-0 peak indicates that the classifier is a nowcaster of community exposure, not a delayed reflector of past complaint-driving events
(which would peak at negative lag) and not a forecaster of future events (which
would peak at positive lag).

\begin{table}[htbp]
\centering
\caption{\textbf{Lagged Pearson cross-correlation
between daily-mean CAIRN predicted tier and daily community complaint count
($n = 89$ days, $\pm 7$-day window).} Lag is in days; lag $> 0$ means the model leads
odour reports. The unique global maximum at lag 0 indicates nowcast alignment at daily resolution.}

\label{tab:si_xcorr}
\scriptsize
\setlength{\tabcolsep}{3.4pt}
\begin{tabular}{@{}cccccccccccccccc@{}}
\toprule
Lag (days) & $-7$ & $-6$ & $-5$ & $-4$ & $-3$ & $-2$ & $-1$ & $\mathbf{0}$ & $+1$ & $+2$ & $+3$ & $+4$ & $+5$ & $+6$ & $+7$ \\
\midrule
Pearson $r$ & 0.34 & 0.24 & 0.21 & 0.15 & 0.18 & 0.39 & 0.54 & \textbf{0.71} & 0.59 & 0.34 & 0.24 & 0.22 & 0.31 & 0.34 & 0.42 \\
\bottomrule
\end{tabular}
\end{table}

\subsection*{Limitations of the tier agreement}
\label{ssec:si_operational}

The intermediate Tiers~1 and~2 carry most of the off-by-one disagreement, with the classifier over-triggering both Tier~1 (recall 0.38, precision 0.21) and  Tier~2 (recall 0.36, precision 0.29; Supplementary Table~\ref{tab:si_tier_per_tier}). These errors are inherited from the marginal precision of the constituent walk-forward classifiers (Supplementary Tables~\ref{tab:si_s4_xgb_weekly}, \ref{tab:si_s4_h2s_ch4_weekly}) and from the difficulty of recovering multi-channel agreement at 15-minute resolution. The headline agreement, its confidence interval, the comparison with a vote-count baseline and the correlation with odour reports are reported in main Results and Supplementary Table~\ref{tab:ed_multisite_tier}.

\clearpage
\section*{Supplementary Methods}
\addcontentsline{toc}{section}{Supplementary Methods}

\noindent This section holds the full mathematical and procedural
detail for the methods summarised in main Methods. Six subsections
correspond to sections of main
Methods: ``Environmental characterisation'', ``Multiscale transfer
entropy'', ``Accumulated Local Effects'', ``S4D state space model'',
``Optimiser schedule'' and ``Walk-forward validation''. All
data-handling, software-environment, hyperparameter and class-imbalance
choices are stated in main Methods and are not repeated here.

\subsection*{Environmental characterisation}
\label{ssec:smethods_envchar}

This subsection holds the per-panel mathematical detail for Fig.~1 of
the main text, summarised in main Methods, ``Environmental
characterisation''. The quantitative results are tabulated in
Supplementary Table~\ref{tab:si_envchar_summary} and discussed panel-by-panel in Supplementary
Note~2.

\paragraph{Conditional probability function (Fig.~1a).}
For each station the full $360^{\circ}$ compass is partitioned into
$n_s = 36$ equal $\Delta\theta = 10^{\circ}$ sectors. Each 15-minute
observation is assigned to the sector $\theta_k$ containing its
recorded wind direction. The CPF for sector $k$ is
\begin{equation}
\mathrm{CPF}(\theta_k) =
\frac{N_{\mathrm{exc}}(\theta_k)}{N(\theta_k)},
\label{eq:s_cpf}
\end{equation}
where $N_{\mathrm{exc}}(\theta_k)$ is the count of observations in
sector $k$ with $\mathrm{H_2S} > C_{\mathrm{thr}} = 5\,\mu\mathrm{g\,m^{-3}}$,
$N(\theta_k)$ is the total count, and sectors with $N(\theta_k)=0$
are assigned $\mathrm{CPF}=0$ \citep{ashbaugh1985principal}. Plots
are rendered on a polar axis with North at $0^{\circ}$ proceeding
clockwise. Site coordinates are converted from WGS84 to British
National Grid (OSGB36, EPSG:27700) via standard transverse Mercator
projection for the receptor map.

\paragraph{Chemical fingerprint (Fig.~1b).}
Pearson's product-moment correlation
\begin{equation}
r = \frac{\sum_{i=1}^{n}(x_i-\bar{x})(y_i-\bar{y})}
        {\sqrt{\sum_i(x_i-\bar{x})^2}\sqrt{\sum_i(y_i-\bar{y})^2}}
\label{eq:s_pearson}
\end{equation}
and Spearman's rank correlation $\rho$ are computed on all paired
H$_2$S--CH$_4$ observations after listwise deletion of missing
values; significance is via two-tailed $t$-test with $n-2$ degrees of
freedom. A 96-sample (24-h) sliding-window rolling Pearson $r(t)$
quantifies temporal stability. Spike co-occurrence on the top-5\%
spike sets $S_{\mathrm{H_2S}}, S_{\mathrm{CH_4}}$ is quantified by
the Jaccard index
\begin{equation}
J = \frac{|S_{\mathrm{H_2S}} \cap S_{\mathrm{CH_4}}|}
       {|S_{\mathrm{H_2S}} \cup S_{\mathrm{CH_4}}|},
\label{eq:s_jaccard}
\end{equation}
with significance via $\chi^2$ test of independence on the
$2\!\times\!2$ contingency table. A power-law
$\mathrm{H_2S} = a\cdot\mathrm{CH_4}^{b}$ is fitted by ordinary
least-squares (OLS) on log-transformed concentrations. Co-emission
evidence is classified as Strong if $r > 0.7$ and $J > 0.5$,
Moderate if $r > 0.4$ and $J > 0.3$, Weak if $r > 0.2$.

\paragraph{Diel hysteresis (Fig.~1c).}
Observations are grouped by hour-of-day $h \in \{0,\dots,23\}$ to
form the diel mean trajectory in $(\mathrm{TEMP}, \mathrm{H_2S})$
phase space:
\begin{equation}
\bar{T}_h = \frac{1}{N_h}\sum_{i:h_i=h} T_i,
\qquad
\overline{\mathrm{H_2S}}_h
= \frac{1}{N_h}\sum_{i:h_i=h} c_i,
\label{eq:s_diurnal_mean}
\end{equation}
with hours containing fewer than $\max(5, n/240)$ observations
excluded. The closed 24-point loop is characterised by its enclosed
area, computed via the Shoelace surveyors' formula
\begin{equation}
A = \tfrac{1}{2}\biggl|\sum_{h=0}^{23}\bigl(\bar{T}_h
\overline{\mathrm{H_2S}}_{h+1}
-\bar{T}_{h+1}\overline{\mathrm{H_2S}}_h\bigr)\biggr|,
\label{eq:s_shoelace}
\end{equation}
with indices modulo 24. The dimensionless normalised area is
\begin{equation}
A^{*} = \frac{A}{(\bar{T}_{\max}-\bar{T}_{\min})\,
                 (\overline{\mathrm{H_2S}}_{\max}-\overline{\mathrm{H_2S}}_{\min})}.
\label{eq:s_norm_area}
\end{equation}
A directional-asymmetry index from cooling- (18:00--06:00) and
warming-limb (06:00--14:00) OLS slopes is
\begin{equation}
H_{\mathrm{index}}
= \frac{|\beta_{\mathrm{cool}} - \beta_{\mathrm{warm}}|}
       {|\beta_{\mathrm{cool}}|+|\beta_{\mathrm{warm}}|+\epsilon},
\quad \epsilon = 10^{-6}.
\label{eq:s_hyst}
\end{equation}
Normalised areas $A^{*}<0.05$, $0.05$--$0.15$, $0.15$--$0.30$, $>0.30$
are classified as none / weak / moderate / strong hysteresis.

\paragraph{Diel profile (Fig.~1d).}
Observations are grouped by hour-of-day; the hourly mean and 95th
percentile are reported, with the shaded envelope spanning mean to
P95. Hourly variation is tested by a 24-group Kruskal--Wallis
$H$-test. The peak hour is the hour of maximum mean concentration.

\paragraph{Sunrise alignment (Fig.~1e).}
Astronomical sunrise times $t_{\mathrm{sr}(d_i)}$ are computed for
each calendar date $d_i$ using the \texttt{astral} Python library
(v2.2) at the site latitude/longitude. For each 15-minute observation
the elapsed time since sunrise is
\begin{equation}
\Delta t_i = t_i - t_{\mathrm{sr}(d_i)},
\label{eq:s_sunrise_lag}
\end{equation}
restricted to $\Delta t_i \in [-2, +6]$\,h. The pre-sunrise baseline
is the mean concentration over $\Delta t \in [-2, 0]$\,h. A one-sided
Mann--Whitney $U$ test
\begin{equation}
U = \sum_{i\in\mathrm{post}}\sum_{j\in\mathrm{pre}}
\mathbf{1}(c_i > c_j)
\label{eq:s_mwu}
\end{equation}
is applied with the alternative hypothesis that post-sunrise
concentrations exceed the pre-sunrise baseline. Evidence is classified Strong if
$p<0.01$ and peak/baseline ratio $>1.5$, Moderate if $p<0.05$ and
ratio $>1.2$, Weak if $p<0.10$.

\paragraph{Bivariate mechanism scatter (Fig.~1f).}
For each predictor $x \in \{\mathrm{WS}, \mathrm{TEMP},
\mathrm{d}T/\mathrm{d}t\}$ and response $c$ (H$_2$S), Pearson's $r$
and an OLS regression
\begin{equation}
\hat{c}_i = \beta_0 + \beta_1 x_i,
\qquad
(\hat{\beta}_0, \hat{\beta}_1) = \arg\min_{\beta}
\sum_i (c_i - \beta_0 - \beta_1 x_i)^2
\label{eq:s_ols}
\end{equation}
are reported. Significance on $\hat{\beta}_1$ is tested via two-tailed
$t$-test with $n-2$ degrees of freedom. All three associations are
predicted a priori to be negative: higher wind speed dilutes the
plume, warmer temperatures promote vertical mixing, and positive
$\mathrm{d}T/\mathrm{d}t$ indicates convective destabilisation of the
boundary layer.

\subsection*{Multiscale transfer entropy}
\label{ssec:smethods_mste}

This subsection holds the full procedural specification and pseudo-code
for the MSTE algorithm summarised in main Methods, ``Causal analysis''.
The information-theoretic framework follows
\citet{runge2019inferring} and \citet{goodwell2017temporal}.

\paragraph{Temporal coarse-graining.}
At each scale $\tau$ with aggregation factor $m = \tau/(15\,\mathrm{min})$,
the coarse-grained series is obtained by non-overlapping block averaging:
\begin{equation}
\bar{x}^{(\tau)}_j
= \frac{1}{m}\sum_{i=(j-1)m+1}^{jm} x_i,
\label{eq:s_coarse_grain}
\end{equation}
with right-closed, right-labelled blocks to prevent information
leakage across aggregation boundaries. Block boundaries at coarsest
scales ($\tau \geq 3$\,h) are aligned to 18:00 UTC to centre the
nocturnal accumulation period within a single block.

\textbf{Gap handling.} The 15-minute record for calendar year 2024 at
the case-study site comprises 35,136 slots. Missing observations are
unevenly distributed across channels: wind direction and wind speed
0.04\% each, temperature 0.63\%, pressure 1.75\% and H$_2$S 1.50\%.
Applying the listwise rule, 1,121 slots (3.19\%) lack at least one of
the five series entering the analysis, in 47 runs exceeding 30 minutes
and with a longest contiguous run of 54.2\,h. Gaps are forward-filled
over at most 30 minutes; longer gaps are left missing and the affected
15-minute slots are excluded. That bound recovers 386 of the 1,121
slots (34.4\%), leaving 735 (2.09\% of the year) unusable. Throughout, the 15-minute series denotes the record after this bounded forward-fill, which determines how many slots enter each estimate.

\textbf{Usable $N$ after masking.} Coarse-graining is applied to
the masked series, so the number of blocks entering each estimate falls
with scale. A block contributes if it contains at least one complete
slot; the stricter requirement that \emph{every} slot be complete is
reported alongside, because it is the quantity that degrades:

\begin{center}\footnotesize
\begin{tabular}{@{}lrrrr@{}}
\toprule
Scale & Blocks & Contributing & \% & All slots complete \\
\midrule
15\,min & 35,136 & 34,015 & 96.8 & 34,015 (96.8\%) \\
30\,min & 17,568 & 17,218 & 98.0 & 16,797 (95.6\%) \\
1\,h    &  8,784 &  8,631 & 98.3 &  8,187 (93.2\%) \\
2\,h    &  4,392 &  4,331 & 98.6 &  3,879 (88.3\%) \\
3\,h    &  2,928 &  2,893 & 98.8 &  2,442 (83.4\%) \\
6\,h    &  1,464 &  1,448 & 98.9 &  1,010 (69.0\%) \\
12\,h   &    732 &    725 & 99.0 &    330 (45.1\%) \\
24\,h   &    366 &    364 & 99.5 &      0 (0\%) \\
48\,h   &    183 &    183 & 100  &      0 (0\%) \\
96\,h   &     91 &     91 & 100  &      0 (0\%) \\
\bottomrule
\end{tabular}
\end{center}

These counts apply the block-mean rule; the multiscale
transfer-entropy analysis additionally requires at least 50\% of a
block's nominal sub-samples to survive, which is the stricter rule under
which the 96\,h scale retains $n = 89$ and is reported as not estimable
(Methods, ``Causal analysis''). At $\tau \geq 24$\,h no block is free of at least one filled or
missing slot, so every coarse-scale block mean is computed over a
partially reconstructed window. At the 6\,h scale, which sets the slow-lane anchor timescale of the S4 nowcaster (main Methods), 31.0\% of blocks are affected. This limits the coarse-scale estimates, and the estimates at $\tau \geq 24$\,h should be read as
descriptive: with 91--366 blocks they are additionally
sample-limited.

\paragraph{Conditional transfer entropy and confounder selection.}
At each scale, the causal influence of driver $X$ on target $Y$
(H$_2$S) is quantified by the conditional transfer entropy
\citep{schreiber2000measuring}, formulated as a conditional mutual
information (CMI):
\begin{equation}
T_{X\to Y\mid \mathbf{Z}}
= I\!\left(Y_{t+1};\, X_t \;\middle|\;
Y_t^{(h)},\, \mathbf{Z}_t\right),
\label{eq:s_te_cmi}
\end{equation}
where $Y_t^{(h)} = (Y_t, \dots, Y_{t-h+1})$ is the target history of
order $h$ and $\mathbf{Z}_t$ is a parsimonious confounder set selected
by greedy mutual-information ranking with a Pearson $|r|<0.85$
redundancy filter (excluding the source itself), with
$|\mathbf{Z}|=1$ throughout. Bivariate transfer entropy
($\mathbf{Z} = \varnothing$) cannot distinguish genuine causal
influence from confounded association in this multivariate
system~\citep[Fig.~2a]{runge2019inferring}; the parsimonious
single-confounder choice balances confounder control against the
$\mathcal{O}(d^2/k)$ scaling of $k$NN density-estimator bias with
conditioning dimension $d$ \citep{kraskov2004estimating}.

\paragraph{History embedding depth.}
The embedding order $h$ is set adaptively to approximate 3\,h of
physical memory:
\begin{equation}
h = \max\!\left(1,\;
\min\!\left(2,\;
\lfloor 3\,\mathrm{h}\,/\,\tau \rfloor\right)\right).
\label{eq:s_history}
\end{equation}
This yields $h=2$ at $\tau \leq 1$\,h (capturing 30--120\,min of
target memory) and $h=1$ at coarser scales. The cap $h\leq 2$ keeps
the joint conditioning dimension at $d\leq 3$, within the regime
where the KSG estimator retains adequate sensitivity (Note~5).

\paragraph{Frenzel--Pompe KSG estimator.}
The CMI in Eq.~\eqref{eq:s_te_cmi} is estimated using the
$k$-nearest-neighbour algorithm of \citet{frenzel2007partial}, an
extension of the KSG estimator \citep{kraskov2004estimating} to
conditional mutual information. For $N$ observations in the joint
space $(\mathbf{x}, \mathbf{y}, \mathbf{z}) = (X_t,\, Y_{t+1},\,
[Y_t^{(h)}, \mathbf{Z}_t])$, the procedure constructs a $k$-d tree
in the full joint space using the Chebyshev ($L^\infty$) norm,
identifies for each point $i$ the $k$-th nearest-neighbour distance
$\varepsilon_i$ (excluding all points $j$ with $|j-i|<W$, the Theiler
window), applies the strict-inequality correction
$\varepsilon_i \leftarrow \varepsilon_i(1-10^{-10})$ (because
$k$-d-tree queries return points within a closed $\leq\varepsilon$
ball whereas the KSG Algorithm requires the open ball
$<\varepsilon$), counts marginal neighbours
$n_{\mathbf{xz}}(i)$, $n_{\mathbf{yz}}(i)$, $n_{\mathbf{z}}(i)$
within distance $\varepsilon_i$ in each subspace, and computes
\begin{equation}
\hat{I}(\mathbf{x}; \mathbf{y}\mid \mathbf{z})
= \psi(k) - \frac{1}{N}\sum_{i=1}^{N}\!\Bigl[
\psi(n_{\mathbf{xz}}(i)+1)
+\psi(n_{\mathbf{yz}}(i)+1)
-\psi(n_{\mathbf{z}}(i)+1)\Bigr],
\label{eq:s_frenzel_pompe}
\end{equation}
where $\psi(\cdot)$ is the digamma function.

\paragraph{Theiler window.}
A Theiler exclusion window~\citep{theiler1986spurious} of $W=4$
time steps (1\,h at 15-min sampling) is applied in both the
$\varepsilon_i$ determination and the marginal neighbour counts.
This excludes neighbours lying within 1\,h of the reference point in time, reducing the upward bias in CMI caused by selecting
nearest neighbours that are close in state space simply because
the atmosphere evolves smoothly between consecutive readings
\citep{runge2019inferring}.

\paragraph{Rank transform and adaptive $k$.}
All variables are mapped to $\mathrm{Uniform}[0,1]$ marginals via the
rank transform $\tilde{x}_i = (\mathrm{rank}(x_i)-0.5)/N$, with a
small $\mathcal{N}(0,10^{-8})$ jitter added to break sensor-quantisation
ties and prevent degenerate zero-radius $k$NN balls
\citep{kraskov2004estimating}. The number of neighbours is
\begin{equation}
k = \max\!\bigl(10,\;
\lfloor N_\tau^{0.4} \rfloor,\;
2d + 6\bigr),
\label{eq:s_adaptive_k}
\end{equation}
with $N_\tau$ the valid sample count at scale $\tau$. The $N^{0.4}$
term keeps the relative ball radius stable as effective sample size
falls at coarser scales; the $2d+6$ floor grows with conditioning
dimensionality, ensuring adequate neighbours for stable digamma
estimates.

\paragraph{Surrogate testing.}
\textbf{Surrogate count and reproducibility of the null.}
Surrogate $b$ uses random seed $b$ ($b = 0, \dots, B-1$) and the
tie-breaking jitter uses a fixed seed, so the surrogate mean is exactly
reproducible for any given $B$, which allows the surrogate count of any stored result to be verified. The significance
underpinning Fig.~\ref{fig:causal_links} is computed at $B = 2{,}000$;
the estimator ablation of Supplementary Note~5 runs at $B = 10$, whose
resolution of $\Delta p = 0.1$ cannot support false-discovery
correction across a family of this size and is reported as a coarse
screen only.

Significance is assessed by $B = 2{,}000$ circular-shift surrogates. For
each surrogate $b$ the source series is shifted by a random offset
$\delta_b \sim \mathrm{Uniform}\{1,\dots,N_\tau-1\}$:
\begin{equation}
X^{(b)}_t = X_{(t+\delta_b)\bmod N_\tau},
\label{eq:s_circular_shift}
\end{equation}
while target, target history and confounder series are held fixed.
Circular shifting (rather than random permutation) preserves the
internal autocorrelation structure, periodicity and marginal
distribution of the source exactly, breaking only its temporal
alignment with the target; this produces a conservative null under
which significance reflects driver--target \emph{timing} rather than
the smoothness of the driver
series~\citep[cf.][]{runge2019inferring}. The same KSG--Theiler
estimator is used for both observed TE and surrogates. Effective
transfer entropy removes the finite-sample positive bias of the
estimator:
\begin{equation}
\mathrm{ETE}_{X\to Y}^{(\tau)}
= \hat{T}_{X\to Y}^{(\tau)}
- \frac{1}{B}\sum_{b=1}^{B} \hat{T}_{X^{(b)}\to Y}^{(\tau)}.
\label{eq:s_ete}
\end{equation}
Negative ETE indicates no detectable coupling. Benjamini--Hochberg
false-discovery-rate correction \citep{benjamini1995controlling} at
$\alpha = 0.05$ is then applied across a single family of all
70 driver--scale tests (seven candidate drivers at ten scales, 15\,min to 96\,h).

\paragraph{Pseudo-code.}
Algorithms~\ref{alg:s_mste} and~\ref{alg:s_ksg} give the full
algorithmic specification. Sensitivity to estimator hyperparameter
choice is assessed by the six-configuration ablation reported in
Note~5 and Supplementary Tables~\ref{tab:si_mste_configs}--\ref{tab:si_mste_summary}.

\begin{algorithm}[t]
\caption{Multiscale Transfer Entropy (MSTE)}
\label{alg:s_mste}
\begin{algorithmic}[1]
\Require Raw multivariate time series $\mathbf{D}$ at $\Delta t = 15$\,min;
  driver set $\mathcal{X}$; target $Y$ (H$_2$S);
  scales $\mathcal{T} = \{15\mathrm{m}, 30\mathrm{m}, 1\mathrm{h}, 2\mathrm{h}, 3\mathrm{h}, 6\mathrm{h}\}$
  with aggregation factors $\mathbf{m} = \{1, 2, 4, 8, 12, 24\}$;
  surrogate count $B$; Theiler window $W$; FDR level $\alpha$
\Ensure ETE and adjusted $p$-value for each $(X, \tau)$
\For{each scale $\tau \in \mathcal{T}$ with aggregation factor $m$}
  \State $\bar{\mathbf{D}}^{(\tau)} \gets \textsc{BlockMean}(\mathbf{D}, m)$
    \Comment{Eq.~\eqref{eq:s_coarse_grain}}
  \State $N_\tau \gets$ valid rows in $\bar{\mathbf{D}}^{(\tau)}$
  \State $h \gets \max(1, \min(2, \lfloor 3\mathrm{h}/\tau \rfloor))$
    \Comment{Eq.~\eqref{eq:s_history}}
  \For{each driver $X \in \mathcal{X}$}
    \State Rank $\mathcal{C} = \mathcal{X} \setminus \{X\}$ by $\hat{I}(C_j; Y)$ descending
    \State $\mathbf{Z} \gets$ top-1 with $|r| < 0.85$ vs.\ already selected
    \State $d \gets h + |\mathbf{Z}|$;
           $k \gets \max(10, \lfloor N_\tau^{0.4} \rfloor, 2d+6)$
      \Comment{Eq.~\eqref{eq:s_adaptive_k}}
    \State Build embeddings; rank-transform; jitter
    \State $\hat{T}_{X\to Y}^{(\tau)} \gets \textsc{KSG-CMI}(
      \tilde{\mathbf{x}}, \tilde{\mathbf{y}}, \tilde{\mathbf{z}}, k, W)$
      \Comment{Algorithm~\ref{alg:s_ksg}}
    \For{$b = 1, \dots, B$}
      \State $\delta_b \sim \mathrm{Uniform}\{1, \dots, N_\tau-1\}$
      \State $\tilde{\mathbf{x}}^{(b)} \gets \textsc{CircularShift}(\tilde{\mathbf{x}}, \delta_b)$
        \Comment{Eq.~\eqref{eq:s_circular_shift}}
      \State $\hat{T}^{(b)} \gets \textsc{KSG-CMI}(
        \tilde{\mathbf{x}}^{(b)}, \tilde{\mathbf{y}}, \tilde{\mathbf{z}}, k, W)$
    \EndFor
    \State $\mathrm{ETE}_X^{(\tau)} \gets
      \hat{T}_{X\to Y}^{(\tau)} - \tfrac{1}{B}\sum_b \hat{T}^{(b)}$
      \Comment{Eq.~\eqref{eq:s_ete}}
    \State $p_X^{(\tau)} \gets \tfrac{1}{B}\sum_b
      \mathbf{1}[\hat{T}^{(b)} \geq \hat{T}_{X\to Y}^{(\tau)}]$
  \EndFor
\EndFor
\State $\{p_{\mathrm{adj}}\} \gets
  \textsc{Benjamini--Hochberg}(\{p_X^{(\tau)}\}, \alpha)$
\State \Return $\{\mathrm{ETE}_X^{(\tau)}, p_{\mathrm{adj},X}^{(\tau)}\}$
\end{algorithmic}
\end{algorithm}

\begin{algorithm}[t]
\caption{\textsc{KSG-CMI}: Frenzel--Pompe CMI with Theiler window}
\label{alg:s_ksg}
\begin{algorithmic}[1]
\Require Rank-transformed
$\tilde{\mathbf{x}}\in\mathbb{R}^N$,
$\tilde{\mathbf{y}}\in\mathbb{R}^N$,
$\tilde{\mathbf{z}}\in\mathbb{R}^{N\times d}$;
neighbours $k$; Theiler window $W$
\Ensure $\hat{I}(\tilde{\mathbf{x}}; \tilde{\mathbf{y}} \mid \tilde{\mathbf{z}})$ in nats
\State Build $k$-d tree $\mathcal{T}_{\mathrm{joint}}$ on
$[\tilde{\mathbf{x}}, \tilde{\mathbf{y}}, \tilde{\mathbf{z}}]$
using Chebyshev norm
\For{$i = 1, \dots, N$}
  \State Query $k + 2W' + 1$ neighbours, $W' = \max(1, W)$
  \State Discard neighbours $j$ with $|j-i| < W'$
  \State $\varepsilon_i \gets$ Chebyshev distance to $k$-th surviving neighbour
  \State $\varepsilon_i \gets \varepsilon_i(1 - 10^{-10})$
    \Comment{open-ball correction}
\EndFor
\State Build marginal trees $\mathcal{T}_{\mathbf{xz}},
  \mathcal{T}_{\mathbf{yz}}, \mathcal{T}_{\mathbf{z}}$
\For{$i = 1, \dots, N$}
  \State $n_{\mathbf{xz}}(i), n_{\mathbf{yz}}(i), n_{\mathbf{z}}(i)
    \gets |\{j : \|\cdot\|_\infty < \varepsilon_i \wedge |j-i| \geq W'\}|$
\EndFor
\State \Return
$\psi(k) - \tfrac{1}{N}\sum_i [\psi(n_{\mathbf{xz}}(i)+1)
+ \psi(n_{\mathbf{yz}}(i)+1) - \psi(n_{\mathbf{z}}(i)+1)]$
  \Comment{Eq.~\eqref{eq:s_frenzel_pompe}}
\end{algorithmic}
\end{algorithm}

\subsection*{Accumulated Local Effects}
\label{ssec:smethods_ale}

This subsection holds the full mathematical derivation for the ALE
interpretability analysis summarised in main Methods, ``XGBoost
seasonal nowcaster''. ALE \citep{apley2020visualizing} quantifies
the marginal effect of each feature on the predicted class
probabilities while avoiding the extrapolation into unobserved feature combinations that biases partial-dependence plots
\citep{friedman2001greedy} when predictors are correlated, as in meteorological feature sets where
wind speed, temperature and stability indices are physically coupled.

\paragraph{Definition.}
For a continuous feature $x_s$, the first-order ALE is the centred
accumulated integral of the local partial derivative of the
prediction function $\hat{f}$ along the feature axis:
\begin{equation}
\widehat{\mathrm{ALE}}(x_s)
= \int_{x_{\min}}^{x_s}\!
\mathbb{E}\!\left[\frac{\partial \hat{f}(\mathbf{X})}{\partial X_s}
\;\middle|\; X_s = z_s\right]\!
\mathrm{d}z_s - \mathrm{const.}
\label{eq:s_ale}
\end{equation}
In practice, the feature range is partitioned into $K$ equal-count
intervals $\{[z_{k-1}, z_k)\}_{k=1}^{K}$, and the local effect within
each interval is estimated by finite differences over samples in
that bin:
\begin{equation}
\widehat{\mathrm{ALE}}(x_s)
= \sum_{k=1}^{k_{x_s}} \frac{1}{n_k}
\sum_{i:\,x_{i,s}\in[z_{k-1}, z_k)}\!
\bigl[\hat{f}(z_k, \mathbf{x}_{i,\setminus s})
- \hat{f}(z_{k-1}, \mathbf{x}_{i,\setminus s})\bigr]
- \mathrm{const.},
\label{eq:s_ale_discrete}
\end{equation}
where $n_k$ is the count of samples in bin $k$, $k_{x_s}$ the bin
index containing $x_s$, and $\mathbf{x}_{i,\setminus s}$ all features
except $x_s$ held at their observed values. The centring constant
ensures $\mathbb{E}[\widehat{\mathrm{ALE}}] = 0$ over the data
distribution. ALE perturbs only within observed intervals and
conditions on the empirical neighbourhood, avoiding extrapolation
into unlikely regions of feature space \citep{apley2020visualizing}.

\paragraph{Multiclass aggregation.}
For the three-class XGBoost classifier, ALE is computed separately
for each class $c \in \{\mathrm{Low}, \mathrm{Medium}, \mathrm{High}\}$
using the class-specific soft probability
$\hat{f}_c(\mathbf{x}) = P(Y = c \mid \mathbf{x})$ from the
\texttt{multi:softprob} XGBoost output, yielding three ALE curves
per feature per seasonal classifier.

\paragraph{Feature importance.}
Importance is the bin-size-weighted variance of the ALE effect:
\begin{equation}
I_s^{(c)} = \sum_{k=1}^{K} \frac{n_k}{N}
\bigl(\widehat{\mathrm{ALE}}_k^{(c)}
- \overline{\mathrm{ALE}}^{(c)}\bigr)^{2},
\quad
\overline{\mathrm{ALE}}^{(c)}
= \sum_k \frac{n_k}{N}\widehat{\mathrm{ALE}}_k^{(c)}.
\label{eq:s_ale_importance}
\end{equation}
The total importance is $I_s = \sum_c I_s^{(c)}$ and the class-mean
importance reported in Fig.~4a is $\bar{I}_s = I_s / 3$. ALE plots
are computed on held-out validation data using the \texttt{PyALE}
library with $K = 20$ quantile-based grid points, independently per
seasonal classifier. For the cross-seasonal comparison in
Fig.~4(a,b), per-class ALE curves and importance values are
averaged across the three classes at each grid point.

\subsection*{S4D state space model}
\label{ssec:smethods_s4d}

This subsection holds the full mathematical foundations of the S4D
state space architecture summarised in main Methods, ``S4 dual-pathway
nowcaster'' (Eqs.~\ref{eq:ssm_kernel}--\ref{eq:dt_anchor} of the main
text). The S4 family was introduced in \citet{gu2021efficiently} and
\citet{gu2020hippo}; we use the diagonal (S4D) simplification of
\citet{gu2022parameterization}.

\paragraph{Continuous-time SSM.}
The S4 architecture represents the mapping from a one-dimensional
input signal $u(t) \in \mathbb{R}$ to an output signal
$y(t) \in \mathbb{R}$ as a continuous-time linear time-invariant
dynamical system \citep{gu2021efficiently}:
\begin{align}
\dot{\mathbf{x}}(t) &= \mathbf{A}\,\mathbf{x}(t) + \mathbf{B}\,u(t),
\label{eq:s_ssm_continuous_x}\\
y(t) &= \mathbf{C}\,\mathbf{x}(t) + D\,u(t),
\label{eq:s_ssm_continuous_y}
\end{align}
with latent state $\mathbf{x}(t) \in \mathbb{C}^{N}$,
$\mathbf{A} \in \mathbb{C}^{N\times N}$,
$\mathbf{B} \in \mathbb{C}^{N\times 1}$,
$\mathbf{C} \in \mathbb{C}^{1\times N}$ and
$D \in \mathbb{R}$.

\paragraph{Zero-order-hold discretisation.}
For a discrete uniformly sampled input $u_k = u(k\Delta)$ at
step $\Delta > 0$, the ZOH discretisation
\citep{gu2021efficiently} of
Eqs.~\eqref{eq:s_ssm_continuous_x}--\eqref{eq:s_ssm_continuous_y}
yields the recurrence
\begin{align}
\mathbf{x}_k &= \bar{\mathbf{A}}\mathbf{x}_{k-1} + \bar{\mathbf{B}}u_k,
\qquad y_k = \mathbf{C}\mathbf{x}_k + D\,u_k,
\label{eq:s_ssm_recurrence}\\
\bar{\mathbf{A}} &= \exp(\Delta\mathbf{A}),
\qquad \bar{\mathbf{B}} = (\Delta\mathbf{A})^{-1}
(\exp(\Delta\mathbf{A}) - \mathbf{I})\,\Delta\,\mathbf{B}.
\label{eq:s_ssm_discretisation}
\end{align}
Unrolling Eq.~\eqref{eq:s_ssm_recurrence} from a zero initial state
gives the equivalent convolutional form
\begin{equation}
y_k = \sum_{j=0}^{k}
\mathbf{C}\,\bar{\mathbf{A}}^{\,j}\,\bar{\mathbf{B}}\, u_{k-j}
+ D\,u_k
= (\bar{\mathbf{K}} \ast u)_k + D\,u_k,
\label{eq:s_ssm_conv}
\end{equation}
with the SSM convolution kernel given in main text
Eq.~\eqref{eq:ssm_kernel}. The kernel decomposition has three
properties exploited directly: (i)~strict causality, since
$j \in \{0, \dots, k\}$; (ii)~$\mathcal{O}(L\log L)$ inference cost
via FFT-based convolution, enabling the $L = 384$-step (96-h)
context without quadratic scaling; (iii)~a continuum of learned
timescales governed by
$\bar{\mathbf{A}} = \exp(\Delta\mathbf{A})$, with slow coordinates
($|\mathrm{Re}(\lambda_i)|\ll 1$) holding information over many
time steps and fast coordinates ($|\mathrm{Re}(\lambda_i)|\gg 1$)
decaying rapidly.

\paragraph{Diagonal (S4D) parameterisation.}
The S4D simplification of \citet{gu2022parameterization} restricts
$\mathbf{A}$ to be diagonal, retaining the HiPPO-LegS spectrum that
underpins long-memory behaviour while eliminating the
normal-plus-low-rank kernel construction. With
$\mathbf{A} = \mathrm{diag}(\lambda_1, \dots, \lambda_N)$,
the kernel reduces to a complex Vandermonde evaluation:
\begin{equation}
\bar{K}_k = 2\,\mathrm{Re}\!\left[\sum_{n=1}^{N/2}
C_n \cdot \frac{\exp(k\Delta\lambda_n) - 1}{\lambda_n}
\cdot \exp((k-1)\Delta\lambda_n)\right].
\label{eq:s_s4d_kernel}
\end{equation}
The eigenvalues are parametrised as
\begin{equation}
\lambda_n = -\exp(\log A^{\mathrm{real}}_n) + \mathrm{i}\pi(n-1),
\quad n = 1, \dots, N/2,
\label{eq:s_lambda_param}
\end{equation}
so that $\mathrm{Re}(\lambda_n) < 0$ is enforced by construction
(stability) and gradients remain well-scaled across many orders of
magnitude in $|A|$. The step size is similarly parametrised as
$\Delta = \exp(\log\Delta)$; $\mathbf{C}$ is stored as
real/imaginary parts with Xavier-scale initialisation
$C_{\mathrm{std}} = 1/\sqrt{HN}$ \citep{glorot2010understanding} to
prevent the readout from dominating at initialisation.
Equation~\eqref{eq:s_s4d_kernel} is assembled with a numerically
stable real-arithmetic factorisation: $\mathrm{expm1}(a) = e^a - 1$
is used to retain precision when $\mathrm{Re}(\Delta\lambda)$ is
small, and the decay/phase accumulation $\exp(k\Delta\lambda)$ is
computed in forced FP32 to prevent TF32 tensor-core truncation from
collapsing slow-lane values.

\paragraph{S4D block.}
Each S4D block wraps the SSM kernel in a pre-normalised, gated
residual structure following the template of contemporary sequence
models \citep{vaswani2017attention, xiong2020layer, dauphin2017language}.
For block input $\mathbf{U} \in \mathbb{R}^{B\times L\times H}$:
\begin{align}
\tilde{\mathbf{U}} &= \mathrm{LayerNorm}(\mathbf{U}),
\label{eq:s_block_ln}\\
\mathbf{Y}_{\mathrm{ssm}}[:,:,h]
&= (\bar{\mathbf{K}}^{(h)} \ast \tilde{\mathbf{U}}[:,:,h])
+ D_h\,\tilde{\mathbf{U}}[:,:,h], \;\; h = 1,\dots,H,
\label{eq:s_block_ssm}\\
\mathbf{Y}_{\mathrm{drop}}
&= \mathrm{Dropout}(\mathbf{Y}_{\mathrm{ssm}}),
\label{eq:s_block_drop}\\
[\mathbf{V};\mathbf{G}]
&= \mathbf{W}_{\mathrm{glu}}\,\mathbf{Y}_{\mathrm{drop}}
+ \mathbf{b}_{\mathrm{glu}},
\label{eq:s_block_glu_proj}\\
\mathbf{Z} &= \mathbf{V} \odot \sigma(\mathbf{G}),
\label{eq:s_block_glu}\\
\mathbf{U}' &= \mathbf{U} +
\mathbf{W}_{\mathrm{out}}\mathbf{Z} + \mathbf{b}_{\mathrm{out}},
\label{eq:s_block_residual}\\
\mathbf{U}'' &= \mathbf{U}' +
\mathrm{FFN}(\mathrm{LayerNorm}(\mathbf{U}')),
\label{eq:s_block_ffn}
\end{align}
where $\sigma$ is the logistic sigmoid, $\odot$ is element-wise
product, and the FFN is a two-layer GELU network
\citep{hendrycks2016gaussian} of inner width $2H$. Pre-Norm rather
than Post-Norm \citep{xiong2020layer} is used for stable
gradient flow in residual stacks and normalises each block's input,
including the long-tailed H$_2$S signal, before it reaches the SSM
kernel. Gated Linear Units with sigmoid gate
\citep{dauphin2017language} preserve amplitude information better
than a pointwise non-linearity, so that slow-lane channels are not suppressed by the gate even when their instantaneous
magnitude is small. Zero-initialised skip ($D_h = 0$) forces the
block, at the start of training, to route information through the
SSM memory pathway rather than through a direct feedthrough. To
regularise the shallow ($n_{\mathrm{layers}} = 3$) stack we apply
stochastic depth \citep{huang2016deep} with linearly increasing
per-layer drop probability
$p_\ell = \ell\cdot p_{\max}/(n_{\mathrm{layers}}-1)$ for
$p_{\max} = 0.15$, scaling the residual update by an independent
Bernoulli mask at each forward pass.

\paragraph{End-to-end model.}
For an input window $\mathbf{X}\in\mathbb{R}^{B\times L\times F}$ of
$L = 384$ steps and $F$ standardised features, the full
architecture is
\begin{align}
\mathbf{H}^{(0)} &=
\mathbf{X}\mathbf{W}_{\mathrm{in}} + \mathbf{b}_{\mathrm{in}},
\quad \mathbf{W}_{\mathrm{in}}\in\mathbb{R}^{F\times H},\\
\mathbf{H}^{(\ell)} &=
\mathbf{H}^{(\ell-1)} + \mathrm{DropPath}_{p_\ell}\!\bigl(
\mathrm{S4DBlock}_\ell(\mathbf{H}^{(\ell-1)}) - \mathbf{H}^{(\ell-1)}\bigr),
\quad \ell = 1, \dots, n_{\mathrm{layers}},\\
\mathbf{h}_{L-1} &=
\mathbf{H}^{(n_{\mathrm{layers}})}_{:,L-1,:}
\in \mathbb{R}^{B\times H},
\label{eq:s_causal_pool}\\
\hat{y}_{\mathrm{reg}} &=
\mathrm{MLP}_{\mathrm{reg}}(\mathbf{h}_{L-1}),
\quad
\hat{\mathbf{p}} = \mathrm{softmax}\!\bigl(
\mathrm{MLP}_{\mathrm{cls}}(\mathbf{h}_{L-1})\bigr).
\label{eq:s_dual_heads}
\end{align}
Both heads are
LayerNorm$\to$Linear$(H, H/2)\to$GELU$\to$Dropout$\to$Linear$(H/2, \cdot)$
with zero-initialised final layers, so the model starts training
with uniform class probabilities and zero-mean regression
output.

\paragraph{Causality guarantees.}
Strict causality is enforced at three points: (i)~the SSM kernel in
Eq.~\eqref{eq:s_s4d_kernel} is evaluated at non-negative time indices
$k = 0, \dots, L-1$ only, with kernel length asserted equal to input
length so no padding can introduce an acausal shift; (ii)~the FFT
convolution in Eq.~\eqref{eq:s_block_ssm} is computed with an
explicit $2L$-length FFT and truncated to the first $L$ outputs,
giving the same result as a direct causal convolution
$y_k = \sum_{j=0}^{k} K_j\,u_{k-j}$ with no bidirectional, circular
or future-padded variant; (iii)~the sequence representation passed
to the heads is the final-timestep slice $\mathbf{h}_{L-1}$
(Eq.~\eqref{eq:s_causal_pool}), not a mean or max over the full
window, preserving a single well-defined ``prediction-time''
receptive field.

\paragraph{Hyperparameters.}
The configuration used for the S4 results in main-text Tables~\ref{tab:ed_s4_six_conditions}
and \ref{tab:ed_s4_period_aggregate} and in Supplementary
Table~\ref{tab:ed_multisite_tier} is:
$H = 128$ (model width),
$N = 256$ (state size),
$n_{\mathrm{layers}} = 3$,
$L = 384$ (15-min steps; 96-h context),
$f = 0.5$ (fast/slow channel split),
$|A^{\mathrm{real}}_{\mathrm{fast}}| = 10.0$,
$|A^{\mathrm{real}}_{\mathrm{slow}}| = 0.5$,
log-space spread $s = 0.5$,
dropout $0.29$,
DropPath $p_{\max} = 0.15$,
focal-loss $\gamma = 2.62$,
class weights
$(\alpha_{\mathrm{Low}}, \alpha_{\mathrm{Med}}, \alpha_{\mathrm{High}})
= (1.00, 11.26, 14.07)$.

\subsection*{Optimiser schedule}
\label{ssec:smethods_optim}

This subsection holds the full optimiser configuration for the
S4 dual-pathway nowcaster summarised in main Methods, ``S4
dual-pathway nowcaster''.

\paragraph{Training loss.}
The training objective is a weighted sum of focal cross-entropy
classification \citep{lin2017focal} and standardised MSE regression:
\begin{equation}
\mathcal{L} = \lambda_{\mathrm{cls}}\,\mathcal{L}_{\mathrm{FL}}(\hat{\mathbf{p}}, y)
+ \lambda_{\mathrm{reg}}\,
\mathcal{L}_{\mathrm{MSE}}(\hat{y}_{\mathrm{reg}},
y_{\mathrm{reg}}^{\mathrm{std}}),
\label{eq:s_joint_loss}
\end{equation}
with $\lambda_{\mathrm{cls}} = 1.0$ and $\lambda_{\mathrm{reg}} = 0.0$
in all reported runs, so training is classification-only: the
regression head receives no gradient, its zero-initialised final layer
outputs a constant, and its output enters no reported result or
interpretability analysis. The focal cross-entropy term is
\begin{equation}
\mathcal{L}_{\mathrm{FL}}(\hat{\mathbf{p}}, y)
= -\frac{1}{B}\sum_{i=1}^{B} \alpha_{y_i}
(1-\hat{p}_{i,y_i})^{\gamma}\,\log\hat{p}_{i,y_i},
\label{eq:s_focal}
\end{equation}
with $\gamma = 2.62$ and per-class weights
$(\alpha_{\mathrm{Low}}, \alpha_{\mathrm{Med}}, \alpha_{\mathrm{High}}) =
(1.00, 11.26, 14.07)$. Weights are imposed manually rather than
derived from inverse frequencies and deliberately allocate more
weight to the High class than its $\sim$9:1 inverse prior, with the
focusing factor down-weighting easy high-confidence examples so that
gradient updates concentrate on borderline samples between Medium
and High. This design targets High-class precision and
recall, the metrics of primary interest for exceedance nowcasting. In parallel, an
inverse-prior weighted minibatch sampler renders the class marginal
approximately uniform within each batch despite the
$\sim$13:2:1 Low:Medium:High imbalance, a two-pronged correction combining sampling and weighting \citep{he2009learning, cui2019class}.

\paragraph{Differential learning-rate groups.}
Optimisation uses AdamW \citep{loshchilov2019decoupled} with five
parameter groups chosen to stabilise the learning of the S4D
timescales. Non-SSM parameters (input projection, FFN layers,
output heads) use base learning rate
$\eta_{\mathrm{base}} = 2.31 \times 10^{-2}$ and weight decay
$2.7 \times 10^{-2}$; the $\mathbf{C}$ readout forms a separate group at
$\eta_{\mathrm{base}}$ without weight decay. SSM log-step ($\log\Delta$) and imaginary-A
parameters use a much smaller $\eta_{\mathrm{s4}} = 10^{-4}$ to
prevent rapid migration of the timescale prior away from the
physics-anchored initialisation. SSM real-A parameters
($\log A^{\mathrm{real}}$) use $\eta_A = 10^{-4}$ (the same rate as the other SSM parameters, with no additional
multiplier).
The skip parameter $D$ uses $\eta_D = 0.1\,\eta_{\mathrm{base}}$ to
prevent it from short-circuiting the SSM memory path. Gradients are
clipped globally at norm 1.0.

\paragraph{Warmup and freeze schedule.}
For the first $E_{\mathrm{w}} = 6$ warmup epochs, $\log A^{\mathrm{real}}$ and
$A^{\mathrm{imag}}$ are frozen, allowing $\mathbf{C}$ to learn to
discriminate the fixed physics-anchored timescales before the $A$
spectrum is allowed to migrate; this prevents the optimiser's
early high-loss gradient from collapsing the slow-lane memory
during the first epoch. When warmup ends at epoch $E_{\mathrm{w}}$, the $A$
matrix is unfrozen and $\eta_{\mathrm{base}}$ is transiently
multiplied by a factor of 3.0 to accelerate
subsequent adaptation of the timescales. A reduce-on-plateau learning-rate
scheduler then monitors inner-probe High-class F$_1$ and halves the
learning rate after $p_{\mathrm{plateau}} = 2$ epochs without
improvement, down to a floor of $10^{-5}$. Early stopping monitors
the same metric with patience 7 and restores the best-epoch weights.
Maximum training is 60 epochs. The inner probe is the seven days immediately preceding the evaluation week, withheld from training, so the evaluation week is never monitored. The four single-site models fused in the multi-site tier (Supplementary Note~6) are instead retrained on the full training window for the number of epochs selected on the inner probe, with the learning-rate schedule of the probe run reapplied epoch by epoch, without early stopping, and with the final-epoch weights retained. The cross-species CH$_4$ model of main Methods is the exception: its learning-rate scheduler and early stopping both monitor the evaluation week, so its reported values are optimistic, as stated in main Methods.

\subsection*{Walk-forward validation}
\label{ssec:smethods_walkforward}

This subsection holds the per-week pipeline, leakage controls,
metric definitions and post-hoc attribution procedures (lane
ablation and feature sensitivity) for the walk-forward evaluation
summarised in main Methods, ``Walk-forward validation and metrics''.

\paragraph{Validation horizon and training cutoff.}
The validation horizon (1 October--31 December 2024) is partitioned
into 13 non-overlapping weekly blocks
$\{\mathcal{V}_w\}_{w=1}^{13}$ with
$t_{w+1}^{\mathrm{start}} = t_w^{\mathrm{end}} + 1\,\mathrm{day}$.
For each block $w$, the training set is the expanding window
\begin{equation}
\mathcal{T}_w = \{t \in \mathcal{D}_{\mathrm{all}} :
t \leq t_w^{\mathrm{start}} - L\,\Delta t_{\mathrm{base}}\},
\label{eq:s_train_cutoff}
\end{equation}
with $\mathcal{D}_{\mathrm{all}}$ the full 15-min record from
2023-09-01. The training cutoff lies $L = 384$ steps
($\approx 96$\,h) before the first validation timestamp; the
intervening $L-1$ steps form a context reserve $\mathcal{C}_w$
used \emph{only as input} to the sliding-window convolution that
produces the first prediction at $t_w^{\mathrm{start}}$. No label in
$\mathcal{C}_w$ contributes to the training loss.

\paragraph{Per-week pipeline.}
For each of the 13 weeks the following procedure is executed
independently:
\begin{enumerate}[leftmargin=*, itemsep=2pt]
\item Build features on $\mathcal{D}_{\mathrm{all}}$ once (hour
sin/cos, WD sin/cos, four Fourier seasonality harmonics), then
partition into $\mathcal{T}_w$ and
$\mathcal{V}_w \cup \mathcal{C}_w$ via Eq.~\eqref{eq:s_train_cutoff}.
\item Drop rows with missing labels or features.
\item Fit a per-feature standardisation on $\mathcal{T}_w$ alone
(zero-mean, unit-variance per feature) and apply to
$\mathcal{V}_w \cup \mathcal{C}_w$. The training regression target
is standardised using $\mathcal{T}_w$ statistics only.
\item Instantiate a fresh model (no carry-over of weights, optimiser
state or LR schedule from week $w-1$) and train for up to 60 epochs
with the objective and optimiser of Supplementary Methods,
``Optimiser schedule'', monitoring F$_1$-High on the inner probe for early stopping (the nowcasters fused in Supplementary Note~6 are instead retrained for a fixed number of epochs selected on the inner probe).
\item Compute predicted class probabilities $\hat{\mathbf{p}}_i$ at
every timestamp $t_i \in \mathcal{V}_w$ and persist alongside true
labels, hard predictions and timestamps.
\item Compute week-$w$ metrics (definitions below).
\end{enumerate}
Period aggregates are computed by pooling the confusion counts over all 8,040 evaluation timesteps (main-text Table~\ref{tab:ed_s4_period_aggregate}); Supplementary Table~\ref{tab:si_s4_xgb_weekly} additionally reports the unweighted mean of the 13 weekly values.

\paragraph{Leakage controls.}
Six leakage pathways are controlled by design:
\begin{itemize}[leftmargin=*, itemsep=2pt]
\item \emph{Label leakage:} Eq.~\eqref{eq:s_train_cutoff} guarantees
that no training sample has a label whose timestamp falls within
$\mathcal{V}_w \cup \mathcal{C}_w$.
\item \emph{Feature leakage:} every feature is either a pointwise
function of the current timestamp (hour sin/cos, Fourier seasonality,
WD sin/cos) or the raw meteorology itself; no rolling mean, lag,
derivative or stagnation index is present to carry information
across the train/validation boundary, and no scaling or target
statistic is fitted on validation data.
\item \emph{Model leakage:} the SSM kernel, FFT convolution and
final-timestep pool are strictly causal; no bidirectional,
attention-based or global-pool operation is used; the convolution kernel length is constrained to equal the input length, which excludes acausal padding.
\item \emph{Hyperparameter leakage:} the hyperparameter set used for
the 13 weekly retrains is fixed in advance, carried from an
independent hyperparameter search on Sep~2023--Sep~2024 data whose
validation period does not overlap with $\bigcup_w \mathcal{V}_w$;
a Week-0 sanity check compares the first weekly retrain against the
hyperopt's held-out test metrics and flags any relative macro-F$_1$
discrepancy exceeding 20\%.
\item \emph{Retrain leakage:} each week's model is instantiated from
scratch (weights, optimiser state and LR schedule all
reinitialised), so no information from week $w$'s validation period
can re-enter week $w' > w$ via carried-over parameters.
\item \emph{Selection leakage:} the stopping epoch, the restored weights and the learning-rate schedule are chosen on the inner probe, which lies inside $\mathcal{T}_w$ and ends before $\mathcal{C}_w$; the nowcasters fused in Supplementary Note~6 are retrained on $\mathcal{T}_w$ for the number of epochs selected on the inner probe, with the same learning-rate schedule, so no training or selection quantity is chosen on $\mathcal{V}_w$. The cross-species CH$_4$ model, which uses checkpoint selection on the evaluation week, is the one exception (main Methods).
\end{itemize}

\paragraph{Metric definitions.}
For each weekly block we report hard-classification metrics
(per-class precision $P_c$, recall $R_c$, F$_1^{(c)}$; macro F$_1$;
false-alarm rate FAR), probabilistic metrics (Brier-High, binary
ranked probability score, log-loss), and conditional-rolling metrics
aligned with the WHO 30-minute averaging convention. Define
$y_i^{(c)} = \mathbf{1}(y_i = c)$ as the one-hot true label and
$\hat{p}_i^{(c)}$ the model's predicted probability of class $c$.

The Brier score for the High class (binary one-vs-rest) is the
mean-squared probabilistic error \citep{brier1950verification}:
\begin{equation}
\mathrm{Brier\text{-}High}
= \frac{1}{N}\sum_{i=1}^{N}
\bigl(\hat{p}_i^{(\mathrm{High})} - y_i^{(\mathrm{High})}\bigr)^{2}.
\label{eq:s_brier_high}
\end{equation}
The binary ranked probability score \citep{epstein1969scoring} for
the ordinal Low~$<$~Medium~$<$~High classes is
\begin{equation}
\mathrm{RPS}_i
= \frac{1}{K-1}\sum_{k=1}^{K-1}
\bigl(\hat{F}_i(k) - F_i(k)\bigr)^{2},
\quad
\hat{F}_i(k) = \sum_{c \leq k} \hat{p}_i^{(c)},
\;\; F_i(k) = \sum_{c \leq k} y_i^{(c)},
\label{eq:s_rps}
\end{equation}
with $K=3$. The multiclass log-loss is
\begin{equation}
\mathrm{LL} = -\frac{1}{N}\sum_{i=1}^{N}\sum_{c=1}^{K}
y_i^{(c)}\log\hat{p}_i^{(c)}.
\label{eq:s_logloss}
\end{equation}
Conditional-rolling F$_1$/precision/recall and FAR are computed in a
4-h moving window aligned with the WHO 30-minute averaging
convention. With window half-width
$T_w = 8$ samples (2\,h on each side, 16 samples total at 15-min
cadence) centred on timestep $t$,
\begin{equation}
M^{\mathrm{roll}}(t) =
M\!\left(\{(y_j, \hat{y}_j)\}_{j: |j-t| \leq T_w}\right),
\label{eq:s_cond_rolling}
\end{equation}
where $M \in \{\mathrm{F}_1, P, R, \mathrm{FAR}\}$ and the function
on the right-hand side is the standard hard-classification metric
applied to the windowed predictions and labels. The rolling metrics
are evaluated only at timesteps for which the window contains
$\geq T_w$ samples. False-alarm rate is
$\mathrm{FAR} = \mathrm{FP}/(\mathrm{FP} + \mathrm{TN})$ for the
High vs.\ non-High binary partition.

\paragraph{Lane ablation.}
To attribute the trained model's High-class skill to each
architectural lane, an inference-time ablation is applied to every
weekly checkpoint. For S4D layer $\ell$, the ablation takes the SSM kernel
$\bar{\mathbf{K}}^{(\ell)} \in \mathbb{R}^{H\times L}$ and zeros the
rows corresponding to the targeted lane:
\begin{equation}
\bar{\mathbf{K}}^{(\ell)}_{\mathrm{ablated}}[h, k]
= \begin{cases}
0, & h \in \mathcal{I}_{\mathrm{lane}} \\
\bar{\mathbf{K}}^{(\ell)}[h, k], & \text{otherwise},
\end{cases}
\label{eq:s_lane_mask}
\end{equation}
where $\mathcal{I}_{\mathrm{fast}} = \{0, \dots, fH - 1\}$ and
$\mathcal{I}_{\mathrm{slow}} = \{fH, \dots, H - 1\}$ are the fixed
channel indices of the two lanes. The mask is applied after kernel
construction but before FFT convolution
(Eq.~\eqref{eq:s_block_ssm}), so the SSM contribution of the masked
lane is suppressed while the $D$-skip path, GLU gate, output
projection and downstream FFN cross-channel mixing remain intact.
This isolates the SSM contribution of the lane rather than the
entire signal-flow pathway.

\textbf{Method dependence of the lane attribution.}
\label{ssec:si_lane_ablation_method}
Under the kernel-path mask of Eq.~\eqref{eq:s_lane_mask} the lane-zeroing ablation apportions $+0.083$ to the fast lane, $+0.074$ to the slow lane and $+0.121$ to their interaction, sample-weighted across the 13 weeks. The two lanes are \emph{not} separately distinguishable: the paired per-fold difference is $-0.002$ (fast leads on 8 of 13 folds; sign test $p = 0.58$, Wilcoxon $p = 1.00$; bootstrap 95\% CI $[-0.079, +0.072]$), the interaction exceeds either marginal contribution, and the ordering reverses between aggregation rules. These values therefore do not separate the contributions of the two lanes and do not attribute a mechanism to either lane.
The attribution above is sensitive to where in the signal path the mask is applied, because the mask location determines the sign of the fast/slow comparison. Equation~\eqref{eq:s_lane_mask} specifies a
\emph{kernel-path} mask: the targeted rows of
$\bar{\mathbf{K}}^{(\ell)}$ are zeroed after kernel construction and
before FFT convolution, leaving the $D$-skip path and GLU gate
intact. An alternative \emph{block-path} formulation, in which the
whole S4 block output for the targeted channels is suppressed,
additionally removes the skip and gate contributions and therefore
attributes to each lane every downstream pathway that lane feeds.
Under the block-path formulation the ordering of the two lanes
reverses relative to the kernel-path values, with the fast lane
carrying the larger contribution. The two formulations answer different questions: the kernel-path mask isolates the SSM contribution, the block-path mask the total pathway contribution. Neither is uniquely correct. The values above should accordingly
be read as a statement about the SSM contribution under the
kernel-path definition, and not as evidence that either lane encodes
a specific atmospheric mechanism.

Four forward passes per checkpoint give the metric values
$M_{\mathrm{full}}, M_{\mathrm{no\_fast}}, M_{\mathrm{no\_slow}},
M_{\mathrm{no\_both}}$. The per-lane contribution is
\begin{equation}
\Delta M_{\mathrm{lane}}
= M_{\mathrm{full}} - M_{\mathrm{no\_lane}},
\label{eq:s_lane_contrib}
\end{equation}
positive when the lane increases $M$, and the interaction term
\begin{equation}
\mathcal{I}_M = (M_{\mathrm{full}} - M_{\mathrm{no\_both}})
- (\Delta M_{\mathrm{fast}} + \Delta M_{\mathrm{slow}})
\label{eq:s_lane_interaction}
\end{equation}
is positive when the two lanes are complementary, zero when they
act independently, and negative when partially redundant. Reported
aggregates are sample-weighted means across
the 13 weeks for F$_1$-High and macro F$_1$.

\paragraph{Feature sensitivity over time.}
Per-feature per-lane sensitivity is computed analytically from each
weekly checkpoint. The S4D kernel
$\bar{\mathbf{K}}^{(\ell)} \in \mathbb{R}^{H\times L}$ is
reconstructed from the saved
$\{\log\Delta^{(\ell)}, \log A^{\mathrm{real},(\ell)},
A^{\mathrm{imag},(\ell)}, \mathbf{C}^{(\ell)}\}$ via
Eq.~\eqref{eq:s_s4d_kernel}, and combined with the input projection
$\mathbf{W}_{\mathrm{in}}$ to give the per-lane sensitivity of
feature $f$ at lag $k$:
\begin{equation}
S_{\mathrm{lane}}(f, k) =
\sum_{h \in \mathcal{I}_{\mathrm{lane}}}
|W_{\mathrm{in}}[h, f]| \cdot |\bar{\mathbf{K}}^{(\ell)}[h, k]|,
\quad k = 0, \dots, L-1.
\label{eq:s_feature_sensitivity}
\end{equation}
Column-normalisation at each lag gives the fractional contribution
of each feature to the lane's response:
\begin{equation}
\widetilde{S}_{\mathrm{lane}}(f, k) =
\frac{S_{\mathrm{lane}}(f, k)}
     {\sum_{f'} S_{\mathrm{lane}}(f', k)} \times 100\%,
\label{eq:s_feature_sensitivity_norm}
\end{equation}
suitable for visualisation as stacked-area charts over the 96-h
input window (Fig.~5c). The formulation is causal ($k=0$ is
present, $k=L-1$ is the earliest input), respects the lane
partition, and is linear in the feature step but multiplicative
through the kernel, highlighting the (feature, lag, lane) combinations with the largest combined input-weight and kernel magnitudes in the trained model. Per-week
sensitivity surfaces are averaged across the 13 walk-forward folds
before plotting to suppress single-checkpoint noise.

\clearpage
\section*{Supplementary Tables}
\addcontentsline{toc}{section}{Supplementary Tables}

\begin{table}[htbp]
\centering
\caption{\textbf{Quantitative summary of the environmental
characterisation of receptor-level H$_2$S exposure (main Fig.~1).}
All statistics are evaluated at the principal receptor (MMF9,
$n = 34{,}609$ 15-minute records with valid H$_2$S across 2024) unless otherwise
indicated.}
\label{tab:si_envchar_summary}
\footnotesize
\setlength{\tabcolsep}{4pt}
\begin{tabular}{@{}llp{6cm}@{}}
\toprule
Panel & Statistic & Value \\
\midrule
\multicolumn{3}{@{}l}{\textit{Source attribution (Fig.~1a)}} \\
& Peak CPF bearing                & $265^{\circ}\ \mathrm{(W)}$ \\
& Peak CPF value                  & $\approx 0.37$ \\
& Peak/opposing-sector ratio      & $4.09\times$ \\
& Sector-mean (W, NW, E)          & $11.2$, $10.6$, $1.1\,\mu\mathrm{g\,m^{-3}}$ \\
& Spike threshold (P95)           & $15.3\,\mu\mathrm{g\,m^{-3}}$ \\
& Circular--linear correlation    & $r = 0.149$, $p < 10^{-3}$ \\
& Spike Fisher's odds ratio       & $1.43$, $p = 2.9 \times 10^{-4}$ \\
\midrule
\multicolumn{3}{@{}l}{\textit{Chemical fingerprint (Fig.~1b)}} \\
& Pearson H$_2$S--CH$_4$          & $r = 0.830$, $p < 10^{-15}$ \\
& Spearman H$_2$S--CH$_4$         & $\rho = 0.683$, $p < 10^{-15}$ \\
& Power-law exponent              & $b = 1.95$ \\
& Jaccard spike index             & $J = 0.654$ \\
& Spike co-occurrence             & 1{,}386 vs 87.7 expected, $\chi^2 = 21{,}298$ \\
& PM$_{10}$ negative control      & $r = -0.028$, $p < 10^{-7}$ \\
& PM$_{2.5}$ negative control     & $r = -0.009$, $p = 0.084$ \\
& NO$_x$ confounded correlation   & $r = 0.377$ (6-h diel phase offset) \\
& Pooled multi-site H$_2$S--CH$_4$ & $r = 0.80$, 95\% CI $0.74$--$0.84$, $n = 4$, $I^2 = 99.8\%$ \\
\midrule
\multicolumn{3}{@{}l}{\textit{Temperature--H$_2$S hysteresis (Fig.~1c)}} \\
& Normalised loop area            & $A^{*} = 0.369$ \\
& Cooling-limb slope              & $\beta_{\mathrm{cool}} = -2.83$ \\
& Warming-limb slope              & $\beta_{\mathrm{warm}} = -2.45$ \\
& Hysteresis-asymmetry index      & $H_{\mathrm{index}} = 0.072$ \\
& Wind-speed--H$_2$S loop area    & $A^{*} = 0.012$ (no hysteresis) \\
\midrule
\multicolumn{3}{@{}l}{\textit{Diel profile (Fig.~1d)}} \\
& Peak (02:00 LST) mean           & $9.69\,\mu\mathrm{g\,m^{-3}}$ \\
& Peak (02:00 LST) P95            & $51.6\,\mu\mathrm{g\,m^{-3}}$ \\
& Trough (12:00--15:00 LST) mean         & $1.31\,\mu\mathrm{g\,m^{-3}}$ \\
& Trough (15:00 LST) P95          & $4.2\,\mu\mathrm{g\,m^{-3}}$ \\
& Mean peak/trough ratio          & $7.4\times$ \\
& P95 peak/trough ratio           & $12.3\times$ \\
& Hourly variation                & Kruskal--Wallis $H = 338.5$, $p = 7.0 \times 10^{-58}$ \\
& Pasquill A--B fraction          & $40.0\%$ \\
& Pasquill E--F fraction          & $26.1\%$ \\
& Feb monthly peak                & $14.97\,\mu\mathrm{g\,m^{-3}}$ \\
& Aug monthly minimum             & $0.99\,\mu\mathrm{g\,m^{-3}}$ \\
& Winter/summer ratio             & $15.1\times$, $p < 10^{-15}$ \\
\midrule
\multicolumn{3}{@{}l}{\textit{Sunrise-aligned inversion break-up (Fig.~1e)}} \\
& Pre-sunrise baseline ($-2$ to $0$ h) & $9.49\,\mu\mathrm{g\,m^{-3}}$ ($n = 2{,}920$) \\
& Maximum at $\Delta t = -2$ h    & $10.42\,\mu\mathrm{g\,m^{-3}}$ \\
& $+1$ h, $+2$ h, $+4$ h          & $2.65$, $1.54$, $1.30\,\mu\mathrm{g\,m^{-3}}$ \\
& Pre-/post-sunrise reduction     & $7.3\times$ \\
& One-sided Mann--Whitney         & $U = 1.15 \times 10^{7}$, $p < 10^{-15}$ \\
\midrule
\multicolumn{3}{@{}l}{\textit{Mechanism scatter (Fig.~1f)}} \\
& WS slope                        & $-1.35\,\mu\mathrm{g\,m^{-3}}$ per m\,s$^{-1}$ ($r = -0.122$) \\
& $1/\mathrm{WS}$ slope           & $10.75\,\mu\mathrm{g\,m^{-3}}$ per (m\,s$^{-1}$)$^{-1}$, 95\% CI $9.4$--$12.1$ \\
& TEMP slope                      & $-0.78\,\mu\mathrm{g\,m^{-3}}$ per $\degree$C ($r = -0.173$) \\
& $\mathrm{d}T/\mathrm{d}t$ slope & $-6.33\,\mu\mathrm{g\,m^{-3}}$ per $\degree$C\,hr$^{-1}$ ($r = -0.143$) \\
& Mean conc.\ at $u < 1$ m\,s$^{-1}$ & $11.8\,\mu\mathrm{g\,m^{-3}}$ \\
& Mean conc.\ at $u > 5$ m\,s$^{-1}$ & $1.8\,\mu\mathrm{g\,m^{-3}}$ \\
& Spike WS vs non-spike WS        & $1.75$ vs $3.67$ m\,s$^{-1}$, $p < 10^{-15}$ \\
& Peak cooling rate               & $-0.6\,\degree$C\,hr$^{-1}$ \\
& Per-mechanism $R^2$ range       & $0.015$--$0.030$ \\
\bottomrule
\end{tabular}
\end{table}

\begin{table}[htbp]
\centering
\caption{\textbf{Optimal hyperparameters selected by Bayesian
Tree-structured Parzen Estimator optimisation for each seasonal
XGBoost classifier.} Each season uses an independent one-week holdout
for hyperparameter validation, temporally isolated from the evaluation
quarter by a one-week gap.}
\label{tab:si_xgb_hyperparams}
\small
\begin{tabular}{lcccc}
\toprule
Parameter & Winter & Spring & Summer & Autumn \\
\midrule
\texttt{max\_depth}         & 3     & 7     & 7     & 5     \\
$\eta$ (learning rate)      & 0.092 & 0.118 & 0.132 & 0.040 \\
\texttt{n\_estimators}      & 150   & 100   & 150   & 300   \\
\texttt{subsample}          & 0.883 & 0.978 & 0.832 & 0.981 \\
\texttt{colsample\_bytree}  & 0.798 & 0.790 & 0.789 & 0.740 \\
$\alpha$ (L$_1$)            & 0.078 & 1.131 & 0.326 & 0.930 \\
$\lambda$ (L$_2$)           & 1.792 & 1.867 & 1.978 & 1.195 \\
\texttt{min\_child\_weight} & 4.505 & 4.957 & 4.664 & 4.758 \\
\midrule
TPE trials             & 12    & 15    & 19    & 14    \\
TPE validation macro-F$_1$        & 0.649 & 0.676 & 0.658 & 0.617 \\
\bottomrule
\end{tabular}
\end{table}

\begin{table}[htbp]
\centering
\caption{\textbf{XGBoost seasonal classification performance on the 2024
calendar-year evaluation periods.} Precision (P), recall (R) and F$_1$
score are reported per class; accuracy summarises overall performance
on the evaluation quarter. Sample counts per class are given in the
lower block; the High-class fraction varies from $5.2\%$ in Summer to
$12.8\%$ in Winter, reflecting seasonal emission and dispersion
patterns.}
\label{tab:si_xgb_perf}
\small
\setlength{\tabcolsep}{4pt}
\begin{tabular}{ll ccc ccc ccc}
\toprule
& & \multicolumn{3}{c}{\textbf{Low}} & \multicolumn{3}{c}{\textbf{Medium}} & \multicolumn{3}{c}{\textbf{High}} \\
\textbf{Season} & \textbf{Acc.} & P & R & F$_1$ & P & R & F$_1$ & P & R & F$_1$ \\
\midrule
Winter  & 0.764 & 0.887 & 0.851 & 0.869 & 0.536 & 0.573 & 0.554 & 0.523 & 0.583 & 0.551 \\
Spring  & 0.817 & 0.962 & 0.848 & 0.902 & 0.567 & 0.739 & 0.642 & 0.402 & 0.644 & 0.495 \\
Summer  & 0.836 & 0.980 & 0.864 & 0.918 & 0.355 & 0.631 & 0.455 & 0.474 & 0.749 & 0.581 \\
Autumn  & 0.778 & 0.963 & 0.826 & 0.889 & 0.386 & 0.591 & 0.467 & 0.392 & 0.600 & 0.474 \\
\midrule
\textbf{Annual (sample-weighted)} & \textbf{0.799} & \multicolumn{9}{c}{see per-class rows above} \\
\bottomrule
\end{tabular}

\vspace{1em}
\begin{tabular}{lcccc c}
\toprule
\textbf{Season} & \textbf{Low} & \textbf{Medium} & \textbf{High} & \textbf{Total} & \textbf{High \%} \\
\midrule
Winter  & 5{,}741 & 1{,}591 & 1{,}073 & 8{,}405 & 12.8\% \\
Spring  & 6{,}578 & 1{,}232 &   666   & 8{,}476 &  7.9\% \\
Summer  & 7{,}339 &   792   &   443   & 8{,}574 &  5.2\% \\
Autumn  & 6{,}438 &   845   &   879   & 8{,}162 & 10.8\% \\
\midrule
\textbf{Annual} & 26{,}096 & 4{,}460 & 3{,}061 & 33{,}617 & 9.1\% \\
\bottomrule
\end{tabular}

\footnotesize Six Autumn samples immediately following a
seven-hour meteorological outage on 23 October 2024 are excluded because the causal derivative formulation requires a leading history that the gap interrupts; all six fall in the Low class, so they do not affect the Medium or High counts.
\end{table}

\begin{table}[htbp]
\centering
\caption{\textbf{Comparison of XGBoost hyperparameter-search validation macro-F$_1$
(one-week holdout) and seasonal evaluation macro-F$_1$ (full quarter).} The
small differences ($|\Delta\mathrm{F}_1| \leq 0.009$) are consistent with little hyperparameter overfitting under the temporal isolation strategy.}
\label{tab:si_xgb_hyperopt_vs_eval}
\small
\begin{tabular}{lccc}
\toprule
\textbf{Season} & \textbf{Search macro-F$_1$} & \textbf{Evaluation macro-F$_1$} & $\Delta$\textbf{F$_1$} \\
\midrule
Winter  & 0.649 & 0.658 & $+$0.009 \\
Spring  & 0.676 & 0.679 & $+$0.003 \\
Summer  & 0.658 & 0.651 & $-$0.006 \\
Autumn  & 0.617 & 0.610 & $-$0.007 \\
\bottomrule
\end{tabular}
\end{table}

\begin{table}[htbp]
\centering
\caption{\textbf{Weekly walk-forward classification performance: S4
SSM versus XGBoost over the Oct--Dec 2024 validation window.}
Recall, precision and FAR are computed by taking the
argmax of the three-class posterior;
F$_1$ is the High-class harmonic mean. The argmax rule is not
equivalent to thresholding $P(\mathrm{High})$ at $0.5$: the latter
would give 531/565/341 rather than the reported
539/612/333
true positives, false positives and false negatives.
Bold entries are the weekly
winner per metric. $n_{\mathrm{H}}$ is the number of true High-class
observations in the week. The weekly counts total 872 (mean 67.1). Period-wide aggregate counts are in
main-text Table~\ref{tab:ed_s4_period_aggregate}. Aggregation convention: the
$\bar{x}$ row is the \emph{unweighted} mean of the 13 weekly values.
main-text Table~\ref{tab:ed_s4_period_aggregate} instead pools the confusion counts over all
8,040 timesteps before computing each metric, so the two tables report 0.528 (unweighted weekly mean) and 0.533 (pooled) for the same quantity. The pooled value is the larger of the two because the weeks with the smallest High-class
support are also the weakest and an unweighted mean gives them
disproportionate influence.}
\label{tab:si_s4_xgb_weekly}
\footnotesize
\setlength{\tabcolsep}{4pt}
\begin{tabular}{@{}cccccccccccc@{}}
\toprule
& \multicolumn{2}{c}{F$_1$ High $\uparrow$}
& \multicolumn{2}{c}{Recall $\uparrow$}
& \multicolumn{2}{c}{Precision $\uparrow$}
& \multicolumn{2}{c}{FAR $\downarrow$}
& \multicolumn{2}{c}{Brier H $\downarrow$} & \\
Wk & S4 & XGB & S4 & XGB & S4 & XGB & S4 & XGB & S4 & XGB & $n_{\mathrm{H}}$ \\
\midrule
0  & \textbf{0.807} & 0.617 & \textbf{0.902} & 0.569 & \textbf{0.730} & 0.674 & 0.028 & \textbf{0.023} & \textbf{0.026} & 0.037 & 51 \\
1  & 0.527 & \textbf{0.538} & \textbf{0.744} & 0.641 & 0.408 & \textbf{0.463} & 0.068 & \textbf{0.047} & 0.048 & \textbf{0.046} & 39 \\
2  & \textbf{0.588} & 0.276 & \textbf{0.625} & 0.500 & \textbf{0.556} & 0.190 & \textbf{0.007} & 0.029 & \textbf{0.011} & 0.021 & 8 \\
3  & 0.506 & \textbf{0.585} & 0.568 & \textbf{0.838} & \textbf{0.457} & 0.449 & \textbf{0.059} & 0.089 & 0.051 & \textbf{0.047} & 37 \\
4  & 0.381 & \textbf{0.485} & 0.679 & \textbf{0.755} & 0.265 & \textbf{0.357} & 0.165 & \textbf{0.119} & 0.088 & \textbf{0.076} & 53 \\
5  & \textbf{0.491} & 0.464 & 0.897 & 0.897 & \textbf{0.338} & 0.313 & \textbf{0.123} & 0.138 & 0.067 & \textbf{0.066} & 29 \\
6  & 0.639 & \textbf{0.680} & \textbf{0.943} & 0.934 & 0.483 & \textbf{0.535} & 0.196 & \textbf{0.158} & 0.094 & \textbf{0.078} & 106 \\
7  & \textbf{0.474} & 0.353 & \textbf{0.419} & 0.349 & \textbf{0.545} & 0.357 & \textbf{0.025} & 0.045 & \textbf{0.053} & 0.057 & 43 \\
8  & \textbf{0.791} & 0.733 & \textbf{0.968} & 0.947 & \textbf{0.669} & 0.597 & \textbf{0.080} & 0.106 & \textbf{0.048} & 0.052 & 94 \\
9  & \textbf{0.417} & 0.296 & \textbf{0.652} & 0.522 & \textbf{0.306} & 0.207 & \textbf{0.054} & 0.073 & \textbf{0.034} & 0.051 & 23 \\
10 & \textbf{0.470} & 0.390 & 0.349 & \textbf{0.356} & \textbf{0.718} & 0.430 & \textbf{0.039} & 0.135 & \textbf{0.140} & 0.150 & 146 \\
11 & \textbf{0.260} & 0.083 & \textbf{0.180} & 0.047 & \textbf{0.466} & 0.368 & 0.062 & \textbf{0.024} & \textbf{0.187} & 0.202 & 150 \\
12 & 0.514 & \textbf{0.545} & \textbf{0.796} & 0.774 & 0.379 & \textbf{0.421} & 0.213 & \textbf{0.174} & 0.111 & \textbf{0.104} & 93 \\
\midrule
$\bar{x}$ & \textbf{0.528} & 0.465 & \textbf{0.671} & 0.625 & \textbf{0.486} & 0.413 & \textbf{0.086} & 0.089 & \textbf{0.074} & 0.076 & 67.1 \\
\bottomrule
\end{tabular}
\end{table}

\begin{table}[htbp]
\centering
\caption{\textbf{Weekly walk-forward performance of the two S4 models
on their respective High-class targets.} CH$_4$ thresholds are obtained from the H$_2$S thresholds by inverting the linear fit
$\mathrm{H_2S} = 5.55\,\mathrm{CH_4} - 8.29$. Bold highlights the
higher per-week value for the species pair on F$_1$-High.
\textbf{Both columns use checkpoint selection on the evaluation week, not the inner-probe protocol of main-text Table~\ref{tab:ed_s4_period_aggregate} and Supplementary Table~\ref{tab:si_s4_xgb_weekly}.} Evaluation-week selection is applied to both models so that the cross-species comparison uses a common protocol.
The H$_2$S values here therefore differ from Supplementary Table~\ref{tab:si_s4_xgb_weekly} --
period-mean F$_1$-High $0.577$ against $0.528$ -- and the difference is attributable to the selection protocol rather than the model. The table shows that the same architecture transfers to a co-emitted species under a common protocol; it is not part of the benchmark comparison.}
\label{tab:si_s4_h2s_ch4_weekly}
\footnotesize
\setlength{\tabcolsep}{4pt}
\begin{tabular}{@{}c|ccccc|ccccc@{}}
\toprule
& \multicolumn{5}{c|}{S4 H$_2$S}
& \multicolumn{5}{c}{S4 CH$_4$} \\
Wk & F$_1$ & Prec & Rec & Brier & LL & F$_1$ & Prec & Rec & Brier & LL \\
\midrule
1  & 0.807 & 0.706 & 0.941 & 0.026 & 0.079 & \textbf{0.870} & 0.803 & 0.950 & 0.023 & 0.076 \\
2  & 0.563 & 0.453 & 0.744 & 0.045 & 0.145 & \textbf{0.622} & 0.472 & 0.911 & 0.065 & 0.192 \\
3  & \textbf{0.588} & 0.556 & 0.625 & 0.011 & 0.037 & 0.576 & 0.436 & 0.850 & 0.026 & 0.079 \\
4  & 0.505 & 0.414 & 0.649 & 0.053 & 0.151 & \textbf{0.769} & 0.625 & 1.000 & 0.034 & 0.111 \\
5  & 0.411 & 0.288 & 0.717 & 0.091 & 0.260 & \textbf{0.621} & 0.465 & 0.935 & 0.084 & 0.246 \\
6  & \textbf{0.677} & 0.636 & 0.724 & 0.037 & 0.110 & 0.625 & 0.500 & 0.833 & 0.055 & 0.166 \\
7  & \textbf{0.711} & 0.581 & 0.915 & 0.078 & 0.243 & 0.705 & 0.587 & 0.882 & 0.118 & 0.360 \\
8  & \textbf{0.487} & 0.543 & 0.442 & 0.055 & 0.224 & 0.477 & 0.337 & 0.817 & 0.135 & 0.397 \\
9  & \textbf{0.841} & 0.750 & 0.957 & 0.040 & 0.134 & 0.818 & 0.701 & 0.981 & 0.047 & 0.154 \\
10  & 0.575 & 0.420 & 0.913 & 0.031 & 0.092 & \textbf{0.609} & 0.483 & 0.824 & 0.035 & 0.112 \\
11 & 0.617 & 0.586 & 0.651 & 0.121 & 0.405 & \textbf{0.701} & 0.571 & 0.910 & 0.111 & 0.329 \\
12 & 0.186 & 0.308 & 0.133 & 0.210 & 0.844 & \textbf{0.455} & 0.327 & 0.750 & 0.083 & 0.260 \\
13 & \textbf{0.533} & 0.403 & 0.785 & 0.098 & 0.291 & 0.447 & 0.303 & 0.852 & 0.101 & 0.292 \\
\midrule
$\bar{x}$ & 0.577 & 0.511 & 0.707 & 0.069 & 0.232 & \textbf{0.638} & 0.508 & 0.884 & 0.074 & 0.236 \\
$\sigma$  & 0.164 & 0.138 & 0.219 & 0.054 & 0.207 & 0.128 & 0.142 & 0.069 & 0.040 & 0.108 \\
\bottomrule
\end{tabular}
\end{table}

\begin{table}[htbp]
\centering
\caption{\textbf{Ablation configurations for the KSG--Theiler conditional
mutual information estimator used in the multiscale transfer entropy
analysis.} Each configuration is additive: subsequent rows inherit all
preceding modifications and change a single additional parameter. Configuration~4
(bold) is the configuration used in main Fig.~2.}
\label{tab:si_mste_configs}
\footnotesize
\begin{tabular}{@{}clcccc@{}}
\toprule
ID & Description & $h_{\max}$ & Confounders & $k$ & $\varepsilon$-correction \\
\midrule
1 & Baseline           & 4 & 1 & 10 (fixed)               & No  \\
2 & + Boundary correction ($\varepsilon$-shrinkage) & 4 & 1 & 10 (fixed)    & Yes \\
3 & + Adaptive $k$                & 4 & 1 & adaptive (Methods)        & Yes \\
\textbf{4} & \textbf{+ Reduced history ($h \leq 2$)} & \textbf{2} & \textbf{1} & \textbf{adaptive (Methods)} & \textbf{Yes} \\
5 & + Minimal conditioning ($h = 1$) & 1 & 1 & adaptive (Methods)     & Yes \\
6 & Bivariate (no conditioning)   & 1 & 0 & adaptive (Methods)        & Yes \\
\bottomrule
\end{tabular}
\end{table}

\begin{table}[htbp]
\centering
\caption{\textbf{Effective transfer entropy (ETE, nats $\times 10^{3}$)
for each driver--scale combination under six estimator configurations.}
Columns labelled 1--6 refer to the configurations defined in
Supplementary Table~\ref{tab:si_mste_configs}. Asterisks mark cells in which zero of
ten circular-shift surrogates exceeded the observed value; at $B = 10$
this is a coarse screen and not a false-discovery-corrected test (see
the section text). Bold
entries in column~4 indicate the configuration used in main
Fig.~2. Dashes denote ETE $\leq 0$ (no detectable coupling).}
\label{tab:si_mste_results}
\footnotesize
\setlength{\tabcolsep}{3pt}
\begin{tabular}{@{}llrrrrrr@{}}
\toprule
Driver & Scale & 1 (B) & 2 (BF) & 3 (AK) & \textbf{4 (RH)} & 5 (MC) & 6 (BV) \\
\midrule
\multicolumn{8}{@{}l}{\textit{Core drivers (screened in nearly all configurations)}} \\
WD$_{\sin}$
  & 15\,min & 17.9* & 20.0* & 12.2* & \textbf{18.1*} & 24.2* & 42.6* \\
  & 1\,h    & 23.5* & 27.3* & 18.6* & \textbf{25.6*} & 33.1* & 51.4* \\
  & 6\,h    & 56.5* & 61.6* & 52.8* & \textbf{52.8*} & 52.8* & 64.0* \\
WS
  & 15\,min & 10.1* & 11.6* &  5.3* &  \textbf{8.2*} & 10.5* &  9.3* \\
  & 1\,h    &  8.9* & 10.1* &  8.3* & \textbf{12.8*} & 18.8* & 22.4* \\
  & 3\,h    & 18.5  & 21.5* & 14.9* & \textbf{14.9*} & 14.9* & 22.9* \\
Pressure
  & 15\,min &  5.9* &  6.4* &  2.6* &  \textbf{2.9*} &  3.7* & 12.6* \\
  & 1\,h    &  7.0* &  9.7* &  9.2* & \textbf{12.6*} & 12.5* & 23.4* \\
  & 6\,h    & 25.2* & 26.6* & 29.2* & \textbf{29.2*} & 29.2* & 24.3* \\
\midrule
\multicolumn{8}{@{}l}{\textit{Rate-of-change drivers (scale-dependent screening)}} \\
$\mathrm{d}P/\mathrm{d}t$
  & 15\,min & ---   &  0.3  & ---   & \textbf{---}   & ---   &  0.6  \\
  & 2\,h    & ---   & ---   &  5.2* &  \textbf{5.2*} &  5.2* & 14.2* \\
  & 6\,h    & 12.6  & 11.6  &  8.5* &  \textbf{8.5*} &  8.5* & 19.5* \\
$\mathrm{d}(\mathrm{WS})/\mathrm{d}t$
  & 15\,min &  4.2* &  4.3* & ---   & \textbf{---}   &  0.1  & ---   \\
  & 2\,h    &  5.1  &  4.9  &  7.9* &  \textbf{7.9*} &  7.9* & 16.2* \\
  & 3\,h    & 20.5* & 18.2* &  6.9* &  \textbf{6.9*} &  6.9* & 13.7* \\
$\mathrm{d}T/\mathrm{d}t$
  & 2\,h    &  9.6* &  8.8* & 11.9* & \textbf{11.9*} & 11.9* &  5.8* \\
  & 6\,h    & ---   & ---   & ---   & \textbf{---}   & ---   & ---   \\
\midrule
\multicolumn{8}{@{}l}{\textit{Slow driver (screened consistently only at the coarsest scale)}} \\
TEMP
  & 15\,min &  2.5* &  3.2* & ---   & \textbf{---}   & ---   & ---   \\
  & 1\,h    &  1.2  & ---   &  1.7  &  \textbf{0.9}  &  0.3  & ---   \\
  & 6\,h    & 12.0* & 15.9* & 17.5* & \textbf{17.5*} & 17.5* & 10.5* \\
\bottomrule
\end{tabular}
\end{table}

\begin{table}[htbp]
\centering
\caption{\textbf{Aggregate ablation statistics across all 42
driver--scale combinations.} ``Screened links'' counts the
cells in which zero of ten circular-shift surrogates exceeded the
observed transfer entropy. At $B = 10$ this is a coarse screen, not a
false-discovery-corrected test; it is reported to compare stability
across estimator settings, and the significance underpinning main
Fig.~2 comes from the main analysis at $B = 2{,}000$.}
\label{tab:si_mste_summary}
\footnotesize
\begin{tabular}{@{}clrrc@{}}
\toprule
ID & Configuration & Mean ETE (nats) & Screened\ links (/42) & Notes \\
\midrule
1 & Baseline                       & $+0.0098$ & 27 & Spurious TEMP significance (fine scales) \\
2 & + Boundary correction & $+0.0108$ & 27 & Marginal improvement \\
3 & + Adaptive $k$                 & $+0.0083$ & 24 & More conservative; fewer screened links \\
\textbf{4} & \textbf{+ Reduced $h$} & \textbf{$+0.0088$} & \textbf{25}
  & \textbf{Selected bias--variance compromise} \\
5 & + Minimal cond.                & $+0.0099$ & 26 & Loses $30$\,min target memory \\
6 & Bivariate                      & $+0.0143$ & 29 & Inflated; confounders uncontrolled \\
\bottomrule
\end{tabular}
\end{table}

\begin{table}[htbp]
\centering
\caption{\textbf{Three-model decomposition isolating the contribution of
memory.} All three models share the walk-forward protocol, the 7-day
adjacent inner probe, the argmax decision rule and the same 8,040
evaluation timesteps (872 High). They differ in one respect each.
\emph{XGB-16} receives CAIRN's sixteen raw inputs, none of which
carries a lag, rolling window or derivative, and is therefore
memoryless by construction; hyperparameters are tuned on the reduced set so the comparison is symmetric. \emph{XGB-24} adds the
engineered feature set, in which Butterworth derivatives and stagnation
indices supply memory by hand. \emph{CAIRN} returns to the raw sixteen
and supplies memory through the state-space kernels. Bold marks the
best value in each row. The ordering is monotone on every High-class
metric and on false alarms, but not on overall accuracy or multiclass
log-loss, where CAIRN is last: its focal loss and High-class weighting
trade specificity for sensitivity (Results).}
\label{tab:si_three_arm}
\small
\setlength{\tabcolsep}{6pt}
\begin{tabular}{@{}lccc@{}}
\toprule
 & XGB-16 & XGB-24 & CAIRN \\
Memory & none & hand-coded & learnt \\
Inputs & 16 raw & 24 engineered & 16 raw \\
\midrule
\multicolumn{4}{@{}l}{\textit{High-class detection}} \\
F$_1$-High        & 0.4625 & 0.5013 & \textbf{0.5329} \\
Precision-High    & 0.4000 & 0.4445 & \textbf{0.4683} \\
Recall-High       & 0.5482 & 0.5745 & \textbf{0.6181} \\
Brier-High        & 0.0835 & 0.0773 & \textbf{0.0748} \\
\midrule
\multicolumn{4}{@{}l}{\textit{Confusion (pooled)}} \\
True positives    & 478 & 501 & \textbf{539} \\
False positives   & 717 & 626 & \textbf{612} \\
False negatives   & 394 & 371 & \textbf{333} \\
\midrule
\multicolumn{4}{@{}l}{\textit{Whole-distribution metrics}} \\
Accuracy          & 0.7692 & \textbf{0.7866} & 0.6944 \\
Log-loss          & 0.5733 & \textbf{0.5372} & 0.8997 \\
\bottomrule
\end{tabular}

\vspace{0.6em}
\begin{tabular}{@{}lcccc@{}}
\toprule
Increment over the memoryless floor & pooled $\Delta$F$_1$ & folds won & Wilcoxon $p$ & bootstrap 95\% CI \\
\midrule
$\Delta_{\mathrm{engineering}}$ = XGB-24 $-$ XGB-16 & $+0.040$ & 10/13 & $0.027$ & $[-0.001, +0.087]$ \\
$\Delta_{\mathrm{architecture}}$ = CAIRN $-$ XGB-16 & $+0.073$ & 10/13 & $0.022$ & $[+0.019, +0.150]$ \\
\midrule
endpoint difference = CAIRN $-$ XGB-24 & $+0.032$ & 8/13 & $0.110$ & $[-0.017, +0.081]$ \\
\bottomrule
\end{tabular}

\vspace{0.4em}
{\footnotesize The paired per-fold Wilcoxon test is the primary test
used throughout this work, and both increments over the memoryless
floor clear it where the endpoint difference does not; neither clears a
two-sided sign test ($p = 0.092$ for both), so the evidence is the pair
of increments rather than either alone. Confidence intervals are block
bootstrap over the 13 weekly folds; a row-wise bootstrap ignores
autocorrelation and returns spuriously narrow intervals.
$\Delta_{\mathrm{architecture}}$ falls to $+0.042$ when the two folds
whose inner probe contains no High-class sample are excluded, while the
ordering XGB-16 $<$ XGB-24 $<$ CAIRN survives every jackknife. A state-space model on the engineered features, which would complete a fully crossed design, is not reported; CAIRN is specified for raw inputs.}
\end{table}

\clearpage
\begin{table}[htbp]
\centering
\caption{\textbf{Headline performance of the meteorology-only
multi-site Bayesian alert-tier classifier.} All metrics evaluated on
the synchronous Jan--Mar 2025 walk-forward window
($n = 8{,}536$ shared-grid 15-min timesteps; 13 weekly retrains per
constituent nowcaster; Methods, ``Multi-site Bayesian alert-tier
classifier''). The predicted tier is constructed purely from meteorology: no measured target-species concentration is used as an input at inference time. Confidence intervals are block
bootstrap ($B = 1{,}000$ resamples, 24-h block length) on the
synchronous timeseries to accommodate within-episode
autocorrelation. Per-tier P/R/F$_1$ values are paired with bootstrap
95\% CIs in Supplementary Table~\ref{tab:si_tier_per_tier}; the full
confusion matrix is in Supplementary
Table~\ref{tab:si_tier_confusion}. The same-architecture vote-count
baseline replaces the Bayesian accumulator with a hard count of
latched per-channel activations and holds all other pipeline stages
fixed. The $\Delta\kappa_w$ interval and $p$-value are computed
from a \emph{paired} block bootstrap, in which the same resampled
blocks are used for both estimators so that shared block-to-block
variation cancels. An unpaired bootstrap, which resamples the two
$\kappa$ values independently, approximately triples the interval width
and is not appropriate for two estimators evaluated on identical
timesteps.
The majority-class baseline is reported because the tier
distribution is dominated by the Normal state, so strict accuracy
alone does not evidence skill; the unweighted $\kappa$ is reported
alongside $\kappa_w$ because quadratic weighting materially raises
the reported agreement when most errors are off-by-one tier.
External validation against community odour reports is
fully held out: the odour-report record was not used in training, hysteresis selection or likelihood-ratio design. The leave-one-week-out
figures quantify the fusion-parameter tuning optimism directly ($\Delta\kappa_w = 0.008$); the constrained-ordering sensitivity is
reported in Supplementary Note~6.}
\label{tab:ed_multisite_tier}
\scriptsize
\setlength{\tabcolsep}{4pt}
\begin{tabular}{@{}p{0.62\textwidth}r@{}}
\toprule
Statistic & Value \\
\midrule
\multicolumn{2}{@{}l}{\textit{Headline tier-match against the deterministic ground truth}} \\
Quadratic-weighted Cohen's kappa $\kappa_w$           & \textbf{0.669}\,\,\,[0.565,\,0.755] \\
Quadratic-weighted $\kappa_w$, leave-one-week-out refit & \textbf{0.661} \\
Odour-report correlation, leave-one-week-out tier          & $+0.698$ \\
Unweighted Cohen's kappa $\kappa$                       & 0.397 \\
Strict element-wise accuracy                          & 0.757 \\
Majority-class baseline accuracy                        & 0.800 \\
\midrule
\multicolumn{2}{@{}l}{\textit{Per-tier F$_1$ (point estimates; full P/R/F$_1$ with CIs in Supp.\ Table~\protect\ref{tab:si_tier_per_tier})}}
 \\
Tier~0 (normal background)        & F$_1 = \mathbf{0.888}$ \\
Tier~1                            & F$_1 = 0.272$ \\
Tier~2                            & F$_1 = 0.318$ \\
Tier~3                            & F$_1 = 0.578$ \\
\midrule
\multicolumn{2}{@{}l}{\textit{Comparison to a same-architecture vote-count baseline}} \\
Vote-count baseline $\kappa_w$                                 & 0.578\,\,\,[0.480,\,0.667] \\
Bayesian-fusion improvement $\Delta\kappa_w$                   & $+0.090$\,\,[0.048,\,0.127] \\
One-sided paired-bootstrap $p$ ($\mathrm{H}_0$: Bayesian $\leq$ baseline) & 0.001 \\
Predicted tier $=$ (vote count $-\,1$), fraction of timesteps & 0.925 \\
\midrule
\multicolumn{2}{@{}l}{\textit{Robustness}} \\
$\pm 25\%$ likelihood-ratio perturbation: max $|\Delta\kappa_w|$ (Supp.\ Table~\protect\ref{tab:si_lr_sensitivity})  & 0.003 \\
$\pm 25\%$ likelihood-ratio perturbation: mean $|\Delta\kappa_w|$                                                   & 0.001 \\
\midrule
\multicolumn{2}{@{}l}{\textit{External validation against daily community odour reports ($n = 89$ days)}} \\
Pearson $r$ (daily-mean predicted tier vs daily odour reports) & \textbf{+0.711}\,\,\,[+0.400,\,+0.839] \\
Coefficient of determination $R^2$                          & 0.506 \\
Ground-truth tier ceiling (Pearson $r$)                     & +0.793\,\,\,[+0.307,\,+0.879] \\
Share of ground-truth explained variance ($R^2$ ratio)               & 80\% \\
Paired difference from the ceiling (moving block) & $+0.095$\,\,\,[$-0.211$,\,$+0.187$] \\
Lagged cross-correlation peak                               & lag $0$, $r = +0.711$ \\
\midrule
\multicolumn{2}{@{}l}{\textit{Constituent walk-forward classifiers (period-aggregate F$_1$-High)}} \\
H$_2$S @ MMF9 (broad-coverage)         & 0.425 \\
H$_2$S @ MMF2 (tight-coverage) & 0.517 \\
CH$_4$ @ MMF9                                 & 0.661 \\
CH$_4$ @ MMF2                                 & 0.616 \\
\midrule
\multicolumn{2}{@{}l}{\textit{Window and protocol}} \\
Synchronous timesteps $n$         & 8{,}536 \\
Weekly walk-forward folds         & 13 (Jan--Mar 2025) \\
Walk-forward retrains (weeks $\times$ constituent classifiers) & 52 ($13 \times 4$) \\
Network prior base rate $\pi$     & 0.10 \\
Tier cut-points $(\theta_1, \theta_2, \theta_3)$ & $(0.15,\, 0.50,\, 0.92)$ \\
Constituent training length & probe-selected epochs (Supp.\ Note~6) \\
\bottomrule
\end{tabular}
\end{table}

\clearpage
\section*{Supplementary Figures}
\addcontentsline{toc}{section}{Supplementary Figures}

\begin{figure}[htp!]
\centering
\includegraphics[width=1\textwidth]{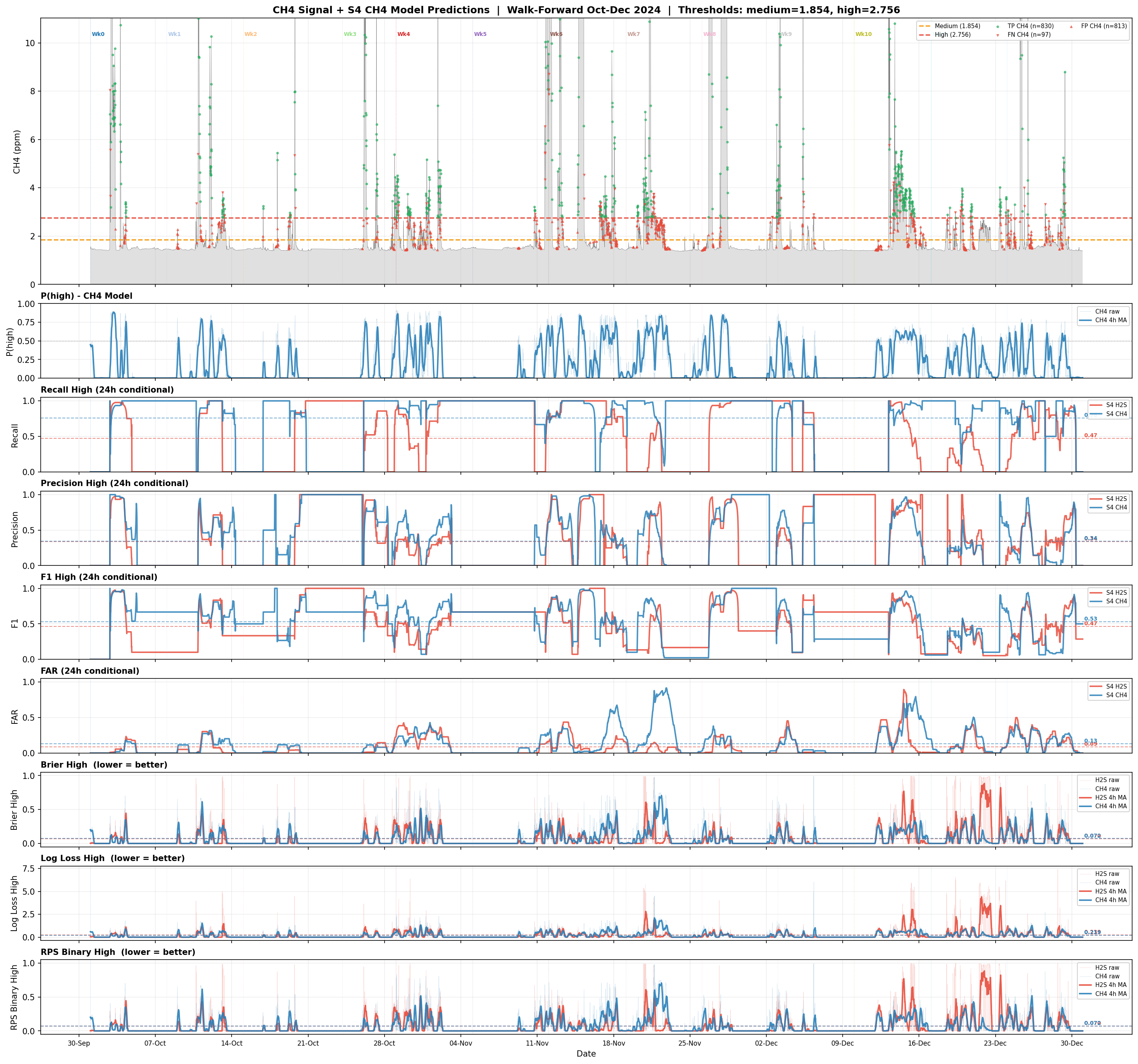}
\caption{\textbf{Co-located 15-minute time series of H$_2$S and CH$_4$
at the principal receptor across the autumn 2024 walk-forward
validation window.} Top: H$_2$S concentration with WHO Medium
($2\,\mu\mathrm{g\,m^{-3}}$) and High ($7\,\mu\mathrm{g\,m^{-3}}$)
thresholds. Bottom: simultaneously measured CH$_4$ concentration on
the class-boundary scale obtained by inverting the linear fit
$\mathrm{H_2S} = 5.55\,\mathrm{CH_4} - 8.29$
(Low $< 1.854$, Medium $1.854$--$2.756$, High $\geq 2.756$\,ppm). The
strong concurrent coupling between the two species is the basis
for the cross-species transfer experiment reported in main Fig.~6,
panels~b (right reliability diagram), c (CH$_4$ cross-correlogram with
odour reports) and d (right panel: cross-model agreement $r = 0.777$).
Methods: ``Cross-species CH$_4$ nowcaster''.}
\label{fig:si_h2s_ch4_timeseries}
\end{figure}

\clearpage
\bibliography{sn-bibliography}


\begin{thebibliography}{106}
\ifx \bisbn   \undefined \def \bisbn  #1{ISBN #1}\fi
\ifx \binits  \undefined \def \binits#1{#1}\fi
\ifx \bauthor  \undefined \def \bauthor#1{#1}\fi
\ifx \batitle  \undefined \def \batitle#1{#1}\fi
\ifx \bjtitle  \undefined \def \bjtitle#1{#1}\fi
\ifx \bvolume  \undefined \def \bvolume#1{\textbf{#1}}\fi
\ifx \byear  \undefined \def \byear#1{#1}\fi
\ifx \bissue  \undefined \def \bissue#1{#1}\fi
\ifx \bfpage  \undefined \def \bfpage#1{#1}\fi
\ifx \blpage  \undefined \def \blpage #1{#1}\fi
\ifx \burl  \undefined \def \burl#1{\textsf{#1}}\fi
\ifx \doiurl  \undefined \def \doiurl#1{\url{https://doi.org/#1}}\fi
\ifx \betal  \undefined \def \betal{\textit{et al.}}\fi
\ifx \binstitute  \undefined \def \binstitute#1{#1}\fi
\ifx \binstitutionaled  \undefined \def \binstitutionaled#1{#1}\fi
\ifx \bctitle  \undefined \def \bctitle#1{#1}\fi
\ifx \beditor  \undefined \def \beditor#1{#1}\fi
\ifx \bpublisher  \undefined \def \bpublisher#1{#1}\fi
\ifx \bbtitle  \undefined \def \bbtitle#1{#1}\fi
\ifx \bedition  \undefined \def \bedition#1{#1}\fi
\ifx \bseriesno  \undefined \def \bseriesno#1{#1}\fi
\ifx \blocation  \undefined \def \blocation#1{#1}\fi
\ifx \bsertitle  \undefined \def \bsertitle#1{#1}\fi
\ifx \bsnm \undefined \def \bsnm#1{#1}\fi
\ifx \bsuffix \undefined \def \bsuffix#1{#1}\fi
\ifx \bparticle \undefined \def \bparticle#1{#1}\fi
\ifx \barticle \undefined \def \barticle#1{#1}\fi
\bibcommenthead
\ifx \bconfdate \undefined \def \bconfdate #1{#1}\fi
\ifx \botherref \undefined \def \botherref #1{#1}\fi
\ifx \url \undefined \def \url#1{\textsf{#1}}\fi
\ifx \bchapter \undefined \def \bchapter#1{#1}\fi
\ifx \bbook \undefined \def \bbook#1{#1}\fi
\ifx \bcomment \undefined \def \bcomment#1{#1}\fi
\ifx \oauthor \undefined \def \oauthor#1{#1}\fi
\ifx \citeauthoryear \undefined \def \citeauthoryear#1{#1}\fi
\ifx \endbibitem  \undefined \def \endbibitem {}\fi
\ifx \bconflocation  \undefined \def \bconflocation#1{#1}\fi
\ifx \arxivurl  \undefined \def \arxivurl#1{\textsf{#1}}\fi
\csname PreBibitemsHook\endcsname

\bibitem[\protect\citeauthoryear{Quist and Johnston}{2024}]{carson2024malodors}
\begin{barticle}
\bauthor{\bsnm{Quist}, \binits{A.J.L.}},
\bauthor{\bsnm{Johnston}, \binits{J.E.}}:
\batitle{Malodors as environmental injustice: health symptoms in the aftermath
  of a hydrogen sulfide emergency in carson, california, usa}.
\bjtitle{Journal of Exposure Science \& Environmental Epidemiology}
\bvolume{34}(\bissue{6}),
\bfpage{935}--\blpage{940}
(\byear{2024})
\doiurl{10.1038/s41370-023-00561-x}
\end{barticle}
\endbibitem

\bibitem[\protect\citeauthoryear{{UK Health Security
  Agency}}{2025}]{ukhsa2025walleys}
\begin{botherref}
\oauthor{\bsnm{{UK Health Security Agency}}}:
Health risk assessment of air quality monitoring results from march 2021 to
  january 2025: Walleys quarry landfill site, silverdale, newcastle-under-lyme.
Technical report,
UKHSA, London
(2025)
\end{botherref}
\endbibitem

\bibitem[\protect\citeauthoryear{Kaza et~al.}{2018}]{kaza2018what}
\begin{bbook}
\bauthor{\bsnm{Kaza}, \binits{S.}}, \betal:
\bbtitle{What a Waste 2.0: A Global Snapshot of Solid Waste Management to
  2050},
pp. \bfpage{17}--\blpage{38}.
\bpublisher{World Bank},
\blocation{Washington, DC}
(\byear{2018}).
\doiurl{10.1596/978-1-4648-1329-0}
\end{bbook}
\endbibitem

\bibitem[\protect\citeauthoryear{Wang et~al.}{2024}]{wang2024methane}
\begin{barticle}
\bauthor{\bsnm{Wang}, \binits{Y.}},
\bauthor{\bsnm{Fang}, \binits{M.}},
\bauthor{\bsnm{Lou}, \binits{Z.}},
\bauthor{\bsnm{He}, \binits{H.}},
\bauthor{\bsnm{Guo}, \binits{Y.}},
\bauthor{\bsnm{Pi}, \binits{X.}},
\bauthor{\bsnm{Wang}, \binits{Y.}},
\bauthor{\bsnm{Yin}, \binits{K.}},
\bauthor{\bsnm{Fei}, \binits{X.}}:
\batitle{Methane emissions from landfills differentially underestimated
  worldwide}.
\bjtitle{Nature Sustainability}
\bvolume{7}(\bissue{4}),
\bfpage{496}--\blpage{507}
(\byear{2024})
\doiurl{10.1038/s41893-024-01307-9}
\end{barticle}
\endbibitem

\bibitem[\protect\citeauthoryear{deSouza et~al.}{2025}]{desouza2026evaluating}
\begin{barticle}
\bauthor{\bsnm{deSouza}, \binits{P.N.}},
\bauthor{\bsnm{Rees}, \binits{A.}},
\bauthor{\bsnm{Oscilowicz}, \binits{E.}},
\bauthor{\bsnm{Lawlor}, \binits{B.}},
\bauthor{\bsnm{Obermann}, \binits{W.}},
\bauthor{\bsnm{Dickinson}, \binits{K.}},
\bauthor{\bsnm{McKenzie}, \binits{L.M.}},
\bauthor{\bsnm{Magzamen}, \binits{S.}},
\bauthor{\bsnm{Miller}, \binits{S.}},
\bauthor{\bsnm{Bell}, \binits{M.L.}}:
\batitle{Evaluating the environmental justice dimensions of odor in denver,
  colorado}.
\bjtitle{Journal of Exposure Science \& Environmental Epidemiology}
\bvolume{36}(\bissue{1}),
\bfpage{67}--\blpage{76}
(\byear{2025})
\doiurl{10.1038/s41370-025-00760-8}
\end{barticle}
\endbibitem

\bibitem[\protect\citeauthoryear{Martuzzi
  et~al.}{2010}]{Martuzzi2010InequalitiesWaste}
\begin{barticle}
\bauthor{\bsnm{Martuzzi}, \binits{M.}},
\bauthor{\bsnm{Mitis}, \binits{F.}},
\bauthor{\bsnm{Forastiere}, \binits{F.}}:
\batitle{Inequalities, inequities, environmental justice in waste management
  and health}.
\bjtitle{European Journal of Public Health}
\bvolume{20}(\bissue{1}),
\bfpage{21}--\blpage{26}
(\byear{2010})
\doiurl{10.1093/eurpub/ckp216}
\end{barticle}
\endbibitem

\bibitem[\protect\citeauthoryear{{Agency for Toxic Substances and Disease
  Registry}}{2001}]{atsdr2001landfillgas}
\begin{botherref}
\oauthor{\bsnm{{Agency for Toxic Substances and Disease Registry}}}:
Landfill gas primer --- an overview for environmental health professionals,
  chapter 2: Landfill gas basics.
Technical report,
U.S. Department of Health and Human Services, Atlanta, GA
(2001)
\end{botherref}
\endbibitem

\bibitem[\protect\citeauthoryear{Shusterman}{1992}]{shusterman1992health}
\begin{barticle}
\bauthor{\bsnm{Shusterman}, \binits{D.}}:
\batitle{Critical review: The health significance of environmental odor
  pollution}.
\bjtitle{Archives of Environmental Health}
\bvolume{47}(\bissue{1}),
\bfpage{76}--\blpage{87}
(\byear{1992})
\doiurl{10.1080/00039896.1992.9935948}
\end{barticle}
\endbibitem

\bibitem[\protect\citeauthoryear{Hirasawa
  et~al.}{2019}]{hirasawa2019subjective}
\begin{barticle}
\bauthor{\bsnm{Hirasawa}, \binits{Y.}},
\bauthor{\bsnm{Shirasu}, \binits{M.}},
\bauthor{\bsnm{Okamoto}, \binits{M.}},
\bauthor{\bsnm{Touhara}, \binits{K.}}:
\batitle{Subjective unpleasantness of malodors induces a stress response}.
\bjtitle{Psychoneuroendocrinology}
\bvolume{106},
\bfpage{206}--\blpage{215}
(\byear{2019})
\doiurl{10.1016/j.psyneuen.2019.03.018}
\end{barticle}
\endbibitem

\bibitem[\protect\citeauthoryear{{European Parliament Intergroup on Climate
  Change, Biodiversity and Sustainable
  Development}}{2021}]{RevisitingOdourPollution2021}
\begin{botherref}
\oauthor{\bsnm{{European Parliament Intergroup on Climate Change, Biodiversity
  and Sustainable Development}}}:
Revisiting Odour Pollution in Europe: Final Agenda.
\url{https://ebcd.org/wp-content/uploads/2021/10/Final-Agenda-Revisiting-Odour-Pollution-in-Europe.pdf}.
Event held 28 October 2021, hosted by MEP Maria Spyraki; co-organized with
  MIO-ECSDE and D-NOSES project
(2021)
\end{botherref}
\endbibitem

\bibitem[\protect\citeauthoryear{Blanes-Vidal
  et~al.}{2012}]{BlanesVidal2012OdorAnnoyance}
\begin{barticle}
\bauthor{\bsnm{Blanes-Vidal}, \binits{V.}},
\bauthor{\bsnm{Nadimi}, \binits{E.S.}},
\bauthor{\bsnm{Ellermann}, \binits{T.}},
\bauthor{\bsnm{Andersen}, \binits{H.V.}},
\bauthor{\bsnm{Løfstrøm}, \binits{P.}}:
\batitle{Perceived annoyance from environmental odors and association with
  atmospheric ammonia levels in non-urban residential communities: a
  cross-sectional study}.
\bjtitle{Environmental Health}
\bvolume{11}(\bissue{1}),
\bfpage{27}
(\byear{2012})
\doiurl{10.1186/1476-069X-11-27}
\end{barticle}
\endbibitem

\bibitem[\protect\citeauthoryear{Wroniszewska and
  Zwoździak}{2020}]{Wroniszewska2020OdorAnnoyance}
\begin{barticle}
\bauthor{\bsnm{Wroniszewska}, \binits{A.}},
\bauthor{\bsnm{Zwoździak}, \binits{J.}}:
\batitle{Odor annoyance assessment by using logistic regression on an example
  of the municipal sector}.
\bjtitle{Sustainability}
\bvolume{12}(\bissue{15}),
\bfpage{6102}
(\byear{2020})
\doiurl{10.3390/su12156102}
\end{barticle}
\endbibitem

\bibitem[\protect\citeauthoryear{{Environmental Protection
  Act}}{1990}]{epa1990sec79}
\begin{botherref}
\oauthor{\bsnm{{Environmental Protection Act}}}:
Section 79
(1990).
\url{https://www.legislation.gov.uk/ukpga/1990/43/section/79}
\end{botherref}
\endbibitem

\bibitem[\protect\citeauthoryear{{Department for Environment, Food \& Rural
  Affairs}}{2017}]{DEFRA2017}
\begin{botherref}
\oauthor{\bsnm{{Department for Environment, Food \& Rural Affairs}}}:
Interaction between Environmental Permitting and local authorities’ statutory
  nuisance duties.
Accessed: 24 June 2026
(2017).
\url{https://www.gov.uk/government/publications/environmental-permitting-guidance-statutory-nuisance/interaction-between-environmental-permitting-and-local-authorities-statutory-nuisance-duties-web-version}
\end{botherref}
\endbibitem

\bibitem[\protect\citeauthoryear{Eykelbosh
  et~al.}{2021}]{eykelbosh2021elucidating}
\begin{barticle}
\bauthor{\bsnm{Eykelbosh}, \binits{A.}},
\bauthor{\bsnm{Maher}, \binits{R.}},
\bauthor{\bsnm{Ferreyro~Monticelli}, \binits{D.}},
\bauthor{\bsnm{Ramkairsingh}, \binits{A.}},
\bauthor{\bsnm{Henderson}, \binits{S.}},
\bauthor{\bsnm{Giang}, \binits{A.}},
\bauthor{\bsnm{Zimmerman}, \binits{N.}}:
\batitle{Elucidating the community health impacts of odours using citizen
  science and mobile monitoring}.
\bjtitle{Environmental Health Review}
\bvolume{64}(\bissue{2}),
\bfpage{24}--\blpage{27}
(\byear{2021})
\doiurl{10.5864/d2021-010}
\end{barticle}
\endbibitem

\bibitem[\protect\citeauthoryear{Pasquill and
  Smith}{1983}]{pasquill1983atmospheric}
\begin{bbook}
\bauthor{\bsnm{Pasquill}, \binits{F.}},
\bauthor{\bsnm{Smith}, \binits{F.B.}}:
\bbtitle{Atmospheric Diffusion},
\bedition{3rd} edn.
\bpublisher{Ellis Horwood},
\blocation{Chichester}
(\byear{1983})
\end{bbook}
\endbibitem

\bibitem[\protect\citeauthoryear{Xu et~al.}{2014}]{xu2014impact}
\begin{barticle}
\bauthor{\bsnm{Xu}, \binits{L.}},
\bauthor{\bsnm{Lin}, \binits{X.}},
\bauthor{\bsnm{Amen}, \binits{J.}},
\bauthor{\bsnm{Welding}, \binits{K.}},
\bauthor{\bsnm{McDermitt}, \binits{D.}}:
\batitle{Impact of changes in barometric pressure on landfill methane
  emission}.
\bjtitle{Global Biogeochemical Cycles}
\bvolume{28}(\bissue{7}),
\bfpage{679}--\blpage{695}
(\byear{2014})
\doiurl{10.1002/2013GB004571}
\end{barticle}
\endbibitem

\bibitem[\protect\citeauthoryear{Poulsen et~al.}{2003}]{young2003relating}
\begin{barticle}
\bauthor{\bsnm{Poulsen}, \binits{T.G.}},
\bauthor{\bsnm{Christophersen}, \binits{M.}},
\bauthor{\bsnm{Moldrup}, \binits{P.}},
\bauthor{\bsnm{Kjeldsen}, \binits{P.}}:
\batitle{Relating landfill gas emissions to atmospheric pressure using
  numerical modelling and state-space analysis}.
\bjtitle{Waste Management \& Research}
\bvolume{21}(\bissue{4}),
\bfpage{356}--\blpage{366}
(\byear{2003})
\doiurl{10.1177/0734242X0302100408}
\end{barticle}
\endbibitem

\bibitem[\protect\citeauthoryear{Forde et~al.}{2019}]{forde2019barometric}
\begin{barticle}
\bauthor{\bsnm{Forde}, \binits{O.N.}},
\bauthor{\bsnm{Cahill}, \binits{A.G.}},
\bauthor{\bsnm{Beckie}, \binits{R.D.}},
\bauthor{\bsnm{Mayer}, \binits{K.U.}}:
\batitle{Barometric-pumping controls fugitive gas emissions from a vadose zone
  natural gas release}.
\bjtitle{Scientific Reports}
\bvolume{9}(\bissue{1}),
\bfpage{14080}
(\byear{2019})
\doiurl{10.1038/s41598-019-50426-3}
\end{barticle}
\endbibitem

\bibitem[\protect\citeauthoryear{Parolari
  et~al.}{2021}]{parolari2021multiscale}
\begin{barticle}
\bauthor{\bsnm{Parolari}, \binits{A.J.}},
\bauthor{\bsnm{Sizemore}, \binits{J.}},
\bauthor{\bsnm{Katul}, \binits{G.G.}}:
\batitle{Multiscale legacy responses of soil gas concentrations to soil
  moisture and temperature fluctuations}.
\bjtitle{Journal of Geophysical Research: Biogeosciences}
\bvolume{126}(\bissue{2}),
\bfpage{2020}--\blpage{005865}
(\byear{2021})
\doiurl{10.1029/2020JG005865}
\end{barticle}
\endbibitem

\bibitem[\protect\citeauthoryear{Mulrow et~al.}{2020}]{wang2019prediction}
\begin{barticle}
\bauthor{\bsnm{Mulrow}, \binits{J.}},
\bauthor{\bsnm{Kshetry}, \binits{N.}},
\bauthor{\bsnm{Brose}, \binits{D.A.}},
\bauthor{\bsnm{Kumar}, \binits{K.}},
\bauthor{\bsnm{Jain}, \binits{D.}},
\bauthor{\bsnm{Shah}, \binits{M.}},
\bauthor{\bsnm{Kunetz}, \binits{T.E.}},
\bauthor{\bsnm{Varshney}, \binits{L.R.}}:
\batitle{Prediction of odor complaints at a large composite reservoir in a
  highly urbanized area: a machine learning approach}.
\bjtitle{Water Environment Research}
\bvolume{92}(\bissue{3}),
\bfpage{418}--\blpage{429}
(\byear{2020})
\doiurl{10.1002/wer.1191}
\end{barticle}
\endbibitem

\bibitem[\protect\citeauthoryear{Rajesh
  et~al.}{2025}]{machinelearning2025realtime}
\begin{barticle}
\bauthor{\bsnm{Rajesh}, \binits{M.}},
\bauthor{\bsnm{Ganesh~Babu}, \binits{R.}},
\bauthor{\bsnm{Moorthy}, \binits{U.}},
\bauthor{\bsnm{Veerappampalayam~Easwaramoorthy}, \binits{S.}}:
\batitle{Machine learning-driven framework for real-time air quality assessment
  and predictive environmental health risk mapping}.
\bjtitle{Scientific Reports}
\bvolume{15}(\bissue{1}),
\bfpage{28801}
(\byear{2025})
\doiurl{10.1038/s41598-025-14214-6}
\end{barticle}
\endbibitem

\bibitem[\protect\citeauthoryear{Gu et~al.}{2022}]{gu2022parameterization}
\begin{bchapter}
\bauthor{\bsnm{Gu}, \binits{A.}},
\bauthor{\bsnm{Gupta}, \binits{A.}},
\bauthor{\bsnm{Goel}, \binits{K.}},
\bauthor{\bsnm{R{\'e}}, \binits{C.}}:
\bctitle{On the parameterization and initialization of diagonal state space
  models}.
(\byear{2022}).
\doiurl{10.48550/arXiv.2206.11893}
\end{bchapter}
\endbibitem

\bibitem[\protect\citeauthoryear{{Environment
  Agency}}{2024}]{ea2024dataadjustment}
\begin{botherref}
\oauthor{\bsnm{{Environment Agency}}}:
Walleys Quarry landfill site, Silverdale: Hydrogen sulphide monitoring data
  adjustment statement.
Statement on calibration issues and adjustment of historic hydrogen sulphide
  monitoring data
(2024).
\url{https://engageenvironmentagency.uk.engagementhq.com/h2s-calibration/widgets/95761/documents}
\end{botherref}
\endbibitem

\bibitem[\protect\citeauthoryear{Runge et~al.}{2019}]{runge2019inferring}
\begin{barticle}
\bauthor{\bsnm{Runge}, \binits{J.}},
\bauthor{\bsnm{Bathiany}, \binits{S.}},
\bauthor{\bsnm{Bollt}, \binits{E.}},
\bauthor{\bsnm{Camps-Valls}, \binits{G.}},
\bauthor{\bsnm{Coumou}, \binits{D.}},
\bauthor{\bsnm{Deyle}, \binits{E.}},
\bauthor{\bsnm{Glymour}, \binits{C.}},
\bauthor{\bsnm{Kretschmer}, \binits{M.}},
\bauthor{\bsnm{Mahecha}, \binits{M.D.}},
\bauthor{\bsnm{Mu{\~n}oz-Mar{\'i}}, \binits{J.}},
\bauthor{\bsnm{Nes}, \binits{E.H.}},
\bauthor{\bsnm{Peters}, \binits{J.}},
\bauthor{\bsnm{Quax}, \binits{R.}},
\bauthor{\bsnm{Reichstein}, \binits{M.}},
\bauthor{\bsnm{Scheffer}, \binits{M.}},
\bauthor{\bsnm{Sch{\"o}lkopf}, \binits{B.}},
\bauthor{\bsnm{Spirtes}, \binits{P.}},
\bauthor{\bsnm{Sugihara}, \binits{G.}},
\bauthor{\bsnm{Sun}, \binits{J.}},
\bauthor{\bsnm{Zhang}, \binits{K.}},
\bauthor{\bsnm{Zscheischler}, \binits{J.}}:
\batitle{Inferring causation from time series in earth system sciences}.
\bjtitle{Nature Communications}
\bvolume{10}(\bissue{1}),
\bfpage{2553}
(\byear{2019})
\doiurl{10.1038/s41467-019-10105-3}
\end{barticle}
\endbibitem

\bibitem[\protect\citeauthoryear{Dai
  et~al.}{2026}]{aldegunde2024pollutionalerts}
\begin{barticle}
\bauthor{\bsnm{Dai}, \binits{Y.}},
\bauthor{\bsnm{Qian}, \binits{J.}},
\bauthor{\bsnm{Yang}, \binits{Y.}},
\bauthor{\bsnm{Liu}, \binits{B.}},
\bauthor{\bsnm{Li}, \binits{S.}},
\bauthor{\bsnm{Zhang}, \binits{K.}},
\bauthor{\bsnm{Xie}, \binits{Q.}},
\bauthor{\bsnm{Tong}, \binits{C.}},
\bauthor{\bsnm{Chen}, \binits{Y.}},
\bauthor{\bsnm{MacKenzie}, \binits{A.R.}},
\bauthor{\bsnm{Shi}, \binits{Z.}}:
\batitle{Significant benefits of pollution alerts for cleaner air and better
  health}.
\bjtitle{PNAS Nexus}
\bvolume{5}(\bissue{3}),
\bfpage{054}
(\byear{2026})
\doiurl{10.1093/pnasnexus/pgag054}
\end{barticle}
\endbibitem

\bibitem[\protect\citeauthoryear{Ashbaugh et~al.}{1985}]{ashbaugh1985principal}
\begin{barticle}
\bauthor{\bsnm{Ashbaugh}, \binits{L.L.}},
\bauthor{\bsnm{Malm}, \binits{W.C.}},
\bauthor{\bsnm{Sadeh}, \binits{W.Z.}}:
\batitle{A residence time probability analysis of sulfur concentrations at
  grand canyon national park}.
\bjtitle{Atmospheric Environment}
\bvolume{19}(\bissue{8}),
\bfpage{1263}--\blpage{1270}
(\byear{1985})
\doiurl{10.1016/0004-6981(85)90256-2}
\end{barticle}
\endbibitem

\bibitem[\protect\citeauthoryear{Stull}{1988}]{stull1988introduction}
\begin{bbook}
\bauthor{\bsnm{Stull}, \binits{R.B.}}:
\bbtitle{An Introduction to Boundary Layer Meteorology}.
\bpublisher{Kluwer Academic Publishers},
\blocation{Dordrecht}
(\byear{1988}).
\doiurl{10.1007/978-94-009-3027-8}
\end{bbook}
\endbibitem

\bibitem[\protect\citeauthoryear{Cawley and Talbot}{2010}]{cawley2010over}
\begin{barticle}
\bauthor{\bsnm{Cawley}, \binits{G.C.}},
\bauthor{\bsnm{Talbot}, \binits{N.L.C.}}:
\batitle{On over-fitting in model selection and subsequent selection bias in
  performance evaluation}.
\bjtitle{Journal of Machine Learning Research}
\bvolume{11},
\bfpage{2079}--\blpage{2107}
(\byear{2010})
\end{barticle}
\endbibitem

\bibitem[\protect\citeauthoryear{Cusworth
  et~al.}{2024}]{cusworth2024quantifying}
\begin{barticle}
\bauthor{\bsnm{Cusworth}, \binits{D.H.}},
\bauthor{\bsnm{Duren}, \binits{R.M.}},
\bauthor{\bsnm{Ayasse}, \binits{A.K.}},
\bauthor{\bsnm{Jiorle}, \binits{R.}},
\bauthor{\bsnm{Howell}, \binits{K.}},
\bauthor{\bsnm{Aubrey}, \binits{A.}},
\bauthor{\bsnm{Green}, \binits{R.O.}},
\bauthor{\bsnm{Eastwood}, \binits{M.L.}},
\bauthor{\bsnm{Chapman}, \binits{J.W.}},
\bauthor{\bsnm{Thorpe}, \binits{A.K.}},
\bauthor{\bsnm{Heckler}, \binits{J.}},
\bauthor{\bsnm{Asner}, \binits{G.P.}},
\bauthor{\bsnm{Smith}, \binits{M.L.}},
\bauthor{\bsnm{Thoma}, \binits{E.}},
\bauthor{\bsnm{Krause}, \binits{M.J.}},
\bauthor{\bsnm{Heins}, \binits{D.}},
\bauthor{\bsnm{Thorneloe}, \binits{S.}}:
\batitle{Quantifying methane emissions from {United States} landfills}.
\bjtitle{Science}
\bvolume{383}(\bissue{6690}),
\bfpage{1499}--\blpage{1504}
(\byear{2024})
\doiurl{10.1126/science.adi7735}
\end{barticle}
\endbibitem

\bibitem[\protect\citeauthoryear{Nesser et~al.}{2024}]{nesser2024high}
\begin{barticle}
\bauthor{\bsnm{Nesser}, \binits{H.}},
\bauthor{\bsnm{Jacob}, \binits{D.J.}},
\bauthor{\bsnm{Maasakkers}, \binits{J.D.}},
\bauthor{\bsnm{Lorente}, \binits{A.}},
\bauthor{\bsnm{Chen}, \binits{Z.}},
\bauthor{\bsnm{Lu}, \binits{X.}},
\bauthor{\bsnm{Shen}, \binits{L.}},
\bauthor{\bsnm{Qu}, \binits{Z.}},
\bauthor{\bsnm{Sulprizio}, \binits{M.P.}},
\bauthor{\bsnm{Winter}, \binits{M.}},
\bauthor{\bsnm{Ma}, \binits{S.}},
\bauthor{\bsnm{Bloom}, \binits{A.A.}},
\bauthor{\bsnm{Worden}, \binits{J.R.}},
\bauthor{\bsnm{Stavins}, \binits{R.N.}},
\bauthor{\bsnm{Randles}, \binits{C.A.}}:
\batitle{High-resolution {US} methane emissions inferred from an inversion of
  2019 {TROPOMI} satellite data: contributions from individual states, urban
  areas, and landfills}.
\bjtitle{Atmospheric Chemistry and Physics}
\bvolume{24}(\bissue{8}),
\bfpage{5069}--\blpage{5091}
(\byear{2024})
\doiurl{10.5194/acp-24-5069-2024}
\end{barticle}
\endbibitem

\bibitem[\protect\citeauthoryear{Guadalupe-Fern{\'a}ndez
  et~al.}{2021}]{guadalupe2021industrial}
\begin{barticle}
\bauthor{\bsnm{Guadalupe-Fern{\'a}ndez}, \binits{V.}},
\bauthor{\bsnm{De~Sario}, \binits{M.}},
\bauthor{\bsnm{Vecchi}, \binits{S.}},
\bauthor{\bsnm{Bauleo}, \binits{L.}},
\bauthor{\bsnm{Michelozzi}, \binits{P.}},
\bauthor{\bsnm{Davoli}, \binits{M.}},
\bauthor{\bsnm{Ancona}, \binits{C.}}:
\batitle{Industrial odour pollution and human health: a systematic review and
  meta-analysis}.
\bjtitle{Environmental Health}
\bvolume{20}(\bissue{1}),
\bfpage{108}
(\byear{2021})
\doiurl{10.1186/s12940-021-00774-3}
\end{barticle}
\endbibitem

\bibitem[\protect\citeauthoryear{Heaney et~al.}{2011}]{heaney2011relation}
\begin{barticle}
\bauthor{\bsnm{Heaney}, \binits{C.D.}},
\bauthor{\bsnm{Wing}, \binits{S.}},
\bauthor{\bsnm{Campbell}, \binits{R.L.}},
\bauthor{\bsnm{Caldwell}, \binits{D.}},
\bauthor{\bsnm{Hopkins}, \binits{B.}},
\bauthor{\bsnm{Richardson}, \binits{D.}},
\bauthor{\bsnm{Yeatts}, \binits{K.}}:
\batitle{Relation between malodor, ambient hydrogen sulfide, and health in a
  community bordering a landfill}.
\bjtitle{Environmental Research}
\bvolume{111}(\bissue{6}),
\bfpage{847}--\blpage{852}
(\byear{2011})
\doiurl{10.1016/j.envres.2011.05.021}
\end{barticle}
\endbibitem

\bibitem[\protect\citeauthoryear{Ko et~al.}{2015}]{ko2015review}
\begin{barticle}
\bauthor{\bsnm{Ko}, \binits{J.H.}},
\bauthor{\bsnm{Xu}, \binits{Q.}},
\bauthor{\bsnm{Jang}, \binits{Y.-C.}}:
\batitle{Emissions and control of hydrogen sulfide at landfills: a review}.
\bjtitle{Critical Reviews in Environmental Science and Technology}
\bvolume{45}(\bissue{19}),
\bfpage{2043}--\blpage{2083}
(\byear{2015})
\doiurl{10.1080/10643389.2015.1010427}
\end{barticle}
\endbibitem

\bibitem[\protect\citeauthoryear{Njoku et~al.}{2025}]{njoku2025landfill}
\begin{barticle}
\bauthor{\bsnm{Njoku}, \binits{P.O.}},
\bauthor{\bsnm{Edokpayi}, \binits{J.N.}},
\bauthor{\bsnm{Makungo}, \binits{R.}}:
\batitle{Assessment of landfill gas dispersion and health risks using {AERMOD}
  and {TROPOMI} satellite data: A case study of the thohoyandou landfill, south
  africa}.
\bjtitle{Atmosphere}
\bvolume{16}(\bissue{12}),
\bfpage{1402}
(\byear{2025})
\doiurl{10.3390/atmos16121402}
\end{barticle}
\endbibitem

\bibitem[\protect\citeauthoryear{M{\'e}ndez et~al.}{2023}]{mendez2023machine}
\begin{barticle}
\bauthor{\bsnm{M{\'e}ndez}, \binits{M.}},
\bauthor{\bsnm{Merayo}, \binits{M.G.}},
\bauthor{\bsnm{N{\'u}{\~n}ez}, \binits{M.}}:
\batitle{Machine learning algorithms to forecast air quality: a survey}.
\bjtitle{Artificial Intelligence Review}
\bvolume{56}(\bissue{9}),
\bfpage{10031}--\blpage{10066}
(\byear{2023})
\doiurl{10.1007/s10462-023-10424-4}
\end{barticle}
\endbibitem

\bibitem[\protect\citeauthoryear{Houdou et~al.}{2024}]{houdou2024interpretable}
\begin{barticle}
\bauthor{\bsnm{Houdou}, \binits{A.}},
\bauthor{\bsnm{El~Badisy}, \binits{I.}},
\bauthor{\bsnm{Khomsi}, \binits{K.}},
\bauthor{\bsnm{Abdala}, \binits{S.A.}},
\bauthor{\bsnm{Abdulla}, \binits{F.}},
\bauthor{\bsnm{Najmi}, \binits{H.}},
\bauthor{\bsnm{Obtel}, \binits{M.}},
\bauthor{\bsnm{Belyamani}, \binits{L.}},
\bauthor{\bsnm{Ibrahimi}, \binits{A.}},
\bauthor{\bsnm{Khalis}, \binits{M.}}:
\batitle{Interpretable machine learning approaches for forecasting and
  predicting air pollution: a systematic review}.
\bjtitle{Aerosol and Air Quality Research}
\bvolume{24}(\bissue{1}),
\bfpage{230151}
(\byear{2024})
\doiurl{10.4209/aaqr.230151}
\end{barticle}
\endbibitem

\bibitem[\protect\citeauthoryear{Prudenza
  et~al.}{2023}]{prudenza2023implementation}
\begin{barticle}
\bauthor{\bsnm{Prudenza}, \binits{S.}},
\bauthor{\bsnm{Bax}, \binits{C.}},
\bauthor{\bsnm{Capelli}, \binits{L.}}:
\batitle{Implementation of an electronic nose for real-time identification of
  odour emission peaks at a wastewater treatment plant}.
\bjtitle{Heliyon}
\bvolume{9}(\bissue{10}),
\bfpage{20437}
(\byear{2023})
\doiurl{10.1016/j.heliyon.2023.e20437}
\end{barticle}
\endbibitem

\bibitem[\protect\citeauthoryear{Brancher and
  De~Melo~Lisboa}{2014}]{brancher2014odour}
\begin{barticle}
\bauthor{\bsnm{Brancher}, \binits{M.}},
\bauthor{\bsnm{De~Melo~Lisboa}, \binits{H.}}:
\batitle{Odour impact assessment by community survey}.
\bjtitle{Chemical Engineering Transactions}
\bvolume{40},
\bfpage{139}--\blpage{144}
(\byear{2014})
\doiurl{10.3303/CET1440024}
\end{barticle}
\endbibitem

\bibitem[\protect\citeauthoryear{Nicell}{2009}]{nicell2009assessment}
\begin{barticle}
\bauthor{\bsnm{Nicell}, \binits{J.A.}}:
\batitle{Assessment and regulation of odour impacts}.
\bjtitle{Atmospheric Environment}
\bvolume{43}(\bissue{1}),
\bfpage{196}--\blpage{206}
(\byear{2009})
\doiurl{10.1016/j.atmosenv.2008.09.033}
\end{barticle}
\endbibitem

\bibitem[\protect\citeauthoryear{Lyons et~al.}{2016}]{lyons2016airaware}
\begin{barticle}
\bauthor{\bsnm{Lyons}, \binits{R.A.}},
\bauthor{\bsnm{Rodgers}, \binits{S.E.}},
\bauthor{\bsnm{Thomas}, \binits{S.}},
\bauthor{\bsnm{Bailey}, \binits{R.}},
\bauthor{\bsnm{Brunt}, \binits{H.}},
\bauthor{\bsnm{Thayer}, \binits{D.}},
\bauthor{\bsnm{Bidmead}, \binits{J.}},
\bauthor{\bsnm{Evans}, \binits{B.A.}},
\bauthor{\bsnm{Harold}, \binits{P.}},
\bauthor{\bsnm{Hooper}, \binits{M.}},
\bauthor{\bsnm{Snooks}, \binits{H.}}:
\batitle{Effects of an air pollution personal alert system on health service
  usage in a high-risk general population: a quasi-experimental study using
  linked data}.
\bjtitle{Journal of Epidemiology and Community Health}
\bvolume{70}(\bissue{12}),
\bfpage{1184}--\blpage{1190}
(\byear{2016})
\doiurl{10.1136/jech-2016-207222}
\end{barticle}
\endbibitem

\bibitem[\protect\citeauthoryear{Dai et~al.}{2026}]{dai2026significant}
\begin{barticle}
\bauthor{\bsnm{Dai}, \binits{Y.}},
\bauthor{\bsnm{Qian}, \binits{J.}},
\bauthor{\bsnm{Yang}, \binits{Y.}},
\bauthor{\bsnm{Liu}, \binits{B.}},
\bauthor{\bsnm{Li}, \binits{S.}},
\bauthor{\bsnm{Zhang}, \binits{K.}},
\bauthor{\bsnm{Xie}, \binits{Q.}},
\bauthor{\bsnm{Tong}, \binits{C.}},
\bauthor{\bsnm{Chen}, \binits{Y.}},
\bauthor{\bsnm{MacKenzie}, \binits{A.R.}},
\bauthor{\bsnm{Shi}, \binits{Z.}}:
\batitle{Significant benefits of pollution alerts for cleaner air and better
  health}.
\bjtitle{PNAS Nexus}
\bvolume{5}(\bissue{3}),
\bfpage{054}
(\byear{2026})
\doiurl{10.1093/pnasnexus/pgag054}
\end{barticle}
\endbibitem

\bibitem[\protect\citeauthoryear{Lam et~al.}{2023}]{lam2023learning}
\begin{barticle}
\bauthor{\bsnm{Lam}, \binits{R.}},
\bauthor{\bsnm{Sanchez-Gonzalez}, \binits{A.}},
\bauthor{\bsnm{Willson}, \binits{M.}},
\bauthor{\bsnm{Wirnsberger}, \binits{P.}},
\bauthor{\bsnm{Fortunato}, \binits{M.}},
\bauthor{\bsnm{Alet}, \binits{F.}},
\bauthor{\bsnm{Ravuri}, \binits{S.}},
\bauthor{\bsnm{Ewalds}, \binits{T.}},
\bauthor{\bsnm{Eaton-Rosen}, \binits{Z.}},
\bauthor{\bsnm{Hu}, \binits{W.}},
\bauthor{\bsnm{Merose}, \binits{A.}},
\bauthor{\bsnm{Hoyer}, \binits{S.}},
\bauthor{\bsnm{Holland}, \binits{G.}},
\bauthor{\bsnm{Vinyals}, \binits{O.}},
\bauthor{\bsnm{Stott}, \binits{J.}},
\bauthor{\bsnm{Pritzel}, \binits{A.}},
\bauthor{\bsnm{Mohamed}, \binits{S.}},
\bauthor{\bsnm{Battaglia}, \binits{P.}}:
\batitle{Learning skillful medium-range global weather forecasting}.
\bjtitle{Science}
\bvolume{382}(\bissue{6677}),
\bfpage{1416}--\blpage{1421}
(\byear{2023})
\doiurl{10.1126/science.adi2336}
\end{barticle}
\endbibitem

\bibitem[\protect\citeauthoryear{Lang et~al.}{2024}]{lang2024aifs}
\begin{botherref}
\oauthor{\bsnm{Lang}, \binits{S.}},
\oauthor{\bsnm{Alexe}, \binits{M.}},
\oauthor{\bsnm{Chantry}, \binits{M.}},
\oauthor{\bsnm{Dramsch}, \binits{J.}},
\oauthor{\bsnm{Pinault}, \binits{F.}},
\oauthor{\bsnm{Raoult}, \binits{B.}},
\oauthor{\bsnm{Clare}, \binits{M.C.A.}},
\oauthor{\bsnm{Lessig}, \binits{C.}},
\oauthor{\bsnm{Maier-Gerber}, \binits{M.}},
\oauthor{\bsnm{Magnusson}, \binits{L.}},
\oauthor{\bsnm{Ben~Bouall{\`e}gue}, \binits{Z.}},
\oauthor{\bsnm{Prieto~Nemesio}, \binits{A.}},
\oauthor{\bsnm{Dueben}, \binits{P.D.}},
\oauthor{\bsnm{Pappenberger}, \binits{F.}},
\oauthor{\bsnm{Rabier}, \binits{F.}}:
{AIFS} -- {ECMWF}'s data-driven forecasting system.
ECMWF Artificial Intelligence Forecasting System; operational at ECMWF since
  February 2025
(2024).
\doiurl{10.48550/arXiv.2406.01465}
\end{botherref}
\endbibitem

\bibitem[\protect\citeauthoryear{Price et~al.}{2025}]{price2024probabilistic}
\begin{barticle}
\bauthor{\bsnm{Price}, \binits{I.}},
\bauthor{\bsnm{Sanchez-Gonzalez}, \binits{A.}},
\bauthor{\bsnm{Alet}, \binits{F.}},
\bauthor{\bsnm{Andersson}, \binits{T.R.}},
\bauthor{\bsnm{El-Kadi}, \binits{A.}},
\bauthor{\bsnm{Masters}, \binits{D.}},
\bauthor{\bsnm{Ewalds}, \binits{T.}},
\bauthor{\bsnm{Stott}, \binits{J.}},
\bauthor{\bsnm{Mohamed}, \binits{S.}},
\bauthor{\bsnm{Battaglia}, \binits{P.}},
\bauthor{\bsnm{Lam}, \binits{R.}},
\bauthor{\bsnm{Willson}, \binits{M.}}:
\batitle{Probabilistic weather forecasting with machine learning}.
\bjtitle{Nature}
\bvolume{637}(\bissue{8044}),
\bfpage{84}--\blpage{90}
(\byear{2025})
\doiurl{10.1038/s41586-024-08252-9} .
\bcomment{GenCast; published online Dec 2024}
\end{barticle}
\endbibitem

\bibitem[\protect\citeauthoryear{Schreiber}{2000}]{schreiber2000measuring}
\begin{barticle}
\bauthor{\bsnm{Schreiber}, \binits{T.}}:
\batitle{Measuring information transfer}.
\bjtitle{Physical Review Letters}
\bvolume{85}(\bissue{2}),
\bfpage{461}--\blpage{464}
(\byear{2000})
\doiurl{10.1103/PhysRevLett.85.461}
\end{barticle}
\endbibitem

\bibitem[\protect\citeauthoryear{Frenzel and Pompe}{2007}]{frenzel2007partial}
\begin{barticle}
\bauthor{\bsnm{Frenzel}, \binits{S.}},
\bauthor{\bsnm{Pompe}, \binits{B.}}:
\batitle{Partial mutual information for coupling analysis of multivariate time
  series}.
\bjtitle{Physical Review Letters}
\bvolume{99}(\bissue{20}),
\bfpage{204101}
(\byear{2007})
\doiurl{10.1103/PhysRevLett.99.204101}
\end{barticle}
\endbibitem

\bibitem[\protect\citeauthoryear{Theiler}{1986}]{theiler1986spurious}
\begin{barticle}
\bauthor{\bsnm{Theiler}, \binits{J.}}:
\batitle{Spurious dimension from correlation algorithms applied to limited
  time-series data}.
\bjtitle{Physical Review A}
\bvolume{34}(\bissue{3}),
\bfpage{2427}--\blpage{2432}
(\byear{1986})
\doiurl{10.1103/PhysRevA.34.2427}
\end{barticle}
\endbibitem

\bibitem[\protect\citeauthoryear{Benjamini and
  Hochberg}{1995}]{benjamini1995controlling}
\begin{barticle}
\bauthor{\bsnm{Benjamini}, \binits{Y.}},
\bauthor{\bsnm{Hochberg}, \binits{Y.}}:
\batitle{Controlling the false discovery rate: a practical and powerful
  approach to multiple testing}.
\bjtitle{Journal of the Royal Statistical Society: Series B}
\bvolume{57}(\bissue{1}),
\bfpage{289}--\blpage{300}
(\byear{1995})
\doiurl{10.1111/j.2517-6161.1995.tb02031.x}
\end{barticle}
\endbibitem

\bibitem[\protect\citeauthoryear{Chen and Guestrin}{2016}]{chen2016xgboost}
\begin{bchapter}
\bauthor{\bsnm{Chen}, \binits{T.}},
\bauthor{\bsnm{Guestrin}, \binits{C.}}:
\bctitle{Xgboost: A scalable tree boosting system}.
In: \bbtitle{Proceedings of the 22nd ACM SIGKDD International Conference on
  Knowledge Discovery and Data Mining},
pp. \bfpage{785}--\blpage{794}
(\byear{2016}).
\doiurl{10.1145/2939672.2939785}
\end{bchapter}
\endbibitem

\bibitem[\protect\citeauthoryear{Oppenheim
  et~al.}{1999}]{oppenheim1999discrete}
\begin{bbook}
\bauthor{\bsnm{Oppenheim}, \binits{A.V.}},
\bauthor{\bsnm{Schafer}, \binits{R.W.}},
\bauthor{\bsnm{Buck}, \binits{J.R.}}:
\bbtitle{Discrete-Time Signal Processing},
\bedition{2nd} edn.
\bpublisher{Prentice Hall},
\blocation{Upper Saddle River, NJ}
(\byear{1999}).
\doiurl{10.21236/ada110902}
\end{bbook}
\endbibitem

\bibitem[\protect\citeauthoryear{Elkan}{2001}]{elkan2001foundations}
\begin{bchapter}
\bauthor{\bsnm{Elkan}, \binits{C.}}:
\bctitle{The foundations of cost-sensitive learning}.
In: \bbtitle{Proceedings of the 17th International Joint Conference on
  Artificial Intelligence (IJCAI)},
pp. \bfpage{973}--\blpage{978}
(\byear{2001})
\end{bchapter}
\endbibitem

\bibitem[\protect\citeauthoryear{Chawla et~al.}{2002}]{chawla2002smote}
\begin{barticle}
\bauthor{\bsnm{Chawla}, \binits{N.V.}},
\bauthor{\bsnm{Bowyer}, \binits{K.W.}},
\bauthor{\bsnm{Hall}, \binits{L.O.}},
\bauthor{\bsnm{Kegelmeyer}, \binits{W.P.}}:
\batitle{Smote: Synthetic minority over-sampling technique}.
\bjtitle{Journal of Artificial Intelligence Research}
\bvolume{16},
\bfpage{321}--\blpage{357}
(\byear{2002})
\doiurl{10.1613/jair.953}
\end{barticle}
\endbibitem

\bibitem[\protect\citeauthoryear{Bergstra
  et~al.}{2011}]{bergstra2011algorithms}
\begin{bchapter}
\bauthor{\bsnm{Bergstra}, \binits{J.}},
\bauthor{\bsnm{Bardenet}, \binits{R.}},
\bauthor{\bsnm{Bengio}, \binits{Y.}},
\bauthor{\bsnm{K\'{e}gl}, \binits{B.}}:
\bctitle{Algorithms for hyper-parameter optimization}.
In: \bbtitle{Advances in Neural Information Processing Systems 24 (NeurIPS)},
pp. \bfpage{2546}--\blpage{2554}
(\byear{2011})
\end{bchapter}
\endbibitem

\bibitem[\protect\citeauthoryear{Apley and Zhu}{2020}]{apley2020visualizing}
\begin{barticle}
\bauthor{\bsnm{Apley}, \binits{D.W.}},
\bauthor{\bsnm{Zhu}, \binits{J.}}:
\batitle{Visualizing the effects of predictor variables in black box supervised
  learning models}.
\bjtitle{Journal of the Royal Statistical Society: Series B (Statistical
  Methodology)}
\bvolume{82}(\bissue{4}),
\bfpage{1059}--\blpage{1086}
(\byear{2020})
\doiurl{10.1111/rssb.12377}
\end{barticle}
\endbibitem

\bibitem[\protect\citeauthoryear{Gu et~al.}{2022}]{gu2021efficiently}
\begin{bchapter}
\bauthor{\bsnm{Gu}, \binits{A.}},
\bauthor{\bsnm{Goel}, \binits{K.}},
\bauthor{\bsnm{R\'{e}}, \binits{C.}}:
\bctitle{Efficiently modeling long sequences with structured state spaces}.
In: \bbtitle{International Conference on Learning Representations (ICLR)}
(\byear{2022})
\end{bchapter}
\endbibitem

\bibitem[\protect\citeauthoryear{Xiong et~al.}{2020}]{xiong2020layer}
\begin{bchapter}
\bauthor{\bsnm{Xiong}, \binits{R.}},
\bauthor{\bsnm{Yang}, \binits{Y.}},
\bauthor{\bsnm{He}, \binits{D.}},
\bauthor{\bsnm{Zheng}, \binits{K.}},
\bauthor{\bsnm{Zheng}, \binits{S.}},
\bauthor{\bsnm{Xing}, \binits{C.}},
\bauthor{\bsnm{Zhang}, \binits{H.}},
\bauthor{\bsnm{Lan}, \binits{Y.}},
\bauthor{\bsnm{Wang}, \binits{L.}},
\bauthor{\bsnm{Liu}, \binits{T.-Y.}}:
\bctitle{On layer normalization in the transformer architecture}.
In: \bbtitle{Proceedings of the 37th International Conference on Machine
  Learning (ICML)},
vol. \bseriesno{PMLR 119},
pp. \bfpage{10524}--\blpage{10533}
(\byear{2020})
\end{bchapter}
\endbibitem

\bibitem[\protect\citeauthoryear{Dauphin et~al.}{2017}]{dauphin2017language}
\begin{bchapter}
\bauthor{\bsnm{Dauphin}, \binits{Y.N.}},
\bauthor{\bsnm{Fan}, \binits{A.}},
\bauthor{\bsnm{Auli}, \binits{M.}},
\bauthor{\bsnm{Grangier}, \binits{D.}}:
\bctitle{Language modeling with gated convolutional networks}.
In: \bbtitle{Proceedings of the 34th International Conference on Machine
  Learning (ICML)},
vol. \bseriesno{PMLR 70},
pp. \bfpage{933}--\blpage{941}
(\byear{2017})
\end{bchapter}
\endbibitem

\bibitem[\protect\citeauthoryear{Hendrycks and
  Gimpel}{2016}]{hendrycks2016gaussian}
\begin{botherref}
\oauthor{\bsnm{Hendrycks}, \binits{D.}},
\oauthor{\bsnm{Gimpel}, \binits{K.}}:
Gaussian error linear units (gelus).
arXiv preprint
\textbf{arXiv:1606.08415}
(2016)
\doiurl{10.48550/arXiv.1606.08415}
\end{botherref}
\endbibitem

\bibitem[\protect\citeauthoryear{Huang et~al.}{2016}]{huang2016deep}
\begin{bchapter}
\bauthor{\bsnm{Huang}, \binits{G.}},
\bauthor{\bsnm{Sun}, \binits{Y.}},
\bauthor{\bsnm{Liu}, \binits{Z.}},
\bauthor{\bsnm{Sedra}, \binits{D.}},
\bauthor{\bsnm{Weinberger}, \binits{K.Q.}}:
\bctitle{Deep networks with stochastic depth}.
In: \bbtitle{Computer Vision -- ECCV 2016}.
\bsertitle{Lecture Notes in Computer Science},
vol. \bseriesno{9908},
pp. \bfpage{646}--\blpage{661}.
\bpublisher{Springer},
\blocation{Cham}
(\byear{2016}).
\doiurl{10.1007/978-3-319-46493-0_39}
\end{bchapter}
\endbibitem

\bibitem[\protect\citeauthoryear{Lin et~al.}{2017}]{lin2017focal}
\begin{bchapter}
\bauthor{\bsnm{Lin}, \binits{T.-Y.}},
\bauthor{\bsnm{Goyal}, \binits{P.}},
\bauthor{\bsnm{Girshick}, \binits{R.}},
\bauthor{\bsnm{He}, \binits{K.}},
\bauthor{\bsnm{Doll{\'a}r}, \binits{P.}}:
\bctitle{Focal loss for dense object detection}.
In: \bbtitle{Proceedings of the IEEE International Conference on Computer
  Vision (ICCV)},
pp. \bfpage{2980}--\blpage{2988}
(\byear{2017}).
\doiurl{10.1109/ICCV.2017.324}
\end{bchapter}
\endbibitem

\bibitem[\protect\citeauthoryear{Loshchilov and
  Hutter}{2019}]{loshchilov2019decoupled}
\begin{bchapter}
\bauthor{\bsnm{Loshchilov}, \binits{I.}},
\bauthor{\bsnm{Hutter}, \binits{F.}}:
\bctitle{Decoupled weight decay regularization}.
In: \bbtitle{International Conference on Learning Representations (ICLR)}
(\byear{2019})
\end{bchapter}
\endbibitem

\bibitem[\protect\citeauthoryear{Brier}{1950}]{brier1950verification}
\begin{barticle}
\bauthor{\bsnm{Brier}, \binits{G.W.}}:
\batitle{Verification of forecasts expressed in terms of probability}.
\bjtitle{Monthly Weather Review}
\bvolume{78}(\bissue{1}),
\bfpage{1}--\blpage{3}
(\byear{1950})
\doiurl{10.1175/1520-0493(1950)078<0001:VOFEIT>2.0.CO;2}
\end{barticle}
\endbibitem

\bibitem[\protect\citeauthoryear{Epstein}{1969}]{epstein1969scoring}
\begin{barticle}
\bauthor{\bsnm{Epstein}, \binits{E.S.}}:
\batitle{A scoring system for probability forecasts of ranked categories}.
\bjtitle{Journal of Applied Meteorology}
\bvolume{8}(\bissue{6}),
\bfpage{985}--\blpage{987}
(\byear{1969})
\doiurl{10.1175/1520-0450(1969)008<0985:ASSFPF>2.0.CO;2}
\end{barticle}
\endbibitem

\bibitem[\protect\citeauthoryear{Cohen}{1968}]{cohen1968weighted}
\begin{barticle}
\bauthor{\bsnm{Cohen}, \binits{J.}}:
\batitle{Weighted kappa: Nominal scale agreement provision for scaled
  disagreement or partial credit}.
\bjtitle{Psychological Bulletin}
\bvolume{70}(\bissue{4}),
\bfpage{213}--\blpage{220}
(\byear{1968})
\doiurl{10.1037/h0026256}
\end{barticle}
\endbibitem

\bibitem[\protect\citeauthoryear{Landis and Koch}{1977}]{landis1977measurement}
\begin{barticle}
\bauthor{\bsnm{Landis}, \binits{J.R.}},
\bauthor{\bsnm{Koch}, \binits{G.G.}}:
\batitle{The measurement of observer agreement for categorical data}.
\bjtitle{Biometrics}
\bvolume{33}(\bissue{1}),
\bfpage{159}--\blpage{174}
(\byear{1977})
\doiurl{10.2307/2529310}
\end{barticle}
\endbibitem

\bibitem[\protect\citeauthoryear{Parker et~al.}{2002}]{parker2002gaseous}
\begin{botherref}
\oauthor{\bsnm{Parker}, \binits{T.}},
\oauthor{\bsnm{Dottridge}, \binits{J.}},
\oauthor{\bsnm{Kelly}, \binits{S.}}:
Investigation of the composition and emissions of trace components in landfill
  gas. r\&d technical report p1-438/tr.
Technical report,
Environment Agency, Bristol
(2002)
\end{botherref}
\endbibitem

\bibitem[\protect\citeauthoryear{Nastev et~al.}{2001}]{nastev2001gas}
\begin{barticle}
\bauthor{\bsnm{Nastev}, \binits{M.}}, \betal:
\batitle{Gas production and migration in landfills and geological materials}.
\bjtitle{Journal of Contaminant Hydrology}
\bvolume{52}(\bissue{1--4}),
\bfpage{187}--\blpage{211}
(\byear{2001})
\doiurl{10.1016/s0169-7722(01)00158-9}
\end{barticle}
\endbibitem

\bibitem[\protect\citeauthoryear{{ENDS Report}}{2024}]{ends2024}
\begin{botherref}
\oauthor{\bsnm{{ENDS Report}}}:
Growing number of persistently poor permitting performers in 2024 --- 15
  permits in Band F over the last two years
(2024).
\url{https://www.endsreport.com/article/1940634/growing-number-persistently-poor-permitting-performers-2024-\%E2\%80\%93-15-permits-band-f-last-two-years}
\end{botherref}
\endbibitem

\bibitem[\protect\citeauthoryear{{Environment Agency}}{2025}]{ea2025chief}
\begin{botherref}
\oauthor{\bsnm{{Environment Agency}}}:
Environment Agency Chief Regulator's Report 2024--25: Supporting Evidence
(2025).
\url{https://www.gov.uk/government/publications/environment-agency-chief-regulators-report-2024-25/environment-agency-chief-regulators-report-2024-25-supporting-evidence}
\end{botherref}
\endbibitem

\bibitem[\protect\citeauthoryear{Lee et~al.}{2006}]{lee2006reduced}
\begin{barticle}
\bauthor{\bsnm{Lee}, \binits{S.}}, \betal:
\batitle{Reduced sulfur compounds in gas from construction and demolition
  debris landfills}.
\bjtitle{Waste Management}
\bvolume{26}(\bissue{5}),
\bfpage{526}--\blpage{533}
(\byear{2006})
\doiurl{10.1016/j.wasman.2005.10.010}
\end{barticle}
\endbibitem

\bibitem[\protect\citeauthoryear{Townsend et~al.}{2004}]{townsend2004heavy}
\begin{barticle}
\bauthor{\bsnm{Townsend}, \binits{T.}}, \betal:
\batitle{Heavy metals in recovered fines from construction and demolition
  debris recycling facilities in florida}.
\bjtitle{Science of the Total Environment}
\bvolume{332}(\bissue{1-3}),
\bfpage{1}--\blpage{11}
(\byear{2004})
\doiurl{10.1016/j.scitotenv.2004.03.011}
\end{barticle}
\endbibitem

\bibitem[\protect\citeauthoryear{Townsend et~al.}{2005}]{townsend2005cd}
\begin{botherref}
\oauthor{\bsnm{Townsend}, \binits{T.}}, et al.:
C\&d waste landfill in florida: Assessment of true impact and exploration of
  innovative control techniques.
Technical report,
Florida Center for Solid \& Hazardous Waste Management
(2005)
\end{botherref}
\endbibitem

\bibitem[\protect\citeauthoryear{Jang and Townsend}{2001}]{jang2001sulfate}
\begin{barticle}
\bauthor{\bsnm{Jang}, \binits{Y.C.}},
\bauthor{\bsnm{Townsend}, \binits{T.}}:
\batitle{Sulfate leaching from recovered construction and demolition debris
  fines}.
\bjtitle{Advances in Environmental Research}
\bvolume{5}(\bissue{3}),
\bfpage{203}--\blpage{217}
(\byear{2001})
\doiurl{10.1016/s1093-0191(00)00056-3}
\end{barticle}
\endbibitem

\bibitem[\protect\citeauthoryear{Asakura}{2015}]{asakura2015sulfate}
\begin{barticle}
\bauthor{\bsnm{Asakura}, \binits{H.}}:
\batitle{Sulfate and organic matter concentration in relation to hydrogen
  sulfide generation at inert solid waste landfill site --- limit value for
  gypsum}.
\bjtitle{Waste Management}
\bvolume{43},
\bfpage{328}--\blpage{334}
(\byear{2015})
\doiurl{10.1016/j.wasman.2015.06.018}
\end{barticle}
\endbibitem

\bibitem[\protect\citeauthoryear{{Agency for Toxic Substances and Disease
  Registry}}{2016}]{ATSDR2016HydrogenSulfideToxGuide}
\begin{botherref}
\oauthor{\bsnm{{Agency for Toxic Substances and Disease Registry}}}:
Hydrogen sulfide and carbonyl sulfide toxguide.
Technical report,
U.S. Department of Health and Human Services, Public Health Service
(December 2016).
Accessed: 2026-07-03.
\url{https://www.atsdr.cdc.gov/toxguides/toxguide-114.pdf}
\end{botherref}
\endbibitem

\bibitem[\protect\citeauthoryear{{Gov.uk}}{}]{govuk_nd_h2s}
\begin{botherref}
\oauthor{\bsnm{{Gov.uk}}}:
Hydrogen sulphide: toxicological overview.
\url{https://www.gov.uk/government/publications/hydrogen-sulphide-properties-incident-management-and-toxicology/hydrogen-sulphide-toxicological-overview}
\end{botherref}
\endbibitem

\bibitem[\protect\citeauthoryear{{National Research Council
  (US)}}{2010}]{nrc2010acute}
\begin{bbook}
\bauthor{\bsnm{{National Research Council (US)}}}:
\bbtitle{Acute Exposure Guideline Levels for Selected Airborne Chemicals:
  Volume 9}.
\bpublisher{National Academies Press},
\blocation{Washington, DC}
(\byear{2010}).
\bcomment{Chap. 4: Hydrogen Sulfide}.
\doiurl{10.17226/12978} .
\burl{https://www.ncbi.nlm.nih.gov/books/NBK208170/}
\end{bbook}
\endbibitem

\bibitem[\protect\citeauthoryear{Gostelow et~al.}{2001}]{gostelow2001odour}
\begin{barticle}
\bauthor{\bsnm{Gostelow}, \binits{P.}},
\bauthor{\bsnm{Parsons}, \binits{S.A.}},
\bauthor{\bsnm{Stuetz}, \binits{R.M.}}:
\batitle{Odour measurements for sewage treatment works}.
\bjtitle{Water Research}
\bvolume{35}(\bissue{3}),
\bfpage{579}--\blpage{597}
(\byear{2001})
\doiurl{10.1016/s0043-1354(00)00313-4}
\end{barticle}
\endbibitem

\bibitem[\protect\citeauthoryear{{National Academies of
  Sciences}}{1979}]{national1979odors}
\begin{bbook}
\bauthor{\bsnm{{National Academies of Sciences}}}:
\bbtitle{Odors from Stationary and Mobile Sources}.
\bpublisher{National Academies Press},
\blocation{Washington, DC}
(\byear{1979}).
\doiurl{10.17226/19818}
\end{bbook}
\endbibitem

\bibitem[\protect\citeauthoryear{Leonardos}{1996}]{leonardos1996review}
\begin{bchapter}
\bauthor{\bsnm{Leonardos}, \binits{G.}}:
\bctitle{Review of odour control regulations in the usa}.
In: \bbtitle{Odors, Indoor and Environmental Air},
pp. \bfpage{73}--\blpage{84}.
\bpublisher{Air \& Waste Management Association},
\blocation{Bloomington}
(\byear{1996})
\end{bchapter}
\endbibitem

\bibitem[\protect\citeauthoryear{Hudon et~al.}{2000}]{hudon2000measurement}
\begin{barticle}
\bauthor{\bsnm{Hudon}, \binits{G.}},
\bauthor{\bsnm{Guy}, \binits{C.}},
\bauthor{\bsnm{Hermia}, \binits{J.}}:
\batitle{Measurement of odor intensity by an electronic nose}.
\bjtitle{Journal of the Air \& Waste Management Association}
\bvolume{50}(\bissue{10}),
\bfpage{1750}--\blpage{1758}
(\byear{2000})
\doiurl{10.1080/10473289.2000.10464202}
\end{barticle}
\endbibitem

\bibitem[\protect\citeauthoryear{Freeman and Cudmore}{}]{freeman_nd_review}
\begin{botherref}
\oauthor{\bsnm{Freeman}, \binits{T.}},
\oauthor{\bsnm{Cudmore}, \binits{R.}}:
Review of Odour Management in New Zealand.
Aurora Pacific Limited
\end{botherref}
\endbibitem

\bibitem[\protect\citeauthoryear{{Environmental Protection
  Act}}{1990}]{epa1990}
\begin{botherref}
\oauthor{\bsnm{{Environmental Protection Act}}}:
Environmental Protection Act 1990, c. 43
(1990).
\url{https://www.legislation.gov.uk/ukpga/1990/43}
\end{botherref}
\endbibitem

\bibitem[\protect\citeauthoryear{{Gov.uk}}{2024}]{govuk2024statutory}
\begin{botherref}
\oauthor{\bsnm{{Gov.uk}}}:
Statutory nuisances: how councils deal with complaints
(2024).
\url{https://www.gov.uk/guidance/statutory-nuisances-how-councils-deal-with-complaints}
\end{botherref}
\endbibitem

\bibitem[\protect\citeauthoryear{{Gov.uk}}{2025}]{govuk2025prep}
\begin{botherref}
\oauthor{\bsnm{{Gov.uk}}}:
Preparation and planning for emergencies: responsibilities of responder
  agencies and others
(2025)
\end{botherref}
\endbibitem

\bibitem[\protect\citeauthoryear{{Gov.uk}}{2015}]{govuk2015nuisance}
\begin{botherref}
\oauthor{\bsnm{{Gov.uk}}}:
Nuisance smells: how councils deal with complaints
(2015).
\url{https://www.gov.uk/guidance/nuisance-smells-how-councils-deal-with-complaints}
\end{botherref}
\endbibitem

\bibitem[\protect\citeauthoryear{{UK Health Security Agency}}{2023}]{ukhsa2023}
\begin{botherref}
\oauthor{\bsnm{{UK Health Security Agency}}}:
Environmental Public Health Surveillance System (EPHSS): Report for 2021--2023
(2023).
\url{https://www.gov.uk/government/publications/environmental-public-health-surveillance-system/environmental-public-health-surveillance-system-ephss-report-for-2021-to-2023}
\end{botherref}
\endbibitem

\bibitem[\protect\citeauthoryear{{Local Government
  Association}}{2024}]{lga2024civil}
\begin{botherref}
\oauthor{\bsnm{{Local Government Association}}}:
Civil resilience and emergency planning
(2024).
\url{https://www.local.gov.uk/topics/community-safety/civil-resilience-and-emergency-planning}
\end{botherref}
\endbibitem

\bibitem[\protect\citeauthoryear{Beychok}{1994}]{beychok1994fundamentals}
\begin{bbook}
\bauthor{\bsnm{Beychok}, \binits{M.R.}}:
\bbtitle{Fundamentals of Stack Gas Dispersion}
vol. \bseriesno{29},
p. \bfpage{3397}.
\bpublisher{Milton R. Beychok},
\blocation{Irvine, CA}
(\byear{1994}).
\doiurl{10.1016/1352-2310(95)90214-7}
\end{bbook}
\endbibitem

\bibitem[\protect\citeauthoryear{Turner}{1994}]{turner1994workbook}
\begin{bbook}
\bauthor{\bsnm{Turner}, \binits{D.B.}}:
\bbtitle{Workbook of Atmospheric Dispersion Estimates: An Introduction to
  Dispersion Modeling},
\bedition{2nd} edn.
\bpublisher{CRC Press},
\blocation{Boca Raton}
(\byear{1994}).
\doiurl{10.1201/9780138733704}
\end{bbook}
\endbibitem

\bibitem[\protect\citeauthoryear{Catena et~al.}{2022}]{plant2022hydrogen}
\begin{barticle}
\bauthor{\bsnm{Catena}, \binits{A.M.}},
\bauthor{\bsnm{Zhang}, \binits{J.}},
\bauthor{\bsnm{Commane}, \binits{R.}},
\bauthor{\bsnm{Murray}, \binits{L.T.}},
\bauthor{\bsnm{Schwab}, \binits{M.J.}},
\bauthor{\bsnm{Leibensperger}, \binits{E.M.}},
\bauthor{\bsnm{Marto}, \binits{J.}},
\bauthor{\bsnm{Smith}, \binits{M.L.}},
\bauthor{\bsnm{Schwab}, \binits{J.J.}}:
\batitle{Hydrogen sulfide emission properties from two large landfills in new
  york state}.
\bjtitle{Atmosphere}
\bvolume{13}(\bissue{8}),
\bfpage{1251}
(\byear{2022})
\doiurl{10.3390/atmos13081251}
\end{barticle}
\endbibitem

\bibitem[\protect\citeauthoryear{{California Air Resources
  Board}}{2020}]{cal2020carb}
\begin{botherref}
\oauthor{\bsnm{{California Air Resources Board}}}:
Estimation and comparison of methane, nitrous oxide, and trace volatile organic
  compound emissions and gas collection system efficiencies at california
  landfills. calpoly final report.
Technical report,
CARB, Sacramento, CA
(2020)
\end{botherref}
\endbibitem

\bibitem[\protect\citeauthoryear{Oke}{1987}]{oke1987boundary}
\begin{bbook}
\bauthor{\bsnm{Oke}, \binits{T.R.}}:
\bbtitle{Boundary Layer Climates},
\bedition{2nd} edn.
\bpublisher{Routledge},
\blocation{London}
(\byear{1987}).
\doiurl{10.4324/9780203407219}
\end{bbook}
\endbibitem

\bibitem[\protect\citeauthoryear{Banerjee et~al.}{2025}]{banerjee2025entropy}
\begin{botherref}
\oauthor{\bsnm{Banerjee}, \binits{S.}},
\oauthor{\bsnm{Ghosh}, \binits{K.}},
\oauthor{\bsnm{De}, \binits{M.}},
\oauthor{\bsnm{Basu}, \binits{U.}},
\oauthor{\bsnm{Basu}, \binits{B.}}:
Entropy-based analysis of urban pollutant--weather correlations.
arXiv:2508.17453; update venue/DOI on journal publication
(2025).
\doiurl{10.48550/arXiv.2508.17453}
\end{botherref}
\endbibitem

\bibitem[\protect\citeauthoryear{Gu and Dao}{2024}]{gu2024mamba}
\begin{bchapter}
\bauthor{\bsnm{Gu}, \binits{A.}},
\bauthor{\bsnm{Dao}, \binits{T.}}:
\bctitle{Mamba: linear-time sequence modeling with selective state spaces}.
In: \bbtitle{First Conference on Language Modeling (COLM)}
(\byear{2024})
\end{bchapter}
\endbibitem

\bibitem[\protect\citeauthoryear{Goodwell and
  Kumar}{2017}]{goodwell2017temporal}
\begin{barticle}
\bauthor{\bsnm{Goodwell}, \binits{A.E.}},
\bauthor{\bsnm{Kumar}, \binits{P.}}:
\batitle{Temporal information partitioning: Characterizing synergy, uniqueness,
  and redundancy in interacting environmental variables}.
\bjtitle{Water Resources Research}
\bvolume{53}(\bissue{7}),
\bfpage{5920}--\blpage{5942}
(\byear{2017})
\doiurl{10.1002/2016WR020216}
\end{barticle}
\endbibitem

\bibitem[\protect\citeauthoryear{{World Health
  Organization}}{2000}]{who2000airquality}
\begin{botherref}
\oauthor{\bsnm{{World Health Organization}}}:
Air quality guidelines for europe.
Technical report,
WHO Regional Office for Europe, Copenhagen
(2000)
\end{botherref}
\endbibitem

\bibitem[\protect\citeauthoryear{Kass and Raftery}{1995}]{kass1995bayesfactors}
\begin{barticle}
\bauthor{\bsnm{Kass}, \binits{R.E.}},
\bauthor{\bsnm{Raftery}, \binits{A.E.}}:
\batitle{Bayes factors}.
\bjtitle{Journal of the American Statistical Association}
\bvolume{90}(\bissue{430}),
\bfpage{773}--\blpage{795}
(\byear{1995})
\doiurl{10.1080/01621459.1995.10476572}
\end{barticle}
\endbibitem

\bibitem[\protect\citeauthoryear{Kraskov et~al.}{2004}]{kraskov2004estimating}
\begin{barticle}
\bauthor{\bsnm{Kraskov}, \binits{A.}},
\bauthor{\bsnm{St{\"o}gbauer}, \binits{H.}},
\bauthor{\bsnm{Grassberger}, \binits{P.}}:
\batitle{Estimating mutual information}.
\bjtitle{Physical Review E}
\bvolume{69}(\bissue{6}),
\bfpage{066138}
(\byear{2004})
\doiurl{10.1103/PhysRevE.69.066138}
\end{barticle}
\endbibitem

\bibitem[\protect\citeauthoryear{Friedman}{2001}]{friedman2001greedy}
\begin{barticle}
\bauthor{\bsnm{Friedman}, \binits{J.H.}}:
\batitle{Greedy function approximation: A gradient boosting machine}.
\bjtitle{Annals of Statistics}
\bvolume{29}(\bissue{5}),
\bfpage{1189}--\blpage{1232}
(\byear{2001})
\doiurl{10.1214/aos/1013203451}
\end{barticle}
\endbibitem

\bibitem[\protect\citeauthoryear{Gu et~al.}{2020}]{gu2020hippo}
\begin{bchapter}
\bauthor{\bsnm{Gu}, \binits{A.}},
\bauthor{\bsnm{Dao}, \binits{T.}},
\bauthor{\bsnm{Ermon}, \binits{S.}},
\bauthor{\bsnm{Rudra}, \binits{A.}},
\bauthor{\bsnm{R{\'e}}, \binits{C.}}:
\bctitle{Hippo: Recurrent memory with optimal polynomial projections}.
In: \bbtitle{Advances in Neural Information Processing Systems},
vol. \bseriesno{33},
pp. \bfpage{1474}--\blpage{1487}.
\bpublisher{Curran Associates, Inc.},
\blocation{Red Hook, NY, USA}
(\byear{2020}).
\burl{https://proceedings.neurips.cc/paper_files/paper/2020/file/102f0bb6efb3a6128a3c750dd16729be-Paper.pdf}
\end{bchapter}
\endbibitem

\bibitem[\protect\citeauthoryear{Glorot and
  Bengio}{2010}]{glorot2010understanding}
\begin{bchapter}
\bauthor{\bsnm{Glorot}, \binits{X.}},
\bauthor{\bsnm{Bengio}, \binits{Y.}}:
\bctitle{Understanding the difficulty of training deep feedforward neural
  networks}.
In: \bbtitle{Proceedings of the Thirteenth International Conference on
  Artificial Intelligence and Statistics (AISTATS)},
vol. \bseriesno{PMLR 9},
pp. \bfpage{249}--\blpage{256}
(\byear{2010})
\end{bchapter}
\endbibitem

\bibitem[\protect\citeauthoryear{Vaswani et~al.}{2017}]{vaswani2017attention}
\begin{bchapter}
\bauthor{\bsnm{Vaswani}, \binits{A.}},
\bauthor{\bsnm{Shazeer}, \binits{N.}},
\bauthor{\bsnm{Parmar}, \binits{N.}},
\bauthor{\bsnm{Uszkoreit}, \binits{J.}},
\bauthor{\bsnm{Jones}, \binits{L.}},
\bauthor{\bsnm{Gomez}, \binits{A.N.}},
\bauthor{\bsnm{Kaiser}, \binits{L.}},
\bauthor{\bsnm{Polosukhin}, \binits{I.}}:
\bctitle{Attention is all you need}.
In: \bbtitle{Advances in Neural Information Processing Systems 30 (NeurIPS)},
pp. \bfpage{5998}--\blpage{6008}.
\bpublisher{Curran Associates, Inc.},
\blocation{Red Hook, NY, USA}
(\byear{2017})
\end{bchapter}
\endbibitem

\bibitem[\protect\citeauthoryear{He and Garcia}{2009}]{he2009learning}
\begin{barticle}
\bauthor{\bsnm{He}, \binits{H.}},
\bauthor{\bsnm{Garcia}, \binits{E.A.}}:
\batitle{Learning from imbalanced data}.
\bjtitle{IEEE Transactions on Knowledge and Data Engineering}
\bvolume{21}(\bissue{9}),
\bfpage{1263}--\blpage{1284}
(\byear{2009})
\doiurl{10.1109/TKDE.2008.239}
\end{barticle}
\endbibitem

\bibitem[\protect\citeauthoryear{Cui et~al.}{2019}]{cui2019class}
\begin{bchapter}
\bauthor{\bsnm{Cui}, \binits{Y.}},
\bauthor{\bsnm{Jia}, \binits{M.}},
\bauthor{\bsnm{Lin}, \binits{T.-Y.}},
\bauthor{\bsnm{Song}, \binits{Y.}},
\bauthor{\bsnm{Belongie}, \binits{S.}}:
\bctitle{Class-balanced loss based on effective number of samples}.
In: \bbtitle{Proceedings of the IEEE Conference on Computer Vision and Pattern
  Recognition (CVPR)},
pp. \bfpage{9268}--\blpage{9277}.
\bpublisher{IEEE},
\blocation{Piscataway, NJ, USA}
(\byear{2019}).
\doiurl{10.1109/CVPR.2019.00949}
\end{bchapter}
\endbibitem

\end{thebibliography}
\end{document}

\end{document}